\documentclass[rmp,aps,amsmath,amsfonts,noshowkeys,noshowpacs,floatfix,fix,twocolumn
]{revtex4}
\usepackage{color,bm}
\usepackage{fancybox,amsmath,amssymb,rotating,pifont}
\usepackage{epic,eepic,calc,array,shadow}
\usepackage{npsfont}
\usepackage{graphicx}
\usepackage{dcolumn}  
\usepackage{subfigure}
\usepackage{multirow}
\newcommand{\subsub}[1]{\goodbreak\goodbreak
\goodbreak\goodbreak\goodbreak\par
\vskip 2mm plus 1mm minus 1mm \setcounter{paragraph}{0} {\bf
\boldmath #1 \unboldmath} \vskip 2mm plus 1mm minus 1mm}
\usepackage{mdwlist}
\makeatletter
\newlength{\earraycolsep}
\def\eqnarray{\stepcounter{equation}\let\@currentlabel%
\theequation \global\@eqnswtrue\m@th
\global\@eqcnt\z@\tabskip\@centering\let\\\@eqncr
$$\halign to\displaywidth\bgroup\@eqnsel\hskip\@centering
$\displaystyle\tabskip\z@{##}$&\global\@eqcnt\@ne \hskip
2\earraycolsep \hfil$\displaystyle{##}$\hfil &\global\@eqcnt\tw@
\hskip 2\earraycolsep $\displaystyle\tabskip\z@{##}$\hfil
\tabskip\@centering&\llap{##}\tabskip\z@\cr} \makeatother

\let\subsub\subsubsection
\begin{document}
\newcommand{\bc}        {\begin{center}}
\newcommand{\ec}        {\end{center}}
\newcommand{\er}{$\pm$}
\newcommand{\be}{\begin{eqnarray}}
\newcommand{\ee}{\end{eqnarray}}
\newcommand{\widt}{\rm\Gamma_{tot}}
\newcommand{\wadd}{$\rm\Gamma_{miss}$}
\newcommand{\gpiN}{$\rm\Gamma_{\pi N}$}
\newcommand{\getN}{$\rm\Gamma_{\eta N}$}
\newcommand{\gkla}{$\rm\Gamma_{K \Lambda}$}
\newcommand{\gksi}{$\rm\Gamma_{K \Sigma}$}
\newcommand{\gNpi}{$\rm\Gamma_{P_{11} \pi}$}
\newcommand{\gDpf}{$\rm\Gamma_{\Delta\pi(L\!<\!J)}$}
\newcommand{\gDps}{$\rm\Gamma_{\Delta\pi(L\!>\!J)}$}
\newcommand{\sqgDpf}{$\rm\sqrt\Gamma_{\Delta\pi(L\!<\!J)}$}
\newcommand{\sqgDps}{$\rm\sqrt\Gamma_{\Delta\pi(L\!>\!J)}$}
\newcommand{\gnsi}{$\rm N\sigma$}
\newcommand{\gNpf}{$\rm\Gamma_{D_{13}\pi}$}
\newcommand{\gNps}{$\Gamma_{D_{13}\pi(L\!>\!J)}$}

\newcommand{\ket}[1]{\ensuremath{\vert\,#1\,\rangle}}

\newcommand{\Nsonea}{$N_{1/2^-}(1535)$}
\newcommand{\Nsoneb}{$N_{1/2^-}(1650)$}
\newcommand{\Nsonec}{$\color{grey0}N_{1/2^-}(1846)$}
\newcommand{\Nsoned}{$\color{grey1}N_{1/2^-}(2090)$}
\newcommand{\Lsonea}{$\Lambda_{1/2^-}(1405)$}
\newcommand{\Lsoneb}{$\Lambda_{1/2^-}(1670)$}
\newcommand{\Lsonec}{$\color{grey3}\Lambda_{1/2^-}(1800)$}
\newcommand{\Ssonea}{$\color{grey2}\Sigma_{1/2^-}(1620)$}
\newcommand{\Ssoneb}{$\color{grey3}\Sigma_{1/2^-}(1750)$}
\newcommand{\Ssonec}{$\color{grey1}\Sigma_{1/2^-}(2000)$}

\newcommand{\Nsoneabf}{$N_{1/2^-}(1535)$}
\newcommand{\Nsonebbf}{$N_{1/2^-}(1650)$}
\newcommand{\Nsonecbf}{$N_{1/2^-}(1905)$}
\newcommand{\Nsonedbf}{$N_{1/2^-}(2090)$}
\newcommand{\Lsoneabf}{$\Lambda_{1/2^-}(1405)$}
\newcommand{\Lsonebbf}{$\Lambda_{1/2^-}(1670)$}
\newcommand{\Lsonecbf}{$\Lambda_{1/2^-}(1800)$}
\newcommand{\Ssoneabf}{$\Sigma_{1/2^-}(1620)$}
\newcommand{\Ssonebbf}{$\Sigma_{1/2^-}(1750)$}
\newcommand{\Ssonecbf}{$\Sigma_{1/2^-}(2000)$}

\newcommand{\Nponea}{$N_{1/2^+}(1440)$}
\newcommand{\Nponeb}{$\color{grey3}N_{1/2^+}(1710)$}
\newcommand{\Nponec}{$\color{grey0}N_{1/2^+}(1880)$}
\newcommand{\Nponed}{$\color{grey1}N_{1/2^+}(2100)$}
\newcommand{\Lponea}{$\Lambda_{1/2^+}(1600)$}
\newcommand{\Lponeb}{$\color{grey3}\Lambda_{1/2^+}(1810)$}
\newcommand{\Sponea}{$\color{grey3}\Sigma_{1/2^+}(1660)$}
\newcommand{\Sponeb}{$\color{grey1}\Sigma_{1/2^+}(1770)$}
\newcommand{\Sponec}{$\color{grey2}\Sigma_{1/2^+}(1880)$}

\newcommand{\Nponeabf}{$N_{1/2^+}(1440)$}
\newcommand{\Nponebbf}{$N_{1/2^+}(1710)$}
\newcommand{\Nponecbf}{$N_{1/2^+}(1880)$}
\newcommand{\Nponedbf}{$N_{1/2^+}(2100)$}
\newcommand{\Lponeabf}{$\Lambda_{1/2^+}(1600)$}
\newcommand{\Lponebbf}{$\Lambda_{1/2^+}(1810)$}
\newcommand{\Sponeabf}{$\Sigma_{1/2^+}(1660)$}
\newcommand{\Sponebbf}{$\Sigma_{1/2^+}(1770)$}
\newcommand{\Sponecbf}{$\Sigma_{1/2^+}(1880)$}

\newcommand{\Npthreea}{$N_{3/2^+}(1720)$}
\newcommand{\Npthreeb}{$\color{grey2}N_{3/2^+}(1900)$}
\newcommand{\Npthreec}{$\color{grey0}N_{1/2^+}(2200)$}
\newcommand{\Lpthreea}{$\Lambda_{3/2^+}(1890)$}
\newcommand{\Spthreea}{$\Sigma_{3/2^+}(1840)$}
\newcommand{\Spthreeb}{$\color{grey2}\Sigma_{3/2^+}(2080)$}

\newcommand{\Npthreeabf}{$N_{3/2^+}(1720)$}
\newcommand{\Npthreebbf}{$N_{3/2^+}(1900)$}
\newcommand{\Npthreecbf}{$N_{1/2^+}(2200)$}
\newcommand{\Lpthreeabf}{$\Lambda_{3/2^+}(1890)$}
\newcommand{\Spthreeabf}{$\Sigma_{3/2^+}(1840)$}
\newcommand{\Spthreebbf}{$\Sigma_{3/2^+}(2080)$}

\newcommand{\Ndthreea}{$N_{3/2^-}(1520)$}
\newcommand{\Ndthreeb}{$\color{grey3}N_{3/2^-}(1700)$}
\newcommand{\Ndthreec}{$\color{grey0}N_{3/2^-}(1875)$}
\newcommand{\Ndthreed}{$\color{grey1}N_{3/2^-}(2080)$}
\newcommand{\Ldthreea}{$\Lambda_{3/2^-}(1520)$}
\newcommand{\Ldthreeb}{$\Lambda_{3/2^-}(1690)$}
\newcommand{\Ldthreec}{$\color{grey1}\Lambda_{3/2^-}(2325)$}
\newcommand{\Sdthreea}{$\color{grey1}\Sigma_{3/2^-}(1580)$}
\newcommand{\Sdthreeb}{$\Sigma_{3/2^-}(1670)$}
\newcommand{\Sdthreec}{$\color{grey3}\Sigma_{3/2^-}(1940)$}

\newcommand{\Ndthreeabf}{$N_{3/2^-}(1520)$}
\newcommand{\Ndthreebbf}{$N_{3/2^-}(1700)$}
\newcommand{\Ndthreecbf}{$N_{3/2^-}(1860)$}
\newcommand{\Ndthreedbf}{$N_{3/2^-}(2080)$}
\newcommand{\Ldthreeabf}{$\Lambda_{3/2^-}(1520)$}
\newcommand{\Ldthreebbf}{$\Lambda_{3/2^-}(1690)$}
\newcommand{\Ldthreecbf}{$\Lambda_{3/2^-}(2325)$}
\newcommand{\Sdthreeabf}{$\Sigma_{3/2^-}(1580)$}
\newcommand{\Sdthreebbf}{$\Sigma_{3/2^-}(1670)$}
\newcommand{\Sdthreecbf}{$\Sigma_{3/2^-}(1940)$}

\newcommand{\Nffivea}{$N_{5/2^+}(1680)$}
\newcommand{\Nffiveb}{$\color{grey2}N_{5/2^+}(2000)$}
\newcommand{\Lffivea}{$\Lambda_{5/2^+}(1820)$}
\newcommand{\Lffiveb}{$\color{grey3}\Lambda_{5/2^+}(2110)$}
\newcommand{\Sffivea}{$\Sigma_{5/2^+}(1915)$}
\newcommand{\Sffiveb}{$\color{grey3}\Sigma_{5/2^+}(2070)$}

\newcommand{\Nffiveabf}{$N_{5/2^+}(1680)$}
\newcommand{\Nffivebbf}{$N_{5/2^+}(1870)$}
\newcommand{\Nffivecbf}{$N_{5/2^+}(2000)$}
\newcommand{\Lffiveabf}{$\Lambda_{5/2^+}(1820)$}
\newcommand{\Lffivebbf}{$\Lambda_{5/2^+}(2110)$}
\newcommand{\Sffiveabf}{$\Sigma_{5/2^+}(1915)$}
\newcommand{\Sffivebbf}{$\Sigma_{5/2^+}(2070)$}

\newcommand{\Ndfivea}{$N_{5/2^-}(1675)$}
\newcommand{\Ndfiveb}{$\color{grey0}N_{5/2^-}(2200)$}
\newcommand{\Ldfivea}{$\Lambda_{5/2^-}(1830)$}
\newcommand{\Sdfivea}{$\color{grey3}\Sigma_{5/2^-}(1775)$}

\newcommand{\Ndfiveabf}{$N_{5/2^-}(1675)$}
\newcommand{\Ndfivebbf}{$N_{5/2^-}(2200)$}
\newcommand{\Ldfiveabf}{$\Lambda_{5/2^-}(1830)$}
\newcommand{\Sdfiveabf}{$\Sigma_{5/2^-}(1775)$}

\newcommand{\Nfsevena}{$\color{grey2}N_{7/2^+}(1990)$}
\newcommand{\Lfsevena}{$\color{grey1}\Lambda_{7/2^+}(2020)$}
\newcommand{\Sfsevena}{$\Sigma_{7/2^+}(2030)$}

\newcommand{\Nfsevenabf}{$N_{7/2^+}(1990)$}
\newcommand{\Lfsevenabf}{$\Lambda_{7/2^+}(2020)$}
\newcommand{\Sfsevenabf}{$\Sigma_{7/2^+}(2030)$}

\newcommand{\Ngsevena}{$N_{7/2^-}(2190)$}
\newcommand{\Lgsevena}{$\Lambda_{7/2^-}(2100)$}
\newcommand{\Sgsevena}{$\color{grey1}\Sigma_{7/2^-}(2100)$}

\newcommand{\Ngsevenabf}{$N_{7/2^-}(2190)$}
\newcommand{\Lgsevenabf}{$\Lambda_{7/2^-}(2100)$}
\newcommand{\Sgsevenabf}{$\Sigma_{7/2^-}(2100)$}

\newcommand{\Nhninea}{$N_{9/2^+}(2220)$}
\newcommand{\Ngninea}{$N_{9/2^-}(2250)$}
\newcommand{\Nielevena}{$\color{grey3}N_{11/2^-}(2600)$}
\newcommand{\Nkthirteena}{$\color{grey3}N_{13/2^+}(2700)$}
\newcommand{\Dsonea}{$\Delta_{1/2^-}(1620)$}
\newcommand{\Dsoneb}{$\color{grey2}\Delta_{1/2^-}(1900)$}
\newcommand{\Dsonec}{$\color{grey1}\Delta_{1/2^-}(2150)$}
\newcommand{\Dponea}{$\color{grey1}\Delta_{1/2^+}(1750)$}
\newcommand{\Dponeb}{$\Delta_{1/2^+}(1910)$}
\newcommand{\Dpthreea}{$\Delta_{3/2^+}(1232)$}
\newcommand{\Dpthreeb}{$\color{grey3}\Delta_{3/2^+}(1600)$}
\newcommand{\Dpthreec}{$\color{grey3}\Delta_{3/2^+}(1920)$}
\newcommand{\Ddthreea}{$\Delta_{3/2^-}(1700)$}
\newcommand{\Ddthreeb}{$\color{grey1}\Delta_{3/2^-}(1940)$}
\newcommand{\Ddthreec}{$\color{grey0}\Delta_{3/2^-}(2350)$}
\newcommand{\Dffivea}{$\Delta_{5/2^+}(1905)$}
\newcommand{\Dffiveb}{$\color{grey2}\Delta_{5/2^+}(2000)$}
\newcommand{\Ddfivea}{$\color{grey2}\Delta_{5/2^-}(1930)$}
\newcommand{\Ddfiveb}{$\color{grey2}\Delta_{5/2^-}(2350)$}
\newcommand{\Dfsevena}{$\Delta_{7/2^+}(1950)$}
\newcommand{\Dfsevenb}{$\color{grey1}\Delta_{7/2^+}(2390)$}
\newcommand{\Dgsevena}{$\color{grey1}\Delta_{7/2^-}(2200)$}
\newcommand{\Dhninea}{$\color{grey3}\Delta_{9/2^+}(2300)$}
\newcommand{\Dgninea}{$\color{grey1}\Delta_{9/2^-}(2400)$}
\newcommand{\Dhelevena}{$\Delta_{11/2^+}(2420)$}
\newcommand{\Dithirteena}{$\color{grey2}\Delta_{13/2^-}(2750)$}
\newcommand{\Dkfifteena}{$\color{grey2}\Delta_{15/2^+}(2950)$}

\newcommand{\Nhnineabf}{$N_{9/2^+}(2220)$}
\newcommand{\Ngnineabf}{$N_{9/2^-}(2250)$}
\newcommand{\Nielevenabf}{$N_{11/2^-}(2600)$}
\newcommand{\Nkthirteenabf}{$N_{13/2^+}(2700)$}
\newcommand{\Dsoneabf}{$\Delta_{1/2^-}(1620)$}
\newcommand{\Dsonebbf}{$\Delta_{1/2^-}(1900)$}
\newcommand{\Dsonecbf}{$\Delta_{1/2^-}(2150)$}
\newcommand{\Dponeabf}{$\Delta_{1/2^+}(1750)$}
\newcommand{\Dponebbf}{$\Delta_{1/2^+}(1910)$}
\newcommand{\Dpthreeabf}{$\Delta_{3/2^+}(1232)$}
\newcommand{\Dpthreebbf}{$\Delta_{3/2^+}(1600)$}
\newcommand{\Dpthreecbf}{$\Delta_{3/2^+}(1920)$}
\newcommand{\Ddthreeabf}{$\Delta_{3/2^-}(1700)$}
\newcommand{\Ddthreebbf}{$\Delta_{3/2^-}(1940)$}
\newcommand{\Ddthreecbf}{$\Delta_{3/2^-}(2350)$}
\newcommand{\Dffiveabf}{$\Delta_{5/2^+}(1905)$}
\newcommand{\Dffivebbf}{$\Delta_{5/2^+}(2000)$}
\newcommand{\Ddfiveabf}{$\Delta_{5/2^-}(1930)$}
\newcommand{\Ddfivebbf}{$\Delta_{5/2^-}(2350)$}
\newcommand{\Dfsevenabf}{$\Delta_{7/2^+}(1950)$}
\newcommand{\Dfsevenbbf}{$\Delta_{7/2^+}(2390)$}
\newcommand{\Dgsevenabf}{$\Delta_{7/2^-}(2200)$}
\newcommand{\Dhnineabf}{$\Delta_{9/2^+}(2300)$}
\newcommand{\Dgnineabf}{$\Delta_{9/2^-}(2400)$}
\newcommand{\Dhelevenabf}{$\Delta_{11/2^+}(2420)$}
\newcommand{\Dithirteenabf}{$\Delta_{13/2^-}(2750)$}
\newcommand{\Dkfifteenabf}{$\Delta_{15/2^+}(2950)$}
\newcommand{\rthe}{A^{1/2}/A^{3/2}}
\newcommand{\amoh}{$A^{1/2}$}
\newcommand{\amth}{$A^{3/2}$}
\newcommand{\broh}{\Gamma^{1/2}_{\gamma p}/\widt}
\newcommand{\brth}{\Gamma^{3/2}_{\gamma p}/\widt}
\newcommand{\btot}{$\Gamma(\gamma p)$}
\newcommand{\Dpi}{\Delta\pi}
\newcommand{\KL}{\Lambda K}
\newcommand{\KS}{\Sigma K}

\newcommand{\lamst}{\Lambda(1530)}

\def\RE{\mathop{\Re{\rm e}}\nolimits}
\def\IM{\mathop{\Im{\rm m}}\nolimits}
\def\vec#1{\boldsymbol{#1}}
\newcommand{\mev}{\mathrm{MeV}}
\newcommand{\gev}{\mathrm{GeV}}
\newcommand{\mevc}{\mathrm{MeV}/c}
\newcommand{\gevc}{\mathrm{GeV}/c}
\newcommand{\mevm}{\mathrm{MeV}/c^2}
\newcommand{\gevm}{\mathrm{GeV}/c^2}

\newcommand{\scstn}{\Sigma_c(2800)^0}
\newcommand{\scstp}{\Sigma_c(2800)^+}
\newcommand{\scstpp}{\Sigma_c(2800)^{++}}
\newcommand{\mnERRo}{515.4 {^{+3.2}_{-3.1}} {^{+2.1}_{-6.0}}}
\newcommand{\mppERRo}{514.5 {^{+3.4}_{-3.1}} {^{+2.8}_{-4.9}}}
\newcommand{\mpERRo}{505.4 {^{+5.8}_{-4.6}} {^{+12.4}_{-\phantom{1}2.0}}}
\newcommand{\gnERRo}{61 {^{+18}_{-13}} {^{+22}_{-13}}}
\newcommand{\gppERRo}{75 {^{+18}_{-13}} {^{+12}_{-11}}}
\newcommand{\gpERRo}{62 {^{+37}_{-23}} {^{+52}_{-38}}}

\newcommand{\br}{\mathcal{B}}\definecolor{mygreen}{rgb}{0.0,0.5,0.0}
\definecolor{myblue}{rgb}{0.0,0.0,0.55}
\definecolor{myblueh}{rgb}{0.0,0.0,0.65}
\definecolor{myred}{rgb}{0.6,0.0,0.}
\definecolor{mycyan}{rgb}{0.0,0.6,0.6}
\definecolor{mylila}{rgb}{0.6,0.,0.6}
\pagestyle{myheadings}
\definecolor{lightyellow}{cmyk}{0,0,0.5,0}
\definecolor{lightred}{rgb}{1,0.5,0.5}
\definecolor{lightgreen}{rgb}{0.5,1,0.5}
\definecolor{lightblue}{rgb}{0.5,0.5,1}
\definecolor{darkred}{rgb}{0.8,0,0}
\definecolor{darkgreen}{rgb}{0,0.4,0}
\definecolor{darkcyan}{cmyk}{1,0.3,0.3,0.3}
\definecolor{darkblue}{rgb}{0,0,0.6}
\definecolor{lightbrown}{rgb}{0.7,0.3,0.3}
\definecolor{darkbrown}{rgb}{0.5,0,0}
\definecolor{bluegreen}{rgb}{0,0.5,0.5}

\definecolor{grey0}{rgb}{0.7,0.7,0.7}
\definecolor{grey1}{rgb}{0.6,0.6,0.6}
\definecolor{grey2}{rgb}{0.4,0.4,0.4}
\definecolor{grey3}{rgb}{0.2,0.2,0.2}

\newcounter{univ_counter}
\setcounter{univ_counter} {0}
\addtocounter{univ_counter} {1}
\edef\HISKP{$^{\arabic{univ_counter}}$ }
\addtocounter{univ_counter}{1}
\edef\LPSC{$^{\arabic{univ_counter}}$ }

\title{BARYON SPECTROSCOPY}
\author{Eberhard Klempt}
\email{klempt@hiskp.uni-bonn.de}
\affiliation{Helmholtz-Institut f\"ur Strahlen- und Kernphysik\\
der Rheinischen Friedrich-Wilhelms Universit\"at\\
 Nu\ss allee 14-16, D--53115 Bonn, Germany}
\author{Jean-Marc Richard}
\email{j-m.richard@ipnl.in2p3.fr}
\affiliation{%
Laboratoire de Physique Subatomique et Cosmologie,\\
Universit\'e Joseph Fourier--CNRS-IN2P3--INPG, Grenoble, France,\\
and\\
Institut de Physique Nucl\'eaire de Lyon, Universit\'e de Lyon,\\
 CNRS-IN2P3-Universit\'e Claude Bernard,
 4, rue Enrico Fermi, F-69622 Villeurbanne, France}

\begin{abstract}
About 120 baryons and baryon resonances are known, from the abundant
nucleon with $u$ and $d$ light-quark constituents up to the
$\Xi_b^-=(bsd)$ which contains one quark of each generation and to
the recently discovered $\Omega_b^-=(bss)$. In spite of this
impressively large number of states, the underlying mechanisms
leading to the excitation spectrum are not yet understood.
Heavy-quark baryons suffer from a lack of known spin-parities. In
the light-quark sector, quark-model calculations have met with
considerable success in explaining the low-mass excitations spectrum
but some important aspects like the mass degeneracy of
positive-parity and negative-parity baryon excitations remain
unclear. At high masses, above 1.8\,GeV, quark models predict a very
high density of resonances per mass interval which is not yet
observed. In this review, issues are identified discriminating
between different views of the resonance spectrum; prospects are
discussed how open questions in baryon spectroscopy may find answers
from photo- and electro-production experiments which are presently
carried out in various laboratories.\\[1ex]
\end{abstract}
\date{\today}
 \maketitle
\clearpage
\markboth{\sl Baryon spectroscopy} {\sl Table of contents} \tableofcontents
\markboth{\sl Baryon spectroscopy }{\sl Introduction: Why baryons?}
\section{\label{Introduction}Introduction}
\subsection{\label{Why baryons?}Why baryons?}
Understanding meson resonances and the search for glueballs, hybrids
and multiquark states has remained an active field of research since
the time when the high-energy frontier brought into light the
existence of the zoo of elementary particles. At that time, baryon
spectroscopy flourished as well; but it came to a still-stand when
the complexity of the three-quark system was realized.

In the recent years, interest in baryon spectroscopy has grown
again. In his memorable closing speech at the workshop on Excited
Nucleons and Hadronic Structure in Newport News, 2000, Nathan Isgur
asked ``Why $N^*$'s\,?" \cite{Isgur:2000ad}, and gave three answers:
``The first is that nucleons are the stuff of which our world is
made. My second reason is that they are the simplest system in which
the quintessentially nonabelian character of QCD is manifest. The
third reason is that history has taught us that, while relatively
simple, baryons are sufficiently complex to reveal physics hidden
from us in the mesons". Indeed, baryons were at the roots of the
development of the quark model.
For refs.~to some early papers, see, e.g., \cite{Gell-Mann:101798,Kokkedee:101899}.
For an introduction to Quantum ChromoDynamics (QCD), see, e.g., \cite{Narison:994162,Yndurain:679369}.

Today, we have a series of precise questions for which we would like
to see answers from experiments which are presently on the floor or
are being planned. While the spectroscopy of baryons with $b$ quarks
is still in its infancy, the number of known charmed baryon
ground-states and resonances has increased substantially in recent years.
But we do not know: \\[-5mm]
\begin{enumerate}\itemsep -1mm
 \item Will baryons with triple charm reveal the genuine
spectroscopy of three color charges bound by gluons, which is
somewhat hidden by the chiral dynamics in light baryons?
 \item Will baryons with two heavy quarks combine a charmonium-like
heavy quark dynamics  and a charmed-meson-like relativistic motion
of a light quark bound around a static color source?
  \item Will single-charm baryons, and their beauty analogs  help
understanding the hierarchy of light-quark excitations and provide
keys to disentangle the pattern of highly-excited nucleon and
$\Delta$ resonances?\\[-5mm]
\suspend{enumerate}

Several questions should be answered by studying light baryons:
\\[-5mm]

\resume{enumerate}\itemsep -1mm

 \item Can we relate the occurrence of Regge trajectories
and the confinement property of QCD?
\item Can high-mass
excitations be described by the dynamics of three quarks (in
symmetric quark models) or do diquark effects play an important
role? Quark models describe baryons as dynamics of three flavored
quarks. Chiral symmetry breaking is supposed to provide constituent
masses; the color-degrees of freedom are integrated out. In spite of
the indisputable success of the quark model, the question needs to
be raised if this type of mean-field theories can be applied to the
full resonance spectrum.
\item Can we identify leading interactions between constituent quarks?
Can we find signatures for the property of flavor independence which
is expected in QCD?
\item Are hyperfine splittings and other spin-dependent effects
generated by an effective one-gluon exchange, even for light quarks?
Or by the exchange of Goldstone bosons? Or are instanton-induced
interactions at work?
\item  What are missing resonances and why
are they missing? Mostly, missing resonances are defined as
resonances which are predicted by symmetric quark models but which
have not (yet) been found. More restricted is a definition where
baryons expected in symmetric but not in diquark models are
considered to be missing resonances. The lowest-mass example of this
type of resonances is the not-well established quartet of nucleon
resonances consisting of $N_{1/2^+}(1880)$, $N_{3/2^+}(1900)$,
$N_{5/2^+}(1890)$, $N_{7/2^+}(1990)$.
\item  The observed spectrum of baryon resonances seems to exhibit a
rather simple pattern. Is this pattern accidental or does it reflect
a phase transition which may occur when baryons are highly
excited?
\item Are high-mass baryons organized in the
form of spin-parity doublets or chiral multiplets, of
mass-degenerate states having identical spin and parity\,?
\item Do we understand baryon decays, or what can be
learned studying decays?
\end{enumerate}

\subsection{\label{The structure of baryons}The structure of baryons}
 From deep inelastic scattering we know that the nucleon has a
complicated structure. The structure functions reveal the
longitudinal momentum distributions of valence and sea quarks;
generalized parton distributions give access to their transverse momenta
and their correlation with the longitudinal momenta. By integration,
a few interesting global features follow. The number $N_v$ of
valence quarks (integrated over Feynman $x$) is $N_v = N_q - 2 N_s =
3$. The nucleon has a strange quark sea with $N_s \approx 0.1
N_{u,d}$. In the infinite momentum frame, gluons carry a large
($\approx 0.5$) fraction of the total momentum. From the
hadronization of $e^+e^-$ pairs it is known that there are three
colors, $ N_c=3$. And the width of the neutral weak interaction
boson $Z^0$ reveals the number of generations $N_\text{G}$ (with at
least one neutrino with mass below 45\,GeV), $N_\text{G} = 3 $.
Time-like and spatial form-factors of protons differ by factor of 2
at $Q^2\approx 10$\,GeV$^2$. Perturbatively, this factor should be
1. The discrepancy teaches us that even at this large momentum
transfer, quark correlations play an important role.

\subsection{\label{Naming scheme}Naming scheme}

The  Particle Data Group (PDG) \cite{Amsler:2008zz} identifies a
baryon by its name and its mass. The particle name is $N$ or
$\Delta$ for baryons having isospin 1/2 or 3/2, respectively, with
three $u, d$ quarks; the name is $\Lambda$ or $\Sigma$ for baryons
having two $u, d$ quarks and one $s$ quark; the two light quarks
couple to isospin 0 or 1, respectively. Particles with one $u$ or
$d$ quark are called $\Xi$, they have isospin 1/2. The $\Omega$ with
no $u$ or $d$ quark has isospin 0. If no suffix is added, the
remaining quarks are strange. Thus, the $\Omega$ has three $s$
quarks. Any $s$ quark can be replaced by a $c$ (or $b$) quark which
is then added as a suffix. Depending on isospin, $\Lambda_c$ or
$\Sigma_c$ (or $\Lambda_b$ or $\Sigma_b$) are formed by replacing
one $s$ quark by a heavy quark. Resonances with one charmed and one
strange quark are called $\Xi_c$, those with two or three charmed
quarks $\Xi_{cc}$ or $\Omega_{ccc}$. The $\Xi_b$ with one $b$, one
$s$, and one $u$ or $d$ quark has already been mentioned.

Resonances are characterized by adding $L_{2I,2J}$ behind the
particle name where $L$ defines the lowest orbital-angular momentum
required when  they  disintegrate into the ground state and a
pseudoscalar meson, $I$ and $J$ are isospin and total angular
momentum, respectively.

We deviate from this definition. E.g., the two particles
$N(1535)S_{11}$ and $N(1520)D_{13}$ derive their name from the fact
that they form an S-wave (D-wave) in $\pi N$ scattering. The first
``1" indicates that they have isospin 1/2 (which is already clear for
a nucleon excitation), the second ``1" defines its total spin to be
$J=1/2$. The parity of the states is deduced from the positive
parity of the orbital angular momentum state and the intrinsic
parities of the ground state baryon (which is $+1$) and of the
pseudoscalar meson (which is $-1$).

We  call these two states $N_{1/2^-}(1535)$ and $N_{3/2^-}(1520)$.
These are the observed states. They can be mixtures of quark model
states. E.g., the $N_{1/2^-}(1535)$  and  $N_{1/2^-}(1650)$ can be written in the form
\begin{eqnarray}\label{qms}
N_{1/2^-}(1535)= \cos\Theta_{1/2^-}\ket{^{2}N_{1/2^-}} -
\sin\Theta_{1/2^-}\ket{^{4}N_{1/2^-}}\nonumber\\
N_{1/2^-}(1650)= \sin\Theta_{1/2^-}\ket{^{2}N_{1/2^-}} +
\cos\Theta_{1/2^-}\ket{^{4}N_{1/2^-}}
\end{eqnarray}
where $^{2}N_{1/2^-}$ has intrinsic quark spin $s=1/2$ while
$^{4}N_{1/2^-}$ belongs to the $s=3/2$ quartet. It is often useful
to classify baryons according to a baryon model in which the
interaction between the (constituent) quarks are approximated by
harmonic oscillators (HO). In the HO approximation, baryons develop
a band structure. Mixing between states belonging to different bands
but having identical external quantum numbers is possible. Further
components to the states in eq.~(\ref{qms}) could come from the
third excitation band with $\textsf{N}=3$. A state
\begin{equation}
\ket{^{2}N_{1/2^-}, D_{56} (L=1)^{P=-1}_{\textsf{N=3}}},
\end{equation}
 is a
spin-doublet quark model state belonging to the third excitation
band with one unit of orbital angular momentum, having a 56-plet
SU(3) flavor structure. Explicit quark model calculations give a
small mixing between different bands and the band structure is
preserved.

\subsection{\label{Guide to the literature}Guide to the literature}

Prime sources of original information is found in the proceedings of
three conference series on the Structure of Baryons and on $N^*$.
The latest conferences were held as tri-annual International
Conference on the Structure of Baryons, Baryons'07, in Seoul, Korea
(2007), and as bi-annual International Conference on Meson-Nucleon
Physics and the Structure of the Nucleon (MENU 2007) in J\"ulich,
Germany, (2007). Irregularly, mostly bi-annual, took place the NSTAR
Workshop (Physics of Excited Nucleons) which, in 2009, was hosted in
Beijing.

Experimentally indispensable is the Review of Particle Properties
published by the Particle Data Group \cite{Amsler:2008zz} which will
be used throughout this review. It includes a few minireviews on
baryons:~\cite{Wohl:2008gw,Wohl:2008st,Hohler:2004gt,Trilling:2006wg}.
Still very useful is the broad review by \cite{Hey:1982aj}. The
advances of the quark model to describe the baryon excitation
spectrum and baryon decays are reviewed  by \cite{Capstick:2000qj}.
Low-energy photoproduction and implications for low-lying resonances
are critically discussed by \cite{Krusche:2003ik}. Not included here
is the physics of cascade resonances: of $\Xi$'s and $\Omega$'s
where little information has been added since the review of
\cite{Hey:1982aj}. There is a proposal to study $\Xi$ resonances at
Jlab,  and first results  demonstrated the feasibility
\cite{Guo:2007dw}. The latest review on $\Xi$ baryons can be found
in \cite{Meadows:1980vr}.

\subsection{\label{Abbreviations}Abbreviations}
For the sake of readability, we collect here abbreviations used in
the text.\\
$\vec{\rho},\,\vec{\lambda}$ are the Jacobi variables for the 3-body problem,\\
$\vec L$ is the orbital angular momentum, $\vec L=\vec l_{\rho}+\vec
l_{\lambda}$,\\
$\vec S=\vec s_1+\vec s_2+\vec s_3$  is the total quark spin,\\
$\vec J=\vec L+\vec S$ is the total angular momentum,\\
$J, L, S, l_{\rho},l_{\lambda}$ are the corresponding quantum numbers,\\
$\mathrm{L}$ is the sum ${\rm L}=l_{\rho}+l_{\lambda}$,\\
$\vec I$ is the isospin having components $I_{k}$,\\
$I$ the isospin quantum number,\\
$\mathrm{S}$ is the strangeness, $\mathrm{Y}$ the hypercharge,\\
$\Upsilon=(b\bar b)$ stands for the bottomonium family,\\
$P$ is the parity, Q the charge, $+e$ is the unit charge,\\
$N(xxx)$ represents a nucleon $N$ with mass $xxx$,\\
$\tt {N} = \tt{n}_{\rho}+\tt{n}_{\lambda}$ is the radial number,\\
\textsf{N} gives the band number,\\
$p$, $n$ represent proton and neutron,\\
$u,d$  are light quarks, $q=u,d,s$ include strangeness, \\
$Q=c,b$ are heavy quarks,\\
$M_{p,n}$ are proton and neutron mass,\\
$\kappa_{p,n}$ their anomalous magnetic moments,\\
$\alpha$, $\alpha_s$ are the electromagnetic and strong couplings.\\
\subsection{\label{Outline}Outline} Exciting new results have been
obtained for heavy baryons containing a charmed or a bottom quark.
The results are reviewed in section \ref{Heavy-quark baryons}. Most
information on light-quark baryons stems from $\pi\,N$ or $K\,N$
elastic or charge exchange scattering but new information is now
added from photo- and electro-production experiments. The progress
is discussed in section \ref{Light-quark baryon resonances}. Section
\ref{se:mod} provides a framework within which baryon excitations
can be discussed and gives an outline of current theoretical ideas.
The rich spectrum of light baryon resonances reveals symmetries and
a mass pattern. Based on these observation, a tentative
interpretation of the baryon spectrum is offered. In the summary
(\ref{se:sum}), conclusions are given to what extent the new
experiments have contributed to baryon spectroscopy and suggestions
for further work are made.

\markboth{\sl Baryon spectroscopy}{\sl Heavy--quark baryons}
\section{\label{Heavy-quark baryons} Heavy-quark baryons}

With the discovery of the $J$ particle \cite{Aubert:1974js} at BNL
and of the $\psi$ \cite{Augustin:1974xw} and $\psi'$
\cite{Abrams:1974yy} at Stanford and their interpretation as
$(c\bar{c})$ bound states, and with the discovery of charmed mesons
\cite{Goldhaber:1976xn}, charmed baryons had of course to exist as
well, and their properties were predicted early
\cite{Gaillard:1974mw,DeRujula:1975ge}. Experimental evidence for
the first charmed baryon was reported at BNL in the reaction
$\nu_{\mu} p\to \mu^- \Lambda \pi^+\pi^+\pi^+\pi^-$ with $\Lambda$
decaying into $p\pi^-$ \cite{Cazzoli:1975et}. None of the $\pi^+$
could be interpreted as $K^+$ and no $\pi^+\pi^-$ pair formed a
$K^0$, hence the event could signal either violation of the $\Delta
S=\Delta Q$ rule, or be due to production of a baryon with charm.
Now we know that a $\Sigma_c^{++}$ was produced.

At present, 34 charmed baryons and 7 beauty baryons are known. For
most of them, spin and parity have not been measured; for some
states the quantum numbers can be deduced from their decay modes or
by comparison of measured masses with the expectation from
quark-models, in particular \cite{Copley:1979wj}.

The study of charmed baryons is mostly pursued by searching for
resonances which decay into $\Lambda_c^+$ plus one (or more)
pion(s). The momenta of the - comparatively slow - pions can be
measured with high precision. Hence the best precision is obtained
for the mass difference to the $\Lambda_c^+$. The $\Lambda_c^+$ is
sometimes reconstructed from up to 15 different decay modes. In
other cases, the most prominent and well measurable modes
$\Lambda_c^+\to p\overline{K}{}^0$ and $\Lambda_c^+\to pK^-\pi^+$ are
sufficient to obtain a significant signal. The study of charmed
baryons was often a by-product: the main aim of the experiments at
Cornell, SLAC or KEK was the study of CP violation in $B$ decays
from $\Upsilon(4S)$ and, perhaps, the study of the $\Upsilon$
family. Charmed baryons are then produced in the $e^+e^-\to q\bar q$
continuum and in $B$ decays.

\subsection{\label{The life time of charmed particles}The life time of charmed particles}

Weak interaction physics is not covered in this review. However, the
finite lifetime of hadrons with heavy flavors plays an important
role in their experimental identification. In
Table~\ref{tab:lifetime} are summarized the measured lifetimes of
flavored mesons and baryons. The precision is truncated to 100\,keV.
\begin{table}[pt]
\caption{\label{tab:lifetime} Lifetime of flavored mesons and
baryons (in s) \cite{Amsler:2008zz}. Lifetimes of $\Xi_b^-$ and
$\Omega_b^-$, see also \cite{Aaltonen:2009ny}.\vspace{-2.6mm}}
\begin{footnotesize}
\renewcommand{\arraystretch}{1.5}
$$\begin{array}{lrclr}
\hline\hline
K^\pm\quad & (123.85\pm 0.24)\times 10^{-10}&&
K^0_S & (0.8953\pm 0.0005)\times 10^{-10}\\
K^0_L&(511.4\pm2.1)\times 10^{-10}&&
D^\pm& (1040\pm7)\times 10^{-15}\\
D^0&(410.1\pm1.5)\times 10^{-15}&&
D_s&(500\pm7)\times 10^{-15}\\
B^\pm&(1638\pm11)\times 10^{-15}&&
B^0&(1530\pm9)\times 10^{-15}\\
B_s&(1466\pm59)\times 10^{-15}&\\[.5ex]
\hline\\[-3.5ex] \Lambda&(2.631\pm0.020)\times 10^{-10}&&
\Sigma^\pm &(0.8018\pm0.0026)\times 10^{-10}\\
\Xi^0&(2.90\pm0.09)\times 10^{-10}&&
\Xi^-&(1.639\pm0.015)\times 10^{-10}\\
\Omega^-&(0.821\pm0.011)\times 10^{-10}&&
\Lambda_c&(200\pm6)\times 10^{-15}\\
\Xi_c^+&(442\pm26)\times 10^{-15}&&
\Xi_c^0&(112\genfrac{}{}{0pt}{}{+13}{-10})\times 10^{-15}\\
\Omega_c^0&(69\pm12)\times 10^{-15}&&
\Lambda_b&(1230\pm74)\times 10^{-15}\\
\Xi_b^-&(1490^{+200}_{-180})\times 10^{-15}&&
\Omega_b^-&1130^{+530}_{-400})\times 10^{-15}\\
\hline\hline
\end{array}$$
\renewcommand{\arraystretch}{1.0}
\end{footnotesize}
\end{table}

Comments are in order:\\[-5mm]
\begin{itemize}\itemsep -1mm
\item
While the lifetimes of particles carrying a $b$ quark are very
similar, this is not the case with strangeness, where more than a
factor of 3 is observed from the most stable hyperon to the
shortest-lived.
\item The differences are even more pronounced for charmed baryons.
When the difference between the charged and the neutral $D$-meson
lifetime was discovered, this was a striking surprise, and it took
some time to realize that besides the simplest mechanism, where the
$c$ quark emits a virtual $W$  boson which dissociates into a lepton
pair or a quark--antiquark pair, there are diagrams in which the $W$
is exchanged.  This is, however, permitted for $D^0$ and $D_s$ but
forbidden for $D^\pm$. The lifetime is also influenced by
interferences. If $c\to s+W^+\to s+u+\bar{d}$, for instance,
initiates some hadronic decay, this $\bar d$ should antisymmetrize
with the $\bar d$ of $D^0$, an effect that does not exist for $D^+$.
In principle, a fusion mechanism such as $c+\bar{s}\to u+\bar{d}$ should also contribute
to the $D_s$ decay.
\item
The analysis was then extended to charmed baryons, with predictions
by  \cite{Guberina:1986gd}; see, also
\cite{Fleck:1990ma,Guberina:2000de}. Some effects are enhanced with
respect to the case of mesons, for instance the role of
antisymmetrization. The fusion mechanism, on the other hand, is
suppressed as requiring an antiquark from the sea. The trend of the
predicted hierarchy is well reproduced by the experimental data, but
the observed differences are even more pronounced.
\item
It would be particularly interesting to measure the lifetime of
double-charm baryons, or heavier baryons with triple charm, or with
charm and beauty. Another effect should be taken into account, that
of the deep binding of the heavy quarks. This is already discussed
for the $B_c$ meson with quark content $(b\bar{c})$.
\item At {\sc\small COMPASS}, LHC, {\sc\small PANDA}, or at a second
generation of $B$-factories, there is the possibility to search for
 weak decays of $\Xi_{cc}(3520)^+$ and $\Xi_{cc}^{++}$ double
charmed baryons into charmless final states \cite{Liu:2007twb}. Such
decays could signal new physics.
\item The lifetimes of charmed
particles are just sufficiently long to identify them by a decay
vertex separated from the interaction vertex. For
$\beta\gamma\approx 1$, the lifetime  of $B$-mesons leads to a
separation of $500\mu$m. Precise vertexing is therefore a major
experimental requirement.
\end{itemize}

\subsection{\label{Summary of heavy baryons}Summary of heavy baryons}
The masses of heavy baryons known so far are summarized in Table
\ref{allcharm}, an account of their discoveries and the most recent
experimental results is given below. For most resonances, the
quantum numbers have not been measured, except for
$\Lambda_c(2593)^+$ with $J^P=1/2^-$ and $\Lambda_c^+(2880)$ for
which $J^P=5/2^+$ is suggested. The quantum numbers of the
lowest-mass states are deduced from the quark model.
\begin{table}[pt]
 \caption{\label{allcharm}Masses (in MeV)  of heavy
baryons quoted from \cite{Amsler:2008zz} except for $\Sigma_b$ and
$\Omega_b$ (see text).  The isospin of
$\Lambda_c^{+}/\Sigma_c^{+}(2765)$ (two faint entries) is not
known.\vspace{-3mm}}
\begin{footnotesize}
\begin{center}
\renewcommand{\arraystretch}{1.5}
\begin{tabular}{lllcll}
\hline\hline
$\Lambda_c^+$&2286.5$\pm$0.2&2595.4$\pm$0.6&2628.1$\pm$0.6&\color{grey2}{2766.6$\pm$2.4}& 2881.5$\pm$0.4\\
$\Sigma_c^{++}$&2454.0$\pm$0.2&2518.4$\pm$0.6&2801$^{+4}_{-6}$&$\qquad\ \ \mid\Lambda_c^+$:&2939.3$\pm$1.4\\
$\Sigma_c^{+}$&2452.9$\pm$0.4&2517.5$\pm$2.3&2792$^{+14}_{-5}$&\color{grey2}{2766.6$\pm$2.4}\\
$\Sigma_c^{+}$&2453.8$\pm$0.2&2518.0$\pm$0.5&2802$^{+4}_{-7}$\\
$\Xi_c^{+}$&2467.9$\pm$0.4&2575.7$\pm$3.1&2646.6$\pm$1.4&2789.2$\pm$3.2&2816.5$\pm$1.2\vspace{-2mm}\\
&&2969.3$\pm$2.8&3054.2$\pm$1.3&3077.0$\pm$0.5&3122.9$\pm$1.3\\
$\Xi_c^{0}$&2471.0$\pm$0.4&2578.0$\pm$2.9&2646.1$\pm$1.2&2791.9$\pm$3.3&2818.2$\pm$2.1\vspace{-2mm}\\
&&2972.9$\pm$4.7&&3079.3$\pm$1.1&\\
$\Omega_c^{0}$&2697.5$\pm$2.6&2768.3$\pm$3.0&&$\qquad\ \ \mid\Xi_{cc}^{+}$:&3518.9$\pm$0.9\\
$\Lambda_b^0$&5620.2$\pm$1.6&&&& \\
$\Sigma_b^{+}$& 5807.8$\pm$2.7&5829.0$\pm$3.4&
$\qquad\ \ \mid\Sigma_b^{-}$:& 5815.2$\pm$2.0&5836.4$\pm$2.8 \\
$\Xi_b^{-}$&5792.4$\pm$2.2 &&
$\qquad\ \ \mid\Omega_b^{-}$:& \multicolumn{2}{l}{6165$\pm$17 or 6054.4$\pm$6.8}\\
\hline\hline
\end{tabular}
\renewcommand{\arraystretch}{1.0}
\end{center}
\end{footnotesize} \vspace{-5mm} \end{table}

\begin{figure}[pb]
\begin{center}
\includegraphics[width=0.25\textwidth,angle=-90,clip]{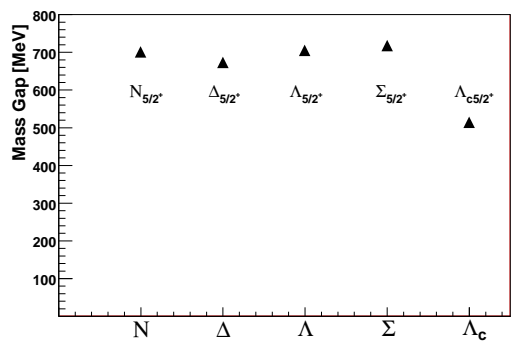}
\end{center}
\caption {\label{massgap}Mass gap from the respective ground states
to the lowest excitation with $J^P=5/2^+$. }
\end{figure}

Figure~\ref{massgap} shows the flavor dependence of the mass
difference between $J^P=5/2^+$ and ground states. The mass gap
between $\Lambda_c^+(2880)$ and $\Lambda_c^+$ is smaller than that
of light-quark baryons. To test this conjecture we compare the
spectrum of all observed $\Lambda_c^+$ baryons with their
light-quark analogue states.

\begin{figure}[pt]
\begin{center}
\includegraphics[width=.45\textwidth,clip]{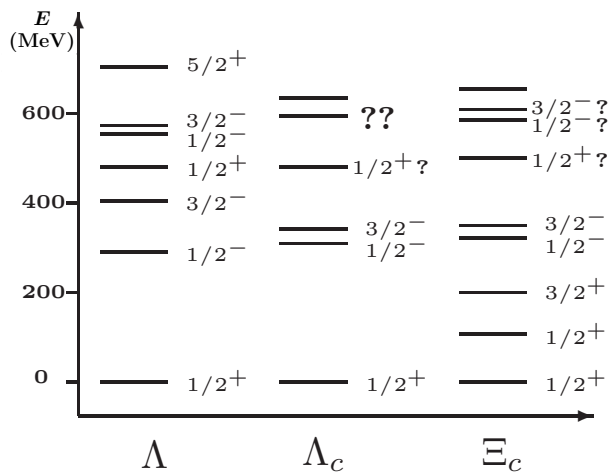}
\vspace{-3mm}
\end{center}
\caption {\label{llcxic}Excitation spectrum of $\Lambda$,
$\Lambda_c^+$, and $\Xi_c$. Between $\Lambda_{3/2^-}(1690)$ and
$\Lambda_{5/2^+}(1690)$ there are two further states which are
omitted for clarity. The quantum number assignments of $\Lambda_c$,
and $\Xi_c$ follow \cite{Amsler:2008zz}, those with question marked
are our tentative assignments. The $\Lambda_c(2880)$, marked {\bf
??} is suggested to have $J^P=5/2^+$
\cite{Abe:2006rz}.\vspace{-3mm}}
\end{figure}

In Fig. \ref{llcxic}, the excitation spectra of $\Lambda$,
$\Lambda_c^+$, and $\Xi_c$ are compared. In the three lowest states,
the light quark pair has spin 0. In the $\Xi_c$ spectrum, there are
two additional states, the $\Xi_c'$ with spin 1/2 and $\Xi_c(2645)$
with spin 3/2, in which the light quark pair has spin 1. These are
forbidden for the isoscalar $\Lambda$ and $\Lambda_c^+$. Above these
states, a doublet of negative-parity states are the lowest
excitations with fully antisymmetric wave functions. In the
$\Lambda$ spectrum, the Roper-like $\Lambda_{1/2^+}(1600)$ follows,
and then a doublet -- $\Lambda_{1/2^-}(1670)$ and
$\Lambda_{3/2^-}(1690)$ -- and a triplet -- $\Lambda_{1/2^-}(1800)$,
$\Lambda_{3/2^-}(xxx)$, and $\Lambda_{5/2^-}(1830)$ -- of negative
parity states. The $\Lambda_{1/2^+}(1810)$, not shown in Fig.
\ref{llcxic}, might be the analogue of $N_{1/2^+}(1710)$ and
$\Delta_{1/2^+}(1750)$.

Far above, a spin doublet $\Lambda_{3/2^+}(1890)$ and
$\Lambda_{5/2^+}(1820)$ is known. It is very tempting to assign
$1/2^+$ quantum numbers to the isolated states in all three spectra,
followed by a doublet of negative-parity states. This scenario is,
however, ruled out by the $5/2^+$ assignment to $\Lambda_c^+(2880)$.
We urge that the quantum number measurement should be repeated;
below we present arguments why the $5/2^+$ assignment is unlikely.
Quite in general, the determination of the quantum numbers of heavy baryons
remains an important task for the future.

\subsection{Major experiments in heavy-baryon spectroscopy}
A large fraction of our knowledge of charmed baryons presented in
Table \ref{allcharm} comes from the {\sc\small CLEO} detector at the
intersecting storage ring CESR. The {\sc\small CLEO} detector was
upgraded continuously. It consisted of a four-layer silicon-strip
vertex detector, a wire drift chamber and a particle identification
system based on Cherenkov ring imaging, time-of-flight counters, a
7800-element CsI electromagnetic calorimeter, a 1.5\,T
superconducting solenoid, iron for flux return and muon
identification, and muon chambers
\cite{Kopp:1996kg,Viehhauser:2000cu}. The integrated luminosity on
the $\Upsilon(4S)$ resonance accumulated in the years 1999-2003 was
$16\rm\,fb^{-1}$.

Of course, the $B$-factories have reached a much higher luminosity;
{\sc\small BaBaR} and {\sc\small BELLE} $700\rm\,fb^{-1}$ both
collected about $1300\rm\,fb^{-1}$. The data shown below are mostly
based on a fraction of the data. Both $B$-factories operated mostly
at the peak cross section for formation of the $\Upsilon(4S)$, at
10.58\,GeV, with energies of the colliding electron and positron
beam of 9 (8)\,GeV and 3.1 (3.5)\,GeV, for {\sc\small BaBaR}
({\sc\small BELLE}) respectively, resulting in a Lorentz boost of
the center of mass of $\beta=0.55$ $(0.425)$.

The inner part of the {\sc\small BaBaR} detector
\cite{Aubert:2001tu} includes tracking, particle identification and
electromagnetic calorimetry. It is surrounded by a superconductive
solenoid providing a magnetic field of 1.5\,T. The tracking system
is composed of a Silicon Vertex Tracker and a drift chamber.  A
40-layer drift chamber is used to measure particle momenta and the
ionization loss $dE/dx$.  Particle identification is provided by the
$dE/dx$ measurement and a ring-imaging detector. The electromagnetic
calorimeter is a finely segmented array of CsI(Tl) crystals with
energy resolution of $\sigma_E/E\approx 2.3\%\cdot E^{-1/4} + 1.9\%$
(E in GeV). The iron return yoke is instrumented with resistive
plate chambers and limited streamer tubes for detection of muons and
neutral hadrons.

Tracking, identification and calorimetric systems of the {\sc\small
BELLE} detector \cite{Iijima:2000cq} at KEKB are placed inside a
1.5\,T superconducting solenoid magnet. Tracking and vertex
measurements are provided by a silicon vertex detector and a central
drift chamber. The central drift chamber has 50 layers of anode
wires for tracking and $dE/dx$ measurements. Particle identification
is achieved using the central drift chamber, time of flight
counters, and aerogel Cherenkov counters. The electromagnetic
calorimeter consists of CsI(Tl) crystals of projective geometry. The
flux return is instrumented with 14 layers of resistive plate
chambers for muon identification and detection of neutral hadrons.

We will mention results obtained by the {\sc\small ARGUS} and
{\sc\small SELEX} collaborations without introducing the detectors
here and refer the reader interested in their performance to two
reports by \cite{Albrecht:1988vy} and \cite{Engelfried:1997rp}. Also
some early bubble chamber results and results from the CERN ISR and
SPS will be mentioned. At Fermilab, the photoproduction experiments
E687, E691, E791 and Focus and {\sc\small SELEX} using a hadron beam
produced interesting results on charmed baryons.

So far, only a few baryons with beauty have been discovered. The
energy of the $B$-factories operating at the $\Upsilon(4S)$ is
obviously not sufficient to produce beauty baryons. These are
however produced abundantly by the Tevatron at Fermilab, in which
antiprotons and protons collide at 1.96\,TeV center-of-mass energy.
Two major experiments, {\sc\small CDF} and {\sc\small D\O},
exploit the physics; the discovery of the top quark, the measurement
of its mass to a precision of nearly 1\%, and the study of $B_s$
oscillations belong to the highlights of the Tevatron results.
Earlier important results on beauty baryons were achieved at the
CERN ISR and at LEP.

The {\sc\small CDF} detector \cite{Acosta:2004yw} consists of multiple layers of
silicon micro-strip detectors, providing for a precise measurement
of a track's impact parameter with respect to the primary vertex,
and a large open-cell drift chamber enclosed in a 1.4\,T
superconducting solenoid, which in turn is surrounded by
calorimeters. The electromagnetic calorimeters use lead-scintillator
sampling, the hadron calorimeters  iron-scintillator sampling.

The inner tracking of {\sc\small D\O} \cite{Abazov:2005pn} is
composed of a silicon microstrip tracker for vertexing and a central
fiber tracker, both located within a 2\,T superconducting solenoidal
magnet. Calorimetry relies on liquid-argon and uranium detectors. An
outer muon system consists of a layer of tracking detectors and
scintillation trigger counters in front of and behind  1.8\,t iron
toroids.

\subsection{Charmed baryons}
 \subsub{The $\Lambda_c$ states}
\paragraph{$\Lambda_c^+$:} The first observation of a charmed baryon, of $\Lambda_c^+$, was
reported two years after the $J/\psi$ discovery \cite{Knapp:1976qw}.
Now, $\Lambda_c^+$ is the best known charmed baryon. Due to its high
mass, it has a large number of decay modes. Among these,
$\Lambda_c^+\to p\overline K\pi,\,  p\overline K\pi\pi$ and $\Lambda\pi^+ \pi,\,\Lambda\pi^+ \pi\pi$ have
the largest decay fractions, summing up to about 20\%. The most
precise mass measurement was made by the {\sc\small BaBaR}
collaboration \cite{Aubert:2005gt} finding
\begin{equation} M_{\Lambda_c} = 2286.46\pm
0.14\,{\rm MeV}.
\end{equation}
The lifetime was measured by  E687, {\sc\small CLEO}, Focus, and
{\sc\small SELEX}.  The lifetimes of all heavy baryons stable
against hadronic decays are collected in Table
\ref{tab:lifetime}.\vspace{-4mm}

\begin{figure}[pt]
\centering
\includegraphics[width=8cm,height=6cm]{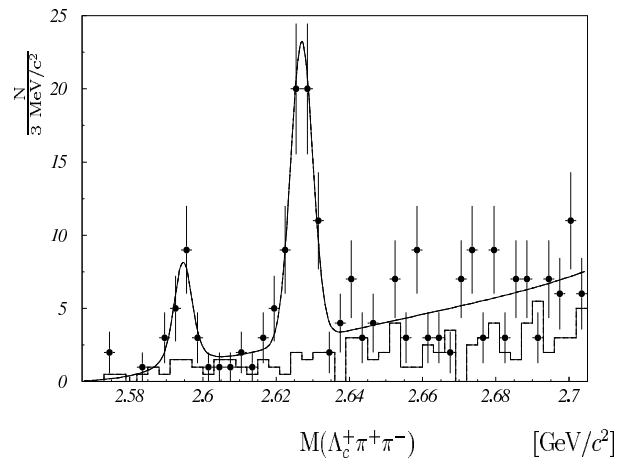}\\[-1ex]
\caption{\label{Argus} The $\Lambda_c^+\pi^+\pi^-$ invariant mass
distribution after a cut on the $\Lambda_c^+$ (reconstructed from
five decay modes) and using side bins (dashed line)
\cite{Albrecht:1997qa}.\\[-3ex]}
\end{figure}

\paragraph{$\Lambda_c(2593)^+$ and $\Lambda_c(2625)^+$:} The
$\Lambda_c(2625)^+$ was discovered by the {\sc\small ARGUS}
collaboration at the $e^+e^-$ storage ring DORIS II at DESY
\cite{Albrecht:1993pt}. Figure~\ref{Argus} shows the
$\Lambda_c^+\pi^+\pi^-$ invariant mass distribution with increased
statistics \cite{Albrecht:1997qa} in which the $\Lambda_c(2593)^+$
is observed as well. The latter state was first observed by
{\sc\small CLEO} \cite{Edwards:1994ar}. Table \ref{lc2593} compares
the results on both states from the {\sc\small ARGUS}
\cite{Albrecht:1997qa}, the {\sc\small CLEO} \cite{Edwards:1994ar},
and the {\sc\small E687} \cite{Frabetti:1993hg,Frabetti:1995sb}
collaborations.

\begin{table}[pt]
\caption{\label{lc2593}Mass and width of the $\Lambda_c(2593)^+$ and
$\Lambda_c(2625)^+$ measured at {\sc\small CLEO}, {\sc\small BaBaR}
and {\sc\small BELLE}.}
\begin{footnotesize}
\begin{center}
\renewcommand{\arraystretch}{1.5}
\begin{tabular}{lccc}
\hline\hline
& & $M,\;\mevm$ & $\Gamma,\;\mevm$ \\
\hline
{\sc\small ARGUS} & \quad$\Lambda_c(2593)$ \quad& $2596.3\pm0.9\pm0.6\;\;$ & $2.9^{+2.9+1.8}_{-2.1-1.4}$ \\
{\sc\small CLEO}  & \quad$\Lambda_c(2593)$ \quad& $2594.0\pm0.4\pm1.0\;\;$ & $3.9^{+1.4+2.0}_{-1.2-1.0}$\\
{\sc\small E687} & \quad$\Lambda_c(2593)$ \quad& $2581.2\pm0.2\pm0.4$ &  \\
\hline
{\sc\small ARGUS} & \quad$\Lambda_c(2625)$ \quad& $2628.5\pm0.5\pm0.5$ & $<3.2$ \\
{\sc\small CLEO}  & \quad$\Lambda_c(2625)$ \quad& $2629.5\pm0.2\pm0.5$ & $<1.9$  \\
{\sc\small E687}& \quad$\Lambda_c(2625)$ \quad& $2627.7\pm0.6\pm0.3$ & \\
\hline\hline
\end{tabular}
\renewcommand{\arraystretch}{1.0}\\[-3ex]
\end{center}
\end{footnotesize}
\end{table}
The $\Lambda_c(2593)^+$ decays with a large fraction ($>$70\%) via
$\Sigma_c\pi$;  the small phase space favors vanishing orbital
angular momentum. The $\Sigma_c$ is the lowest mass charmed
isovector state and is thus expected to have $J^P=1/2^+$. Then,
$J^P=1/2^-$ follows for the $\Lambda_c(2593)^+$. Most likely, the
$\Lambda_c(2625)^+$ is its $J^P=3/2^-$ companion and the two states
correspond to $\Lambda_{1/2^-}(1405)$ and $\Lambda_{3/2^-}(1520)$.
See section \ref{se:sub-band} for further discussion.\vspace{-4mm}

\paragraph{$\Lambda_c(2765)^+$ (or $\Sigma_c(2765)^+$), $\Lambda_c(2880)^+$ and
$\Lambda_c(2940)^+$:} The {\sc\small CLEO} Collaboration reported
two peaks in the $\Lambda_c^+\pi^+\pi^-$ final
state~\cite{Artuso:2000xy} which could be $\Lambda_c^+$ or
$\Sigma_c^+$ excitations. One of them is found 480\,MeV above the
$\Lambda_c^+$ baryon and is rather broad, $\Gamma\approx 50$\,MeV;
the other one is narrow, $\Gamma < 8$\,MeV, and its mass lies
$596\pm1\pm2$\,MeV above the $\Lambda_c^+$.

\begin{figure}[pt]
\centering
\begin{picture}(550,160)
\put(5,0){\includegraphics[width=8cm,height=6cm]{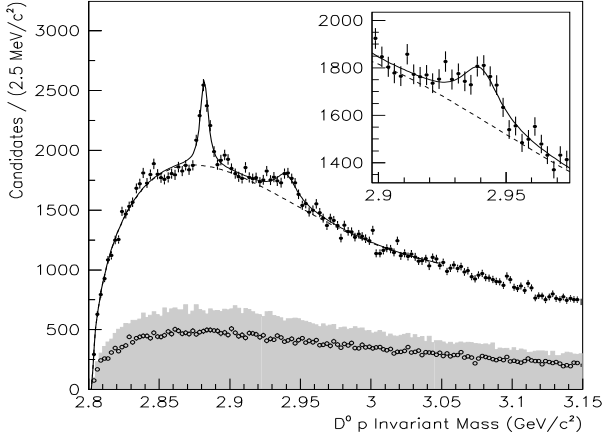}}
\end{picture}
\caption{Invariant mass distribution for $D^0p$ candidates at
  {\sc\small BaBaR}~\cite{Aubert:2006sp}. Also shown are the contributions from $D^0$
  sidebands (grey) and wrong-sign combinations (open dots).\vspace{-3mm}}
\label{fit_both}
\end{figure}

The {\sc\small BaBaR} Collaboration observed two peaks in the $D^0p$
invariant mass distribution (see
Fig.~\ref{fit_both})~\cite{Aubert:2006sp}. It is the first
observation of a heavy baryon disintegration into a heavy-quark
meson and a light-quark baryon. Due to the kinematics, the larger
part of the released energy is carried away by the baryon. The
$D^+p$ final state shows no peaks; thus the isospin of the heavy
baryon must be zero which identifies the peaks as
$\Lambda_c(2880)^+$ and $\Lambda_c(2940)^+$ (and not belonging to
the $\Sigma_c^+$ series). The former one coincides with the narrow
state observed by \cite{Artuso:2000xy}, called $\Lambda_c(2880)^+$.

The {\sc\small BELLE} Collaboration confirmed the
$\Lambda_c(2940)^+$ in $\Lambda_c^+\pi^+\pi^-$. The decay proceeds
via formation of $\Sigma_c(2455)^{++}$ or $\Sigma_c(2455)^0$
resonances in the intermediate state~\cite{Abe:2006rz}. The
$\Lambda_c(2880)^+$ and $\Lambda_c(2940)^+$ mass and width measured
by {\sc\small BaBaR} and {\sc\small BELLE} are consistent (see
Table~\ref{tab:CLEO-bb}).\vspace{-6mm}

\begin{table}[htbp]
\caption{\label{tab:CLEO-bb}Mass and width of the $\Lambda_c(2880)$
and $\Lambda_c(2940)$ measured at {\sc\small CLEO}
\cite{Artuso:2000xy}, {\sc\small BaBaR~\cite{Aubert:2006sp}} and
{\sc\small BELLE}~\cite{Abe:2006rz}.}
\begin{footnotesize}
\begin{center}
\renewcommand{\arraystretch}{1.5}
\begin{tabular}{llll}
\hline\hline
& &\quad  $M,\;\mevm$ &\quad $\Gamma,\;\mevm$ \\
\hline
{\sc\small CLEO}  & $\Lambda_c(2880)$ & $2882.5\pm1\pm2\;\;$ & $<8$ \\
{\sc\small BaBaR} & $\Lambda_c(2880)$ & $2881.9\pm0.1\pm0.5\;\;$ & $5.8\pm1.5\pm1.1$ \\
{\sc\small BELLE} & $\Lambda_c(2880)$ & $2881.2\pm0.2\pm0.4$ & $5.8\pm0.7\pm1.1$ \\
\hline
{\sc\small BaBaR} & $\Lambda_c(2940)$ & $2939.8\pm1.3\pm1.0$ & $17.5\pm5.2\pm5.9$ \\
{\sc\small BELLE} & $\Lambda_c(2940)$ & $2938.0\pm1.3^{+2.0}_{-4.0}$ & $13^{+8}_{-5}{^{+27}_{-\phantom{2}7}}$ \\
\hline\hline
\end{tabular}
\renewcommand{\arraystretch}{1.0}\vspace{-3mm}
\end{center}
\end{footnotesize}
\end{table}

\begin{figure}[pt]
\centering
\includegraphics[width=6cm,height=4cm]{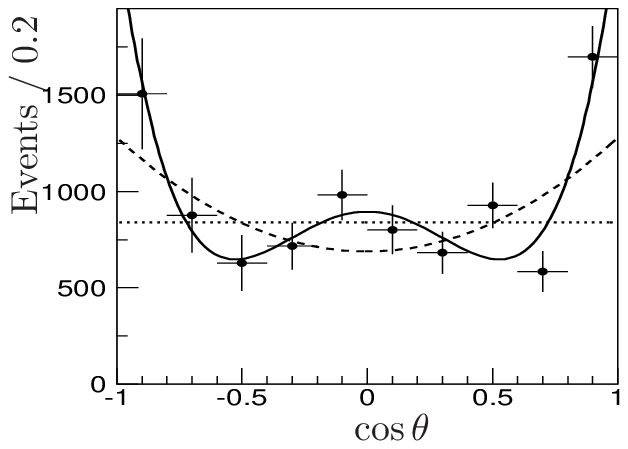}
\caption{\label{angular_2880}The yield of
$\Lambda_c(2880)^+\to\Sigma_c(2455)^{0}\pi^{+}$ and
$\Sigma_c(2455)^{++}\pi^{-}$ decays as a function of the helicity
angle. The fits correspond to $\Lambda_c(2880)^+$ spin hypotheses
$J=1/2$ (dotted line), $3/2$ (dashed curve), $5/2$ (solid curve),
respectively \cite{Abe:2006rz}.\vspace{-2mm}}
\end{figure}

The two sequential decay modes improve the sensitivity to study the
quantum numbers of the resonance.  As shown in \cite{Abe:2006rz},
the angular distribution of the
$\Lambda_c(2880)^+\to\Sigma_c(2455)\pi$ decay favors  high spin and
is compatible with $J=5/2$ (see Fig.~\ref{angular_2880}).  The
experimental ratio of the $\Lambda_c(2880)^+$ partial widths
$\Gamma[\Sigma_c(2520)\pi]/\Gamma[\Sigma_c(2455)\pi]=0.23\pm0.06\pm0.03$
is calculated in the framework of heavy-quark symmetry to be 1.45
for $J^P=5/2^-$ and 0.23 for $J^P=5/2^+$
\cite{Isgur:1991wq,Cheng:2006dk}. Thus the spin-parity assignment
$5/2^+$ is favored over $5/2^-$. Note that this assignment requires
angular momentum $L=3$ between $\Sigma_c(2455)$ and $\pi$ at a decay
momentum 370\,MeV/c while $L=1$ is sufficient for the suppressed
$\Sigma_c(2520)\pi$ decay mode. The $D^0p$ decay mode of
$\Lambda_c(2880)^+$ poses a further problem. Again, $L=3$ is
required for $J^P=5/2^+$, now at 320\,MeV decay momentum. When
$J^P=1/2^-$ is assigned to $\Lambda_c(2880)^+$, the
$\Sigma_c(2455)\pi$ and $D^0p$ decay mode proceed via S-wave while
the suppressed $\Sigma_c(2520)\pi$ decay requires D-wave.

Based on the spin-parity assignment $5/2^+$ for the
$\Lambda_c(2880)^+$ and on the Mass Load Flux Tube Model
\cite{LaCourse:1988cu}, the series of $\Lambda_c^+$ states in the
first line of Table~\ref{allcharm} is suggested to have quantum
numbers $1/2^+$, $1/2^-$, $3/2^-$, $3/2^+$, $5/2^+$, and $5/2^-$
\cite{Cheng:2009tm}. The spin-parity assignment $5/2^+$ for the
$\Lambda_c(2880)^+$ is constitutive for this interpretation of the
spectrum.

 Finally we notice that the mass of the
$\Lambda_c(2940)^+$ is at the $D^*p$ threshold, a fact which invites
interpretations of this state as a $D^*p$
molecule~\cite{He:2006is}.\vspace{-2mm}
 \subsub{The $\Sigma_c$ states}
\begin{figure}[pb]
\includegraphics[width=0.36\textwidth,height=0.28\textwidth]{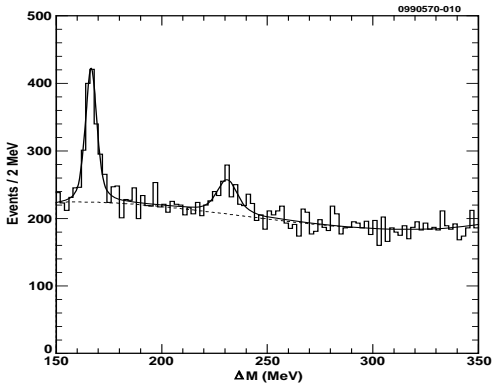}
\caption{\label{Sigma_ground_states}Mass difference spectrum,
$M(\Lambda_c^+\pi^0)-M(\Lambda_c^+)$ from CLEO \cite{Ammar:2000uh}.
The solid line fit is to a third-order polynomial background shape
and two $P$-wave Breit--Wigner functions smeared by Gaussian
resolution functions for the two signal shapes. The dashed line
shows the background function.}
\end{figure}
\paragraph{$\Sigma_c(2455)$  and $\Sigma_c(2520)$:} These two states
have been observed in a large number of experiments; here we show
only the results of the most recent publication of the {\sc\small
CLEO} collaboration. $\Sigma_c^+$ and $\Sigma_c^{*+}$ were observed
in their $\Lambda_c^+\pi^0$ decay \cite{Ammar:2000uh}, and
$\Sigma_c^{*++}$ and $\Sigma_c^{*0}$ in their decay into
$\Lambda_c^+\pi^{\pm}$ \cite{Athar:2004ni}. The data of
\cite{Athar:2004ni} cover the $e^+e^-$ energy range 9.4 to 11.5\,GeV
while \cite{Ammar:2000uh} used data at the $\Upsilon(4S)$. But $B$
decays were suppressed by kinematic cuts and in both cases, the
$\Sigma_c^{*}$ baryons are likely produced from the $e^+e^-\to q\bar
q$ continuum. Figure~\ref{Sigma_ground_states} shows the momentum of
pions recoiling against the $\Lambda_c^+$ which defines the mass gap
between $\Sigma_c$ or $\Sigma_c^{*}$ and $\Lambda_c^+$. From the
angular distribution of the $B^- \to \Sigma_{c}(2455)^{0} \bar{p}$
decays, the spin of the $\Sigma_{c}(2455)^{0}$ baryon is determined
to be 1/2 \cite{Aubert:2008if} while the $\Sigma_c(2520)$ quantum
numbers $J^P=3/2^+$ are quark-model assignments. The numerical
results on masses and widths are reproduced in Table
\ref{tab:CLEO-sigma}.\vspace{-4mm}

\paragraph{$\Sigma_c(2800)^+$:} The {\sc\small BELLE} Collaboration
observed an isotriplet of charmed baryons decaying to the
$\Lambda_c^+\pi$ final state at 2800\,MeV \cite{Mizuk:2004yu}. An
additional peak at $\Delta M\sim0.42\,\gevm$, visible in the
$\Lambda_c^+\pi^+$ and $\Lambda_c^+\pi^-$ invariant mass
distributions, was identified as a reflection from the
$\Lambda_c(2880)^+\to\Sigma_c(2455)\pi\to\Lambda_c^+\pi^+\pi^-$
decays. The parameters of all isospin partners are consistent (see
Table~\ref{tab:2800}). Based on the mass and width, the $3/2^-$
assignment for these states was proposed~\cite{Mizuk:2004yu}.
\cite{Aubert:2008if} observe the state at $(2846\pm 8\pm 10)$\,MeV
and with a width of $(86^{+33}_{-22}\pm 12)$\,MeV.\\[-2ex]
\begin{table}[pt]
\caption{\label{tab:CLEO-sigma}Mass and width of the
$\Sigma_c(2455)$ and $\Sigma_c(2520)$ measured at {\sc\small CLEO}.}
\begin{footnotesize}
\begin{center}
\renewcommand{\arraystretch}{1.5}
\begin{tabular}{lccc}
\hline\hline
& & $M,\;\mevm$ & $\Gamma,\;\mevm$ \\
\hline
$\Sigma_c(2455)$& $M(\Sigma_c^{++})-M(\Lambda_c^+)$&$167.4\pm 0.1\pm 0.2$&$2.3\pm0.2\pm0.3$\\
& $M(\Sigma_c^{+})-M(\Lambda_c^+)$&$166.4\pm 0.2\pm 0.3$&$<4.6$\\
& $M(\Sigma_c^{0})-M(\Lambda_c^+)$&$167.2\pm 0.1\pm 0.2$&$2.5\pm0.2\pm0.3$\\
$\Sigma_c(2520)$& $M(\Sigma_c^{*++})-M(\Lambda_c^+)$&$231.5\pm0.4\pm0.3$&$14.4^{+1.6}_{-1.5}\pm1.4$\\
& $M(\Sigma_c^{*+})-M(\Lambda_c^+)$&$231.0\pm 1.1\pm 2.0$&$<17$\\
& $M(\Sigma_c^{*0})-M(\Lambda_c^+)$&$231.4\pm0.5\pm0.3$&$16.6^{+1.9}_{-1.7}\pm1.4$\\
\hline\hline
\end{tabular}
\renewcommand{\arraystretch}{1.0}
\end{center}
\end{footnotesize}
\caption{\label{tab:2800}Mass and
width of the $\Sigma_c(2800)$
 measured at {\sc\small CLEO}.\vspace{-4mm}}
\begin{footnotesize}
\begin{center}
\renewcommand{\arraystretch}{1.5}
\begin{tabular}{lcll}
\hline\hline
& & $M,\;\mevm$ & $\Gamma,\;\mevm$ \\
\hline
 $\Sigma_c(2800)$&$M(\scstpp)-M(\Lambda_c^+)$& $\mppERRo$& $\gppERRo$\\
&$M(\scstp)-M(\Lambda_c^+)$ & $\mpERRo\;\;$ & $\gpERRo$ \\
&$M(\scstn)-M(\Lambda_c^+)$ & $\mnERRo$ & $\gnERRo\;\;$ \\
\hline\hline
\end{tabular}\vspace{-3mm}
\renewcommand{\arraystretch}{1.0}
\end{center}
\end{footnotesize}
\end{table}
 \subsub{The $\Xi_c$ states}\vspace{-2mm}
\paragraph{$\Xi_c$\ and $\Xi_c'$:} The $\Xi_c^+$ was discovered by
\cite{Biagi:1983en} at the CERN SPS hyperon beam in $\Sigma^-$
nucleon collisions, $\Sigma^- + {\rm Be} \to (\Lambda K^-\pi^+\pi^+)
+ X$, its isospin partner $\Xi_c^0$ by the {\sc\small CLEO}
collaboration \cite{Avery:1988uh} through its decay to $\Xi^-\pi^+$.
Both states were studied in different production and decay modes.
The PDG quotes
\begin{equation}
\begin{aligned}
&\Xi_c^+\quad &M=2467.9\pm 0.4\,{\rm MeV}, &\quad \tau=442\pm26\,{\rm fs},
 \\
&\Xi_c^0\quad &M=2471.0\pm 0.4\,{\rm MeV}, &\quad
\tau=112^{+13}_{-10}\,{\rm fs}.
\end{aligned}
\end{equation}

The $\Xi_c(2645)$:\qquad The spin wave-function of the isospin
doublet $\Xi_c^+$, and $\Xi_c^0$ contains a pair of light quarks,
$[su]$ and $[sd]$, mostly in a spin $S=0$ state. There should exist
a second doublet in which the light quark pair is mostly in spin
triplet $S=1$. This pair is denoted $\Xi_c^{0,+\prime}$.

\begin{figure}[pt]
\vspace{2mm}
\begin{tabular}{cc}
\hspace{-2mm}\includegraphics[width=0.25\textwidth,height=0.32\textwidth]{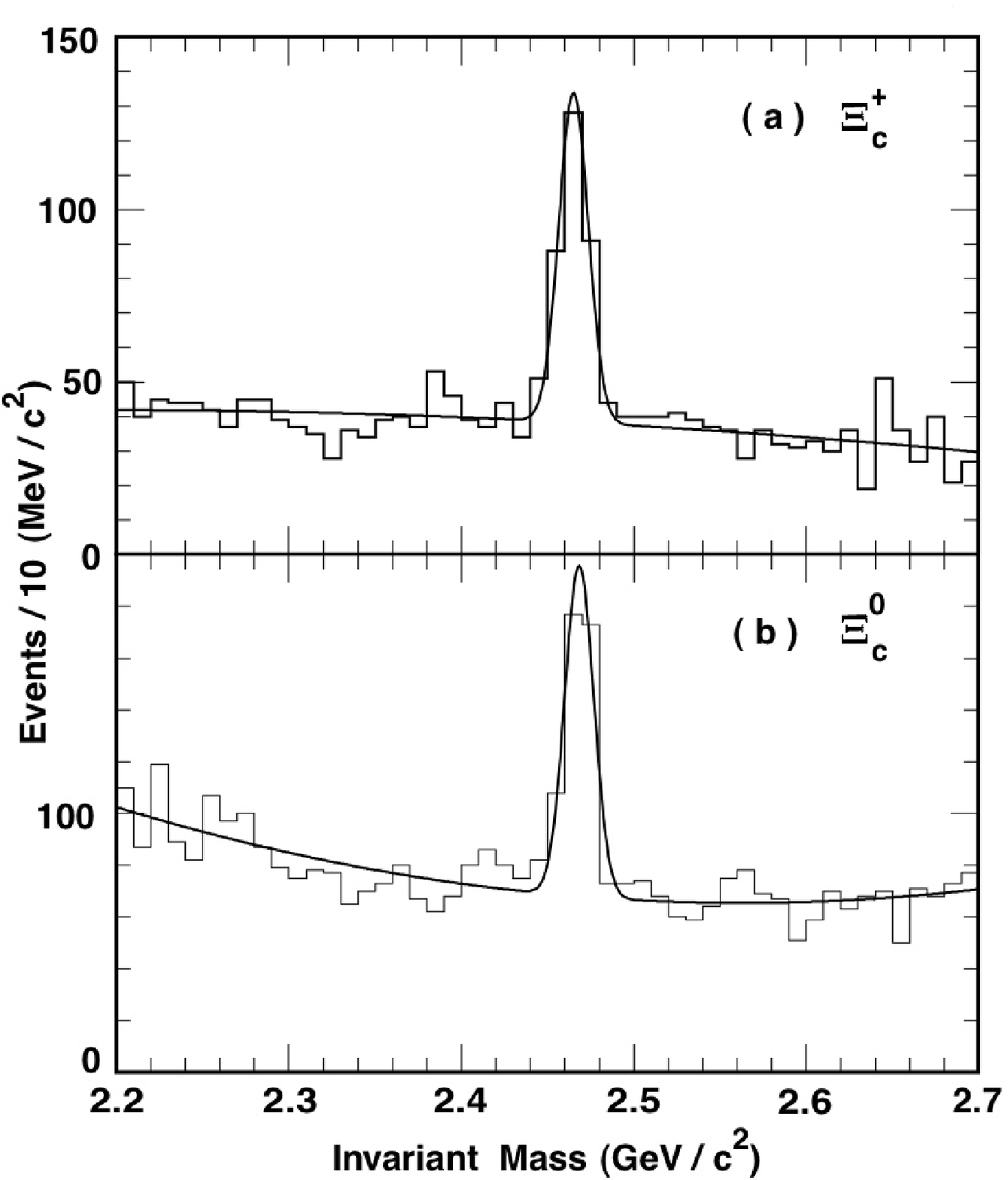}&
\hspace{-2mm}\includegraphics[width=0.25\textwidth,height=0.32\textwidth]{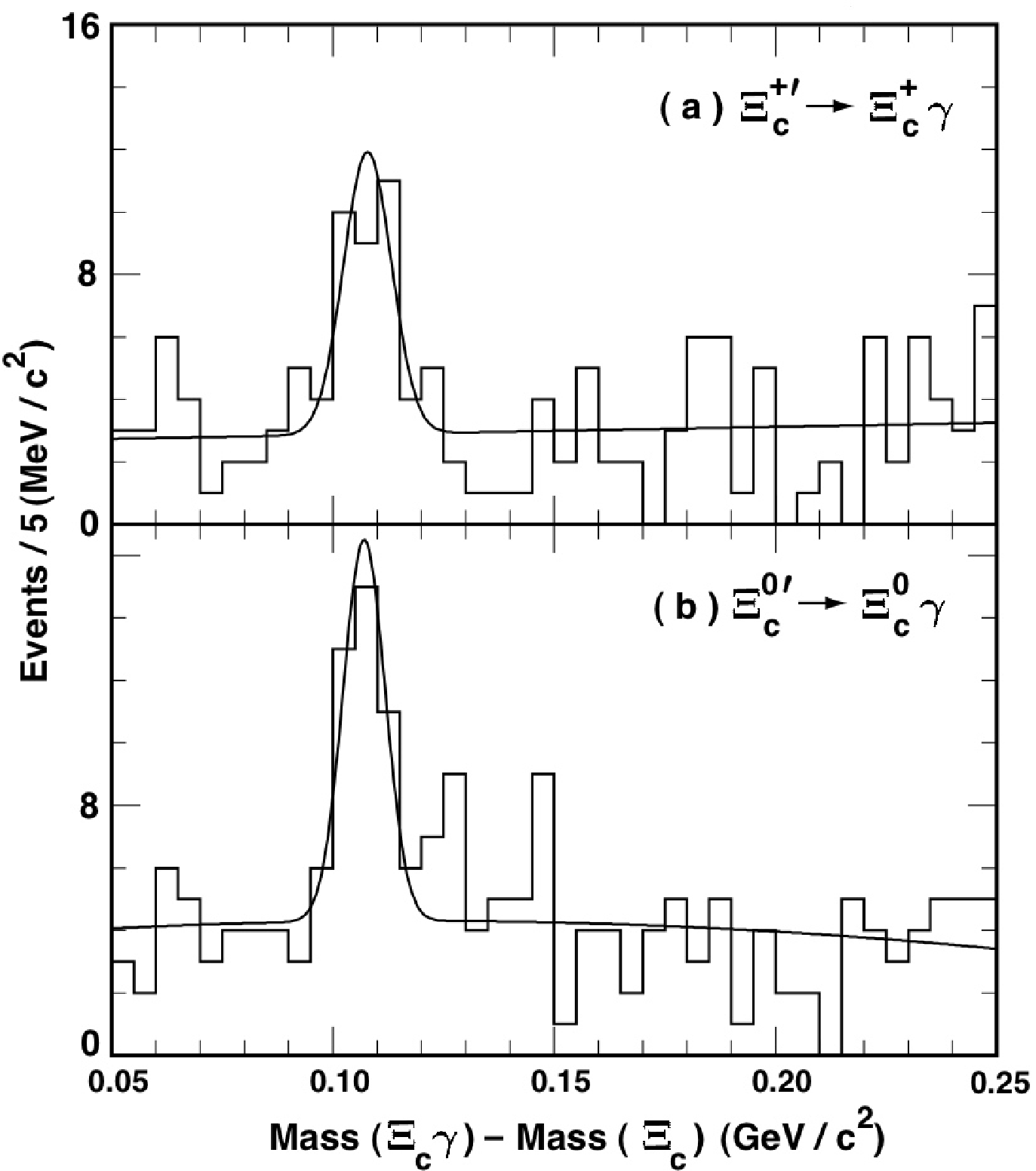}
\end{tabular}
\caption{\label{CLEO_xi_prime} Left: (a) Summed invariant mass
distributions for $\Xi^{-}\pi^{+}\pi^{+}$ and
$\Xi^{0}\pi^{+}\pi^{0}$ combinations with $x_p>0.5$ and 0.6,
respectively, and (b) for $\Xi^{-}\pi^{+}$, $\Xi^{-}\pi^{+}\pi^{0}$,
$\Omega^{-}K^{+}$, and $\Xi^{0}\pi^{+}\pi^{-}$ combinations. Right:
Invariant mass difference $\Delta M(\Xi_c\gamma - \Xi_c)$
distributions for $\Xi_c^+\gamma$ and $\Xi_c^0\gamma$, where
contributions from the different $\Xi_c$ decay modes have been
summed in each case \cite{Jessop:1998wt}.\vspace{-3mm} }\end{figure}

The latter two states were discovered by the {\sc\small CLEO}
collaboration \cite{Jessop:1998wt}. In a first step, the two
ground-state $\Xi_c$ baryons were reconstructed using several decay
modes (see Fig.~\ref{CLEO_xi_prime}). The ground-state $\Xi_c$
baryons were observed jointly with a low-energetic photon. The
$\Xi_c^+\gamma$ and $\Xi_c^0\gamma$ invariant masses show signals
which were interpreted as the missing $\Xi_c^{+,0\prime}$ partners
of the ground state $\Xi_c^{+,0}$ baryons. The mass differences
$M(\Xi_c^{+\prime})-M(\Xi_c^+)$ and $M(\Xi_c^{0\prime})-M(\Xi_c^0)$
were measured to be $107.8\pm 1.7\pm 2.5$ and $107.0\pm 1.4\pm
2.5\;\mathrm{MeV}/c^2$, respectively.

{\sc\small BaBaR} confirmed the existence of the $\Xi_c'$ and found
that the rate of $\Xi_c'$ production over $\Xi_c$ is about 18\% in
the $e^+e^-$ continuum but about 1/3 in $B$ decays. The angular
distribution of  $\Xi_c'\to\Xi_c\gamma$ decays was found  to be
consistent with the prediction for $J^P = 1/2^+$ even though higher
spins cannot yet be ruled out \cite{Aubert:2006rv}. {\sc\small
Belle} determined the $\Xi_c(2645)^+$ mass to be $2645.6\pm
0.2^{+0.6}_{-0.8}$ and the $\Xi_c(2645)^0$ $2645.7\pm
0.2^{+0.6}_{-0.7}$, respectively \cite{Lesiak:2008wz}.

\vspace{-4mm}

\paragraph{$\Xi_c(2790)$ and $\Xi_c(2815)$:} In \cite{Csorna:2000hw},
decays of $\Xi_c$ resonances to $\Xi_c^{\prime}$ plus a pion were
observed. Mass differences for the two states to the $\Xi_c^{+,0}$
ground states are given in Table \ref{tab:2790}. The precision for
the $\Xi_c(2815)$ mass was improved by \cite{Lesiak:2008wz} to
$2817.0\pm 1.2^{+0.7}_{-0.8}$ and $2820.4\pm 1.4^{+0.9}_{-1.0}$ for
the neutral and charged state, respectively. These observations
complement an earlier observation of the {\sc\small CLEO}
collaboration \cite{Alexander:1999ud} in which a doublet of $\Xi_c$
resonances was observed, one decaying into $\Xi_c^+\pi^+\pi^-$ via
an intermediate $\Xi_c^{*0}$, and its isospin partner decaying into
$\Xi_c^0\pi^+\pi^-$ via an intermediate $\Xi_c^{*+}$. Mass
differences and widths are again collected in Table \ref{tab:2790}.
These resonances are interpreted as the $J^P = {1/2}^-$ and
${3/2}^-$ $\Xi_{c}$ particles, the charmed-strange analogues of the
$\Lambda_{c}^+(2593)$ and $\Lambda_{c}^+(2625)$, or of the
light-quark $\Lambda_{1/2^-}(1405)$ and $\Lambda_{3/2^-}(1520)$
pair.

\begin{table}[pt]
\caption{\label{tab:2790}Mass and width of the $\Xi_c(2790)$ and
$\Xi_c(2815)$ measured at {\sc\small CLEO}.}
\begin{footnotesize}
\begin{center}
\renewcommand{\arraystretch}{1.5}
\begin{tabular}{lcll}
\hline\hline
& & $M,\;\mevm$ & $\Gamma,\;\mevm$ \\
\hline
$\Xi_c(2790)$ &$M(\Xi_c^0 \gamma \pi^+)-M(\Xi_c^0)$& $318.2\pm1.3\pm2.9$& $<15$\\
              &$M(\Xi_c^+ \gamma \pi^-)-M(\Xi_c^+)$& $324.0\pm1.3\pm3.0$& $<12$\\
$\Xi_c(2815)$ &$M(\Xi_c^0\pi^+\pi^-)-M(\Xi_c^0)$&$347.2\pm0.7\pm2.0$ & $<6.5$\\
              &$M(\Xi_c^+\pi^+\pi^-)-M(\Xi_c^+)$&$348.6\pm0.6\pm1.0$& $<3.5$\\
\hline\hline
\end{tabular}
\renewcommand{\arraystretch}{1.0}
\end{center}\vspace{-5mm}
\end{footnotesize}
\end{table}

\vspace{-5mm}

\begin{figure}[pt]
\begin{tabular}{cc}
\hspace{-2mm}\includegraphics[width=0.24\textwidth,height=0.16\textwidth]{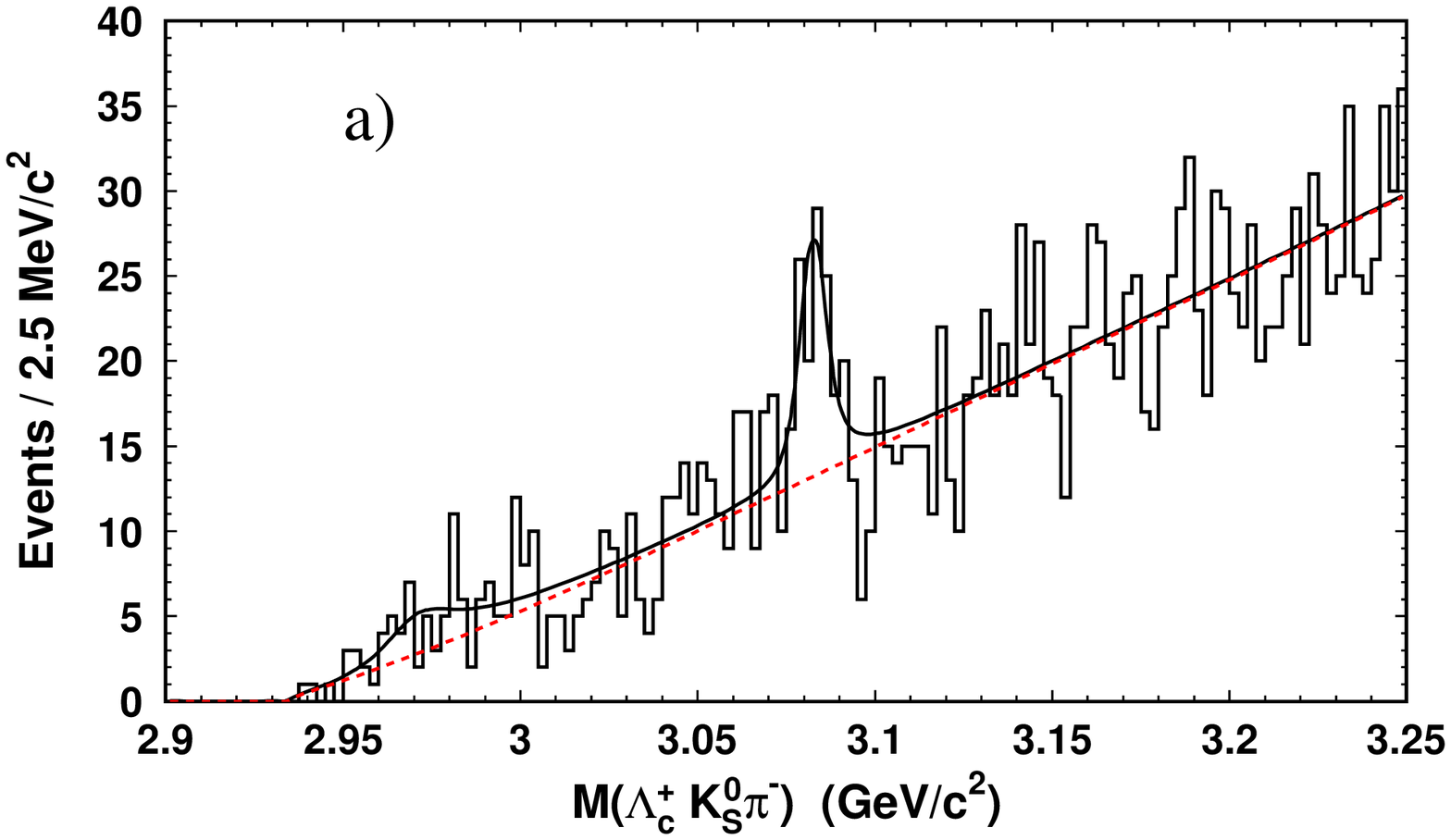}\vspace{-30mm}\\
\hspace{-2mm}\includegraphics[width=0.24\textwidth,height=0.16\textwidth]{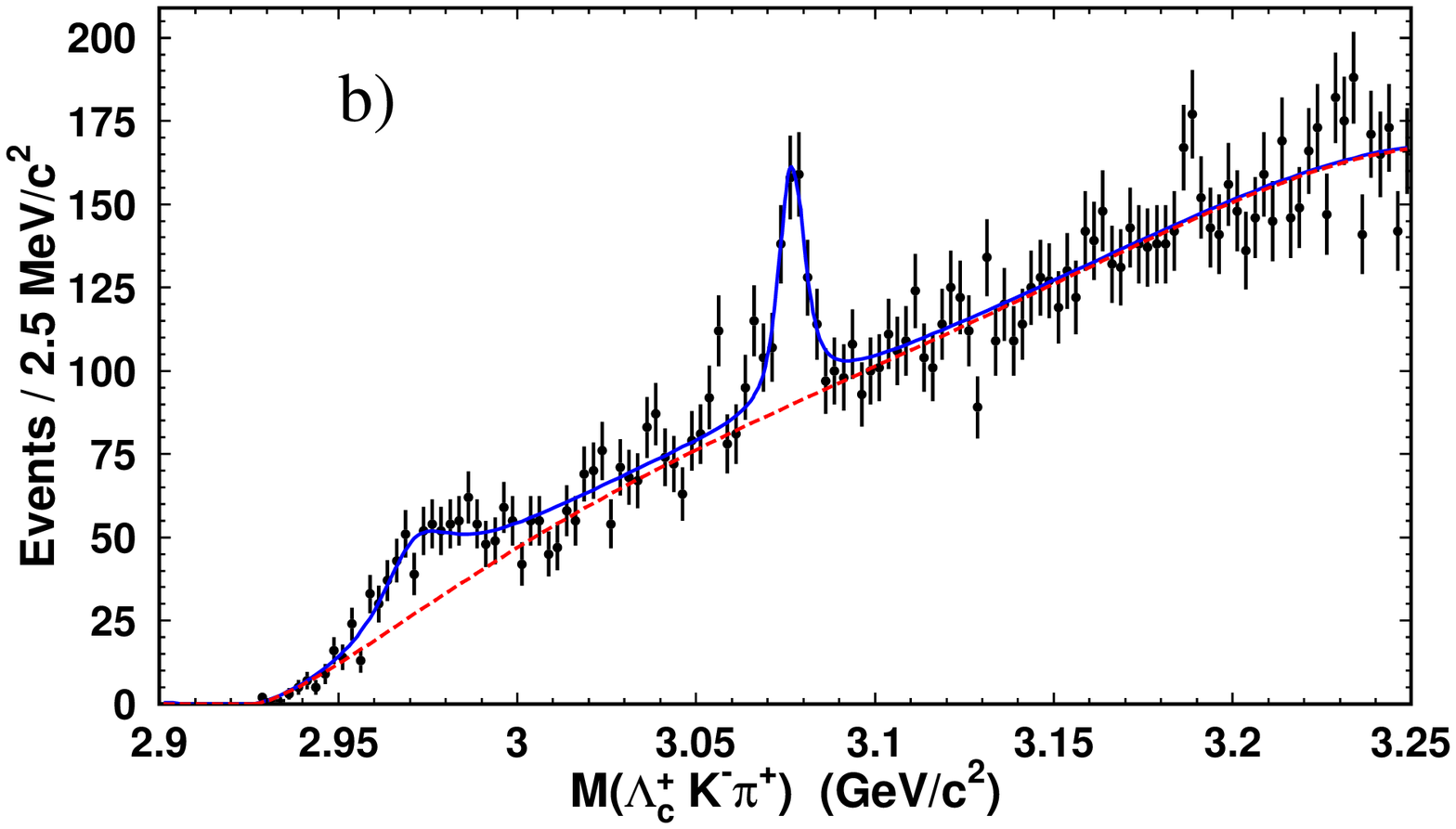}&
\hspace{-1mm}\includegraphics[width=0.25\textwidth,height=0.32\textwidth]{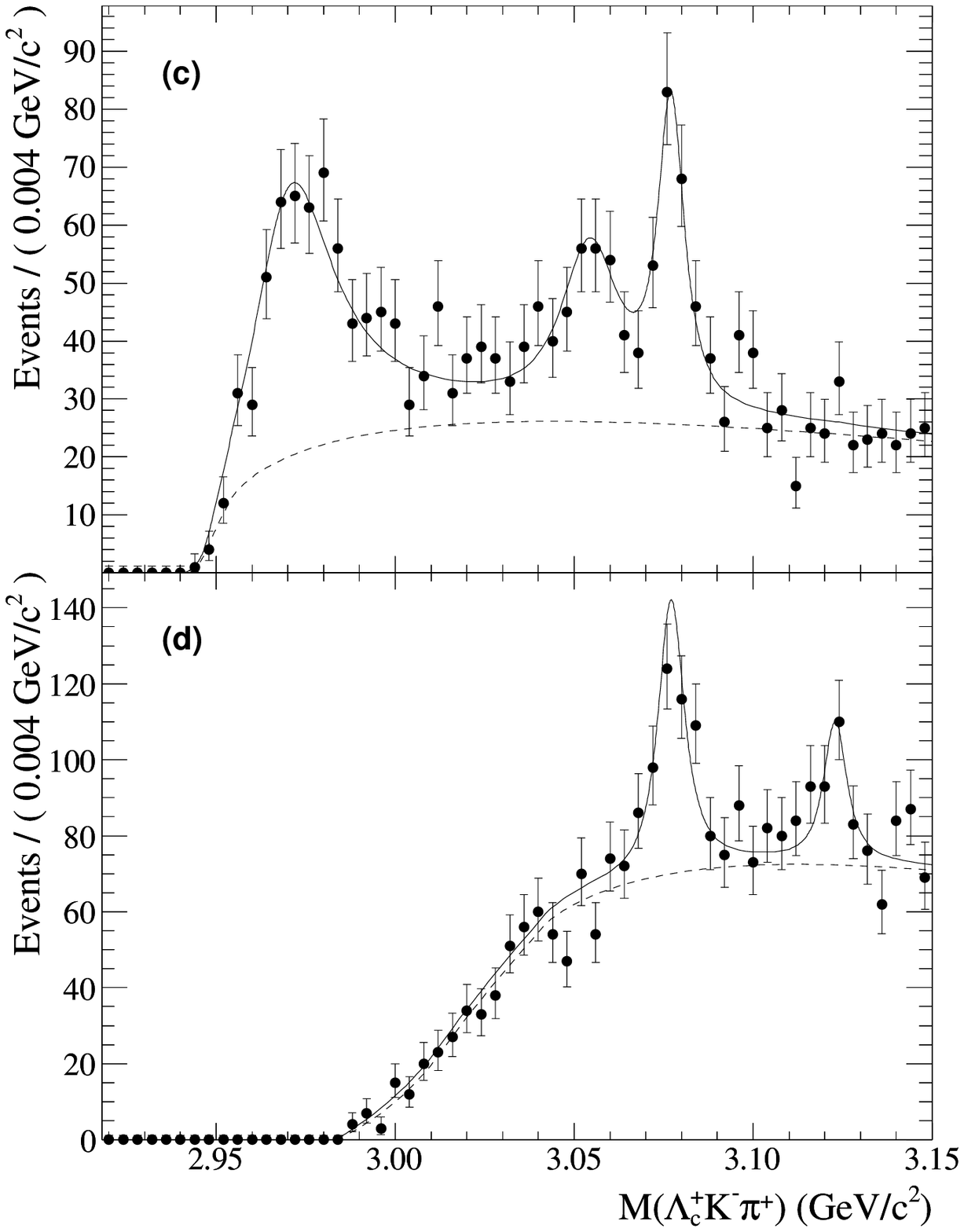}
\end{tabular} \caption{\label{BaBaR_belle} (a)
$M(\Lambda_c^+ K^-\pi^+)$ and (b) $M(\Lambda_c^+ K^0_S\pi^-)$
distribution at {\sc\small BELLE}~\cite{Chistov:2006zj}. (c) The
$\Lambda_c^+K^-\pi^+$ invariant mass distribution for
   $M(\Lambda_c^+\pi^+)$ consistent  with the $\Sigma_c(2455)$ and
   (d) with the $\Sigma_c(2520)$, measured at
   {\sc\small BaBaR}~\cite{Aubert:2006uw,Aubert:2007dt}.}
\end{figure}

\paragraph{$\Xi_c(2980)$ and $\Xi_c(3080)$:} The {\sc\small BELLE}
Collaboration observed two new $\Xi_c$ states, the $\Xi_c(2980)$ and
$\Xi_c(3080)$, decaying to $\Lambda_c^+K^-\pi^+$ and
$\Lambda_c^+K_S\pi^-$ \cite{Chistov:2006zj}, see
Fig.~\ref{BaBaR_belle}a,b. In contrast to other $\Xi_c$ decay modes,
the $c$ and $s$ quark separate, thus forming a charmed baryon and a
strange meson. (Likewise, decays into $\Lambda D^+$ are allowed
above 3\,GeV and could be searched for.) The broader of the two
states was measured to have a mass of $2978.5\pm 2.1\pm 2.0$
MeV/$c^2$ and a width of $43.5\pm 7.5\pm 7.0$ MeV/$c^2$. The mass
and width of the narrow state are measured to be $3076.7\pm 0.9\pm
0.5$ MeV/$c^2$ and $6.2\pm 1.2\pm 0.8$ MeV/$c^2$, respectively. A
search for the isospin partner decaying into $\Lambda_c^+
K_S^0\pi^-$ yielded evidence for a signal at the mass of $3082.8\pm
1.8\pm 1.5$ MeV/$c^2$; the broader low-mass baryon is just visible.

The {\sc\small BaBaR} Collaboration confirmed observations of the
$\Xi_c(2980)$ and $\Xi_c(3080)$~\cite{Aubert:2006uw} by studying the
$\Lambda_c^+K^0_{\scriptscriptstyle S}$, $\Lambda_c^+K^-$,
$\Lambda_c^+K^-\pi^+$, $\Lambda_c^+K^0_{\scriptscriptstyle S}\pi^-$,
$\Lambda_c^+K^0_{\scriptscriptstyle S}\pi^-\pi^+$, and
$\Lambda_c^+K^-\pi^+\pi^-$ mass distributions (see
Fig.~\ref{BaBaR_belle}c).  In addition, {\sc\small BaBaR} studied
the resonant structure of the $\Lambda_c^+K^-\pi^+$ final state
\cite{Aubert:2007dt}, see Fig.~\ref{BaBaR_belle}d. The $\Xi_c(3080)$
was found to decay through the intermediate $\Sigma_c(2455)$ and
$\Sigma_c(2520)$ states, with roughly equal probability. The
$\Xi_c(2980)$ was found to decay through the intermediate
$\Sigma_c(2455)\overline K$; the $\Sigma_c(2455)\overline K$ mass
distribution show an additional signal establishing $\Xi_c(3055)^+$.
The $\Sigma_c(2455)\overline K$ mass distribution shows evidence for
$\Xi_c(2980)$ as strong threshold enhancement, for $\Xi_c(3080)$ and
for a third signal at $\Xi_c(3123)$. The {\sc\small BELLE} and
{\sc\small BaBaR} parameters for the new $\Xi_c$ states are
summarized in Table~\ref{tab:BaBaR-belle}.
\begin{table}[pt]
\caption{\label{tab:BaBaR-belle}Mass and width of the $\Xi_c(2790)$
and $\Xi_c(2815)$ measured at {\sc\small CLEO} \cite{Chistov:2006zj}
and {\sc\small BaBaR} \cite{Aubert:2007dt}.}
\begin{footnotesize}
\begin{center}
\renewcommand{\arraystretch}{1.5}
\begin{tabular}{lccc}
\hline\hline
& &$M,\;\mevm$ & $\Gamma,\;\mevm$ \\
\hline
{\sc\small BELLE} & $\Xi_c(2980)^+\;$ & $2978.5\pm2.1\pm2.0\;$ & $\;43.5\pm7.5\pm7.0\;$ \\
{\sc\small BaBaR} & $\Xi_c(2980)^+$ & $2969.3\pm2.2\pm1.7$ & $27\pm8\pm2$ \\
\hline
{\sc\small BaBaR} &$\Xi_c(3055)^+$ & $3054.2\pm1.2\pm0.5$ & $17\pm6\pm11$  \\
\hline
{\sc\small BELLE} & $\Xi_c(2980)^0\;$ & $2977.1\pm8.8\pm3.5$ & $43.5$ (fixed) \\
{\sc\small BaBaR} & $\Xi_c(2980)^0$ & $2972.9\pm4.4\pm1.6$ & $31\pm7\pm8$ \\
\hline
{\sc\small BELLE} & $\Xi_c(3080)^+$ & $3076.7\pm0.9\pm0.5$ & $6.2\pm1.2\pm0.8$ \\
{\sc\small BaBaR} & $\Xi_c(3080)^+$ & $3077.0\pm0.4\pm0.2$ & $5.5\pm1.3\pm0.6$ \\
\hline
{\sc\small BELLE} & $\Xi_c(3080)^0$ & $3082.8\pm1.8\pm1.5$ & $5.2\pm3.1\pm1.8$ \\
{\sc\small BaBaR} & $\Xi_c(3080)^0$ & $3079.3\pm1.1\pm0.2$ & $5.9\pm2.3\pm1.5$ \\
\hline {\sc\small BaBaR} &$\Xi_c(3123)^+$&$3122.9\pm1.3\pm0.3$&$4.4\pm3.4\pm1.7$\\
\hline\hline
\end{tabular}
\renewcommand{\arraystretch}{1.0}
\end{center}
\end{footnotesize}\vspace{-3mm}
\end{table}

Based on their mass and width, the $\Xi_c(3080)$ state is proposed
to be a strange partner of the spin-parity $J^P=5/2^+$
$\Lambda_c(2880)^+$ resonance, while the $\Xi_c(2980)$ should have
$J^P=1/2^+$ or $3/2^+$
\cite{Cheng:2006dk,Rosner:2006jz,Garcilazo:2007eh,Ebert:2007nw}.\vspace{-2mm}
 \subsub{The $\Omega_c$ states}
\paragraph{$\Omega_c$:} The discovery of the $\Omega_c$ ($=csd$) marked
a milestone; it completed the number of stable single-charmed
baryons. The first evidence for it was reported in
\cite{Biagi:1984mu} and confirmed in several experiments. We quote
here its mass \cite{Amsler:2008zz}
\begin{eqnarray}
M_{\Omega_c}= 2697.5\pm2.6\,{\rm MeV}.
\end{eqnarray}

The $\Omega_c$ lifetime (see Table \ref{tab:lifetime}) was measured
by the {\sc\small WA89} collaboration at  CERN and, recently, by the
{\sc\small FOCUS} and {\sc\small SELEX} experiments at Fermilab. The
{\sc\small SELEX} (E781) experiment used 600 GeV/c $\Sigma^-, \pi^-$
and $p$ beams \cite{Iori:2007pw} while WA89 and Focus are
photoproduction experiments. All three experiments reconstructed
about 75 $\Omega_c^0$ in the $\Omega^-\pi^-\pi^+\pi^+$ and
$\Omega^-\pi^+$ decay modes.\vspace{-4mm}

\paragraph{$\Omega_c^*$:} Recently, an excited $\Omega_c$ state has been
suggested by the {\sc\small BaBaR} collaboration; it was introduced
as $\Omega_c^*$. It was produced inclusively in the process $e^+e^-
\to \Omega_c^* X$, where $X$ denotes the remainder of the event. The
$\Omega_c^*$ was observed in its radiative decay to the $\Omega_c$
ground state. The latter was constructed from one of the $\Omega_c$
decay sequences
\begin{equation}
\begin{split}
\Omega_c^0\to\Omega^-\pi^+, \ &\Omega^-\pi^+\pi^0, \
\Omega^-\pi^+\pi^+\pi^-, \ \Omega^-\to\Lambda K^-\\
\mathrm{or}\quad &\Omega_c^0\to\Xi^-K^-\pi^+\pi^+, \ \Xi^-\to
\Lambda\pi^-
\end{split}
\end{equation}

\begin{figure}[pt]
\bc
\includegraphics[width=0.48\textwidth]{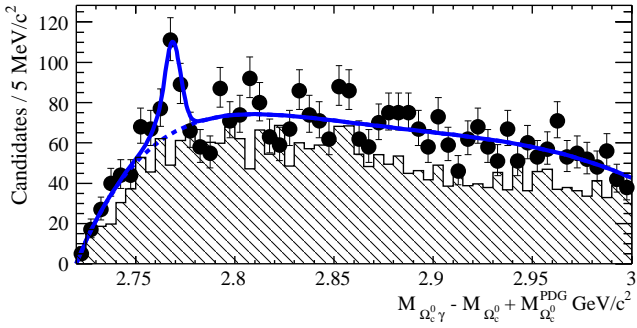}
\vspace{-3mm}\ec \caption{\label{Omega_bb} The invariant mass
distributions of $\Omega_c^{0}{}\gamma$ candidates, with
$\Omega_c^{0}$ reconstructed in various decay modes. The
$M_{\Omega_c^{0}\gamma}$ mass is corrected for the difference
between the reconstructed $\Omega_c^{0}$ mass and the nominal value
$M_{\Omega_c^{0}}^{\rm\tiny PDG}$. The shaded histograms represent
the mass distribution expected from the mass sideband of
$\Omega_c^{*0}$ \cite{Aubert:2006je}. \vspace{-3mm}}
\end{figure}

Figure~\ref{Omega_bb} shows the $\Omega_c^0\gamma$ invariant mass
after all $\Omega_c$ decay modes were added up. A significant
enhancement (with $5.2\,\sigma$) is observed above a smooth
background. It is identified with the $J^P=3/2^+$ excitation of the
$\Omega_c$ ground state. Its mass was found to be $70.8\pm 1.5$\,MeV
above the ground state. The observation was confirmed by {\sc\small
Belle} \cite{Solovieva:2008fw} reporting a mass difference to the
ground state of $(70.7 \pm 0.9 {+0.1 \atop
-0.9})$\,MeV.\vspace{-2mm}
 \subsub{Double-charm baryons}\vspace{-2mm}

The {\sc\small SELEX} Collaboration reported a statistically
significant signal in the $\Lambda_c^+ K^- \pi^+$ invariant mass
distribution at $3519\pm 1$\,MeV, a lifetime of less than 33\,fs at
90\% confidence level \cite{Mattson:2002vu}, and produced in a 600
GeV/c charged hyperon beam. Due to its decay mode, the signal is
assigned to production of a doubly charmed baryon, $\Xi_{cc}^+$. The
state was confirmed by {\sc\small SELEX} in the $\Xi_{cc}^+\to p D^+
K^-$ decay mode \cite{Ocherashvili:2004hi}. In spite of intense
searches, the state failed to be observed in the photoproduction
experiment FOCUS \cite{Ratti:2003ez} although they observe 19,500
$\Lambda_c^+$ baryons, compared to 1.650 observed at {\sc\small
SELEX}. {\sc\small BaBaR} reports  $\approx 600$\,k reconstructed
$\Lambda_c^+$ baryons but only upper limits for $\Xi_{cc}^{+}$ and
$\Xi_{cc}^{++}$ \cite{Aubert:2006qw}. Of course, {\sc\small SELEX}
starts with a hyperon beam which may be better suited to produce
double-charm baryons. But doubts remain concerning the evidence
reported by {\sc\small SELEX}.

The lack of double charm baryons at $B$-factories is surprising. In
these experiments, double charm production is abundant, leading in
particular to $e^+e^-\to J/\psi + X$ and the discovery of the
$\eta_c'$ in the missing-mass spectrum. One could thus expect
double-charm production should hadronize also into
baryon--antibaryon pairs, $\Xi_{cc} + \overline{\Xi}{}_{cc}$, or
$\Xi_{cc} + \overline{\Lambda}{}_{c}+\overline{D}$, etc. In general,
baryon production is suppressed by one order of magnitude as
compared to mesons. In $J/\psi$ decays, e.g., events with baryons in
the final state constitute about 5\% of all hadronic
decays.\vspace{-2mm}
\begin{figure}[pt]
\includegraphics[width=0.45\textwidth,height=0.53\textwidth]{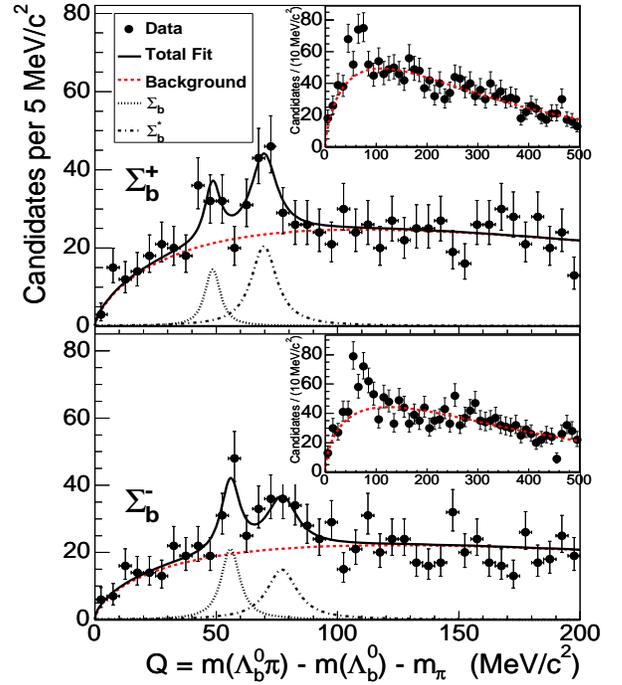}
 \caption{ The invariant mass distributions for the $\Lambda_b^0 \pi^+$
   (top) and $\Lambda_b^0 \pi^-$ (bottom) combinations at
   {\sc\small CDF} \cite{Aaltonen:2007rw}. \vspace{-3mm}}
  \label{sigb}
\end{figure}
\subsection{Beautiful baryons}\vspace{-1mm}
 \subsub{The $\Lambda_b$ states}\vspace{-2mm}
The $\Lambda_b$ was discovered early at the CERN ISR
\cite{Bari:1991in,Bari:1991ty} and later reported by several
collaborations. We give here only the PDG values for its mass
\cite{Amsler:2008zz}\vspace{-1mm}
\begin{eqnarray}
M_{\Lambda_b}= 5620.2\pm1.6\,{\rm MeV};\vspace{-1mm}
\end{eqnarray}
its lifetime is given in Table \ref{tab:lifetime}.\vspace{-2mm}
 \subsub{The $\Sigma_b$ states}
\paragraph{$\Sigma_b$ and $\Sigma_b^*$:} The $\Sigma_b$
baryon with $J^P=1/2^+$ and a low-mass excitation identified as
$J^P=3/2^+$ $\Sigma_b^{*}$ were discovered recently at Fermilab
\cite{Aaltonen:2007rw} by the {\sc\small CDF} Collaboration in the
$\Lambda_b^0\pi^+$ and $\Lambda_b^0\pi^-$ final states (see
Fig.~\ref{sigb}).

The signal region exhibits a clear excess of events even though the
statistics is not sufficient to determine mass and widths of the
expected $\Sigma_b$ and $\Sigma_b^{*}$. Therefore the
$M(\Sigma_b^{*+})-M(\Sigma_b^+)$ and
$M(\Sigma_b^{*-})-M(\Sigma_b^-)$ mass differences were assumed to
the same and the widths of the Breit--Wigner resonances were fixed
to predictions based on the Heavy Quark
Symmetry~\cite{Korner:1994nh}.
\begin{table}[pt]\vspace{-4mm}
\caption{Results of the $\Sigma_b^{(*)}$ fit.\vspace{-3mm}}
\label{tab_sigb}
\begin{footnotesize}
\begin{center}
\renewcommand{\arraystretch}{1.5}
\begin{tabular}{l}
\hline\hline
$m(\Sigma_b^+)-m(\Lambda_b^0)=188.1^{+2.0}_{-2.2}{^{+0.2}_{-0.3}}\,\mevm$ \\
$m(\Sigma_b^-)-m(\Lambda_b^0)=195.5\pm1.0\pm0.2\,\mevm$ \\
\hline
$m(\Sigma_b^{*})-m(\Sigma_b)=21.2^{+2.0}_{-1.9}{^{+0.4}_{-0.3}}\,\mevm$ \\
\hline\hline
\end{tabular}
\renewcommand{\arraystretch}{1.0}
\end{center}\vspace{-3mm}
\end{footnotesize}\end{table}
Both the shape and the normalization of the background were
determined from Monte-Carlo simulations.  The results of the fit are
given in Table~\ref{tab_sigb}. The significance of the four-peak
structure relative to the background-only hypothesis is
$5.2\,\sigma$ (for 7 degrees of freedom). The significance of every
individual peak is about $3\,\sigma$.\vspace{-2mm}
 \subsub{The $\Xi_b$ states}
\paragraph{$\Xi_b$:} A further baryon with beauty, the $\Xi_b$, contains a $b$, $s$,
and a $d$ quark and thus a negatively charged quark from each
family. It was discovered at Fermilab
\cite{Abazov:2007ub,Aaltonen:2007un}. Its history will be outlined
shortly.

\begin{figure}[pt]
\begin{center}
\includegraphics[width=0.38\textwidth]{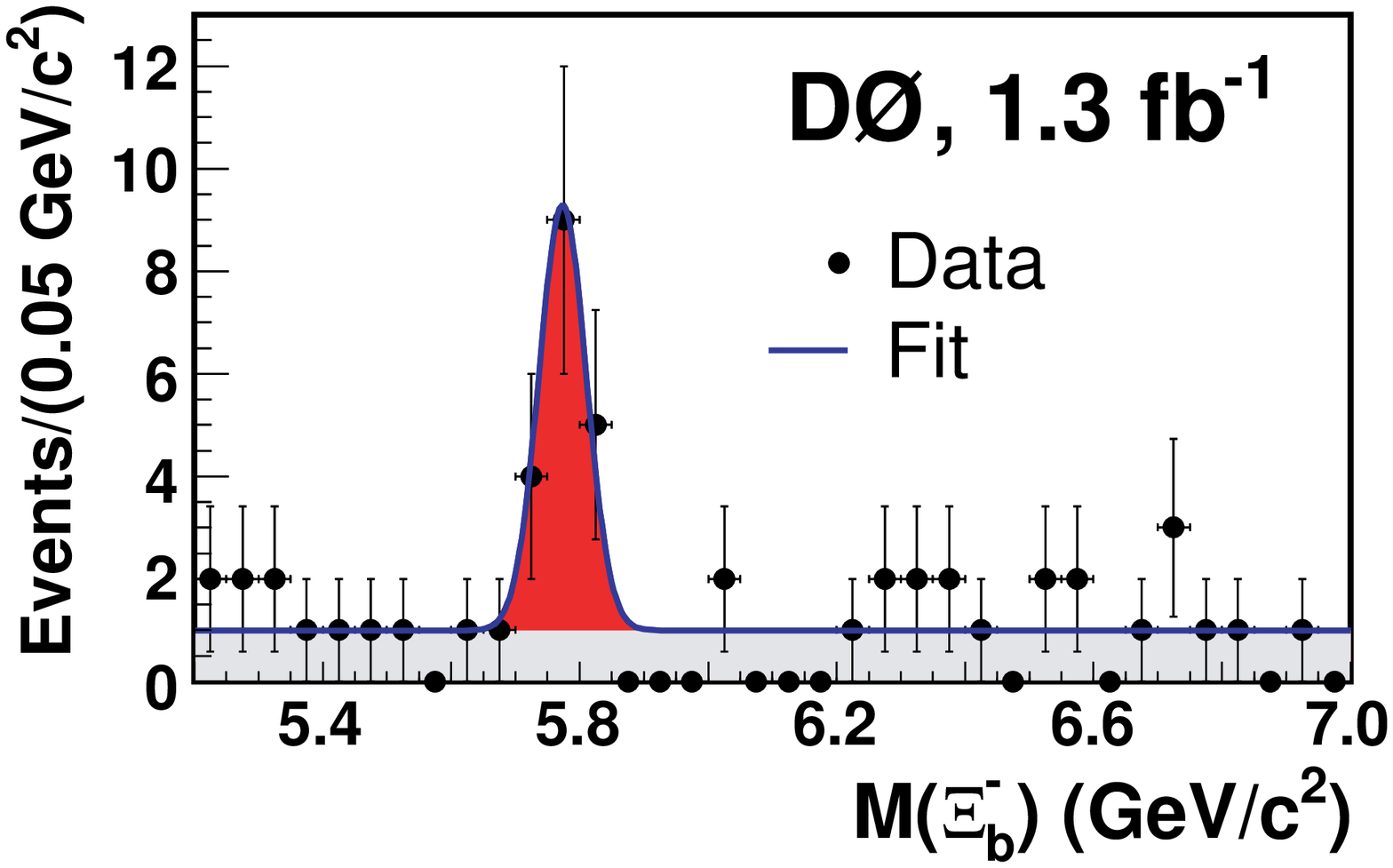}\\
\includegraphics[width=0.385\textwidth]{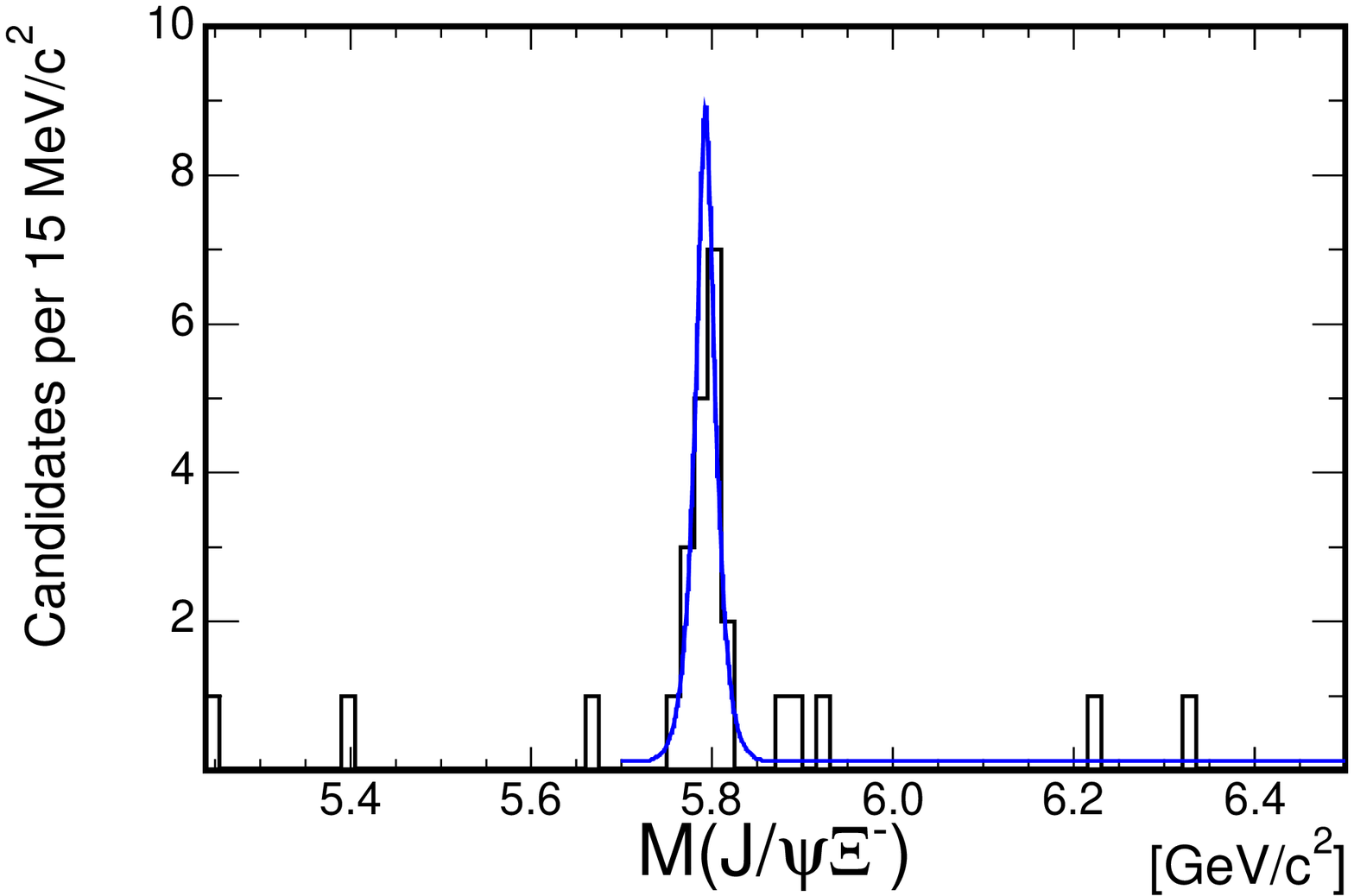}
\end{center}
 \caption{ The invariant mass distributions of the $J/\psi\,\Xi^-$
   combinations at {\sc\small D\O} (top)~\cite{Abazov:2007ub}
   and {\sc\small CDF} (bottom)~\cite{Aaltonen:2007un}. \vspace{-2mm}}
  \label{mxib}
\end{figure}

\begin{figure}[pt]
\scalebox{.9}{ \unitlength=1mm
\begin{picture}(70,96)(0,0)
\put(0,46){\mbox{
\includegraphics[width=70mm,height=50mm]{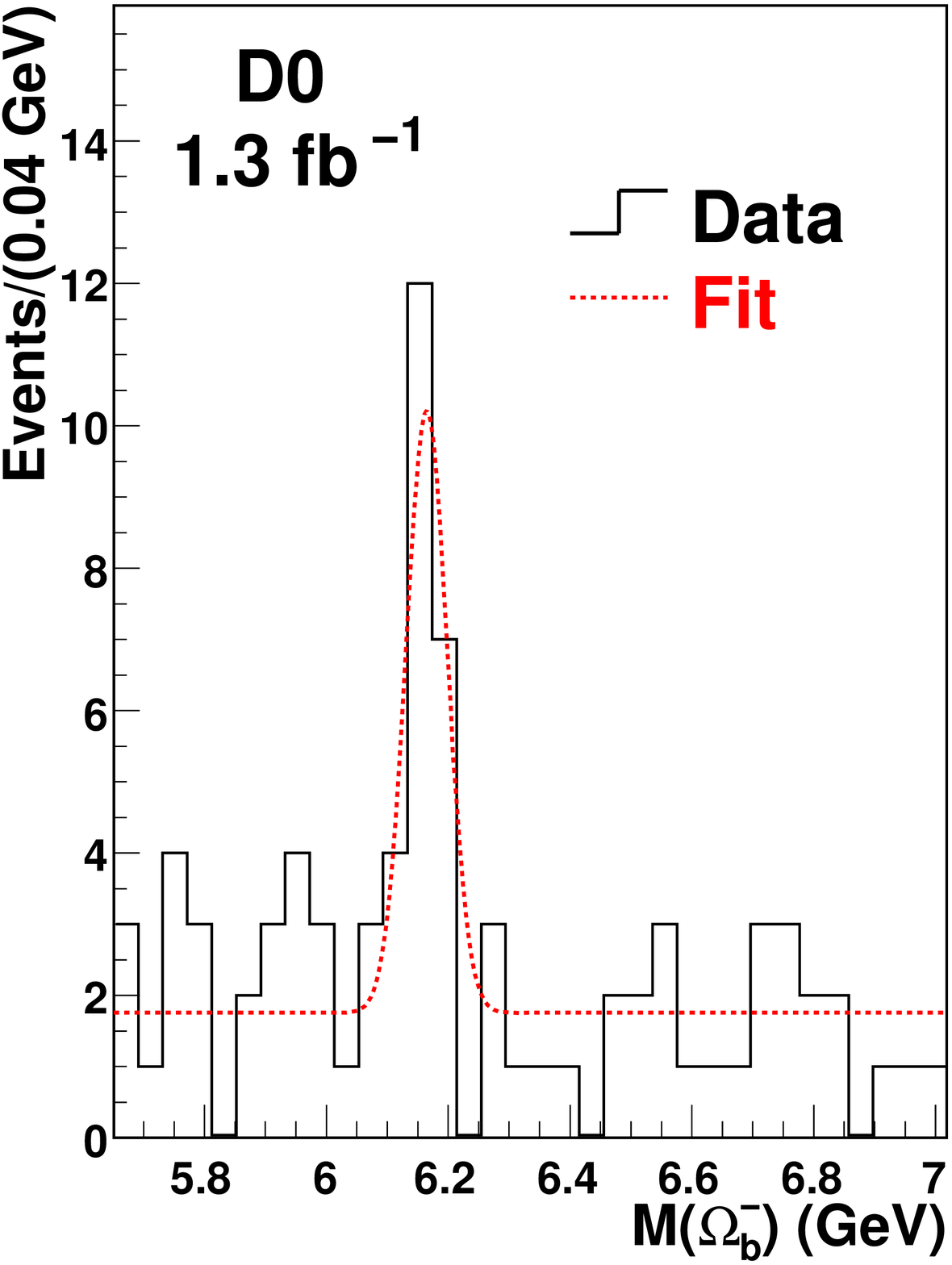}}}
\put(5.0,0){\mbox{\includegraphics[width=56.4mm,height=45mm]{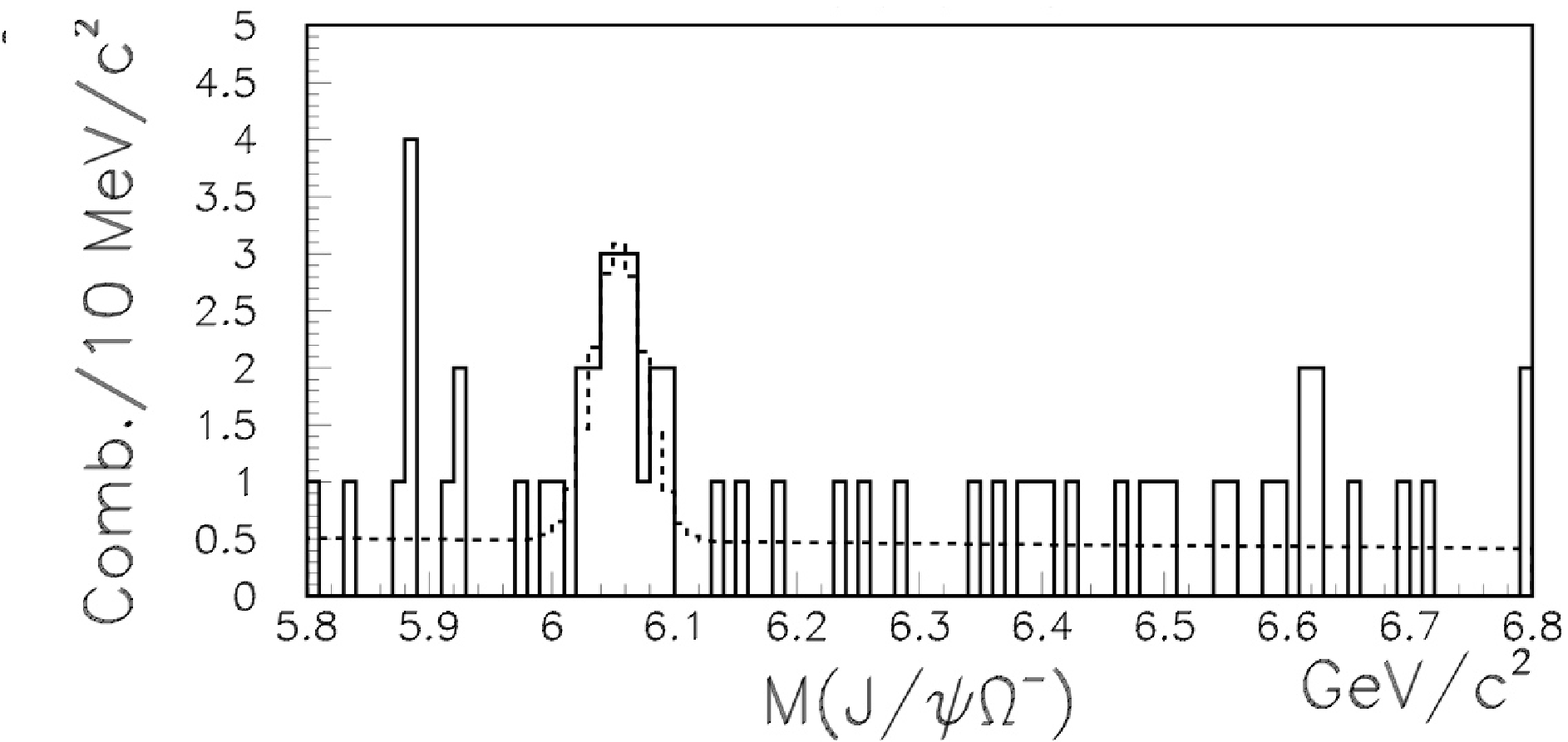}}}
\end{picture}}
\caption{\label{fig:OmegabMass} Top: {\sc\small D\O} data
\cite{Abazov:2008qm}. The $M(\Omega_b^-)$ distribution of the
$\Omega_b^-$ candidates after all selection criteria. The dotted
curve is an unbinned likelihood fit to the model of a constant
background plus a Gaussian signal. Bottom: {\sc\small CDF} data
\cite{Aaltonen:2009ny}. $J/\Psi\Omega^-$ mass distribution.}
\end{figure}

Indirect evidence for the $\Xi_b^-$ baryon based on an excess of
same-sign $\Xi^- \ell^-$ events in jets was observed from
experiments at the CERN LEP $e^+e^-$ collider but no exclusively
measured candidate was reported. The first direct observation of the
strange $b$ baryon $\Xi_b^-\thinspace (\overline{\Xi}{}_b^+)$ was
achieved at Fermilab \cite{Abazov:2007ub} by the {\sc\small D\O}
collaboration by reconstruction  of the decay sequence
$\Xi_b^-\thinspace \to J/\psi\thinspace\Xi^-$, with
$J/\psi\to\mu^+\mu^-$, and $\Xi^-\to\Lambda\pi^-\to p\pi^-\pi^-$
(Fig.~\ref{mxib}, top). The {\sc\small CDF} collaboration reported a
more precise mass value. Their $J/\psi\thinspace\Xi^-$ invariant
mass distribution exhibits a significant peak \cite{Aaltonen:2007un}
at a mass of
\begin{eqnarray}
M_{\Xi_b}= 5792.9\pm2.5\pm1.7\,{\rm MeV}
\end{eqnarray}
which is presented in Fig.~\ref{mxib}, bottom. The mass and number
of $\Xi_b^-$ events observed by \cite{Abazov:2007ub,Aaltonen:2007un}
are given in Table~\ref{tab_xib}, the lifetime in Table
\ref{tab:lifetime}. The results of {\sc\small D\O} and {\sc\small
CDF} are consistent.\vspace{-2mm}
 \subsub{The $\Omega_b$}\vspace{-2mm}
Figure~\ref{fig:OmegabMass} (top) shows evidence for the
$\Omega_b^-$ baryon reported by the {\sc\small D\O} collaboration.
It was reconstructed from the decay sequence $\Omega_b^-\to
J/\psi\thinspace\Omega^-$, with $J/\psi\to\mu^+\mu^-$,
$\Omega^-\to\Lambda K^-$ and $\Lambda\to p\pi^-$. The signal has a
statistical significance exceeding $5\sigma$. Its mass was reported
to be \cite{Abazov:2008qm}
\begin{equation}
M_{\Omega_b}= 6.165\pm 0.010\pm 0.013\,{\rm GeV}.
\end{equation}
It  is unexpectedly high, see section \ref{sub:se-phegs}. Recently,
the $\Omega_b$ has been seen (Fig.~\ref{fig:OmegabMass}, bottom) by
the  {\sc\small CDF} collaboration \cite{Aaltonen:2009ny}; their
result is
\begin{eqnarray}
M_{\Omega_b}= 6.054\pm 0.007\pm 0.013\,{\rm GeV},
\end{eqnarray}
closer to most theoretical predictions. For the $\Omega_b^-$
lifetime, see Table \ref{tab:lifetime}.

\begin{table}[pt]
\caption{The parameters of the $\Xi_b^-$ measured by {\sc\small
D\O} and {\sc\small CDF}.}
\label{tab_xib}
\renewcommand{\arraystretch}{1.4}
\begin{tabular}{llcc}\hline\hline
&\hspace{2mm} Yield & Mass, $\mevm$ & Significance \\
\hline
{\sc\small D\O} & $15.2\pm4.4^{+1.9}_{-0.4}\;\;$ & $5774\pm11\pm15$ & 5.5$\,\sigma$ \\
{\sc\small CDF}$\;\;$ & $17.5\pm4.3$ & $5792.9\pm2.5\pm1.7$ & 7.7$\,\sigma$ \\
\hline\hline\end{tabular}
\end{table}
 \subsection{A future at LHC}
The {\sc\small CDF} and {\sc\small D{\O}} experiments have
demonstrated the potential of hadron machines for the discovery of
new baryon resonances. At LHC, double charmed baryons should be
produced abundantly, a total number of $10^9$ is estimated by
\cite{Berezhnoi:1998aa},
and one may even dream of $(ccc)$
baryons. Baryons (and mesons) with $b$ quarks and their excitations
will also be produced; such events should not be thrown away at the
trigger level.

\markboth{\sl Baryon spectroscopy}{\sl Light--quark baryon resonances}
\section{\label{Light-quark baryon resonances}Light-quark baryon resonances}
In this section we give a survey of data which have been reported in
recent years and give an outline of partial wave analysis methods
used to extract the physical content from the data. The light-baryon
excitation spectrum is discussed.
\subsection{\label{Pion- (Kaon-)nucleon elastic and charge exchange scattering}
Pion- (kaon-) nucleon elastic and charge exchange scattering}
\begin{figure}[pb]
\bc
\includegraphics*[width=.48\textwidth]{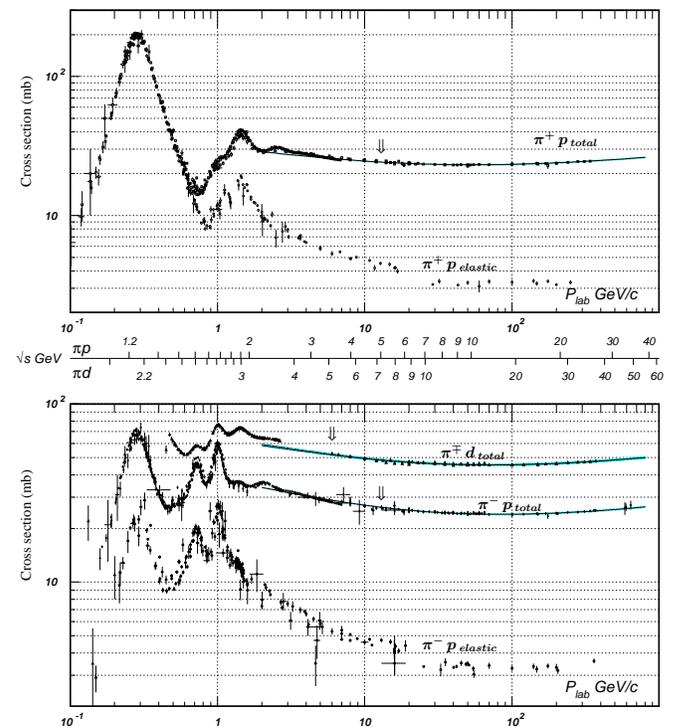}
\ec
\caption{\label{elasticX}The total and elastic cross sections for
$\pi^{\pm}$ scattering off protons from
\tt{http://pdg.lbl.gov/current/xsect/}, courtesy of the COMPAS
group, IHEP, Protvino. }
\end{figure}

\begin{figure}[pt]
\vspace{-4mm}\bc
\includegraphics*[width=.48\textwidth]{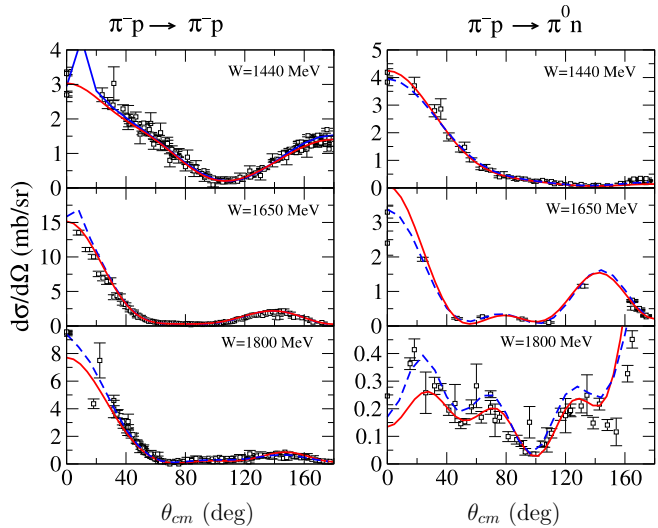}
\vspace{-4mm}
\ec
\caption{\label{differential} Differential cross section for several
different center of mass energies. Solid and dashed curves
correspond to different SAID \text{http://gwdac.phys.gwu.edu/}
solutions.\vspace{-2mm}}
\end{figure}

 \subsub{Cross sections}
 \setcounter{paragraph}{0}
The dynamical degrees of freedom of three quarks bound in a baryon
lead to a very rich excitation spectrum. It is obviously impossible
to observe them all as individual resonances but a sufficiently
large number of states should be known to identify the proper
degrees of freedom and their effective interactions. First insight
into the experimental difficulties can be gained by inspecting, in
Fig.~\ref{elasticX}, the total cross section for elastic $\pi^{\pm}$
scattering off protons. The $\pi^{+}p$ cross section is dominated by
the well-known $\Delta_{3/2^+}(1232)$ resonance. A faint structure
appears at 1.7\,GeV, slightly better visible in the elastic cross
section, a second bump can be identified at 1.9 to 2\,GeV in mass,
and a small enhancement is seen at 2.4\,GeV. Above this mass, the
spectrum becomes structureless. The total cross section for
$\pi^{-}p$ scattering exhibits three distinctive peaks at the
$\Delta_{3/2^+}(1232)$, at 1.5\,GeV and at 1.7\,GeV; a fourth
enhancement at 1.9\,GeV is faint, a further peak at 2.2\,GeV leads
into the continuum.
The gradual disappearance of the resonant structures suggests that
at least part of the problem is due to the increasingly smaller
elastic width of resonances when their masses increase: more and
more inelastic channels open, and the resonances decouple from the
elastic scattering amplitude. A second problem are overlapping
resonances and their large widths. The peaks in Fig.~\ref{elasticX}
may contain several resonances. Hence a partial wave decomposition
is required to determine the amplitudes which contribute to a
particular energy bin. Very high statistics and polarization data
are required to disentangle the different partial waves. At present,
it is an open issue up to which mass baryon resonances can be
identified. A second and even more exciting  question is whether QCD
really supports the full spectrum of three-quark models. In the
literature, diquark models are very popular; the experimental
resonance spectrum has features which are easily understood assuming
quasi-stable diquark configurations within a baryon; however, there
are also resonances - albeit with one or two star classification -
which require three quarks to participate in the dynamics. Less
familiar in this context are two dynamical arguments: an extended
object has three axes but the object rotates only around the two
axes having minimal/maximal moments of inertia. And, surprisingly, a
series of coupled resonators with approximately equal resonance
frequencies resonate coherently after some swinging-in period even
if the oscillators start with random phases and amplitudes. Hence
there may be restrictions concerning the observable spectrum of
baryon resonances.\vspace{-2mm}
 \subsub{Angular distributions}\vspace{-2mm}
 \setcounter{paragraph}{0}
Most of the peaks in Fig.~\ref{elasticX} house several resonances
with similar masses but different angular momenta. The differential
cross sections $d\sigma/d\Omega$ in Fig.~\ref{differential} allow
for a first insight into the dynamics of the scattering process.

\begin{figure}[pt]
\begin{center}
\includegraphics[width=0.15\textwidth]{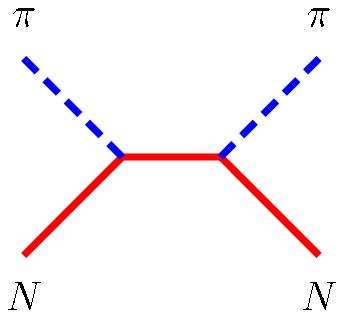}
\includegraphics[width=0.15\textwidth]{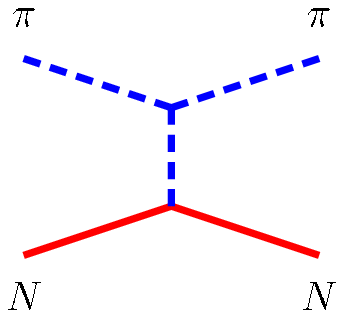}
\includegraphics[width=0.15\textwidth]{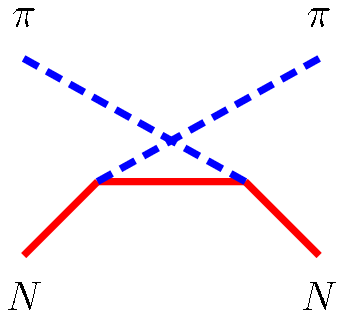}\\
(a)\hspace{23mm}(b)\hspace{23mm}(c)
\end{center}
\caption{\label{diagrams} Pion--nucleon scattering: a) $s$-channel
exchange; b) $t$-channel exchange; c) $u$-channel exchange.}
\end{figure}

The first striking effect seen from the data is the preference for
forward angles ($\theta\leq 40^{\circ}$) of the scattered pion. The
preference for forward pion scattering at low energies reflects the
large role   of background processes like $t$-channel exchange with
a $\rho$ meson (or a $\rho$ Regge trajectory) transmitting
four-momentum from pion to proton. Formation of resonances produces
a symmetry between forward and backward scattering, at least at the
amplitude level; interference between amplitudes can of course lead
to forward-backward asymmetries. Here, it is useful to compare the
Clebsch--Gordan coefficients for different reactions:
\bc
\renewcommand{\arraystretch}{1.4}
\begin{footnotesize}\begin{tabular}{ccc}\hline\hline
                  &\quad$\pi^-p\to\pi^-p$\quad&\quad$\pi^-p\to\pi^0n$\\
\hline$s,u$-channel $N$   & 2/3 &1/3$\sqrt{2}$\\
$s,u$-channel $\Delta$\quad&1/3&1/3$\sqrt{2}$\\
$t$-channel $\rho$ &  1&1\\\hline\hline
\end{tabular}
\end{footnotesize}
\renewcommand{\arraystretch}{1.0}
\ec
The forward cross section for elastic and charge exchange (CEX) have
nearly the same size and the interpretation of the forward peak is
supported. The backward peak at 1440\,MeV is stronger in elastic
than in charge exchange scattering suggesting strong isospin 1/2
contribution in the $s$-channel (via $N(1440)P_{11}$ formation)
and/or $u$-channel nucleon exchange. At $W=1800$\,MeV, there is no
CEX forward peak; a complex distribution evolves indicating
contributions from high-spin $s$-channel resonances. The elastic
cross section continues to exhibit a strong forward peak due to the
exchange of isoscalar mesons, e.g. of the Pomeron. The three
processes $s$-, $t$-, and $u$-channel exchange are visualized in
Fig.~\ref{diagrams}. The data were obtained through the Scattering
Analysis Interactive Dial-in (SAID) online applications
\text{http://gwdac.phys.gwu.edu/}.
\begin{figure}[pb]
\bc
\includegraphics[width=0.3\textwidth]{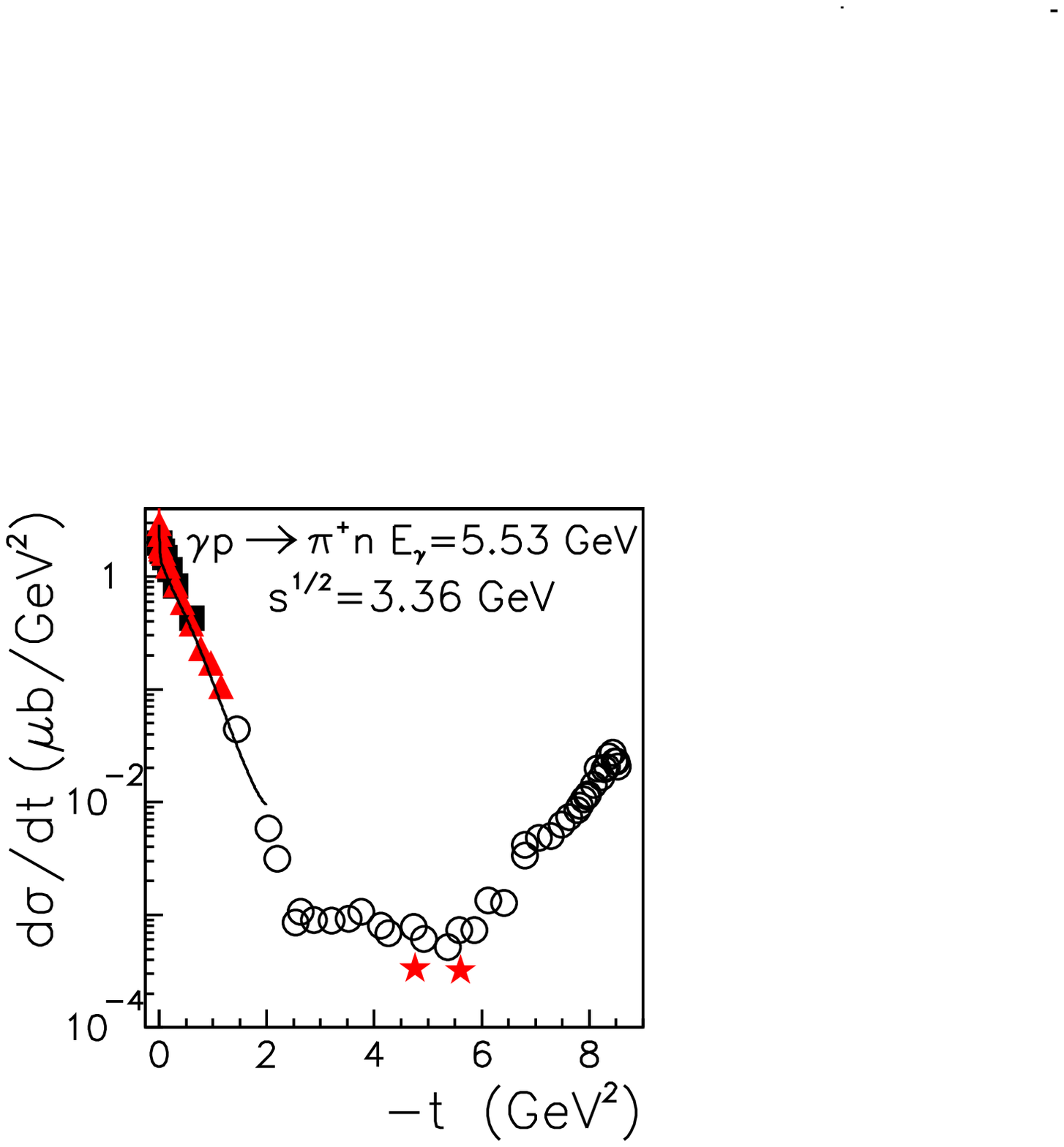}
\vspace{-4mm}\ec \caption{\label{uandt}The $\gamma p\to n\pi^+$
differential cross section as a function of $-t$ for
$E_{\gamma}=5.53$\,GeV \cite{Sibirtsev:2007wk}. The data are from
\cite{Anderson:1969jw,Anderson:1976ph,Zhu:2004dy}. }
\end{figure}
A beautiful example illustrating the effect of $t$- and $u$-channels
exchanges is shown in Fig.~\ref{uandt}. For forward pions, the
four-momentum transfer $t=-q^2$ to the proton is small; a
diffractive-like decrease of the cross section as a function of $t$
is observed. The peak is due to meson exchange in the $t$-channel,
mostly of $\rho$ and $\omega$; in analyses, the exchange is
reggeized to include higher mass $\rho$ and $\omega$ excitations.
The slope corresponds to the $\rho$/$\omega$ mass. For very large
(negative) $t=-2k^2(1-\cos\theta)$, $u=-2k^2(1+\cos\theta)$ becomes
a small number. The slope is smaller and corresponds to the nucleon
mass.

The differential cross sections $\sigma$ are related to the
transversity scattering amplitudes
\begin{eqnarray} \label{e:1}
\sigma & = & |f^+|^2 + |f^-|^2
\end{eqnarray}
which can be decomposed into the nucleon spin-flip amplitude $g$ and
the non-flip amplitude $h$, $f^+=g+ih$, $f^-=g-ih$. The latter
amplitudes can be expanded into the partial waves
\begin{subequations}\label{gamp}
\begin{eqnarray}
g(k,\theta) &=& \frac{1}{k}\sum_l \left[ (l+1) a_{l^{+}} + l
a_{l^{-}} \right] P_l(\cos\theta)\\\label{hamp} h(k,\theta) &=&
\frac{1}{k}\sum_l \left[  a_{l^{+}} -  a_{l^{-}} \right]
\sin\theta\,P_l'(\cos\theta)
\end{eqnarray}
\end{subequations}
where $k$ is the momentum and $\theta$ the scattering angle in
the center-of-mass system. The expansion into Legendre polynomials
extends over all angular momenta $l$, the $\pm$ sign indicates that
the total angular momentum is $J=l\pm 1/2$. The dimensionless
partial wave amplitudes
$a_{l^{\pm}}=[\eta_{l^{\pm}}\,\exp(2i\delta_{l^{\pm}})]/2i$ are
related to the inelasticities $\eta_{l^{\pm}}$ and the phase shifts
$\delta_{l^{\pm}}$.

\begin{figure}[pt]
\includegraphics[width=0.48\textwidth]{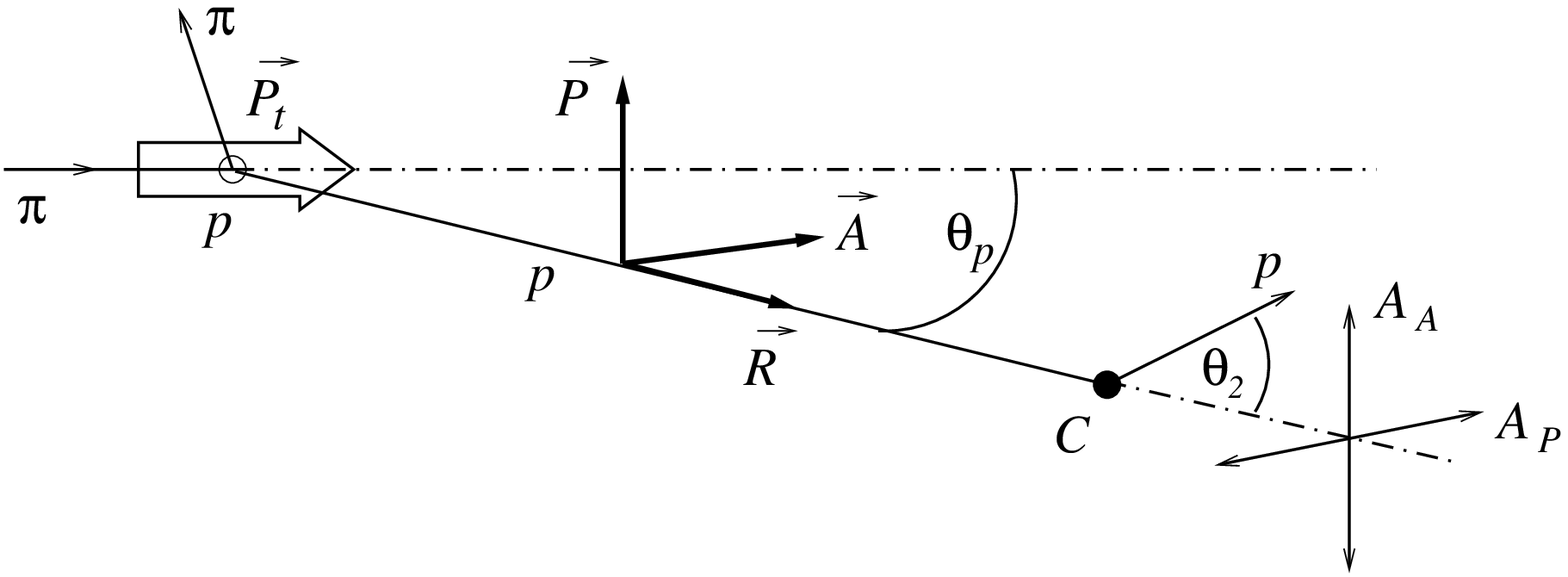}
\caption{\label{pol_def}Definition of polarization variables
\cite{Alekseev:2005zr}.\vspace{-4mm}}
\end{figure}

It is obvious that the two amplitudes cannot be deduced from the
differential cross sections alone. Polarization observables need to
be measured. We discuss the polarization $P$ and the two spin
rotation parameters $A$ and $R$.\vspace{-2mm}
 \subsub{Polarization variables}
 \vskip -2mm plus 1mm minus 1mm \setcounter{paragraph}{0} \vskip
  2mm plus 1mm minus 1mm \vspace{-2mm}
The polarization variable $P$ can be measured using a polarized
target. If the proton polarization vector is parallel to the
decay-plane normal, there is, at any laboratory scattering angle
$\theta$, a left-right asymmetry of the number of scattered pions
which defines $P$. The polarization of the scattered proton does not
need to be known. Thus large data sets exist where $P$ was
determined, from Rutherford \cite{Cox:1968yg}, \cite{Martin:1974rt},
\cite{Brown:1978hw} and from CERN
\cite{Albrow:1971xa,Albrow:1972ky}, among other places. $P$
constrains the amplitudes but does not yet yield a unique solution:
\begin{eqnarray}
P\,\sigma_{\rm tot} & = & |f^+|^2 - |f^-|^2  \label{e:2}
\end{eqnarray}
Further variables need to be measured. Figure \ref{pol_def} shows
the definitions of polarization variables which can be deduced in
$\pi N$ elastic scattering off longitudinally polarized protons. The
proton is deflected by an angle $\theta_{\rm p}$ in the laboratory
system. The proton polarization vector now has a component $P$ which
is perpendicular to the scattering plane, a component $R$ along its
direction of flight, and a component $A$ along the third orthogonal
direction. The components $A$ and $P$ can be measured by scattering
the recoil proton off a Carbon foil as indicated in
Fig.~\ref{pol_def}. The analyzing power of the $\pi$ Carbon
scattering process leads to a left-right asymmetry of the proton
count rate $A_P$ in the scattering plane; analogously, the $A_A$
parameter can be determined by measuring the up-down asymmetry of
proton count rate. The relation between $R,A$ and the scattering
amplitudes are given by
\begin{equation}
(R+iA)\,\sigma_{\rm tot} = f^+ f^-
\exp[-i(\theta_{cm}-\theta_{p})].
\label{e:3}
\end{equation}
The polarization parameters obey the relation
\begin{equation}
P^2+A^2+R^2=1. \label{e:4}
\end{equation}

As can been seen from eqs.~(\ref{e:1}) and (\ref{e:2}), a
measurement of the differential cross-section and of the
polarization $P$ are not sufficient to reconstruct the complex
amplitudes $f^+$ and $f^-$ but only their absolute values. Recoil
polarization data require a secondary interaction of the scattered
nucleon. Such experiments have been performed at Gatchina
\cite{Alekseev:1989cy,Alekseev:1995nw,Alekseev:1996gs,Alekseev:2000nk,Alekseev:2005zr},
at Los Alamos \cite{Mokhtari:1985cb,Mokhtari:1987iy,Seftor:1989er}
and a few other laboratories but only over a limited energy range.
An unbiased energy-independent partial wave analysis is therefore
not possible. Constraints from dispersion relations are necessary to
extract meaningful partial wave amplitudes. For baryon masses and
widths, the PDG refers mostly to five analyses which we call the
reference analyses. Other results are mostly not used to calculate
averages.

\begin{figure}[pt]
\begin{center}
\includegraphics*[width=.48\textwidth]{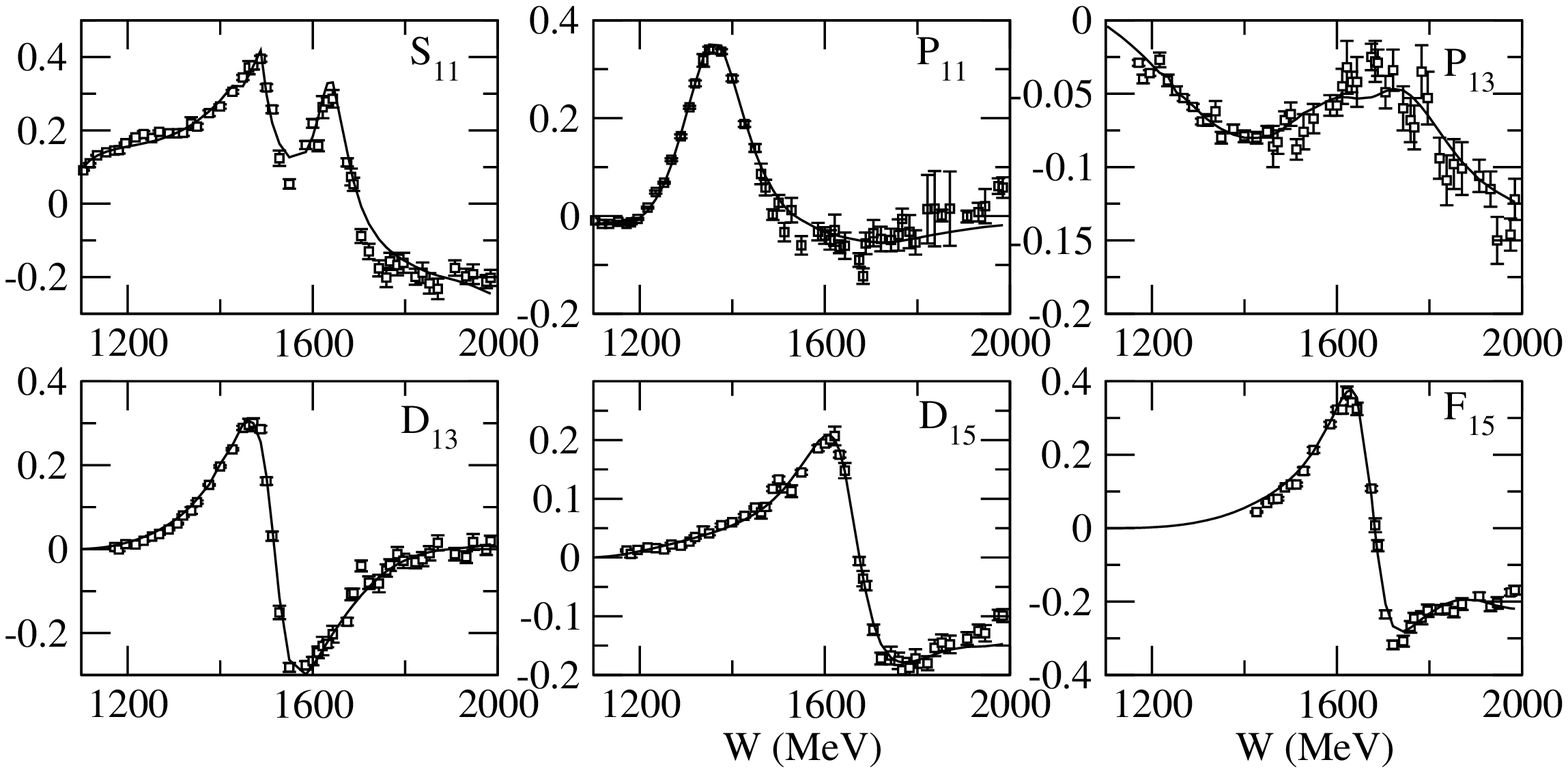}
\vspace{6mm}
\includegraphics*[width=.48\textwidth]{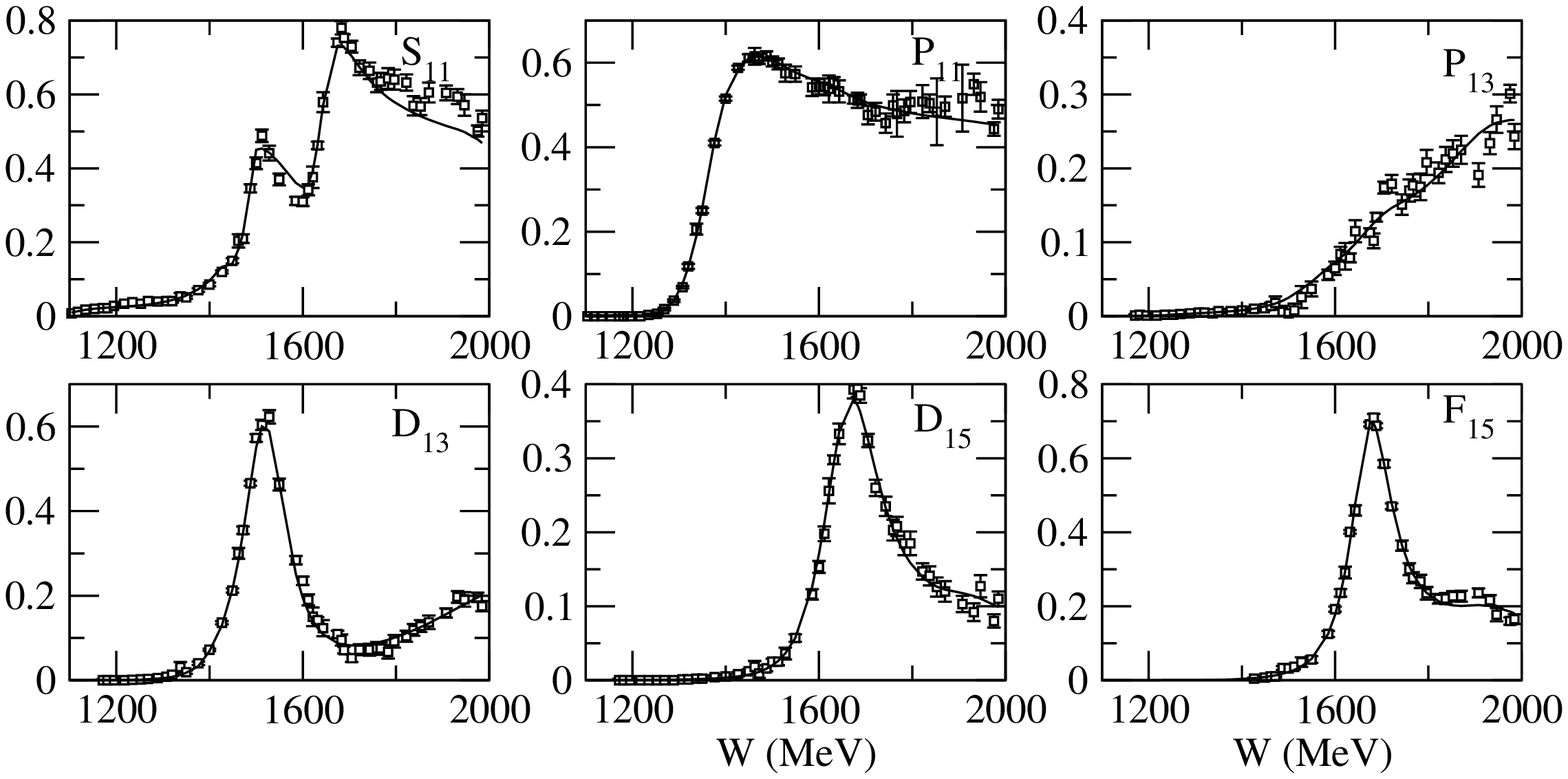}
\vspace{-7mm} \caption{\label{amplitude}Fit to the $I=\frac{1}{2}$
$Re(T_{\pi N,\pi N})$ and  $Im(T_{\pi N,\pi N})$ of SAID
\text{http://gwdac.phys.gwu.edu/}.\vspace{-4mm}}
\end{center}
\end{figure}

 The analyses of the Karlsruhe--Helsinki (KH) and  Carnegie--Mellon
(CM) groups were published in 1979 and 1980, respectively; still
today, they contain the largest body of our knowledge on $N^*$ and
$\Delta^*$ as listed by the PDG. The Kent group made a systematic
study of the inelastic reactions $\pi N\to N\pi\pi$. Hendry
presented data taken on elastic $\pi N$ scattering at 14 momenta in
the range from 1.6\,GeV to 10\,GeV and extracted resonance
contributions. The Virginia Tech Partial-Wave Analysis Facility
(SAID) (which moved to the George Washington University ten years
ago) included more and more data on $\pi N$ scattering, in
particular from Gatchina, Los Alamos, PSI, and TRIUMF, and publishes
regularly updated solutions. In a first step, energy-independent
partial wave amplitudes are constructed, and then energy dependent
partial-wave fits are performed using  a coupled-channel
Chew--Mandelstam K-matrix. The results may not yet satisfy all of
the requirements imposed by analyticity and crossing symmetry. These
requirements are then addressed at fixed four-momentum transfer $t$
by a complete set of fixed-$t$ dispersion relations, which are
handled iteratively with the data fitting. Figure~\ref{amplitude}
shows the reconstructed amplitudes for some partial
waves.\vspace{-2mm}
 \subsub{$K$-nucleon elastic scattering}\vspace{-2mm}
  \vskip 2mm plus 1mm minus 1mm \vspace{-2mm}
\setcounter{paragraph}{0}
Kaon--nucleon scattering remains at a standstill since 1980; a
survey of achievements up to 1980 was presented by
\cite{Gopal:1980ur}. For this reason, we do not elaborate on hyperon
spectroscopy in this review. We will just mention a few recent
results from a low-momentum kaon beam at BNL in which differential
and total cross sections and the induced hyperon polarization have
been measured.\vspace{-1mm}
\boldmath\subsection{\label{sec:inelas}Inelastic pion and kaon
nucleon scattering and other reactions}\unboldmath
Inelastic reactions like $\pi^- p\to n\pi^+\pi^-$ and $\pi^- p\to
p\pi^0\pi^-$ and similar kaon induced reactions require large
solid-angle coverage of the detector. The Large Aperture
Superconducting Solenoid (LASS) spectrometer at SLAC was the last
experiment having an intense 11 GeV/c kaon beam at its disposal. The
main results are reviewed in \cite{Aston:1990ys}. The experiment had
a very significant impact on the spectroscopy of mesons with open or
hidden strangeness. At that time the focus of the community was on
glueballs and hybrids, and the LASS data were important as reference
guide for quarkonium states. The data contained information on
strange baryons as well \cite{Wright:1995em}. Lack of interest and
shortage of manpower prevented an analysis of this unique data set.
Only evidence for one baryon resonance was reported, an $\Omega^*$
at $2474\pm12$\,MeV mass and $72\pm33$ MeV width
\cite{Aston:1988yn}, in its $\Omega\pi^+\pi^-$ decay.

The absence of appropriate beams and detectors gave a long
scientific lifetime to results obtained by the use of bubble
chambers in the sixties and seventieth. The most important results
were reviewed by \cite{Manley:1984jz} who fitted data and provided
amplitudes for the most important isobars. At low energies, data
were recorded by the {\sc\small OMICRON} collaboration at the CERN
synchro-cyclotron \cite{Kernel:1988sua,Kernel:1989yf,Kernel:1990qx}
and TRIMF and Los Alamos
\cite{Sevior:1990yp,Lowe:1991gd,Pocanic:1993mp}.\vspace{-2mm}

\begin{figure}[pb]
\begin{tabular}{c}
\hspace{-2mm}\includegraphics[width=0.4\textwidth]{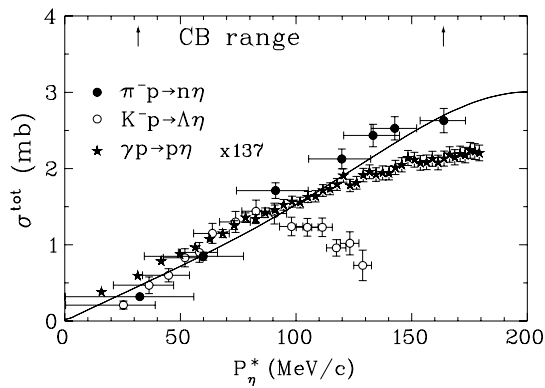}
\vspace{-2mm} \end{tabular} \caption{\label{cball}Total cross
section for $\pi^-p\to n\eta$, $K^-p\to \Lambda\eta$ and $\gamma
p\to p\eta$ \cite{Prakhov:2005qb}. The later cross section is scaled
by a factor 137.}
\end{figure}

 \vskip 2mm plus 1mm minus 1mm
 \subsub{Experiments at BNL}\vspace{-2mm}
 \setcounter{paragraph}{0}
 \vskip 2mm plus 1mm minus 1mm
The Crystal Ball detector has an animated history. It started
operation in 1978 at SPEAR with studies of radiative transitions
between charmonium states \cite{Gaiser:1985ix}. In 1982 it was moved
to DESY for spectroscopy of the $\Upsilon$ family and two-photon
physics \cite{Bienlein:1981rn}. In the late 90's it was transferred
to BNL where it was exposed to $\pi^-$ and $K^-$ beams, and is
presently installed at MAMI for photoproduction experiments (see
section \ref{Photoproduction experiments, a survey}). The ball
consists of 672 NaI detectors covering $\approx94$\% of $4\pi$. The
main results from BNL will be summarized in this section.
\vspace{-3mm}
\paragraph{$\pi^- p\to n\pi^0$ and $n\eta$:}
The Crystal Ball collaboration measured the reaction $\pi^- p\to
n\eta$ from threshold to 747\,MeV/c pion momentum
\cite{Kozlenko:2003hu,Prakhov:2005qb} (see Fig.~\ref{cball}).
Angular distributions with nearly full angular coverage were
reported for seven $\pi^-$ momenta. The total cross section
d$\sigma_{tot}$ was obtained by integration of d$\sigma$/d$\Omega$.
The rapid increase of the cross section and the rather flat angular
distributions indicate that $N_{1/2^-}(1535)$ is formed as
intermediate state. A small quadratic term  reveals contributions
from the $N\eta$ $D$-wave due to $N_{3/2^-}(1520)$. The effect of
the $\eta$ production-threshold can be seen in pion charge exchange
$\pi^-p\to n\pi^0$ \cite{Starostin:2005pd} in the form of a small
cusp. For the latter reaction, the Crystal Ball collaboration
measured precise differential cross section in the momentum interval
$p_{\pi} = 649-752$\,MeV/c. The cusp is rather weak and not as
dramatic as in pion photoproduction. The $\Delta$ region was studied
with full solid angle coverage using eight different momenta
\cite{Sadler:2004yq}.\vspace{-2mm}
\paragraph{$K^- p\to \Lambda\pi^0$, $ \Sigma^0\pi^0$, and $
\Lambda\eta$:} The reaction $K^- p \to \Lambda \pi^0$ was studied in
the mass range from 1565 to 1600\,MeV \cite{Olmsted:2003is}.
Differential cross sections and induced $\Lambda$ polarization were
reported for three $K^-$ momenta. The data were shown to be
incompatible with the claimed existence of $\Sigma_{3/2^-}(1580)$, a
one-star candidate with properties not fitting into expectations
based on SU(3)$_{\rm f}$ symmetry. An interpretation of the hyperon
spectrum including this state is proposed by \cite{Melde:2008yr}.

Differential distributions and hyperon recoil polarization were also
reported for the reaction $K^- p \to \Sigma^0 \pi^0$ at eight beam
momenta between 514 and 750 MeV/c. The (forthcoming) partial wave
analysis could have a significant impact on low-mass $\Lambda$
states \cite{Manweiler:2008zz}.

Particularly interesting is the reaction $K^- p\to \Lambda\eta$
\cite{Manley:2002ue}. The cross section rises steeply from threshold
and reaches a maximum of about $1.4\,mb$ at $W\sim
1.675$\,GeV/c$^2$. The data show a remarkable similarity to the
SU(3)$_{\rm f}$ flavor-related $\pi^-p\to p\eta$ cross section. The
latter is dominated by $N_{1/2-}(1535)$, the former by formation of
the intermediate $\Lambda_{1/2-}(1670)$ state, for which mass and
width, respectively, of $M = 1673\pm2$\,MeV, $\Gamma  =
23\pm6$\,MeV, and an elasticity $x = 0.37\pm0.07$ were measured. The
fraction with which $\Lambda_{1/2-}(1670)$ decays to $\Lambda\eta$
is determined to $(16\pm6)$\%. Resonance parameters and decay modes
were found in striking agreement with the quark-model predictions of
Koniuk and Isgur \cite{Koniuk:1979vy} but disagree with the results
of an analysis using a Bethe--Salpeter coupled-channel formalism
incorporating Chiral Symmetry \cite{GarciaRecio:2002td}. The latter
analysis finds a $\Lambda\eta$ decay fraction of $(68\pm1)$\% and an
inelasticity of $(24\pm1)$\%.

\begin{figure}[pt]
\begin{center}
 \includegraphics*[width=0.35\textwidth,height=0.25\textwidth]{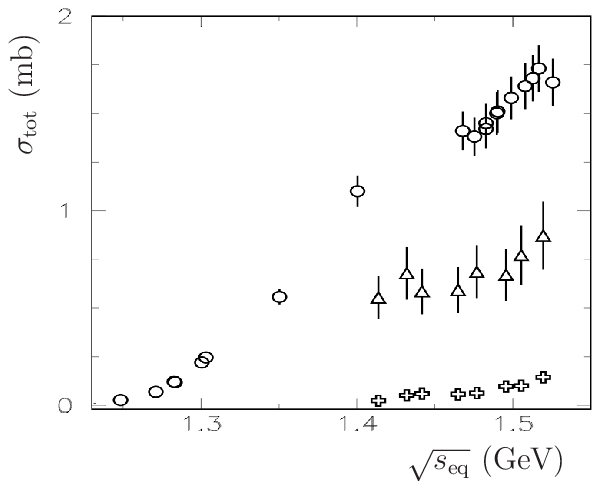}
\vspace{-3mm}\end{center}
\caption{\label{2pitot}The total cross sections as functions of the
\emph{equivalent total energy} $\sqrt{s_{eq}}$, defined as the
standard $s$ for pions and as $\sqrt{s_{eq}} \equiv \sqrt{s} - (m_s
- m_d)$  for incident kaons \cite{Nefkens:2002rz}. Circles:
$\sigma_{tot}(\pi^-p\to\pi^0\pi^0n)$. Triangles:
$\sigma_{tot}(K^-p\to\pi^0\pi^0\Lambda)$. Crosses:
$\sigma_{tot}(K^-p\to\pi^0\pi^0\Sigma^0)$.}
\end{figure}

In both cases, the branching ratio of $\Lambda_{1/2-}(1670)$ for
decays into $\Lambda\eta$ is much larger than that of other
resonances. In Table \ref{etadecay} we list the branching ratios of
negative-parity spin-1/2 resonances for decays into $\eta$ mesons.
We notice that for $N_{1/2^-}$, the lower mass state (mainly
$S=1/2$) has a strong coupling the $N\eta$ while it is smaller by
about one-order-of-magnitude for the higher-mass state (mainly
$s=3/2$). The situation is similar for $\Lambda_{1/2-}$ but opposite
for $\Sigma_{1/2^-}$. We note that in $\Lambda_{1/2-}$, the $ud$
diquark has isospin zero while for $\Sigma_{1/2-}$ $I=1$. The
connection is not yet understood.
\begin{table}
\caption{\label{etadecay}Decay branching ratios to baryon plus
$\eta$ of spin-1/2 negative parity baryons.}
\renewcommand{\arraystretch}{1.4}
\begin{footnotesize}
\begin{tabular}{ccccc}
\hline\hline Decay mode & Fraction &\qquad& Decay mode  & Fraction
\\                                                       \hline
$N_{1/2^-}(1535)\to N\eta$ & 45-60\%               & &
$N_{1/2^-}(1650)\to N\eta$& 3-10\%                 \\
$\Lambda_{1/2^-}(1670)\to\Lambda\eta$ & 10-25\% &&
$\Lambda_{1/2^-}(1800)\to\Lambda\eta$ & not seen  \\
$\Sigma_{1/2^-}(1620)\to\Sigma\eta$ & not seen &&
$\Sigma_{1/2^-}(1750)\to\Sigma\eta$ & 15-55\%  \\
\hline\hline\end{tabular}
\end{footnotesize}
\renewcommand{\arraystretch}{1.0}
\end{table}
 \paragraph{$\pi^- p\to n 2\pi^0$, $K^- p\to \Lambda 2\pi^0$ and to $\Sigma
2\pi^0$:} Three reactions leading to $2\pi^0$ in the final state
were studied; $\pi^- p\to n 2\pi^0$ from threshold to 750\,MeV/c
\cite{Craig:2003yd,Prakhov:2004zv}, $K^- p \to \pi^0\pi^0\Lambda$
and $K^- p \to \pi^0\pi^0\Sigma^0$ for p$_{K^-} = 514$\,MeV/c to
750\,MeV/c \cite{Prakhov:2004ri,Prakhov:2004an}. The cross sections
for the three reactions reveal a few interesting patterns
\cite{Nefkens:2002rz}, see Fig.~\ref{2pitot}. The cross section for
$K^- p\to \Lambda 2\pi^0$ is smaller than that for $\pi^- p\to n
2\pi^0$ by a factor 2. A reduction due to strangeness production is
not unexpected. But the cross section for $K^- p\to \Sigma 2\pi^0$
is much smaller than the other ones. This requires a dynamical
interpretation.

If the reactions would produce $\sigma$ (=$f_0(600)$) at a sizable
rate, one should expect similar cross sections for all three
reactions. This is not the case; at least the two reactions $\pi^-
p\to n 2\pi^0$ and $K^- p\to \Lambda 2\pi^0$ must be dominated by
production of baryon resonances. A partial wave analysis of the
former data revealed a very large contribution of $N_{1/2^+}(1440)$
interfering with $N_{1/2^-}(1535)$ and $N_{3/2^-}(1520)$
\cite{Sarantsev:2007bk} where $N_{1/2^+}(1440)$ decays via
$\Delta\pi$ and via $N\sigma$. The broad shoulder in the $K^- p\to
\Lambda 2\pi^0$ cross section is tentatively interpreted as evidence
for $\Lambda_{1/2^+}(1600)$ decaying via
$\Sigma_{3/2^+}^0(1385)\pi^0$ as intermediate state
\cite{Prakhov:2004ri}. A partial wave analysis of the data has not
been performed.\vspace{-2mm}
 \subsub{Baryon excitations from $J/\psi$ and $\psi'$ decays}\vspace{-2mm}
 \setcounter{paragraph}{0}
 \vskip 2mm plus 1mm minus 1mm
Baryon resonances can be searched for in final states from $J/\psi$
and $\psi'$ decays into a baryon, an antibaryon and at least one
meson. In Table \ref{psitobbbarplus},
\begin{table}[pt]\vspace{-4mm}
\caption{\label{psitobbbarplus}$J/\psi$ and $\psi'$ branching ratios
for decays into final states containing mesons and baryons.}
\begin{footnotesize}
\vspace{-2mm}
\begin{center}
\renewcommand{\arraystretch}{1.3}
\begin{tabular}{ccccccc}
 \hline\hline
 & \quad &\multicolumn{2}{c}{$J/\psi$} & \quad & \multicolumn{2}{c}{$\psi'$} \\
 \hline
$N\bar N\pi$ &&$(9.7\pm0.6)$&$10^{-3}$&&$(7.6\pm0.6)$&$10^{-4}$\\
$p\bar p\pi^+\pi^-$ &&$(6.0\pm0.5)$&$10^{-3}$&&$(7.6\pm0.6)$&$10^{-4}$\\
$N\bar N\eta$ &&$(4.18\pm0.36)$&$10^{-3}$&&$(0.58\pm0.13)$&$10^{-4}$\\
$\Lambda\bar\Lambda\,\eta$&&$(0.26\pm0.08)$&$10^{-3}$&&$<1.2$&$10^{-4}$\\
$pK^-\bar\Lambda$&&$(0.9\pm0.2)$&$10^{-3}$&&&\\
$pK^-\bar\Sigma^0$&&$(0.29\pm0.08)$&$10^{-3}$&&&\\
$\Sigma\,\bar\Lambda\pi$&&$(0.23\pm0.03)$&$10^{-3}$&&&\\
 \hline\hline
\end{tabular}
\renewcommand{\arraystretch}{1.0}
\end{center}
\end{footnotesize}
\vspace{-3mm}
\end{table}
\begin{figure}[pt]
\begin{tabular}{cc}
\hspace{-6mm}\includegraphics[width=0.26\textwidth]{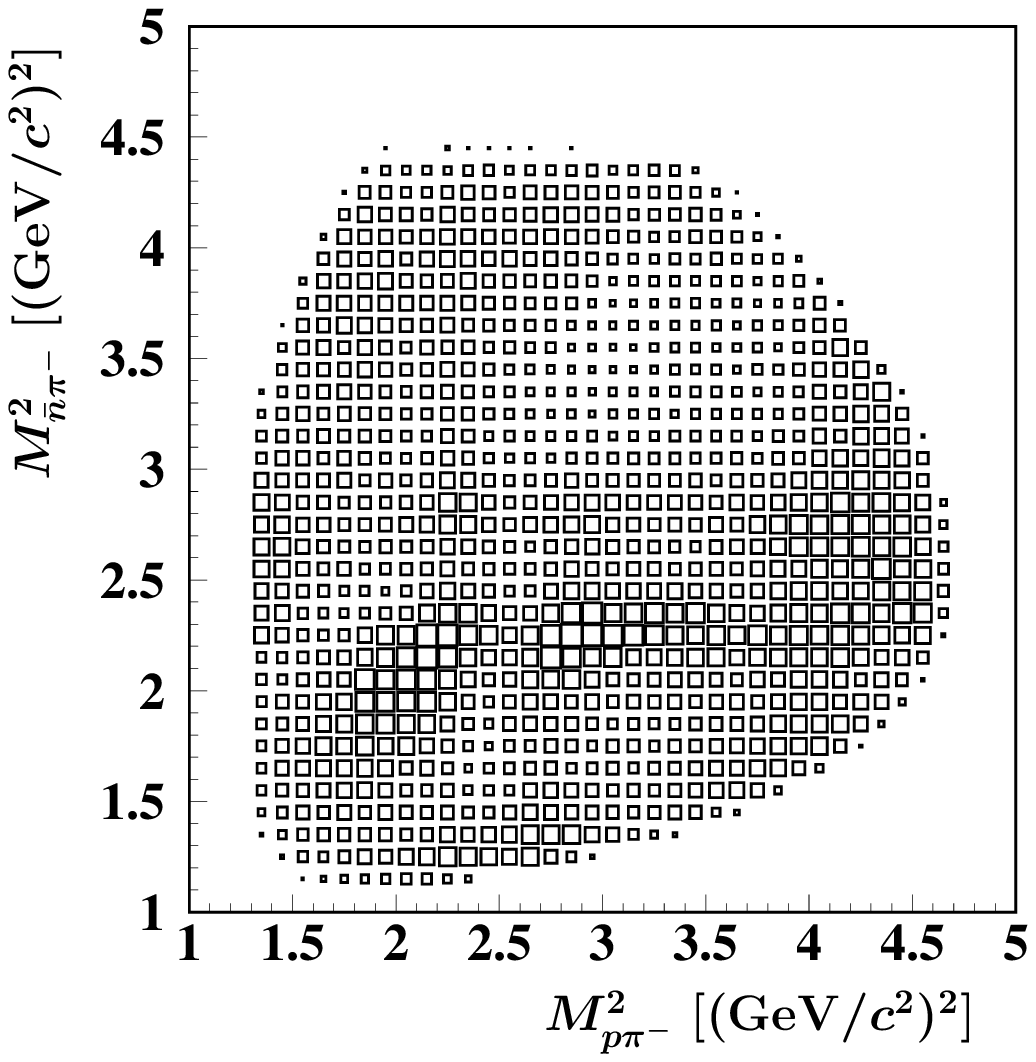}&
\hspace{0mm}\includegraphics[width=0.26\textwidth]{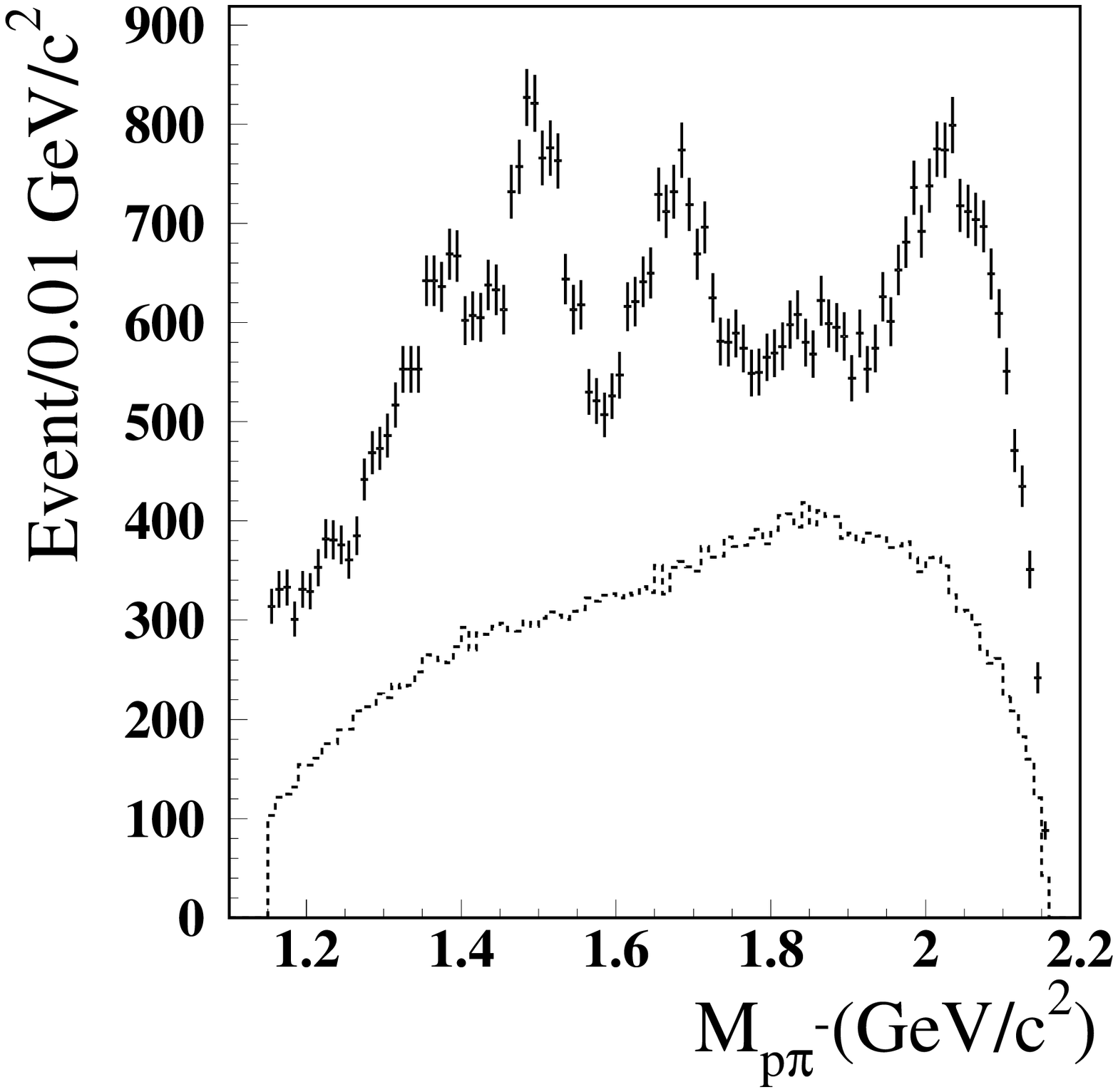}
\end{tabular}
\caption{\label{nnbarpidp}Dalitz plots of $M^2_{n\pi}$ vs.
$M^2_{p\pi}$ for $J/\psi\to p\pi^-\bar n$ and $p\pi^-$ invariant
mass spectrum \cite{LI:2009iw}.}
\end{figure}
relevant branching fractions are given, demonstrating the discovery
potential of $J/\psi$ decays for baryon spectroscopy. In particular
resonances recoiling against $\Lambda$, $\Sigma$,
$\Sigma_{3/2^+}(1385)$, $\Xi$, $\Xi_{3/2^+}(1530)$ are rewarding. In
other reactions, there is no real means to decide if, e.g.,
$\Sigma_{1/2^-}(1750)$ belongs to an SU(3)$_{\rm f}$ octet or
decuplet, or if it a mixture. Observation of $\Sigma_{1/2^-}(1750)$
recoiling against $\Sigma$ and/or $\Sigma_{3/2^+}(1385)$ in $\psi'$
decays would identify its SU(3)$_{\rm f}$ nature.

As example for the use of $J/\psi$ decays in baryon spectroscopy we
show in Fig.~\ref{nnbarpidp} the Dalitz plot $M^2_{n\pi}$ vs.
$M^2_{p\pi}$ for $J/\psi\to p\pi^-\bar n$ decays, and the $p\pi^-$
mass projection \cite{Ablikim:2004ug}. Four peaks can be identified.
A partial wave analysis \cite{LI:2009iw} assigns the first peak to
$N(1440)P_{11}$ with Breit--Wigner mass and width of $1358\pm 6 \pm
16$\,MeV and $179\pm 26\pm 50$\,MeV; the $N^*$ peaks at 1500\,MeV
and 1670\,MeV are identified with the well known second and third
resonance region, and the fourth peak is interpreted as a new $N^*$
resonance with $2040^{+3}_{-4}\pm 25$\,MeV mass and width of $230\pm
8\pm 52$\,MeV. The fit prefers $P_{13}$ quantum numbers.
\subsection{\label{Photoproduction experiments, a survey}Photoproduction experiments, a survey}
 \subsub{Aims of photoproduction experiments}
 \setcounter{paragraph}{0}
\paragraph{How many baryon resonances are known?} Baryon
spectroscopy defined by $\pi N$ elastic scattering is at a
bifurcation point. The listings of the PDG give a large number of
baryon resonances which were reported by the analyses of the
Karlsruhe--Helsinki group \cite{Hohler:1979yr} and of the
Carnegie-Mellon group \cite{Cutkosky:1980rh}, with star ratings from
1-star to 4-star. In the most recent analysis of the
George-Washington group \cite{Arndt:2006bf} including a large number
of additional data sets from pion factories (even though mostly at
low energy), practically only the 4-star resonances are confirmed. A
very decisive question is therefore if H\"ohler is right in his
critique of the GWU analysis that the method used by the GWU group
suppresses weak higher-mass resonances \cite{Hohler:2004pr}. The
confirmation of a few resonances found by \cite{Hohler:1979yr} and
\cite{Cutkosky:1980rh} and questioned by \cite{Arndt:2006bf} would
already help to give credit to the old analyses.\vspace{-4mm}
\paragraph{How many baryon resonances are expected?} Quark models
predict a very large number of baryon resonances. Experimentally,
the density of states in the mass region above 1.8\,GeV is much
smaller than expected. A reason might be \cite{Koniuk:1979vw} that
these \emph{missing resonances} decouple from the $\pi N$ channel.
Then they escape detection in $\pi N$ elastic scattering. These
resonances are expected to have no anomalously low helicity
amplitudes; then they must show up in photoproduction of
multi-particle final states.\vspace{-4mm}
\paragraph{What is the structure of baryon resonances?}
Electroproduction of baryon resonances provides additional
information, inaccessible to $\pi N$ scattering. Helicity
amplitudes, form factors, (generalized) polarizabilities can be
extracted. Intense experimental and theoretical efforts have, e.g.,
been devoted to determinations of the $E_2/M_1$ (electric quadruple
versus magnetic dipole) and $C_2/M_1$ (longitudinal electric
quadruple versus magnetic dipole) ratio for the $N\to \Delta(1232)$
transition amplitude. For a review of the hadron structure at low
$Q^2$, see \cite{Drechsel:2007sq}.\vspace{-2mm}
 \subsub{Experimental facilities}
 \setcounter{paragraph}{0}
 \paragraph{Bubble chambers:} Very early, in the late 1960s, photoproduction was
studied in bubble chamber experiments. Results at DESY were
summarized by \cite{Erbe:1968ke}, those from SLAC by
\cite{Ballam:1971yd,Ballam:1972eq}.\vspace{-4mm}

 \paragraph{NINA:} The electron synchrotron NINA at Dares-bury was
 used to study photoproduction reactions. We quote here only two of
 their late publications \cite{Barber:1981fj,Barber:1985fr} where
 references to earlier work can be found.\vspace{-4mm}

\paragraph{BONN SYNCHROTON:} In Bonn, a 2.5\,GeV electron synchrotron started
operation in 1967 and was used for photoproduction experiments. The
latest publication reported on production of positive pions at large
angles \cite{Dannhausen:2001yz}. The accelerator is now used to feed
ELSA.\vspace{-4mm}

\paragraph{ELSA:} The electron stretcher ring ELSA, in operation
since 1987, serves either as storage ring producing synchrotron
radiation or as post-accelerator and pulse stretcher delivering a
continuous electron beam (1\,nA, duty factor $\approx 70$\%) with up
to 3.5\,GeV energy. A few detectors were installed at ELSA:
{\sc\small Phoenics} \cite{Bock:1998rk}, {\sc\small ELAN}
\cite{Kalleicher:1997qf}, {\sc\small GDH} \cite{Naumann:2003vf},
{\sc\small SAPHIR}, and {\sc\small CBELSA} in different
configurations. {\sc\small SAPHIR} was a magnetic detector with a
central drift chamber (CDC), with a magnetic field perpendicular to
the beam axis and the target placed in the center of the CDC.
Forward hodoscopes in coincidence with the tagging system gave a
fast trigger and provided particle identification by measuring the
time of flight \cite{Schwille:1994vg}. It was dismantled in 1999.
The {\sc\small CBELSA}  experiment is based on the $4\pi$ photon
detector Crystal Barrel \cite{Aker:1992ny} which had been moved in
1997 from LEAR/CERN to Bonn. An inner scintillating fiber detector
is used for charged particle detection and trigger purposes
\cite{vanPee:2007tw}. Later, the forward direction was covered
\cite{Elsner:2007hm} by the TAPS \cite{Gabler:1994ay} or a MiniTaps
detector.\vspace{-4mm}

\paragraph{Jlab:} The continuous electron beam accelerator facility
at the Department of Energy's Thomas Jefferson National Accelerator
Facility (Jlab) delivers a 6\,GeV primary electron beam into three
different experimental areas, Halls A, B, and C, for simultaneous
experiments. Halls A and C both have two spectrometers; in Hall A,
two identical high-resolution spectrometer covering a maximum
momentum of 4 GeV/c are installed while in Hall C one is dedicated
to analyze high-momentum particles, the other has a short path
length for the detection of decay particles. Hall B houses the Jlab
Large Acceptance Spectrometer ({\sc\small CLAS}), the detector most
relevant for baryon spectroscopy. The {\sc\small CLAS} detector is
based on a six-coil toroidal magnet which provides a largely
azimuthal field distribution. Particle trajectories are
reconstructed, using drift chambers, with a momentum resolution of
0.5\% at forward angles. Cherenkov counters, time-of-flight
scintillators, and electromagnetic calorimeters provide good
particle identification \cite{Mecking:2003zu}.\vspace{-4mm}

\paragraph{ESFR:} The  {\sc\small GRAAL}  experiment was installed at the
European Synchrotron Radiation Facility (ESRF) in Grenoble (France).
The tagged and polarized $\gamma$-ray beam was produced by Compton
scattering of laser photons off the 6\,GeV electrons circulating in
the storage ring. The shortest UV wave length of 351\,nm yielded a
maximal $\gamma$-ray energy of 1.5\,GeV. The tagging system used 128
silicon microstrips with a pitch of 300\,$\mu$m. The proton track
was measured by two cylindrical Multi-Wire Proportional Chambers
with striped cathodes and two forward planar chambers. Charged
particles were identified by $dE/dx$ and time-of-flight measurement.
Photons coming from neutral decay channels of $\pi^0$ and $\eta$
were detected in 480 21-radiation-lengths BGO crystals supplemented
by a lead-scintillator sandwich time-of-flight wall in forward
direction \cite{Bartalini:2005wx}.\vspace{-5mm}

\paragraph{SPring-8:} The {\sc\small LEPS} (laser electron photons
at {\sc\small SPring-8} ) detector uses backscattered photons from
the 8\,GeV stored electron beam producing a tagged $\gamma$-ray beam
of up to 2.4\,GeV.  The {\sc\small LEPS} spectrometer consists of a
wide-gap dipole magnet with charged-particle tracking detectors. An
array of scintillator bars 4 meters downstream of target and
scintillators just behind the target provided a time-of-flight
information. Electron--positron pairs are vetoed by an aerogel
Cherenkov detector.\vspace{-4mm}

\paragraph{MAMI:}
The electron accelerator MAMI consists of three cascaded racetrack
microtrons and a harmonic double-sided microtron for final
acceleration \cite{Blomqvist:1998xn}. A linear accelerator provides
a 4\,MeV beam, the racetrack microtrons 15, 180 and 855\,MeV,
respectively. The maximum energy at the end of the new fifth stage
is 1.5\,GeV, with a beam current of up to 100\,$\mu$A. Photons can
be provided with linear or circular polarization. The development of
a polarized target is finalized.

\begin{figure*}[pt]
\includegraphics[width=0.7\textwidth]{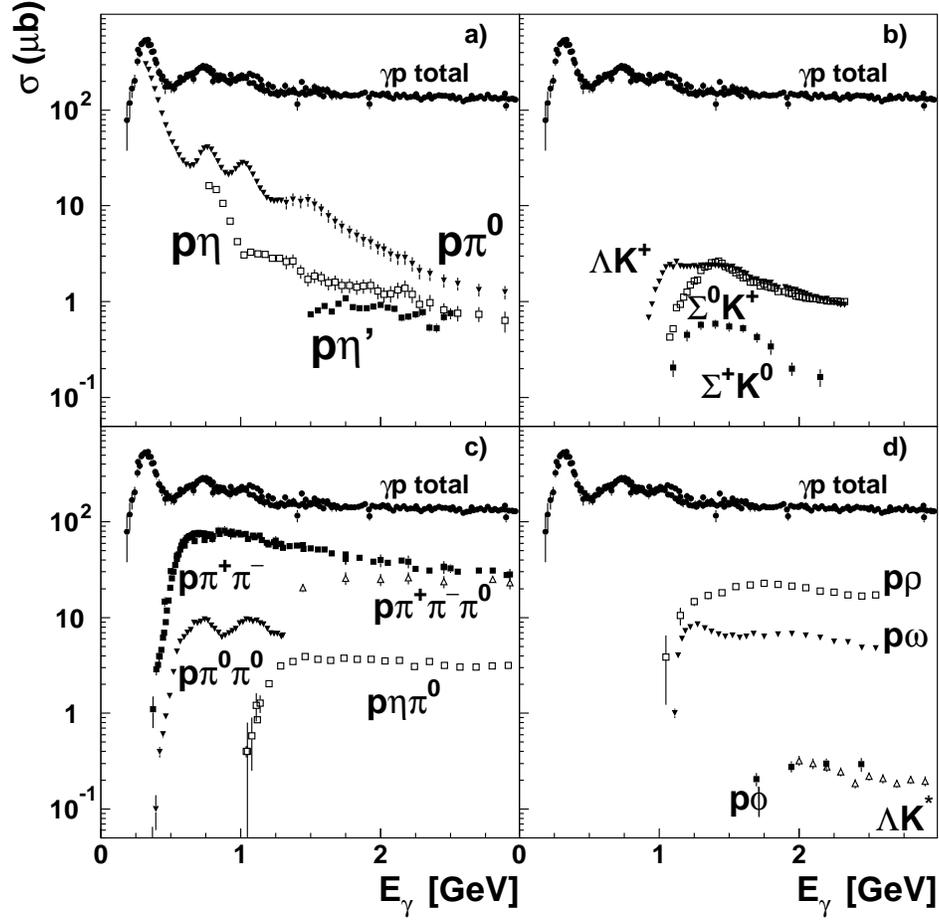}
\caption{\label{allwqs}Total photo-absorption cross section and
exclusive cross sections for single- and multi-meson production. a:
total, $p\pi^0$, $p\eta$, $p\eta^{\prime}$; b: total, $\Lambda K^+$,
$\Sigma^0 K^+$, $\Sigma^+ K^0$; c: $p\rho$, $p\omega$, $p\phi$,
$\Lambda K^{*+}$, $\Sigma^+ K^{*0}$, $\Sigma^0 K^{*+}$; d:
$p\pi^+\pi^-$, $p\pi^0\pi^0$, $p\pi^0\eta$, $p\pi^+\pi^-\pi^0$,
$pK^+K^-$.}
\end{figure*}

A major installation for baryon spectroscopy is the Crystal Ball
detector  (see {\it Experiments at BNL} in section
\ref{sec:inelas}). The detector capabilities are strengthened by a
forward-wall TAPS consisting of 510 hexagonally shaped BaF$_2$
detectors.\vspace{-2mm}
 \subsub{Total cross sections for photo-induced reactions}
 \setcounter{paragraph}{0}
The total photo-absorption cross section shown in Fig.~\ref{allwqs}
exhibits a large peak ($\approx 500\,\mu$b) due to $\Delta(1232)$
production, shows some structures in the second and third resonance
region and levels off at about $150\,\mu$b at a few GeV. At very
high energies, the photon splits into a $q\bar q$ pair with
vector-meson quantum numbers and the interaction between proton and
photon is dominated by Pomeron exchange exhibiting the typical
relativistic rise in the multi-GeV energy range. The structure of
the photon and its interaction with protons, a central issue at
{\sc\small H1} and {\sc\small ZEUS}, is beyond the scope of this
article; we refer the reader to a review by
\cite{Butterworth:2005aq}.\vspace{-4mm}
 \subsub{The GDH sum rule}
 \setcounter{paragraph}{0}
The photoproduction cross section depends on the helicity of proton
and photon. The total helicity may be 3/2 or 1/2; the fractional
difference
\begin{equation}
E=\frac{\sigma_{3/2}-\sigma_{3/2}}{\sigma_{3/2}+\sigma_{3/2}},
\end{equation}
is an important quantity. Such measurements require circularly
polarized photons and a target of polarized protons.

The development of techniques to produce polarized targets and
photons has a long history. The most recent driving force for this
development was the chance to test the Gerasimov--Drell--Hearn sum
rule \cite{Gerasimov:1965et,Drell:1966jv}
\begin{equation}
\int_0^\infty \frac{dE_{\gamma}}{E_{\gamma}}
\left[\sigma_{3/2}(E_{\gamma}) - \sigma_{1/2}(E_{\gamma})\right] =
\frac{2\pi^2\alpha}{M_p^2}\kappa^2_p~
\label{eqn:gdh}
\end{equation}
which relates the integrated helicity-difference cross-section to
the anomalous magnetic moment $\kappa_p$.

\begin{figure}[pb]\vspace{5mm} \centering
\includegraphics[width=0.49\textwidth,height=0.39\textwidth]{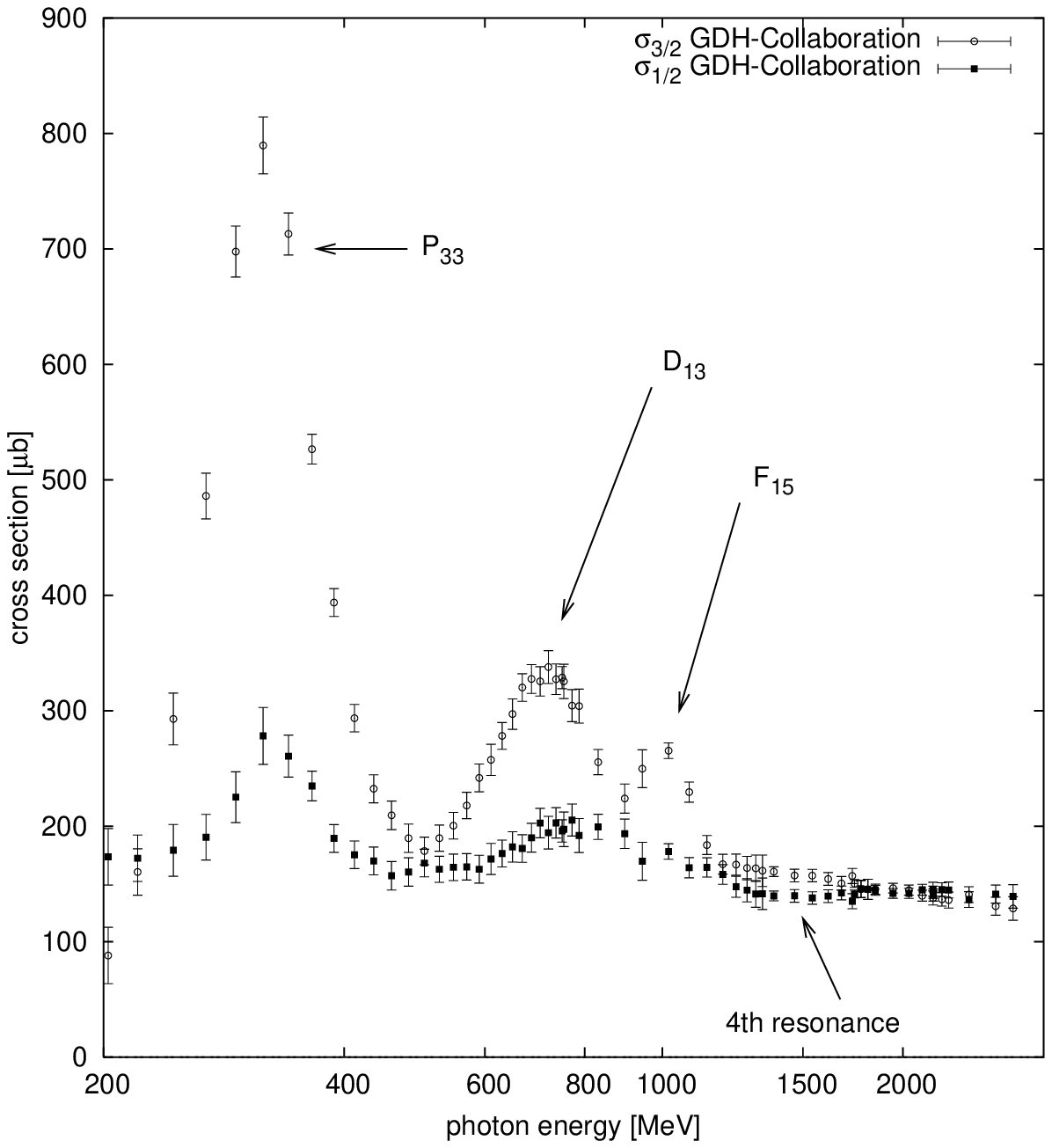}
\caption{Separate helicity state total cross sections $\sigma_{3/2}$
and $\sigma_{1/2}$ of the proton
\cite{Dutz:2003mm,Ahrens:2000bc,Ahrens:2001qt,Ahrens:2006yx}.}
\label{fig:resonances}
\end{figure}

Figure~\ref{fig:resonances} shows the separate helicity
contributions to the total cross section, measured at ELSA
\cite{Dutz:2003mm} and MAMI
\cite{Ahrens:2000bc,Ahrens:2001qt,Ahrens:2006yx}. Obviously, most of
the resonance strength of the first three resonances originates from
the $3/2$ helicity channel. The integrated difference, weighted with
$1/E_{\gamma}$, needs to be corrected for the unmeasured regions.
The low-energy part can be estimated using MAID (Mainz Analysis
Interactive Dial-in) predictions, the integral from 2.9\,GeV up to
$\infty$ using deep inelastic scattering data. The comparison of
calculated $205$\,$\mu$b and measured $212\pm 6\pm 16$\,$\mu$b value
shows remarkable agreement \cite{Helbing:2006zp}.

First measurements of the helicity difference on exclusive final
states have been published recently
\cite{Ahrens:2006gp,Ahrens:2007zzj}; these measurements provide an
important input to partial wave analyses.\vspace{-4mm}
\subsection{\label{Photo-production of pseudoscalar mesons}
Photo-production of pseudoscalar mesons}
 \subsub{Polarization observables}
 \setcounter{paragraph}{0}
The differential cross section for electro-production of
pseudoscalar mesons off nucleons is given by the product of the flux
of the virtual photon field - with longitudinal ($L$) and transverse
($T$) polarization - and the virtual differential cross section
which depends on 6 response functions ($R_i = R_{T}, R_{L}, R_{TL},
R_{TT}, R_{TL'}, R_{TT'}$). The response functions depend on two
additional indices characterizing the target polarization and the
recoil polarization of the final-state baryon. The response
functions can be written as CGNL \cite{Chew:1957tf} or helicity
amplitudes. The formalism is tedious; a derivation of formulas and a
comprehensive compendium of the relations between the different
schemes can be found in \cite{Knochlein:1995qz}. In photoproduction,
the longi\-tudinal component of the photon polarization vector
vanishes, and the problem is easier to handle. From the four CGNL
amplitudes, \emph{sixteen} bilinear products can be formed which
define the measurable quantities. The differential cross sections
can be divided into three classes, for experiments with polarized
photons and polarized target (BT, \ref{pt}) and experiments
measuring the baryon recoil polarization and using either polarized
photons (BR, \ref{pr}) or a polarized target~(TR, \ref{tr}).
\begin{subequations}\begin{eqnarray}\label{pt}
  \sigma  =
  \sigma_0 \bigl\{ 1 &-& p_{\bot} \Sigma \cos 2 \varphi
    + t_x \left( - p_{\bot} H \sin 2 \varphi + p_{\odot} F\right)
\nonumber\\
  &-& t_y \left( - T + p_{\bot} P \cos 2 \varphi \right) \\
&-&t_z \left( - p_{\bot} G \sin 2 \varphi + p_{\odot} E
\right)\bigr\},\nonumber\\
\label{pr}
\sigma  = \sigma_0  \bigl\{  1 &-& p_{{\bot}} \Sigma \cos 2 \varphi+
\sigma_{x'} \left( - p_{\bot} O_{x'} \sin 2 \varphi  -
p_{\odot} C_{x'} \right)\nonumber \\
&-& \sigma_{y'} \left( - P + p_{\bot} T \cos 2 \varphi\right)\\
&-&\sigma_{z'} \left( p_{\bot} O_{z'} \sin 2 \varphi + P_{\odot}
C_{z'} \right) \bigr\},\nonumber
\\
\label{tr}
\sigma  =  \sigma_0 \bigl\{ 1 &+& \sigma_{y'} P
+ t_x \left( \sigma_{x'} T_{x'} + \sigma_{z'} T_{z'} \right)\nonumber\\
&+& t_{y} \left( T + \sigma_{y'} \Sigma \right)\\
&-& t_{z} \left( \sigma_{x'} L_{x'} - \sigma_{z'} L_{z'}
\right)\bigr\}~.\nonumber
\end{eqnarray}
\end{subequations}
We use $\sigma=2\rho_f d \sigma/d \Omega$ where $\rho_f$ denotes the
density matrix for the final state baryon, $\sigma_0$ the
unpolarized differential cross section, $p_{\bot}$ the degree of
linear photon polarization, and $\varphi$ the angle between photon
polarization vector and reaction plane, $p_{\odot}$ the circular
photon polarization. The target polarization vector is represented
by $(t_x,t_y,t_z)$ with $z$ chosen as photon beam direction and $y$
as normal of the reaction plane. The Pauli matrices
$(\sigma_x',\sigma_y',\sigma_z')$ referring to the recoiling baryon
are defined in a frame with the momentum vector of the outgoing
meson as $z'$-axis and where the $y'$-axis is the same as the
$y$-axis. The $x$ and $x'$ axes are defined by orthogonality.

The quantities defined by capital letters (and, of course, the
differential cross section $\sigma_0$) are those to be determined.
Some have traditional names; we mention the beam and target
asymmetries $\Sigma$ and $T$, the recoil polarization $P$ and the
helicity difference of the cross section
$E\,\sigma=\sigma_{1/2}-\sigma_{3/2}$. The spin correlation
coefficients $C_{x'}, C_{z'}$ ($L_{x'}, L_{z'}$) define the transfer
of circular (oblique) polarization to a recoiling baryon.

Not all 16 observables need to be measured to arrive at a unique
solution (up to an overall phase); relations between the observables
reduce the number of required experiments. Seven appropriately
chosen experiments can be sufficient but may lead to discrete
ambiguities of the solution. Hence a minimum of up to 8 functions
need to be measured \cite{Barker:1975bp,Chiang:1996em}. The minimum
contains experiments with polarization of photons, target and
recoiling baryon. This number may be smaller due to inequalities
among observables \cite{Artru:2008cp}. If, e.g., $|A|^2+|B|^2\leq
1$, and if a first measurement gives $A\approx -1$, then a
measurement of $B$ is not anymore needed.

A set of data which allows for an energy-independent full
reconstruction of the amplitude is commonly referred to as a
``complete" experiment. Of course, a complete experiment requires
the measurement of isospin related channels, and it remains open if
the goal can be reached in practice
\cite{Workman:1998vh}.\vspace{-2mm}

\begin{figure}[pt]
\bc
\includegraphics[width=.48\textwidth,height=.70\textheight]{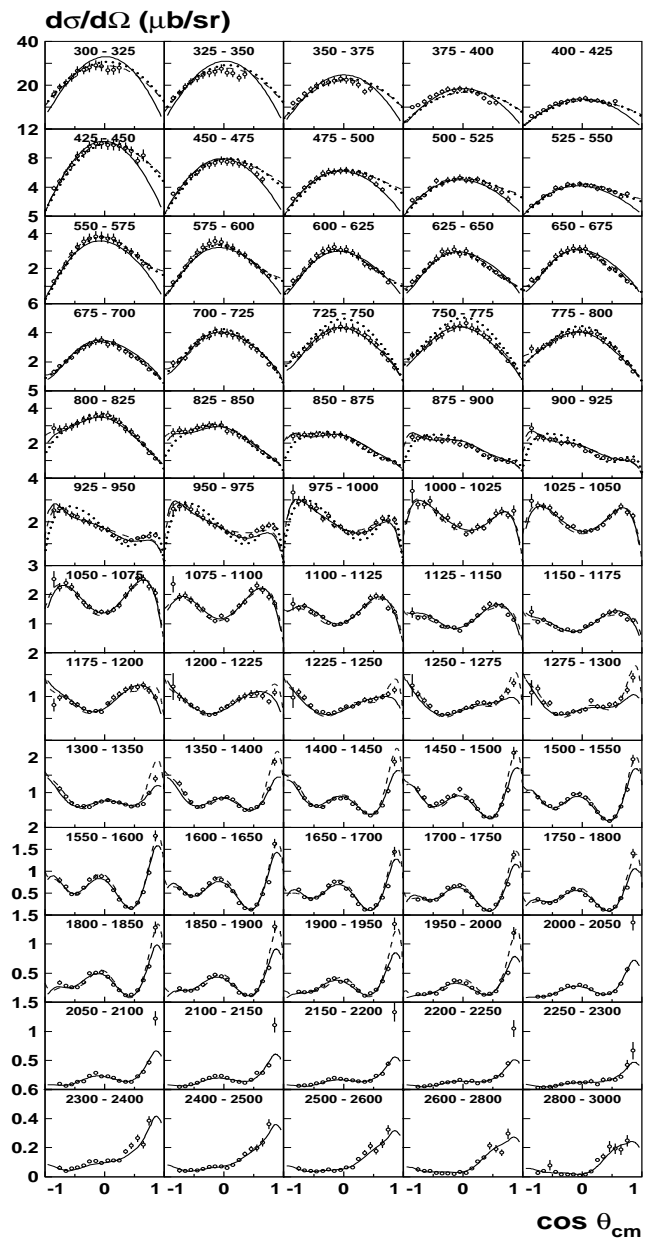}
\ec \caption{Differential cross sections for $\gamma
p\to p\pi^0$. The solid line represents the BnGa, the dashed line
the SAID (SM05), and the dotted line the MAID model. }
\label{FigurePi0DCS}
\end{figure}

 \vskip 2mm plus 1mm minus 1mm
 \subsub{Photoproduction of pions}\vspace{-2mm}
 \setcounter{paragraph}{0}
 \vskip 2mm plus 1mm minus 1mm
The structures observed in the total photo-absorption cross section
are much more pronounced in single-$\pi^0$ photo-production
(Fig.~\ref{allwqs}a); the cross section reaches $400\,\mu$b at the
$\Delta(1232)$ position, $40\,\mu$b at the second and $26\,\mu$b at
the third resonance peak. There are indications for the fourth
resonance region; then, the cross section decreases rapidly. The
cross section for $\pi^0$ production has been derived by integration
over differential cross sections $d\sigma/d\cos\theta$ where
$\theta$ is the angle of the $\pi^0$ meson with respect to the
direction of the photon in the $\gamma p$ rest frame. Most recent
data from Jlab \cite{Dugger:2007bt} and ELSA
\cite{Bartholomy:2004uz,vanPee:2007tw} cover a large energy and
angular range. References to earlier data are listed in
\cite{vanPee:2007tw}. The agreement between the data is remarkable;
at high energy, small discrepancies in the forward direction show up
between the ELSA data (which are shown in Fig. \ref{FigurePi0DCS})
and the Jlab data. The Crystal Barrel collaboration has new data in
the extreme forward angle which will hopefully resolve this
discrepancy.

\begin{figure}[pt]
\bc
\hspace*{-1mm}\includegraphics[width=.45\textwidth,height=.55\textwidth
]{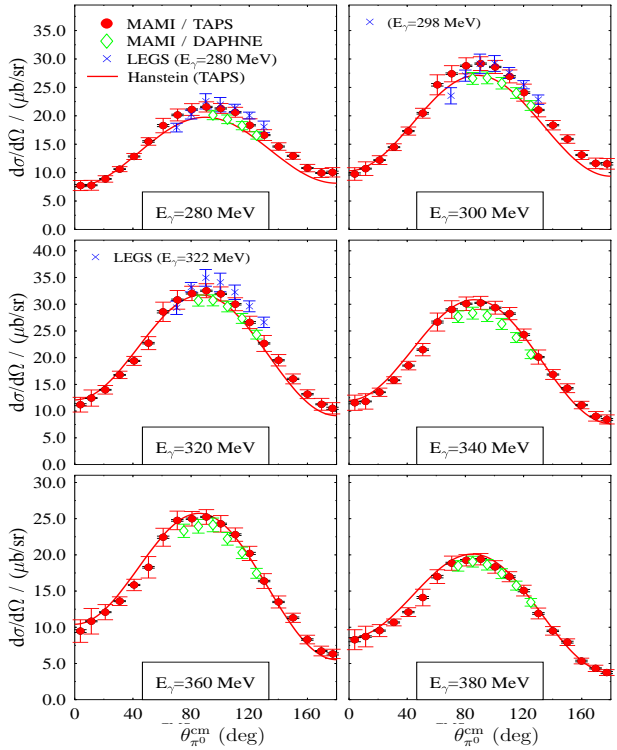}
\ec
\caption{\label{RB}(Color online) Photon asymmetry $\Sigma$ in the $\Delta$
resonance region for $\gamma p\to p\pi^0$ \cite{Beck:2006prc}. The
solid line is the MAID model.}
\end{figure}

The beam asymmetry is available from MAMI in the low-energy region
\cite{Beck:2006prc} (shown in Fig.~\ref{RB}), from GRAAL
\cite{Bartalini:2005wx} and from ELSA \cite{Elsner:2008sn}. Some
data on target and proton recoil polarization and a few data on
double polarization can be found at the GWU Data Analysis Center
\text{http://gwdac.phys.gwu.edu/}. Data on the related reaction
$\gamma p\to n\pi^+$ for the low energy region are given in
\cite{MacCormick:1996jz}, angular distributions and beam asymmetry
in \cite{Bartalini:2002cj}. Recently, differential cross sections
for $\gamma p \to n \pi^+$ have been measured by the {\sc\small
CLAS} collaboration for energies from 0.725 to 2.875 GeV
\cite{Dugger:2009pn}. The results are consistent with previously
published results. For the photon energies ranged from 1.1 to 5.5
GeV, cross sections for $\gamma n \to p\pi^-$ and $\gamma p \to
n\pi^+$ were measured at Jlab (for selected scattering angles) with
the aim to test ideas in perturbative QCD \cite{Zhu:2002su}. Further
details and references to earlier data can be found in
\cite{Zhu:2004dy}. The  beam asymmetry for photoproduction of
neutral pions from quasi-free nucleons in a deuteron target was
measured with the GRAAL detector for photon energies between 0.60
and 1.50\,GeV \cite{DiSalvo:2009zz}. The asymmetries for quasi-free
protons and quasi-free neutrons were found equal up to 0.8\,GeV and
substantially different at higher energies.

\begin{figure}[pb]
\centering
\includegraphics[clip,width=0.4\textwidth]{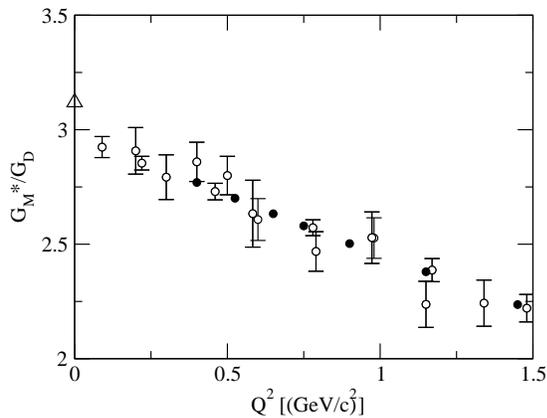}
\caption{\label{JooEbac}$G_M^\ast$ normalized by the dipole factor
$G_D=[1+Q^2/0.71\,({\rm GeV/c})^2]^{-1}$. }
\end{figure}

Electro-production of pions is sensitive to the $Q^2$ dependence of
electromagnetic transition operators and provides the possibility to
determine additional amplitudes; in particular the interference
between real and imaginary amplitudes can be determined. The
longitudinal amplitude $L_{l\pm}$ and the scalar amplitude
$S_{l\pm}$ are related due to gauge invariance and only $S_{l\pm}$
needs to be determined. The reaction $e^-\,p\to e^-\,p\pi^0$ was
studied in the $\Delta$ region at four-momentum transfers $Q^2 =
0.2$ \cite{Elsner:2005cz}, $2.8$ and $4.0$\,GeV$^2$
\cite{Frolov:1998pw}, and ratios of multipoles $S_{0+}/M_{1+}$,
$S_{1+}/M_{1+}$, and $E{_1+}/M_{1+}$ were extracted from decay
angular distributions. The related $e^-\,p\to e^-\,n\pi^+$ reaction
was investigated in the first and second nucleon resonance regions
in the $0.25 < Q^2 < 0.65$\,GeV$^2$ range
\cite{Joo:2001tw,Joo:2003uc,Joo:2005gs,Egiyan:2006ks}. The data were
used by {\sc\small EBAC} \cite{JuliaDiaz:2009ww} to extract the
dependence of the helicity amplitudes on the (squared) momentum
transfer $Q^2$, and ''dressed form factors" were determined. Figure
\ref{JooEbac} shows the resulting magnetic transition form factor
$G_M^\ast$ normalized to the conventional dipole form factor for the
$N\to\Delta$ transition. The $Q^2$ dependence serves as fix-point
for comparison of higher-mass excitations.

At higher invariant masses, electro-production of single pions can
be discussed within the frame of generalized parton distributions or
by extending the Regge formalism to high photon virtualities
\cite{Avakian:2003pk,Ungaro:2006df,DeMasi:2007id}. Recently,
electro-production of pions was studied using a polarized ($\rm
^{15}NH_3$) target. The data, recorded in the first and second
nucleon resonance regions in a $Q^2$ range from 0.187 to
0.770\,GeV$^2$ \cite{Biselli:2008ug}, is expected to place strong
constraints on the electro-coupling amplitudes $A_{1/2}$ and
$S_{1/2}$ for the $N_{1/2^+}(1440)$, $N_{1/2^-}(1535)$, and
$N_{3/2^-}(1520)$ resonances. The {\sc\small CLAS} collaboration
also performed a measurement of semi-inclusive $\pi^+$
electroproduction in the $Q^2$ range from 1.4 to 5.7\,(GeV/c)$^2$
with broad coverage in all other kinematic variables
\cite{Osipenko:2008rv}. The results suggest a similarity between the
spectator diquark fragmentation in $\gamma^*p\to n\pi^+$ and the
anti-quark fragmentation in $e^+e^-$ collisions.

Electro-production of $\pi^0$ mesons in the threshold region,
including the $\pi^+$ production threshold, was studied at very low
$Q^2$ at MAMI \cite{Weis:2007kf}.\vspace{-2mm}

 \subsub{Photoproduction of $\eta$- and $\eta^{\prime}$-mesons}
 \setcounter{paragraph}{0}\vspace{-2mm}
\begin{figure}[pb]\centering
\includegraphics[width=0.99\columnwidth]{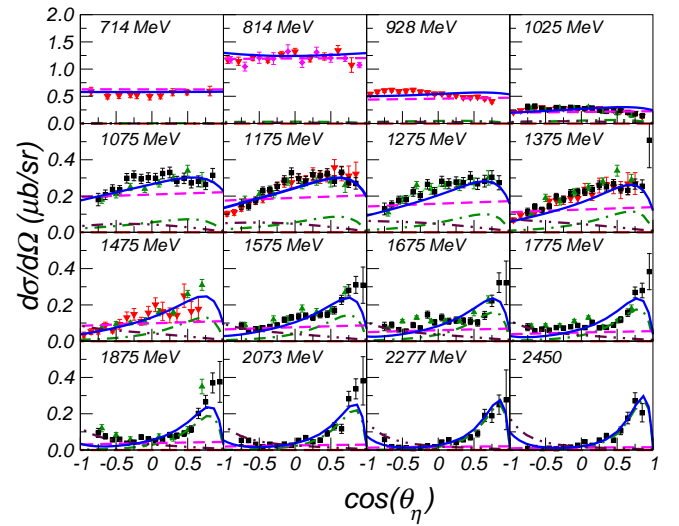}
\caption{\label{fig:photo4c}(Color online) Differential cross
sections for the reaction $\gamma p \to p \eta$ from {\sc\small
CBELSA} \cite{Bartholomy:2007zz} and {\sc\small CLAS}
\cite{Dugger:2002ft} for different invariant masses and fit results
\cite{Nakayama:2008tg}. The dashed line represents the $S_{11}$, the
dash-dot-dot line the $D_{13}$, the dashed-dotted line the
meson-exchange contribution; their sum is given as solid line.
\vspace{-2mm}}
\end{figure}
The cross section for photo-induced production of $\eta$-mesons is
sizable reaching $16\,\mu$b just above its threshold, see
Fig.~\ref{allwqs}a. The most recent data can be found in
\cite{Dugger:2002ft,Crede:2003ax,Bartalini:2007fg,Bartholomy:2007zz}.
\cite{Bartholomy:2007zz} contains a survey of older data. At 1\,GeV
photon energy, a small dip is observed but otherwise, the cross
section does not show any significant structures. (The anomaly in
the {\sc\small GRAAL}  data at 1\,GeV does not show up when the
angular distributions are fitted with the BnGa amplitudes; hence the
anomaly is likely due to the polynomial extrapolation of the angular
distribution into an uncovered region.) At $E_{\gamma}=2$\,GeV, the
$\eta$ cross section is smaller than the $\pi^0$ cross section by a
factor 3. The {\sc\small GRAAL}  beam asymmetry
\cite{Bartalini:2007fg} is confirmed and extended in range by
\cite{Elsner:2007hm}. Very recently, the {\sc\small CLAS}
\cite{Williams:2009yj} and the {\sc\small CB-ELSA/TAPS}
\cite{Crede:2009zz} collaborations reported new data on $\eta$ and
$\eta^{\prime}$ photoproduction. In the high energy region and for
forward angles, the {\sc\small CB-ELSA/TAPS} cross section is
significantly larger than {\sc\small CLAS}. We note that {\sc\small
CB-ELSA/TAPS} data are based on two fully reconstructed $\eta$ decay
modes; both with very little background and a high detection
efficiency.

Photoproduction of $\eta$-mesons off neutrons gives access to the
helicity amplitudes $A_{1/2}^n, A_{3/2}^n$ of $N_{1/2^-}(1535)$
coupling to $N\eta$. The reaction has recently attracted
considerable additional interest due to the possibility that a
narrow $J^P=1/2^+$ nucleon resonance at $\approx 1680$\,MeV may have
been found \cite{Kuznetsov:2007gr,Kuznetsov:2008hj}. For a more
detailed discussion, see paragraph \ref{pentaq} below. Very
recently, precise angular distributions \cite{Jaegle:2008ux} and
beam asymmetries \cite{Fantini:2008zz} have been reported.

Electro-production of $\eta$-mesons was reported in
\cite{Denizli:2007tq} for total center of mass energy
$W=1.5-2.3$\,GeV and invariant squared momentum transfer
$Q^2=0.13-3.3$\,GeV$^2$, and photo-couplings and $\eta N$ coupling
strengths of baryon resonances were deduced. A structure was seen at
$W\sim$\,1.7\,GeV. The shape of the differential cross section is
indicative of the presence of a $P$-wave resonance that persists to
high $Q^2$. The data are extended by \cite{Dalton:2008ff} to $Q^2
\sim 5.7$ and $7.0$\,GeV$^2$ for center-of-mass energies from
threshold to 1.8\,GeV. A first double polarization experiment on
$\eta$ electro-production was reported by \cite{Merkel:2007ig}.

The photoproduction cross section for $\eta^{\prime}$-mesons,
reported by \cite{Dugger:2005my}, rises slowly from threshold,
reaches a maximum of about  $1\,\mu$b at $E_{\gamma}=1.9$\,MeV; at
large energies, its cross section falls below the $\eta$ cross
section by a factor $\approx 2$, likely because of the twice smaller
$u\bar u+d\bar d$ component in the $\eta^{\prime}$ wave
function.\vspace{-2mm}
 \subsub{The reactions $\gamma p \to K^+\Lambda, K^+\Sigma^0$, and
 $K^0\Sigma^+$}\vspace{-2mm}
 \setcounter{paragraph}{0}
Figure \ref{allwqs}b show cross sections for photo-production of
final states with strangeness. For $\Lambda K^+$ and $\Sigma^0K^+$
the cross sections reach about 2.5\,$\mu$b; for $\Sigma^+K^0$, it is
a factor 4 smaller. The ratio for decays of nucleon resonances into
$\Sigma^+K^0$ or $\Sigma^0K^+$ is 1/2, for $\Delta$ resonances it is
2. The $\Sigma^0K^+$ cross section is larger than that for
$\Sigma^+K^0$; the former reaction receives contributions from kaon
exchange which is forbidden for the latter reaction. In partial wave
analyses \cite{Castelijns:2007qt}, the $N_{1/2^+}(1880)$ resonance
is seen to make a significant contribution to final states with open
strangeness.

\begin{figure}[pt]
\bc
\includegraphics[width=0.4\textwidth,height=0.35\textheight]{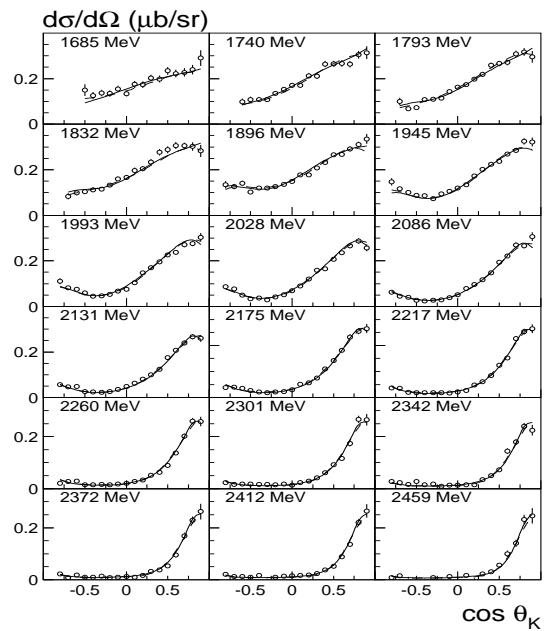}
\vspace{-1mm}\ec \caption{\label{dcs_klam}Differential cross
sections for $\gamma p \rightarrow K^+\Lambda$
\cite{Bradford:2005pt}. The solid curves represent  a Bonn--Gatchina
fit.\vspace{-2mm}}
\end{figure}

Differential distributions for $\gamma p\to K^+ \Lambda$, $K^+
\Sigma^0$ and $K^0 \Sigma^+$ have been measured at ELSA with
{\sc\small SAPHIR} \cite{Glander:2003jw,Lawall:2005np} and
{\sc\small CBELSA/TAPS} \cite{Castelijns:2007qt}, {\sc\small GRAAL}
\cite{Lleres:2007tx}, at Jlab with the {\sc\small CLAS} detector
\cite{Bradford:2005pt}, and by {\sc\small LEPS} at {\sc\small
SPring-8} \cite{Zegers:2003ux,Sumihama:2005er,Hicks:2007zz}. The
data of \cite{Bradford:2005pt} are shown in Fig.~\ref{dcs_klam}. The
reconstruction of the hyperon decay defines its polarization status.
At {\sc\small GRAAL} and {\sc\small SPring-8} , the $\gamma$-ray
beam is created by rescattering of optical photons which are easily
polarized; in these measurements, the beam asymmetry is determined
as well.

\begin{figure}[pt]\vspace{-5mm}
\hspace{35mm}\includegraphics[width=.53\textwidth,height=0.38\textheight]{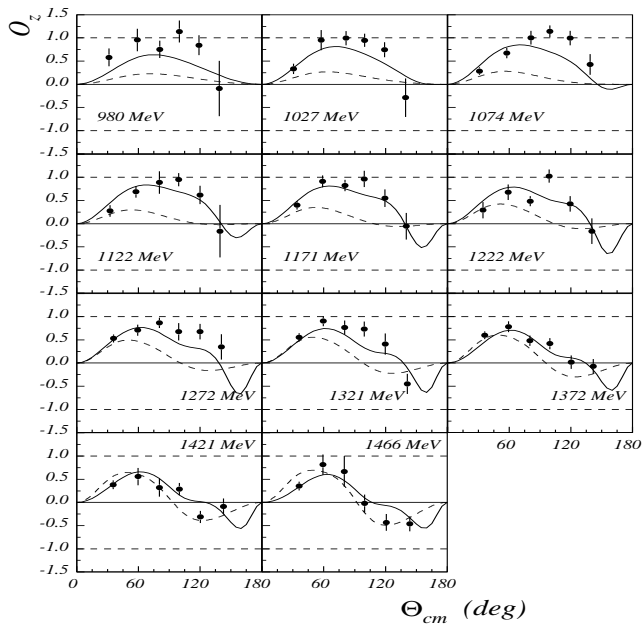}
\caption{\label{ozkl}Angular distributions of the beam recoil
observable $O_z$ \cite{Lleres:2008em}. Data are compared with
predictions of the Bonn--Gatchina (solid line, private
communications from to the GRAAL collaboration) and the
Regge-plus-resonance model of \cite{Corthals:2007kc}.\vspace{-2mm}}
\end{figure}
Recently, spin transfer from linearly and circularly polarized
photons to final-state hyperons has been measured at {\sc\small
GRAAL} \cite{Lleres:2008em}, see Fig.~\ref{ozkl} and Jlab
\cite{Bradford:2006ba}. The data exhibit a striking transfer of the
photon polarization to the $\Lambda$ \cite{Schumacher:2006ii}; the
data mark an important step towards a complete experiment.

Electro-production of $K^+\Lambda$ and $K^+\Sigma^0$ final states
from a proton target was studied at Jlab using the {\sc\small CLAS}
detector. The separated structure functions $\sigma_T$, $\sigma_L$,
$\sigma_{TT}$, and $\sigma_{LT}$ were extracted for momentum
transfers from $0.5\leq Q^2\leq 2.8$ GeV$^2$ and invariant energy
from $1.6\leq W\leq 2.4$ GeV, while spanning nearly the full
center-of-mass angular range of the kaon \cite{Ambrozewicz:2006zj}.
The polarized structure function $\sigma_{LT'}$ was measured for the
reaction $p(\overrightarrow{e},e'K^+)\Lambda$ in the nucleon
resonance region from threshold up to $W$=2.05 GeV for central
values of $Q^2$ of 0.65 and 1.00 GeV$^2$ \cite{Nasseripour:2008fz}.
The separated structure functions reveal clear differences between
the production dynamics for the $\Lambda$ and $\Sigma^0$ hyperons.

The polarization transferred from virtual photons to $\Lambda$ and
$\Sigma^0$ hyperons was measured using the {\sc\small CLAS}
spectrometer at beam energies of  2.567, 4.261, and 5.754 GeV
\cite{Carman:2002se,Carman:2009fi} spanning momentum transfers up to
5.4 GeV$^2$. The data suggest that the $\Lambda$ polarization is
maximal along the virtual photon direction. The large polarization
effects -- as also observed in photoproduction \cite{Lleres:2008em,%
Bradford:2006ba} -- call for a simple interpretation accounting for
the dynamics of quarks and gluons in a domain thought to be
dominated by meson/baryon degrees of freedom. Two possible scenarios
are discussed in \cite{Carman:2009fi}.

\subsection{\label{Photo-production of multi-mesonic final states}
Photo-production of multi-mesonic final states}
\begin{figure}[pb]\vspace{3mm}
\begin{center}
 \subfigure[]{ \includegraphics[width=0.11\textwidth]{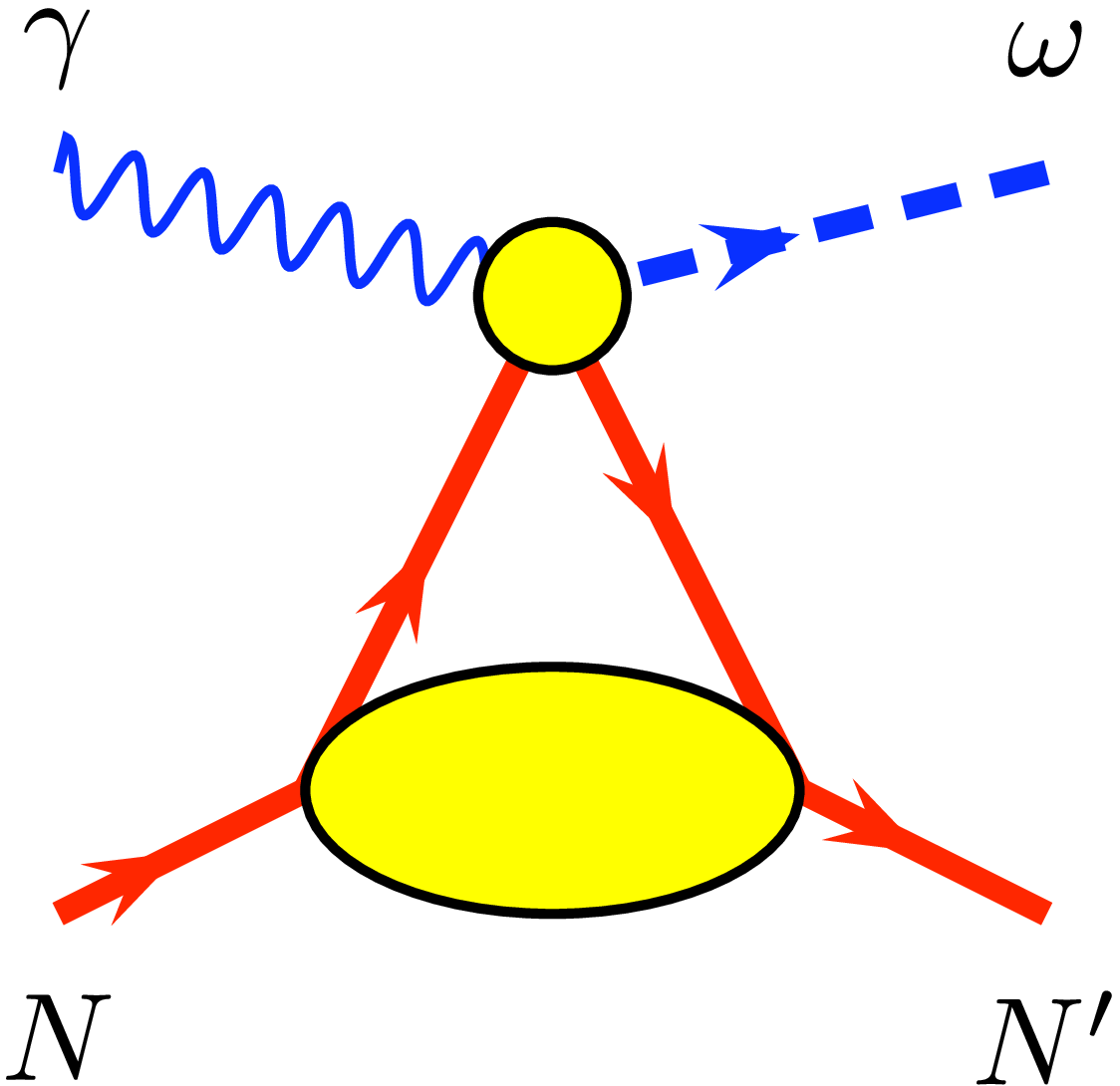}}\
 \subfigure[]{ \includegraphics[width=0.11\textwidth]{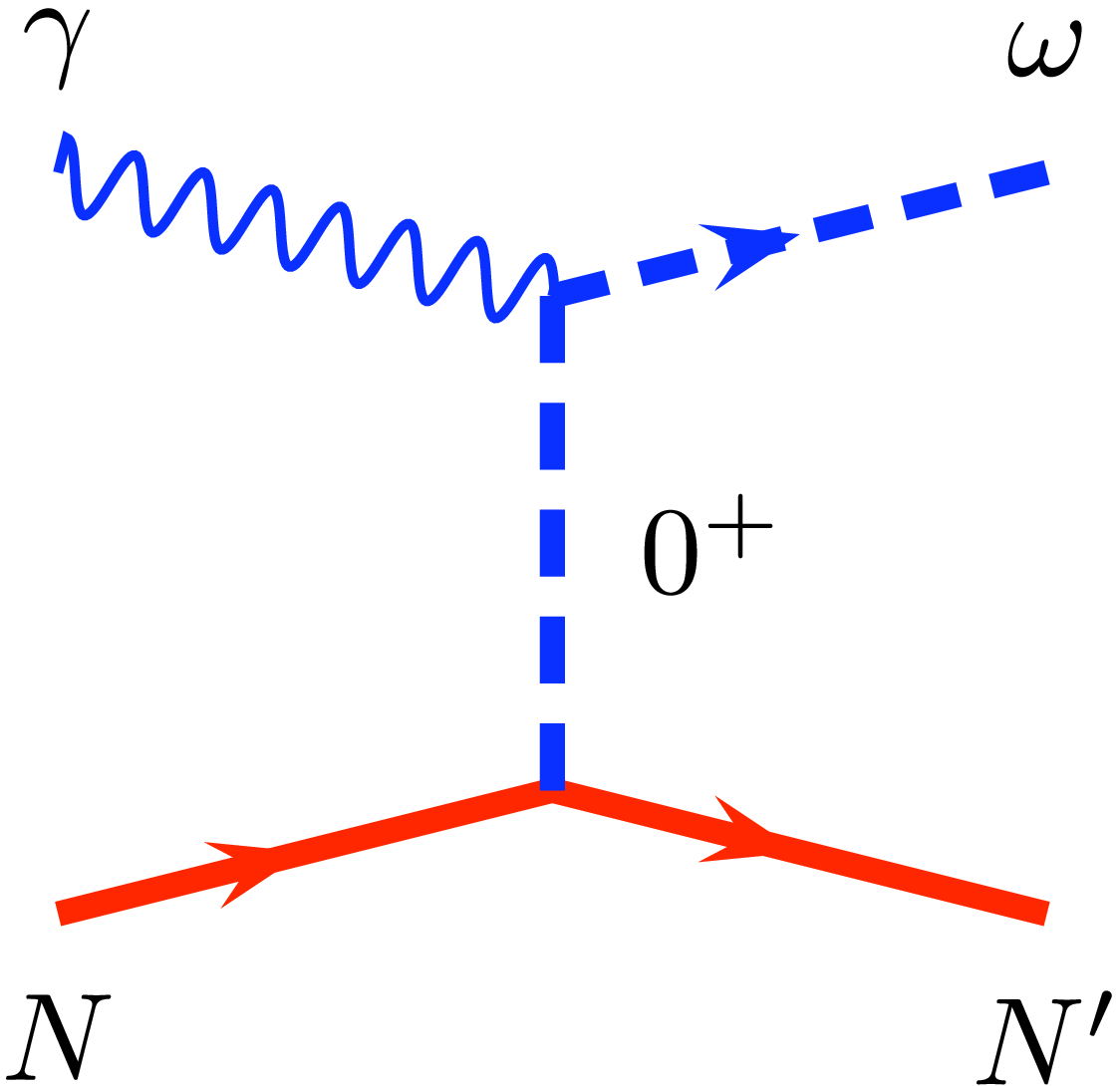}}\
 \subfigure[]{ \includegraphics[width=0.11\textwidth]{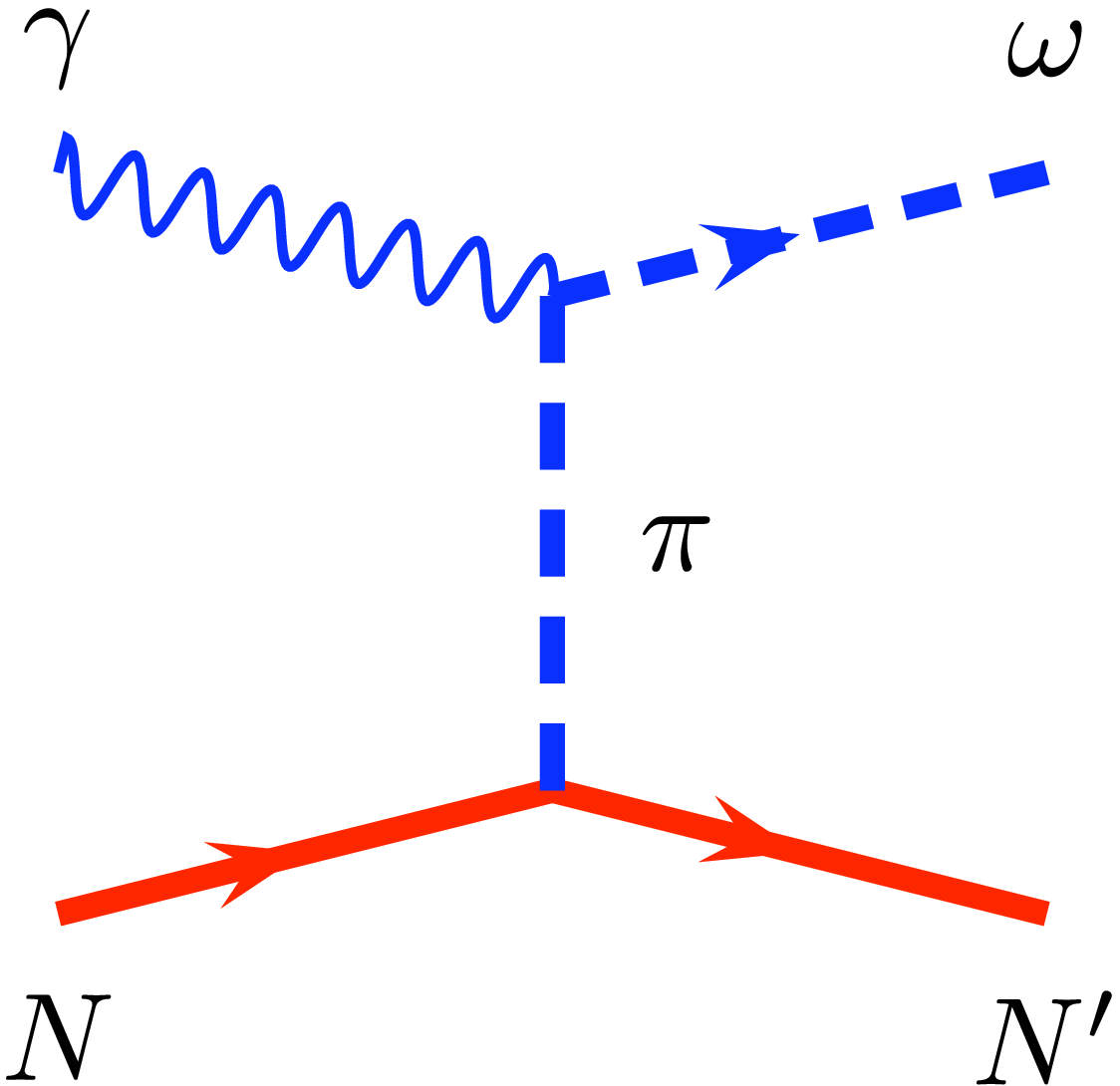}}\\
 \subfigure[]{ \includegraphics[width=0.11\textwidth]{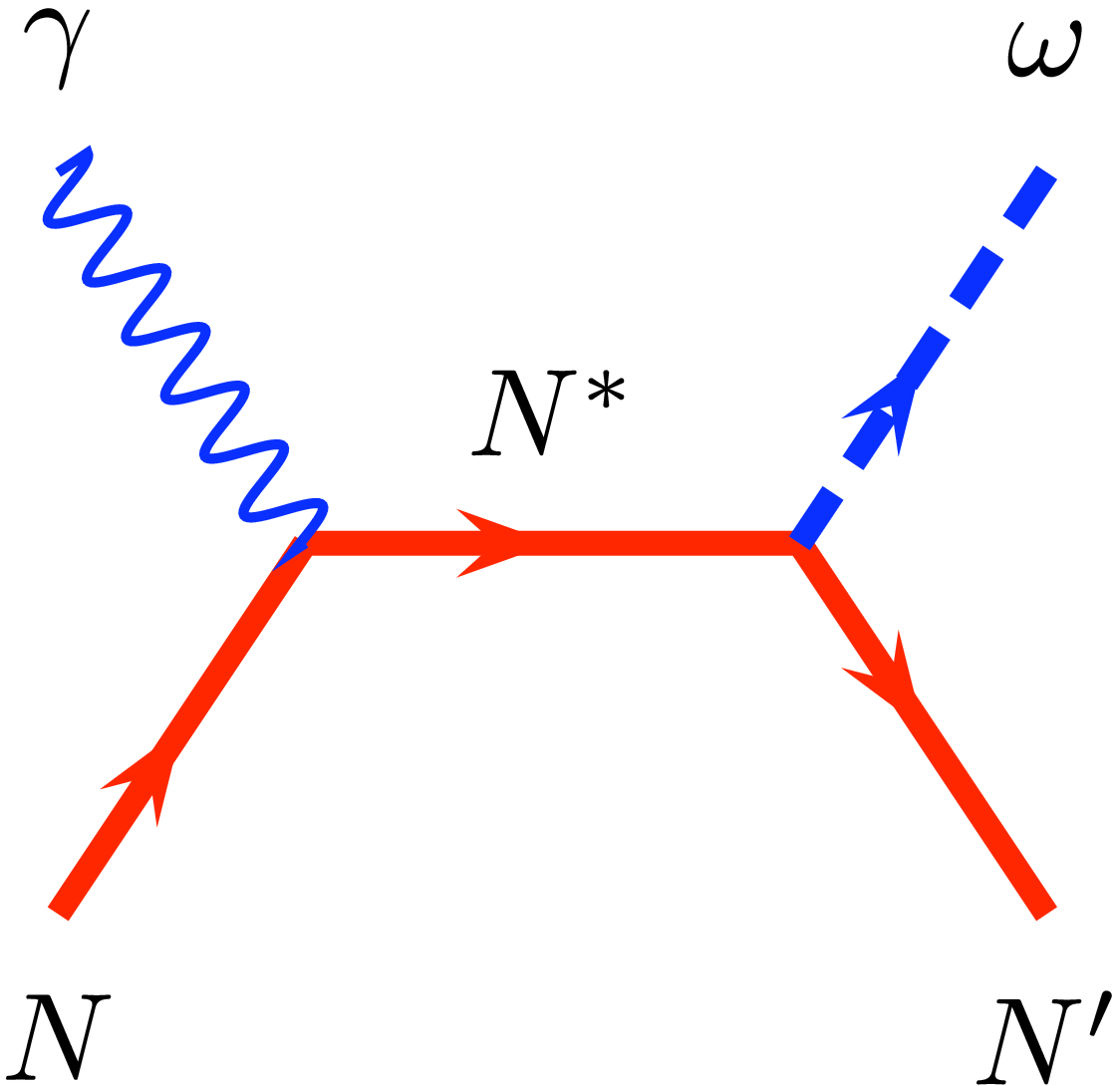}}\quad
 \subfigure[]{ \includegraphics[width=0.11\textwidth]{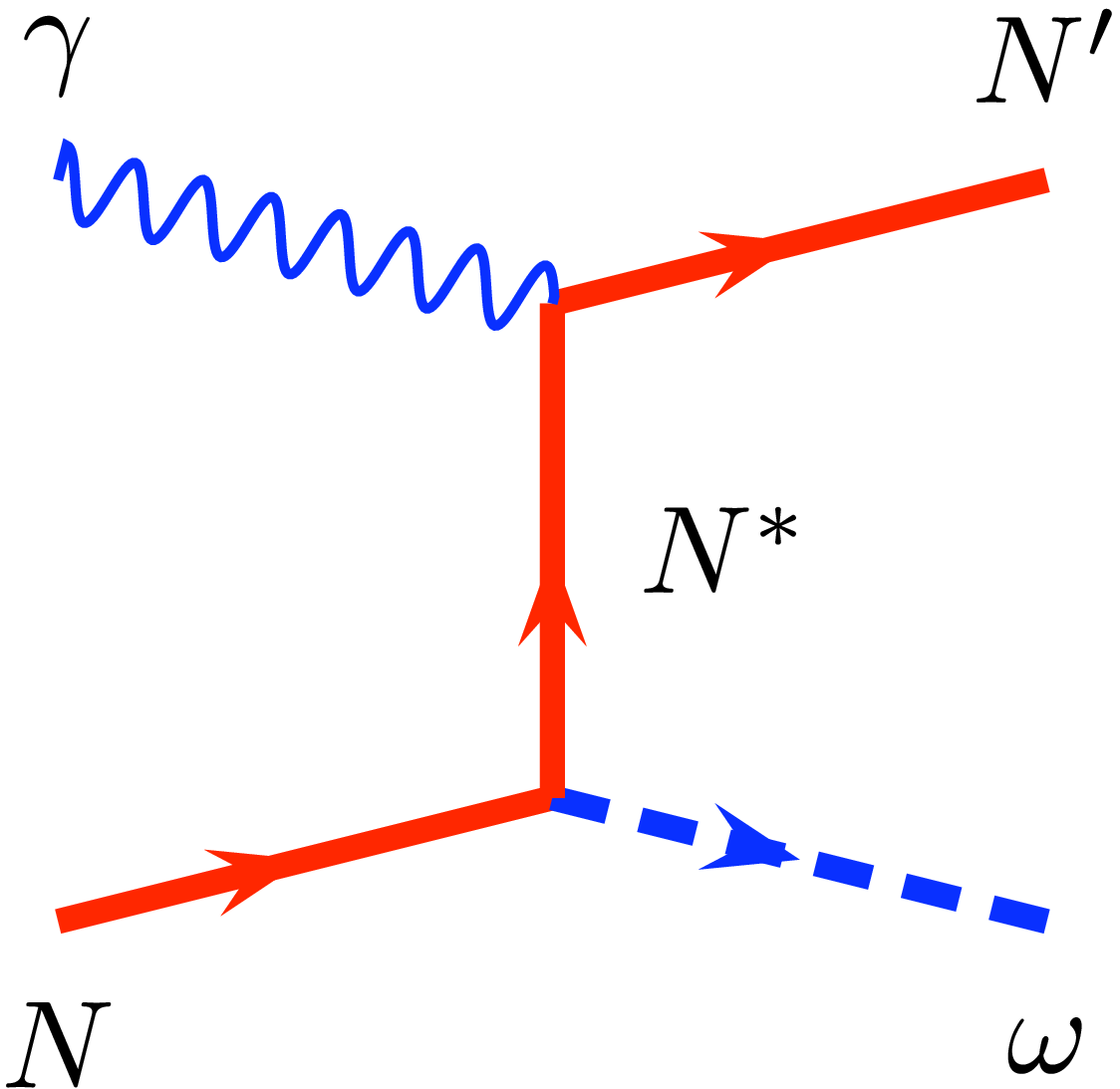}}
\end{center}
\vspace{-3mm} \caption{\label{fig:mechanism} Contributions to
$\omega$ photoproduction: (a): The handbag diagram for hard photo- and
electro-production. The large blob represents the generalized parton
distribution of the nucleon. At lower energies, processes b,c,d are
more appropriate to describe the reaction. (b): Natural parity
$t$-channel exchange and (c): $t$-channel exchange via the pion
trajectory, (d): $s$-channel intermediate resonance excitation. The
same diagrams contribute to $\rho$ production while $\phi$ are
produced dominantly via (b). For $K^*$ production, a kaon trajectory
is exchanged, the outgoing $N'$ is replaced by a hyperon. (d) and (e):
baryon pole in the $s$ and $u$ channel. }
\end{figure}

 \subsub{Vector mesons}
 \setcounter{paragraph}{0}
Photons and unflavored vector mesons share the same quantum numbers.
In soft vector-meson production by real photons, natural-parity
(Pomeron) exchange provides the leading term to the cross section.
The cross section falls off exponentially with the squared recoil
momentum $t$ characteristic for ``diffractive" production. At low
energies, a significant pion (kaon) exchange contribution is
expected because of the large $(\rho, \omega) \to\pi^0\gamma$
($K^*\to K\gamma$) coupling. Most interesting in the context of this
review are contributions from $N^*$ production since quark models
predict for some $N^*$ resonances large couplings to $N\omega$ and
to $N\rho$. Figure~\ref{fig:mechanism} depicts the different reaction
mechanisms.

\begin{figure}[pt]
\bc
\includegraphics[width=.3\textwidth]{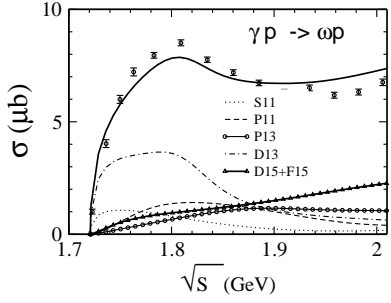}
\vspace{-8mm}
\ec
\caption{\label{barth}The total cross section for
$\omega$ photoproduction \cite{Barth:2003kv} and decomposition into
partial waves by \cite{Shklyar:2004ba}.}
\end{figure}
Photoproduction of $\rho$ mesons was studied by the {\sc\small CLAS}
\cite{Battaglieri:2001xv} and {\sc\small SAPHIR} \cite{Wu:2005wf}
collaborations, $\omega$ mesons by {\sc\small CLAS}
\cite{Battaglieri:2002pr}, {\sc\small SAPHIR} -- these data are
shown in Fig.~\ref{barth} -- \cite{Barth:2003kv}, {\sc\small GRAAL}
\cite{Ajaka:2006bn} and {\sc\small CBELSA/TAPS} \cite{Klein:2008gs}.
Large statistics data on differential cross sections and spin
density matrix elements for $\gamma p \to p \omega$ have been
reported recently by {\sc\small CLAS} \cite{Williams:2009rb}, for
energies from threshold up to $W=2.84$\,GeV.  $N_{5/2^+}(1680)$ and
$N_{3/2^-}(1700)$ near threshold, $N_{7/2^-}(2190)$ and possibly a
$N_{5/2^+}$ around 2\,GeV were determined as leading contributions
in an event-based partial wave analysis \cite{Williams:2009rc}:.

Photoproduction of $\phi$-mesons was reported by {\sc\small CLAS}
\cite{Anciant:2000az}, {\sc\small SAPHIR} \cite{Barth:2003bq} and
{\sc\small LEPS} \cite{Mibe:2005er}; the reactions $\gamma p \to
K^{*0}\Lambda$ and $\gamma p \to K^{*0}\Sigma$ were reported by
{\sc\small CLAS} \cite{Hleiqawi:2007ad}, $\gamma p \to K^{*0}
\Sigma^+$ by {\sc\small CBELSA/TAPS} \cite{Nanova:2008kr}. The size
of the cross section is about 24\,$\mu$b for $\rho$, 8\,$\mu$b for
$\omega$, 0.2\,$\mu$b for $\phi$ production (see Fig.~\ref{allwqs}c)
while ratios 9:1:2 would be expected from the direct
photon-vector-meson couplings. For pion exchange, the $\omega$ cross
sections should exceed the $\rho$ cross section while $\phi$
production would vanish.

In the multi-GeV range, electro-production is sensitive to the
transition between the low energy hadronic and high energy partonic
domains; at sufficiently large energies, generalized parton
distributions can be determined (see, e.g.,
\cite{Goloskokov:2007fd}). However, there is so far no attempt to
use the data for baryon spectroscopy. Here, we give reference to
recent {\sc\small CLAS} papers on electro-production of $\rho$-
\cite{Morrow:2008ek}, $\omega$- \cite{Morand:2005ex}, and
$\phi$-mesons \cite{Santoro:2008ai}.\vspace{-2mm}
 \vskip 2mm plus 1mm minus 5mm
 \subsub{$\gamma N\to N\pi\pi$ and $N\pi\eta$}
 \vskip 2mm plus 1mm minus 5mm
 \setcounter{paragraph}{0}\vspace{-2mm}
Multi-meson production collects an increasing fraction of the cross
section, see Fig.~\ref{allwqs}d. The most important channels are
$\gamma p\to p\pi^+\pi^-$ \cite{Wu:2005wf}; above 2\,GeV, $\gamma
p\to p\pi^+\pi^-\pi^0$ reaches a similar strength
\cite{Barth:2003kv}. In the low-energy region, the different isospin
channels of two-pion photoproduction \cite{Zabrodin:1999sq} can be
used to study chiral dynamics
\cite{GomezTejedor:1995pe,Nacher:2000eq}. Differences in
$\pi^+\pi^0$ and $\pi^0\pi^0$ invariant mass distributions were
assigned to a $N\rho$ decay branch of the $N_{3/2^-}(1520)$ nucleon
resonance \cite{Langgartner:2001sg}. In the resonance region,
photoproduction of two charged pions is dominated by diffractive
$\rho$ production and the direct production $\gamma p\to
\pi^-\Delta(1232)^{++}$; $\gamma p\to \pi^+\Delta(1232)^{0}$ plays a
less important role. The {\sc\small CLAS} collaboration reported a
study of the moments of the di-pion decay angular distributions and
extracted S, P, and D-waves in the 0.4 - 1.4\,GeV $\pi\pi$ mass
range \cite{Battaglieri:2008ps,Battaglieri:2009fq}

\begin{figure}[pt]
\begin{center}
\includegraphics[width=.22\textwidth]{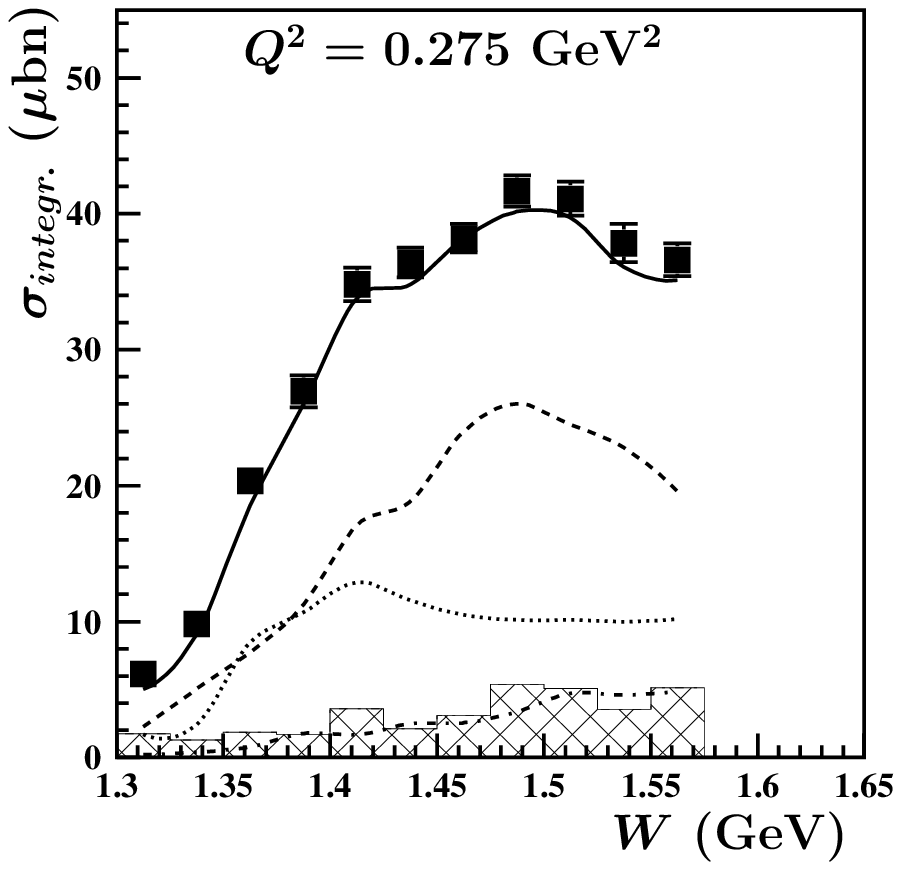}\quad
\includegraphics[width=.22\textwidth]{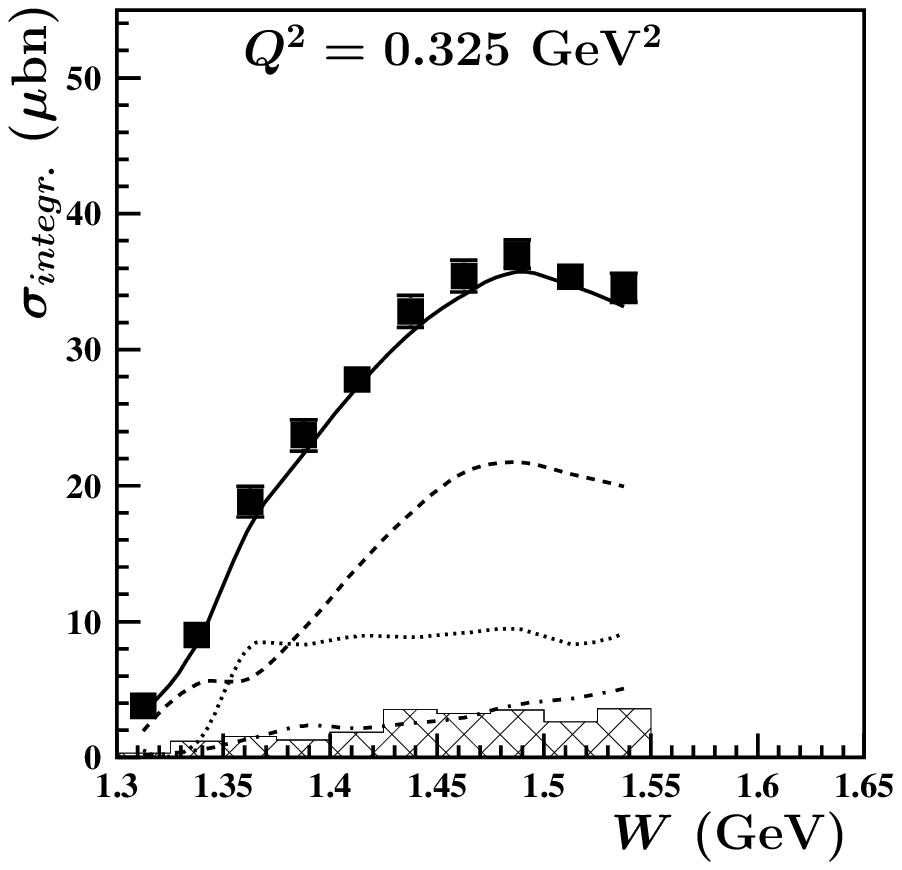}\\
\includegraphics[width=.22\textwidth]{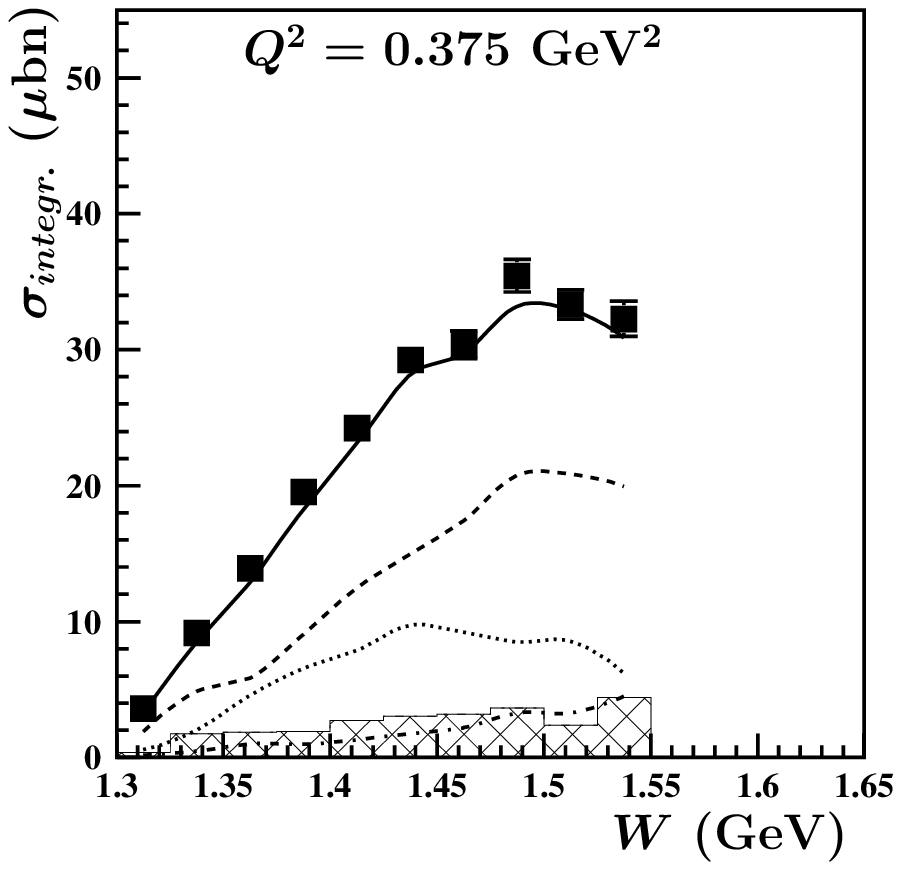}\quad
\includegraphics[width=.22\textwidth]{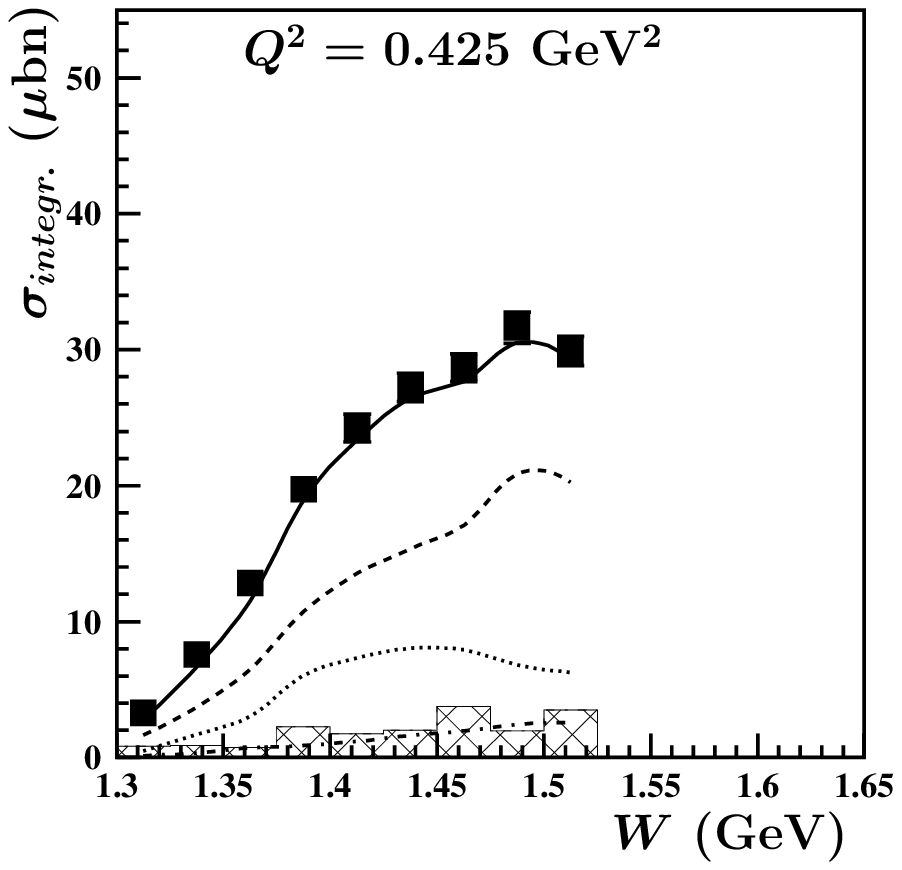}\\
\includegraphics[width=.22\textwidth]{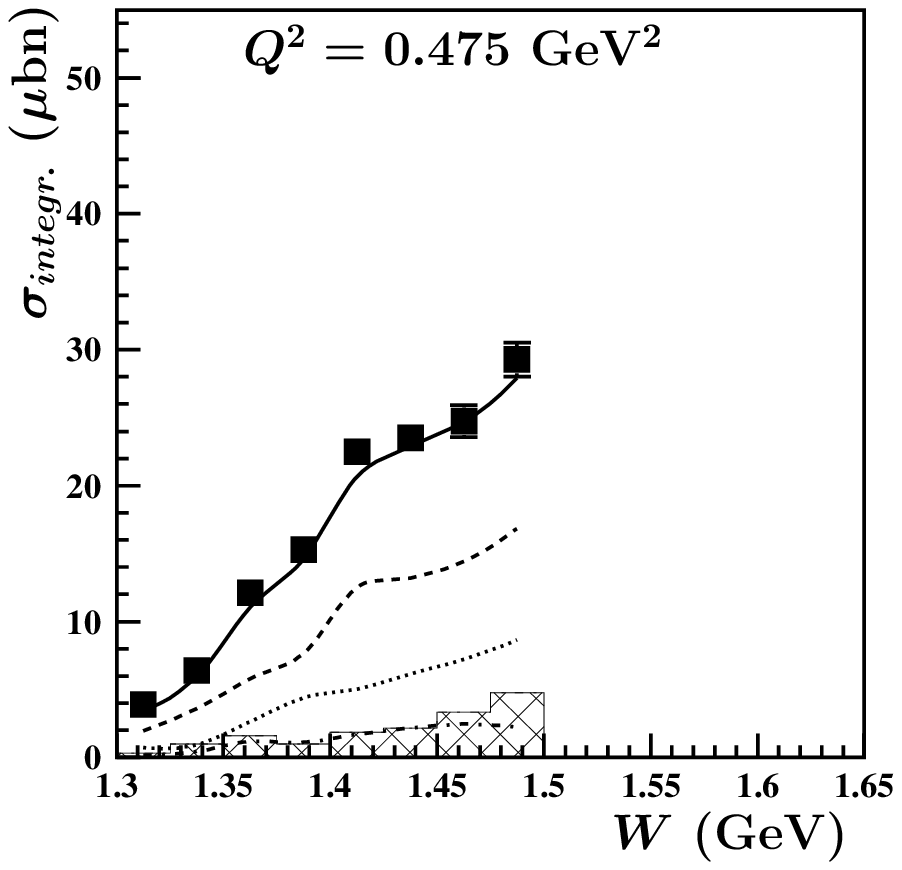}\quad
\includegraphics[width=.22\textwidth]{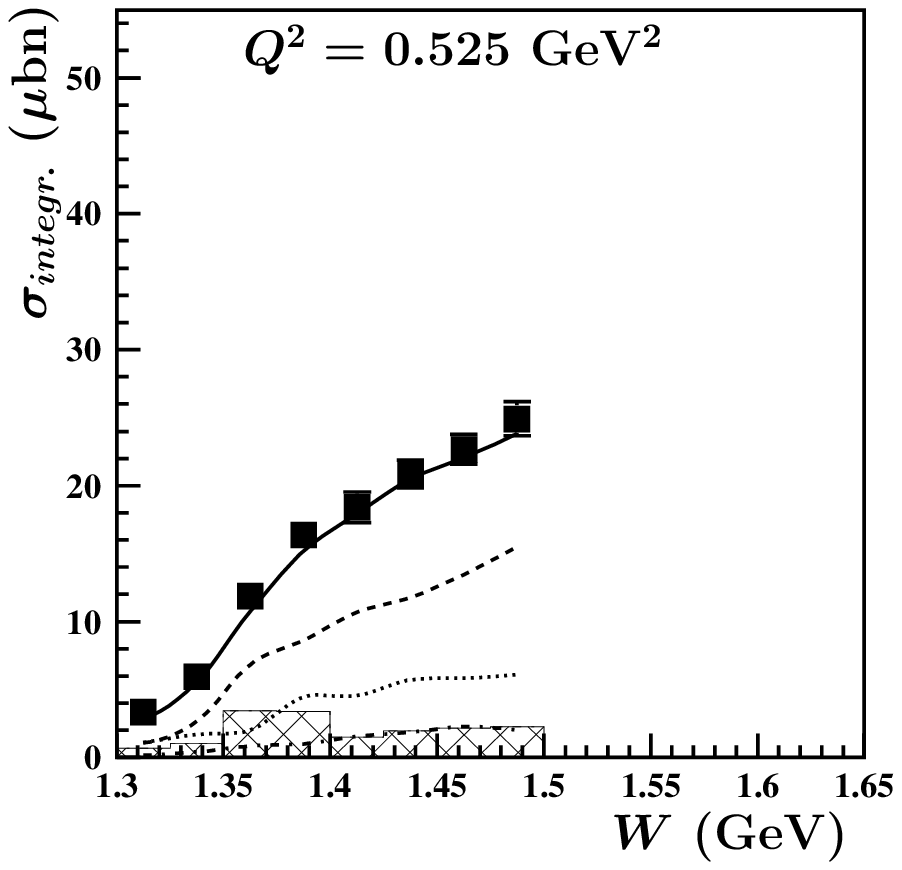}\vspace{-2mm}
\end{center}
\caption{\label{systemerr}Electro-production of $p\pi^+\pi^-$ after
integration over the full dynamics. The cross sections are
decomposed into the dominant isobar channels. The recent {\sc\small
CLAS} data \cite{Fedotov:2008gy} are shown by full symbols. Shadowed
areas represent the systematical uncertainties. The solid lines
correspond to an {\sc\small EBAC} fit (JM06) to the six 1-fold
differential cross sections \cite{Mokeev:2008iw}. The contributions
from $\pi^{-} \Delta^{++}$, $\pi^{+} \Delta^{0}$ channels are shown
by dashed and dot-dashed lines, the contributions from direct 2$\pi$
production by dotted lines, respectively.\vspace{-2mm}}
\end{figure}
Intermediate baryon resonances are much stronger in photoproduction
of two neutral pions
\cite{Ahrens:2005ia,Assafiri:2003mv,Thoma:2007bm}. The helicity
dependence of the $\gamma p\to p\pi^+\pi^-$ total cross-section was
measured at MAMI for photon energies from 400 to 800\,MeV
\cite{Ahrens:2007zzj,Krambrich:2009te}. At higher energies,
beam-helicity asymmetries were studied at Jlab
\cite{Strauch:2005cs}. Two-pion electro-production from Jlab was
reported by \cite{Ripani:2002ss}, \cite{Hadjidakis:2004zm} and, with
very high statistics, by \cite{Fedotov:2008gy}. The pion pair was
produced at photon virtualities ranging in $Q^{2}$ from 0.2 to
0.6\,GeV$^{2}$ and invariant mass $W$ from 1.3 to 1.57\,GeV. A
phenomenological analysis found non-resonant mechanisms to provide
the most significant part of the cross-section. Within the
{\sc\small EBAC} model, electrocouplings of  the $N(1440)P_{11}$ and
$N(1520)D_{13}$ states can be extracted. The present state-of-art of
the fits is described in \cite{Mokeev:2008iw}. A fraction of the
data and the most significant isobar contributions are shown in
Fig.~\ref{systemerr}.

\begin{figure}[pt]
\centerline{
\includegraphics[width=0.24\textwidth,height=0.23\textwidth,clip]{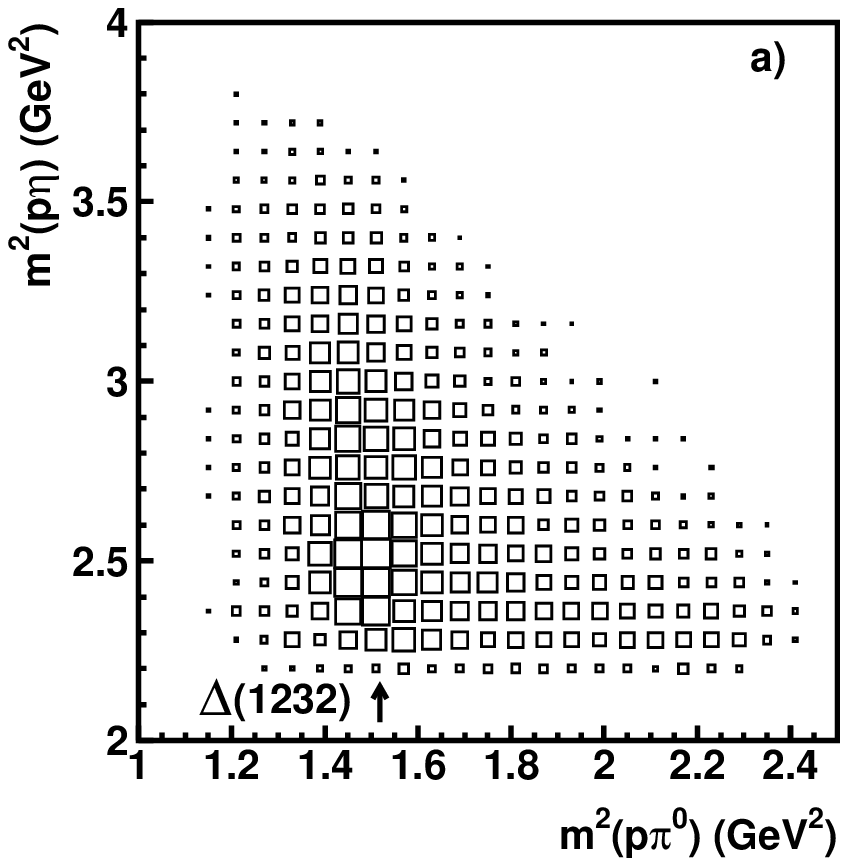}
\includegraphics[width=0.24\textwidth,height=0.23\textwidth,clip]{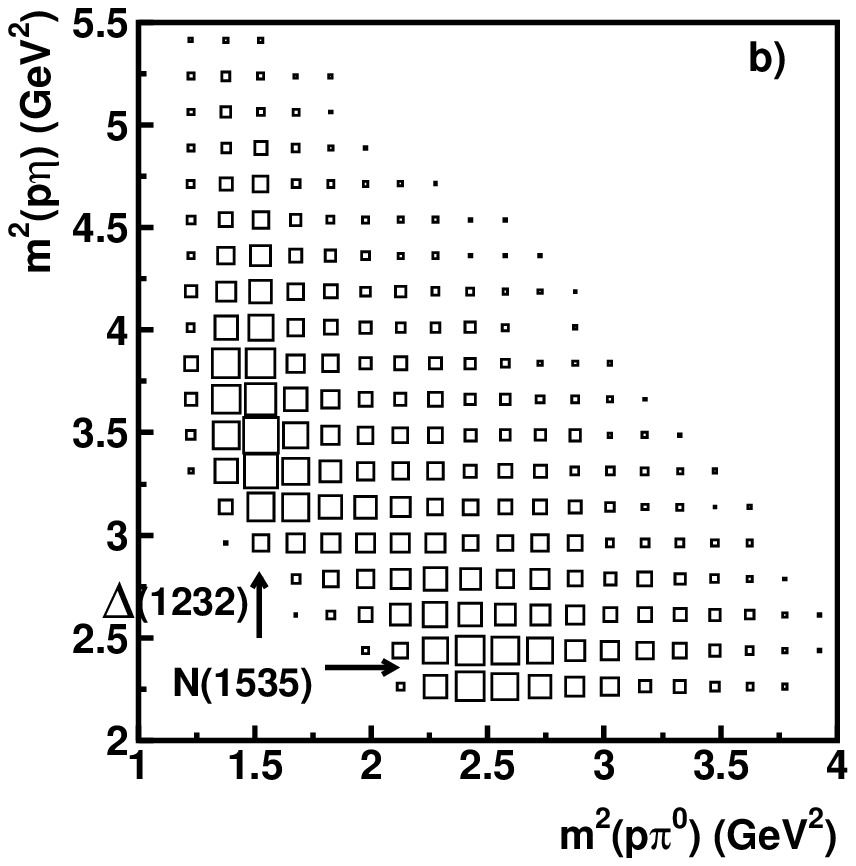}
}
 \caption{\label{Dalitz}Dalitz plot (Crystal Barrel at ELSA) for
the reaction $\gamma p\to p\,\pi^0\eta$ for $E_{\gamma}<1.9$\,GeV
(a) and $E_{\gamma}>1.9$\,GeV (b). $\Delta(1232)$ is seen in both
Dalitz plots; $N(1535)$ is visible only for high photon energies
even though the $N(1535)\pi$ production threshold ($\sim1.0$\,GeV)
is lower than the $\Delta(1232)\eta$ production threshold
($\sim1.2$\,GeV). }
\end{figure}

The reaction $\gamma p\to p\pi^0\eta$ gives access to resonances
in\hspace{-1mm} the\hspace{-1mm} $\Delta\eta$\hspace{-1mm}
system.\hspace{-2mm} The\hspace{-1mm} reaction\hspace{-1mm}
was\hspace{-1mm} studied\hspace{-1mm} at\hspace{-1mm} {\sc\small
SPring-8} \cite{Nakabayashi:2006ut}, at {\sc\small GRAAL}
\cite{Ajaka:2008zz}, at ELSA
\cite{Horn:2007pp,Horn:2008qv,Gutz:2008zz}, and at MAMI
\cite{Kashevarov:2009ww}. The $p\,\pi^0\eta$ Dalitz plots for two
different photon energy ranges are shown in Fig. \ref{Dalitz}, with
$\Delta(1232)$ and $N(1535)$ as intermediate resonances in $\gamma
p\to \left(\Delta(1232)\eta; N(1535)\pi\right)\to p\,\pi^0\eta$
cascade decays. Likewise, $\gamma p\to p\pi^0\omega$ can be used to
study the $\Delta\omega$ system. However, so far data are scarce
\cite{Junkersfeld:2007yr}.\vspace{-2mm}
 \vskip 2mm plus 1mm minus 3mm
 \subsub{Hyperon resonances and the $\Theta(1540)^+$}
 \vskip 2mm plus 1mm minus 3mm\vspace{-2mm}
 \setcounter{paragraph}{0}
In 2003, evidence for a narrow baryon resonance with positive
strangeness $\Theta(1540)^+$, i.e., with a constituent $\bar
s$-quark, was reported by four different laboratories
\cite{Nakano:2003qx,Stepanyan:2003qr,Barth:2003es,Barmin:2003vv}
with properties as predicted in a chiral soliton model
\cite{Diakonov:1997mm}. A broad search was initiated to confirm or
disprove these findings, including the search for related phenomena
like $\Phi(1860)$ ($=ssdd\bar u$) \cite{Alt:2003vb} and
$\Theta_c(3100)$ ($=uudd\bar c$) \cite{Chekanov:2004kn}. The
evidence for pentaquarks has now faded away \cite{Danilov:2008zz}
even though some evidence persists \cite{Nakano:2008ee}; memorable
remarks on the coherence of experimental findings and results from
lattice QCD and QCD sum rules can be found in \cite{Tariq:2007ck}.

Our knowledge of excited strange baryons rests nearly entirely on
$K\,N$ scattering data which are not reviewed here. The
$\Lambda_{3/2^-}(1520)$ hyperon was studied in photoproduction by
{\sc\small LEPS} \cite{Kohri:2009xe} in the threshold region, by
{\sc\small SAPHIR} \cite{Wieland:2010aaa} for photon energies below
2.65\,GeV, and in electro-production by {\sc\small CLAS} at electron
beam energies of 4.05, 4.25, and 4.46\,GeV \cite{Barrow:2001ds}. The
decay angular distributions suggest resonant contributions at low
energies, and at high energy dominance of {\it t}-channel diagrams
with either $K^+$ exchange or longitudinal coupling to an exchanged
$K^*$. The $Q^2$ dependence of the $\Lambda_{3/2^-}(1520)$
production cross section is very similar to the one observed for
photo- and electro-production of the $\Lambda$ hyperon. The reaction
$\gamma p \to K^{*0} \Sigma^+$ provides hints for a significant role
of $K_0(900)$ exchange \cite{Hleiqawi:2007ad}.

Differential cross sections for $\gamma p \to
K^+\Lambda_{1/2^-}(1405)$ and $\gamma p \to K^+\Sigma^0(1385)$ for
forward $K^+$ scattering angles have been reported for photon
energies ranging from 1.5 to 2.4\,GeV \cite{Hicks:2008yn}. The
$\Lambda_{1/2^-}(1405)$ to $\Sigma^0(1385)$ production ratio of
decreased with increasing photon energy possibly suggesting
different internal structures \cite{Niiyama:2008rt}. Cross sections
and beam asymmetries for $K^+\Sigma^{*-}$ photoproduction from the
deuteron at 1.5-2.4\,GeV were reported by \cite{Hicks:2008yn}.

\subsection{\label{Partial wave analyses} Partial wave analyses}

A discussion of problems, principles and achievements of partial
wave analysis goes beyond the scope of this paper which rather
concentrates on a review of the data which have been gathered and
the physical significance of the results. Partial wave analyses are
performed at a number of places, using different methods. Even
though small groups or individuals have made significant
contributions to the field, most partial wave analyses are performed
at a few places only.\vspace{-4mm}

\paragraph{SAID and MAID:}\qquad The longest continuous tradition is
held by the SAID group. The group maintains and updates analyses of
the elastic $\pi N$, (including $\pi d$), $KN$, $NN$ databases and
on photo- and electro-production of pseudoscalar mesons. The web
page \text{http://gwdac.phys.gwu.edu/} provides access to the data,
to partial wave amplitudes, and to current energy-dependent
predictions for observable quantities. A similar page is found at
Mainz \text{http://www.kph.uni-mainz.de/MAID/}. The most recent
solutions for $\pi N$ elastic scattering were obtained by
\cite{Arndt:2006bf}, for $KN$ elastic scattering by
\cite{Hyslop:1992cs}, for photoproduction of pions, jointly with the
most recent {\sc\small CLAS} data by \cite{Dugger:2007bt}.
Amplitudes for photoproduction of $\eta$ and $\eta'$ were determined
by \cite{Chiang:2002vq} and \cite{Briscoe:2005ek}, those for Kaon
photoproduction  by \cite{Mart:2006dk}. Principles of multi-channel
analyses are discussed by \cite{Vrana:1999nt}. The latter results
differ significantly from those of single-channel fits emphasizing
the need to include inelasticity explicitly. Electro-production
amplitudes ({\sc\small MAID-07}) were reported by
\cite{Drechsel:2007if}. The MAID and SAID data bases provide
indispensable tools for physicists working in the field. Both groups
determine masses, widths and quantum numbers mostly from $\pi N$
scattering; photoproduction data complement the information by
providing helicity amplitudes.\vspace{-4mm}

\paragraph{EBAC:}\qquad The Excited Baryon Analysis Center ({\sc\small EBAC}) has
developed a model to study nucleon resonances pion- and
photon-induced reactions \cite{Matsuyama:2006rp}. The model is based
on an energy-in\-de\-pendent Hamiltonian derived from an interaction
Lagrangian. Main results on $\pi N\to N\pi$ were communicated by
\cite{JuliaDiaz:2007kz}, on $\pi N\to N\eta$ \cite{Durand:2008es},
and on $\pi N\to N\pi\pi$ by \cite{Kamano:2008gr}. Photoproduction
of pions was studied by \cite{JuliaDiaz:2007fa} and
\cite{Sibirtsev:2007wk,Sibirtsev:2009bj,Sibirtsev:2009kw},
electroproduction of $\eta$ and $\eta'$ by \cite{JuliaDiaz:2009ww}
and of pion pairs by \cite{Mokeev:2008iw}. A common analysis of
single- and double-pion photoproduction is presented in
\cite{Kamano:2009im}. A review of the method and recent achievements
was presented by \cite{Lee:2009zzo}.\vspace{-4mm}

\paragraph{The Giessen model:}\qquad The Giessen group analyses simultaneously
pion- and photon-induced data on $\gamma N$ and $\pi N$ to $\pi N$,
$2\pi N$, $\eta N$, $K\Lambda$, $K\Sigma,$ and $\omega N$ for
energies from the nucleon mass up to $\sqrt s = 2$\,GeV. The method
is based on a unitary coupled-channel effective Lagrangian model.
The results of the partial wave analyses were reported by
\cite{Penner:2002ma,Penner:2002md},
\cite{Feuster:1997pq,Feuster:1998cj},
\cite{Shklyar:2004ba,Shklyar:2005xg,Shklyar:2006xw}, and
\cite{Shyam:2009za}.\vspace{-4mm}

\paragraph{The Bonn--Gatchina model:}\qquad The Bonn--Gatchina group
analyses large data sets, including the most recent results from
photoproduction of Kaons and multiparticle final states like
$p\pi^0\pi^0$ and $p\pi^0\eta$. The latter data are included in
event-based likelihood fits which exploit fully the information
contained in the correlations between the different variables.
Methods are described by
\cite{Anisovich:2004zz,Anisovich:2007zz,Klempt:2006sa} and results
by \cite{Anisovich:2005tf,Anisovich:2007bq,Anisovich:2008wd}, and
\cite{Thoma:2007bm,Nikonov:2007br,Sarantsev:2005tg,Sarantsev:2007bk}.
Data and on meson and baryon spectroscopy, the BnGa $\pi N$ partial
wave amplitudes, photoproduction multipoles and predictions for
observables can be found on the web page
\text{http://pwa.hiskp.uni-bonn.de/}.\vspace{-4mm}

\paragraph{Other approaches:}\qquad We further mention the analysis
of  the Gent group which describes photo- and electro-production of
hyperons in a Regge-plus-resonance approach
\cite{Corthals:2005ce,Corthals:2006nz,Corthals:2007kc}. \vspace{4mm}

A few words should be added as general remarks. Partial wave
amplitudes are constrained by a number of theoretical
considerations. First, amplitudes need to be analytic function in
the complex $s$ plane; left-hand cuts due to threshold singularities
can be treated using the N/D formalism. Amplitudes should obey
crossing symmetry; ideally, amplitudes should be defined as
functions of $s$, $t$, and $u$. In elastic $\pi N$ scattering, these
requirements are met approximately by forcing amplitudes to satisfy
fixed-$t$ dispersion relations. Amplitudes should respect chiral
symmetry. This requirement can be enforced in models using a chiral
Lagrangian or by including the $\pi N$ scattering amplitudes in the
fits. Finally, amplitudes have to preserve unitarity; the number of
incoming particles in a given partial wave, e.g. $\pi N$ in the
$J^P=3/2^+$ wave, has to be preserved. This requirement can be met
using a K-matrix in which background amplitudes and resonances can
be added in a unitarity-preserving way.

Even when the scattering amplitudes are known, the extraction of
resonance parameters from meson-nucleon and photoinduced reactions
is not easy. The physical quantity which should not depend on the
reaction mechanism is (supposedly) the pole position. Masses and
widths can be determined, e.g. in the $\pi N$ elastic scattering, by
the speed-plot or the time-delay method \cite{Suzuki:2008rp},
methods which may be more stable than parameters deduced from
Breit--Wigner parameterizations. An alternative method to define
Breit--Wigner parameters \cite{Thoma:2007bm} is to construct a
Breit--Wigner amplitude as a function of $s$ which reproduces the
pole position of the scattering amplitude. \cite{Ceci:2006jj}
suggest to derive resonance parameter from the trace of K- and
T-matrices.

Coupling constants for decays of a resonance into $A+b$ can be
determined as residues of pole of the $A+b\to A+b$ scattering
amplitude in the complex $s$-plane. The partial decay width is
usually defined as $\Gamma_{A\,b}=\rho_{A\,b}\,g_{A\,b}^2$ where
$\rho_{A\,b}$ is the phase space (including centrifugal barrier and
Blatt--Weisskopf corrections \cite{Anisovich:2005tf}), calculated at
the nominal mass and $g_{A\,b}^2$ the squared coupling constant,
again at the nominal mass. The definition has the non-intuitive
consequence that the partial decay width of a subthreshold resonance
vanishes identically even though the decay is possible via the tails
of the mother (and/or daughter) resonance. More intuitive, but in
practice less well defined, is a definition where the ratio of
partial to total width is given by the ratio of the intensity in one
channel to the intensity in all channels. One particular case are
the $N\gamma$ decays or the $A_{1/2}$ and $A_{3/2}$ helicity
amplitudes, describing the nucleon-photon coupling for a total spin
1/2 and 3/2, respectively. A thorough discussion of these
amplitudes, including the longitudinal helicity amplitude $S_{1/2}$
is given in \cite{Aznauryan:2008us}. With the definition of a
partial decay width as residue of a pole in the $\gamma N\to
N\gamma$ amplitude, helicity amplitudes become complex quantities.

The coupling of a resonance to a decay channel has an impact on its
mass. Quark model calculations usually give masses of ``stable"
baryons, of baryons before they are ``dressed with a meson cloud".
The {\sc\small EBAC} group makes the attempt to determine bare
baryon masses, masses a resonance might have before it dresses
itself with a meson cloud. In meson spectroscopy, the Gatchina group
\cite{Gatchina:2008aaa} identified the undressed states with the
K-matrix poles. However, in a dedicated study, \cite{Workman:2008iv}
did not find a simple association between K-matrix and T-matrix
poles. We believe bare masses to be highly model-dependent
quantities; the determination of the T-matrix poles is easy once the
amplitudes are known, and they should be given, at least in
addition. Finally, it is the T-matrix pole position which is given
by the PDG and which can be compared to other analyses. The future
will have to decide if dressing of quark model states or undressing
of observed resonances may become a useful concept.

\subsection{\label{Summary of N* and Delta* resonances} Summary of
$N^*$ and $\Delta^*$ resonances                   }

\begin{table*}[pt]
\caption{\label{all-light-N}Breit-Wigner masses W$_R$ and widths
$\Gamma$ (in MeV) of $N$ and $\Delta$ resonances. \vspace{-2mm}}
\begin{footnotesize}
\begin{center}
\renewcommand{\arraystretch}{1.3}
\begin{tabular}{cccccccc}
\hline\hline
Resonance &Our&\hspace{-1mm}Our\vspace{-1mm}& KH&CM&Kent&GWU&BnGa\\
&estimate&\hspace{-1mm}rating\hspace{-1mm}&\\
\hline \Nponeabf     &1450$\pm$32;\,300$\pm$100&\hspace{-1mm}****\hspace{-2mm}%
              &1410$\pm$12; 135$\pm$10&1440$\pm$30; 340$\pm$70%
              &1462$\pm$10; 391$\pm$34&1485\,$\pm$\,1; 284$\pm$18\hspace{-5mm}%
              &1440$\pm$12; 335$\pm$50   \\
\Ndthreeabf   &1522\,$\pm$\,4; 115$\pm10$&\hspace{-1mm}****\hspace{-2mm}%
              &1519\,$\pm$\,4; 114\,$\pm$\,7&1525$\pm$10; 120$\pm$15%
              &1524\,$\pm$\,4; 124\,$\pm$\,8&1516$\pm$ 1; 99$\pm$ 3\hspace{-5mm}%
              &1524\,$\pm$\,4; 117\,$\pm$\,6  \\
\Nsoneabf     &1538$\pm$10; 175$\pm$45&\hspace{-1mm}****\hspace{-2mm}%
              &1526\,$\pm$\,7; 120$\pm$20&1550$\pm$40; 240$\pm$80%
              &1534\,$\pm$\,7; 151$\pm$27&1547\,$\pm$\,1; 188$\pm$ 4\hspace{-5mm}%
              &1535$\pm$20; 170$\pm$35  \\
\Nsonebbf     &1660$\pm$18; 165$\pm$25&\hspace{-1mm}****\hspace{-2mm}%
              &1670\,$\pm$\,8; 180$\pm$20&1650$\pm$30; 150$\pm$40%
              &1659\,$\pm$\,9; 170$\pm$12&1635\,$\pm$\,1; 115\,$\pm$\,3\hspace{-5mm}%
              &1680$\pm$40; 170$\pm$45 \\
\Ndfiveabf    &1675\,$\pm$\,5; 153$\pm$22&\hspace{-1mm}****\hspace{-2mm}%
              &1679\,$\pm$\,8; 120$\pm$15&1675$\pm$10; 160$\pm$20%
              &1676\,$\pm$\,2; 159$\pm$ 7&1674\,$\pm$\,1; 147\,$\pm$\,1\hspace{-5mm}%
              &1678\,$\pm$\,5; 177$\pm$15  \\
\Nffiveabf    &1683\,$\pm$\,3; 126\,$\pm$\,9 &\hspace{-1mm}****\hspace{-2mm}%
              &1684\,$\pm$\,3; 128$\pm$ 8&1680$\pm$10; 120$\pm$10%
              &1684\,$\pm$\,4; 139$\pm$ 8&1680$\,\pm$\,1; 128\,$\pm$\,1\hspace{-5mm}%
              &1685\,$\pm$\,5; 117$\pm$12 \\
\Ndthreebbf   &1725$\pm$50; 190$\pm$110&\hspace{-1mm}***\hspace{-2mm}%
              &1731$\pm$15; 110$\pm$30&1675$\pm$25; 90$\pm$40%
              &1737$\pm$44; 250$\pm$230
              &-
              &1730$\pm$40; 310$\pm$60  \\
\Nponebbf &1713$\pm$12;220$\pm$180&\hspace{-1mm}***\hspace{-2mm}%
              &1723\,$\pm$\,9; 120$\pm$15&1700$\pm$50; 90$\pm$30%
              &1717$\pm$28;\,480$\pm$330& -
              &1725$\pm$25;200$\pm$35\\
\Npthreeabf   &1730$\pm$30; 320$\pm$210&\hspace{-1mm}****\hspace{-2mm}%
              &1710$\pm$20; 190$\pm$30&1700$\pm$50; 125$\pm$70%
              &1717$\pm$31; 380$\pm$180&1750\,$\pm$\,5; 256$\pm$22\hspace{-5mm}%
              &1770$\pm$100; 650$\pm$120  \\
\underline{\Ndthreecbf}&1850$\pm$40;\,260$\pm$170&\hspace{-1mm}**\hspace{-1mm}&-
              &\hspace{-1mm}1880$\pm$100;\,180$\pm$60%
              &1804$\pm$55; 450$\pm$185&-&1870$\pm$25;150$\pm$40 \\
\underline{\Nffivebbf}&1880$\pm$40;\,270$\pm$180&\hspace{-1mm}$\overline{**}$\hspace{-1mm}%
              &1882$\pm$10; 95$\pm$20&-&1903$\pm$87;\,490$\pm$310%
              &1818; 118&1910$\pm$50; 360$\pm$80 \\
\underline{\Nponecbf}&1890$\pm$50;\,210$\pm$100&*\hspace{-1mm}%
              &-&-&1885$\pm$30;\,113$\pm$44&-&1900$\pm$30; 300$\pm$40 \\
\Npthreebbf   &1940$\pm$50;\,340$\pm$150&*\hspace{-1mm}%
              &-&-&1879$\pm$17; 498$\pm$78&-
              &1960$\pm$30; 185$\pm$40\\
\underline{\Nsonecbf}&1905$\pm$50;\,250$\pm$150&*\hspace{-1mm}%
              &1880$\pm$20; 95$\pm$30&-&1928$\pm$59;\,414$\pm$157%
              &-&- \\
\Nfsevenabf   &2020$\pm$60;\,410$\pm$110&\hspace{-1mm}**\hspace{-1mm}%
              &2005$\pm$150;\,350$\pm$100&1970$\pm$50;\,350$\pm$120%
              &2086$\pm$28;\,535$\pm$120& - & -\\
\Ndthreedbf   &2100$\pm$55;\,310$\pm$110%
              &\hspace{-1mm}$\overline{**}$\hspace{-2mm}&2080$\pm$20; 265$\pm$40&2060$\pm$80;\,300$\pm$100%
              &-& - &2160$\pm$35; 370$\pm$50 \\
\Nsonedbf     &&%
              &&2180$\pm$80;\,350$\pm$100%
              &-&-& -                      \\
\Nponedbf     &\hspace{-1mm}2090$\pm$100;\,230$\pm$200%
              &*&2050$\pm$20; 200$\pm$30%
              &2125$\pm$75;\,260$\pm$100&-&-& - \\
\Ndfivebbf    &2160$\pm$85;\,350$\pm$50%
              &**\hspace{-1mm}&2228$\pm$30; 310$\pm$50&2180$\pm$80;\,400$\pm$100%
              &-&-&2065$\pm$25; 340$\pm$40\\
              \hline
              &&& KH&CM&Kent&GWU&Hendry\vspace{1mm}\\ \hline
\Ngsevena     &2150$\pm$30; 440$\pm$110%
              &\hspace{-1mm}****\hspace{-2mm}&2140$\pm$12; 390$\pm$30&2200$\pm$70;\,500$\pm$150%
              &2127$\pm$ 9; 550$\pm$50& 2152$\pm$2; 484$\pm13$%
              &2140$\pm$ 40; 270$\pm$ 50\\
\Nhnineabf    &2260$\pm$60;\,490$\pm$115
              &\hspace{-1mm}****\hspace{-2mm}&2205$\pm$10; 365$\pm$30&2230$\pm$80;\,500$\pm$150%
              &-&2316$\pm$3;\,633$\pm$17&2300$\pm$100;\,450$\pm$150 \\
\Ngnineabf    &2255$\pm$50;\,420$\pm$150&\hspace{-1mm}****\hspace{-2mm}%
              &2268$\pm$15;\,300$\pm$40&2250$\pm$80; 400$\pm$120%
              &-&2302$\pm$6;\,628$\pm$28%
              &2200$\pm$100;\,350$\pm$100 \\
\hspace{-1mm}\Nielevenabf  &\hspace{-1mm}2630$\pm$120;\,650$\pm$250%
              &\hspace{-1mm}**\hspace{-2mm}&2577$\pm$50;400$\pm$100&-&-&-
              &2700$\pm$100;\:900$\pm$100 \\
\hspace{-1mm}\Nkthirteenabf&\hspace{-1mm}2800$\pm$160;\,600$\pm$300%
              &\hspace{-1mm}**\hspace{-1mm}\hspace{-2mm}&2612$\pm$45; 350$\pm$50&-&-&-
              &3000$\pm$100;\,900$\pm$150 \\
\hline
       &&      & KH&CM&Kent&GWU&BnGa\\            \hline
\Dpthreeabf    &1232\;$\pm$\;1; 118\;$\pm$\;2&\hspace{-1mm}****\hspace{-2mm}%
               &1232\;$\pm$\;3; 116$\;\pm$\;5&1232\;$\pm$\;2; 120\;$\pm$\;5%
               &1231\;$\pm$\;1; 118\;$\pm$\;4&1233\,$\pm$\,1; 119\,$\pm$\,1%
               &1230\;$\pm$\;2; 112\;$\pm$\;4\\
\Dpthreebbf    &1615$\pm$80;\,360$\pm$120&\hspace{-1mm}***\hspace{-2mm}%
               &1522$\pm$15; 220$\pm$40&1600$\pm$50;\,300$\pm$100%
               &1706$\pm$10; 430$\pm$73               &-
               &1640$\pm$40;\,480$\pm$100\\
\Dsoneabf      &1626$\pm$23; 130\,$\pm$\,45&\hspace{-1mm}****\hspace{-2mm}%
               &1610\,$\pm$\,7; 139$\pm$18&1620$\pm$20; 140$\pm$20%
               &1672\,$\pm$\,7; 154$\pm$37&1614\,$\pm$\,1; 71\,$\pm$\,3%
               &1625$\pm$10; 148$\pm$15\\
\Ddthreeabf    &1720$\pm$50;\,370$\pm$200&\hspace{-1mm}****\hspace{-2mm}%
               &1680$\pm$70; 230$\pm$80&1710$\pm$30; 280$\pm$80%
               &1762$\pm$44; 600$\pm$250&1688\,$\pm$\,3; 182\,$\pm$\,8%
               &1780$\pm$40; 580$\pm$120\\
\Dponeabf&&    &-&-
               &1744$\pm$36;\,300$\pm$120& -
               &-\\
\Dsonebbf      &1910$\pm$50; 190$\pm$100&\hspace{-1mm}**\hspace{-1mm}%
               &1908$\pm$30; 140$\pm$40&1890$\pm$50; 170$\pm$50%
               &1920$\pm$24; 263$\pm$39& -
               &-\\
\Dffiveabf   &1885$\pm$25; 330\,$\pm$\,50&\hspace{-1mm}****\hspace{-2mm}%
               &1905$\pm$20; 260$\pm$20&1910$\pm$30;\,400$\pm$100%
               &1881$\pm$18; 327$\pm$51& 1856\,$\pm$\,2; 321$\pm$ 9%
               &1870$\pm$32;\,340$\pm$32\\
\Dponebbf      &1935$\pm$90; 280$\pm$150&\hspace{-1mm}****\hspace{-2mm}%
               &1888$\pm$20; 280$\pm$50&1910$\pm$40; 225$\pm$50%
               &1882$\pm$10; 229$\pm$25&2068\,$\pm$\,2; 543$\pm$10%
               &-\\
\Dpthreecbf    &1950$\pm$70; 260$\pm$100&\hspace{-1mm}***\hspace{-2mm}%
               &1868$\pm$10; 220$\pm$80&1920$\pm$80;\,300$\pm$100%
               &2014$\pm$16; 152$\pm$55& -
               &1995$\pm$40; 360$\pm$50\\
\Ddfiveabf     &1930$\pm$30;\,350$\pm$170 &\underline{**}%
               &1901$\pm$15; 195$\pm$60&1940$\pm$30; 320$\pm$60%
               &1956$\pm$22;\,530$\pm$140& -
               &-\\
\Ddthreebbf    &1995$\pm$60;\,340$\pm$130&\hspace{-1mm}$\overline{**}$\hspace{-2mm}%
               &-&\hspace{-1mm}1940$\pm$100;200$\pm$100%
               &\hspace{-1mm}2057$\pm$110;\,460$\pm$320\hspace{1mm}& -
               &1995$\pm$40; 360\,$\pm$\,50\\
\Dfsevenabf    &1930$\pm$16; 285\,$\pm$\,45&\hspace{-1mm}****\hspace{-2mm}%
               &1913\,$\pm$\,8; 224$\pm$10&1950$\pm$15; 340$\pm$50%
               &1945\,$\pm$\,2; 300\,$\pm$\,7& 1921$\pm$ 1; 271\,$\pm$\,1%
               &1928$\pm$8; 290$\pm$14\\
\Dffivebbf     &&&\hspace{-1mm}2200$\pm$125;400$\pm$125&-&1752$\pm$32; 251$\pm$93&- &-\\
\Dsonecbf      &&&-&\hspace{-1mm}2200$\pm$100;\,200$\pm$100 & -
&-&-\\\hline
               &&& KH&CM&Kent&GWU&Hendry\vspace{1mm}\\ \hline
\Dgsevenabf    &2230\,$\pm$\,50;\,420$\pm$100&\hspace{-1mm}**\hspace{-1mm}%
               &2215$\pm$10;\,400$\pm$100&2200$\pm$80;\,450$\pm$100%
               &-&-
               &2280$\pm$80;\,400$\pm$150\\
\Dhnineabf     &2360$\pm$125;\,420$\pm$200&\hspace{-1mm}**\hspace{-1mm}%
               &2217$\pm$80;\,300$\pm$100&2400$\pm$125;\,425$\pm$150%
               &-&-      &\hspace{-1mm}2450$\pm$100;\,500$\pm$200\\
\Ddfivebbf     &2310\,$\pm$\,85; 490$\pm$250&\hspace{-1mm}***\hspace{-2mm}%
               &2305$\pm$26; 300\,$\pm$\,70&2400$\pm$125;\,400$\pm$150%
               &-&\hspace{-5mm}2233$\pm$53;\,773$\pm$187\hspace{-3mm}
               &-\\
\Dfsevenbbf    &\hspace{-1mm}2390$\pm$100;\,300$\pm$200&*%
               &2425\,$\pm$\,60;\,300\,$\pm$\,80&2350$\pm$100;\,300$\pm$100%
               &-& -&-\\
\Dgnineabf     &\hspace{-1mm}2400$\pm$190;\,530$\pm$300&\hspace{-1mm}**\hspace{-1mm}%
               &2468$\pm$50;480$\pm$100&2300$\pm$100;\,330$\pm$100%
               &-&\hspace{-5mm}2643$\pm$141;\,895$\pm$432\hspace{-2mm}
               &2200$\pm$100;\,450$\pm$200 \\
\hspace{-1mm}\Dhelevenabf
&2462$\pm$120;\,490$\pm$150&\hspace{-1mm}***\hspace{-2mm}%
               &2416\,$\pm$\,17; 340$\pm$28&2400$\pm$125;\,450$\pm$150%
               &-&\hspace{-5mm}2633$\pm$29; 692\,$\pm$\,47\hspace{-3mm}
               &2400$\pm$60;\,460$\pm$100\\
\hspace{-1mm}\Dithirteenabf
&\hspace{-1mm}2720$\pm$100;\,420$\pm$200&\hspace{-1mm}**\hspace{-1mm}%
               &2794$\pm$80;\,350$\pm$100&-
               &-& -               &2650$\pm$100;\,500$\pm$100\\
\hspace{-1mm}\Dkfifteenabf
&\hspace{-1mm}2920$\pm$100;\,500$\pm$200&\hspace{-1mm}**\hspace{-1mm}%
               &\hspace{-1mm}2990$\pm$100;\,330$\pm$100&-&-
               &-&2850$\pm$100;\,700$\pm$200\\
\hline\hline
\end{tabular}
\renewcommand{\arraystretch}{1.0}
\end{center}
\end{footnotesize} \vspace{-5mm}
\end{table*}
The Review of Particle Properties of the PDG \cite{Amsler:2008zz} is
indispensable for any physicist working in nuclear and particle
physics, and also in this review frequent use has been made of it.
In baryon spectroscopy, listings of main properties of resonances
are given and a selection is made which data are used to define the
properties, which data are listed but not used for averaging and
which results to not warrant to be mentioned. Based on these
results, a status is defined, with 4 stars given to a resonance with
certain existence and fairly well defined properties, 3-star
resonances are almost certain but some parameters are less well
defined. A resonance is given 2 stars if the evidence for its
existence is fair and 1 star, if it is poor. The judgement is
dominantly based on analyses from \cite{Hohler:1979yr},
\cite{Cutkosky:1979fy} -- updated in \cite{Cutkosky:1980rh} --,
\cite{Manley:1992yb}, and \cite{Arndt:2006bf}.

We suggest here ``our own" version of the PDG Listings by including
the results of the Bonn--Gatchina analysis \cite{Anisovich:2009zy}.
So far, results from photoproduction were not yet used to estimate
the status of a resonance or to determine mass or width. The reason
for this decision is the following one: unlike $\pi N$ elastic
scattering, it is -- at least so far -- not possible to derive
energy-independent partial wave amplitudes from photoproduction
data. For an independent observer, it is very difficult to judge how
reliable a fit to data is, and if alternative solutions exists in
which a particular resonance is not needed. However, in the most
recent analysis of the Bonn--Gatchina group, the same amplitudes are
used as in \cite{Arndt:2006bf}. The BnGa differs by constraining the
amplitudes of the SM06 solution by data on photoproduction. In
previous analyses, the inelasticity of baryon resonances are mostly
unknown and are fitted as free unconstrained parameters of the fit.
Constraining the SM06 amplitudes by known inelasticities can only
improve our knowledge.

In Table \ref{all-light-N} we list the $N^*$ and $\Delta^*$
resonances, give our estimate for mass and width and our rating.
Results from five analyses are given. Four new resonances are
suggested which are underlined.
\begin{enumerate}
\item The \Ndthreecbf\ is found in the PDG listings under the entry
$N(2080)D_{13}$ (\Ndthreedbf). It is observed at this mass in the KH
analysis; CM suggest two states, here we list both under the two
headings. Kent confirmed the lower-mass state at 1804\,MeV. In the
BnGa analysis, it assumes a mass of 1875\,MeV. \Ndthreecbf\ is not
seen by KH nor by GWU and we give it a 2-star status.
\vspace{-2mm}\item A second newly introduced resonance is \Nponecbf.
Evidence comes from the Kent and BnGa analyses. \vspace{-2mm}\item
\Nffivebbf\ replaces the PDG entry $N(2000)F_{15}$ (\Nffivecbf). It
is seen in all but the CM analysis and we rate it with 3 stars.
\vspace{-2mm}\item \Nsonecbf\ was reported by KH and Kent. In PDG,
the two results are combined with the CM result (2180\,MeV) to give
$N(2090)S_{11}$.
\end{enumerate}

The five analyses listed in Table \ref{all-light-N} are used to
determine our rating. Resonances get 4 stars if seen in four
experiments, including the GWU analysis. One star is subtracted, if
it is not seen in the GWU analysis; two stars are assigned if seen
in three, one star if seen by two analyses. Resonances included in
the PDG which are seen only by one of the five analyses, are kept in
Table \ref{all-light-N} but with no star. For those, no mass or
width estimate is given, and they are not considered in section
\ref{se:mod}. In some cases, the ratings differ from PDG; in case of
up- (down-) graded resonances, the star rating is over- (under-)
lined. The mass region above 2.5\,GeV was studied in the KH and
Hendry analysis only; we keep their PDG rating.

Mass and width are estimated from the spread of results rather than
from the quoted errors. As a rule, we do not give extra weight to
analyses quoting smaller errors. Mostly, small errors indicate that
correlations with other variables are not sufficiently explored. For
two-star resonances we give a minimum error of $\pm$3\% on the mass,
for one-star resonances of $\pm$5\%. The width error we assign is
minimally about twice larger than the error in mass.

\markboth{\sl Baryon spectroscopy } {\sl Models and phenomenology}
\section{Models and phenomenology}
\label{se:mod}
\subsection{Historical perspectives}
\subsub{SU(3) symmetry}
The main concern of baryon spectroscopy in the late sixties was to
analyze the meson--baryon interaction and to understand the pattern
of the many nucleon and $\Delta$ resonances, and the relation
between these baryons and the strange baryons, $\Lambda$, $\Sigma$,
$\Xi$, and their excitations. The dynamical mechanism proposed to
generate these resonances was the meson--nucleon interaction: it
accounted, e.g., for the $\Delta$ resonance in the $\pi-N$ system,
but failed to predict most of the other states.

Then came flavor symmetry, based on  the group SU(3), from now on
called SU(3)$_\text{f}$, and  its ``eightfold way''  version. The
lowest mass baryons, with spin $S=1/2$, form an octet $(N,\,
\Lambda,\, \Sigma,\,\Xi$). The baryons with $S=3/2$ are in a
decuplet which, in 1962, included $\Delta(1232),\,
\Sigma_{3/2^+}(1385)$ and $\Xi_{3/2^+}(1530)$ (named $\Sigma^*$ and
$\Xi^*$ at that time). One state was missing. The regular mass
spacing between $\Delta(1232), \Sigma_{3/2^+}(1385)$ and
$\Xi_{3/2^+}(1530)$ was used to predict the existence and the mass
of the $\Omega(1672)$ baryon \cite{GellMann:1962xb}, with
strangeness $\mathrm{S}=-3$. Its experimental discovery
\cite{Barnes:1964pd} was a triumph for SU(3)$_\text{f}$.

It was then realized that, if SU(3)$_\text{f}$ is taken seriously, there
are three states in the fundamental representation, $3$, named
quarks,  and the actual baryons correspond to the flavor
representations found in the $3\times3\times 3$ product.  This was
the beginning of the quark model, first a tool for building the
SU(3)$_\text{f}$ representations, and then becoming a dynamical model.

Today, SU(3)$_\text{f}$  is understood from the universal character
of the quark interaction (flavor independence) and the approximate
equality of the masses of light and strange quarks. SU(3)$_\text{f}$
remains a valuable tool to correlate data in different flavor
sectors and organize the hadron multiplets.\vspace{-2mm}

\subsub{SU(6) symmetry}\vspace{-2mm}
The group SU(6) combines SU(3)$_\text{f}$ with the spin group SU(2).
For instance, the octet baryons with $S=1/2$ and the decuplet
baryons with $S=3/2$ form a 56 representation of  SU(6).  This SU(6)
symmetry emerges automatically in potential models with flavor
independent forces, in the limit where the strange quark mass $m_s$
is equal to that or ordinary quarks, and the spin-dependent forces
are neglected.

Further symmetry schemes have been proposed to analyze the baryon spectrum and  properties. See, for instance, \cite{Bijker:1994yr,Bijker:2000gq,Kirchbach:2001de}.

\subsub{Early models}
The harmonic oscillator model, to be discussed shortly as well as
some of its many refinements, enables to account explicitly for
SU(3)$_\text{f}$ and SU(6) symmetry and their violation, and was crucial to
assess the quark model not only as a mathematical tool to generate
the actual representation out of the fundamental ones, but to
understand the pattern of radial and orbital excitations. More
refined constituent models were proposed later.

More recently, attempts were made to derive the baryon masses and
properties directly from QCD, by sum rules or lattice simulations:
the results are very encouraging, but often restricted to the lowest
levels.
\subsub{Heavier flavors}\vspace{-2mm}
The discovery of charm and beauty enriched significantly the
spectrum of hadrons. The quark model gained in credibility by the
success of potentials fitting the $J/\psi$ and $\Upsilon$
excitations. The problem was to combine these new states in the
existing schemes.

The extension of SU(3)$_\mathrm{f}$ to SU(4)$_\mathrm{f}$ or beyond is straightforward but not
very useful, as the symmetry is largely broken.  However, with the
advent of QCD, the ideas have evolved. The basic coupling, that of
gluons to quarks, is linked to the color, not to the flavor. Hence,
at least in the static limit, the quark--quark interaction should be
\emph{flavor independent} in the same way as in the physics of
exotic atoms, the very same Coulomb potential binds electrons,
muons, kaons and antiprotons.

Flavor independence is probed in various ways: the same ``funnel''
potential (Coulomb $+$ linear) simultaneously fits the charmonium
and bottomonium spectrum in the meson sectors. For baryons,
regularities are also observed, which supports a picture with a
flavor independent  confinement and flavor symmetry broken through
the quark masses entering the kinetic energy and the spin-dependent
corrections. For instance, there is a very smooth evolution of
hyperfine splittings from $\Delta-N$ to $\Sigma_b^*-\Sigma_b$.

It would of course be very appealing to describe all baryons within
in a universal model, the light quark requiring only  relativistic
corrections due to their light mass. This is for instance the spirit
of the work by \cite{Godfrey:1985xj,Capstick:1986bm}. The success of
this model is almost embarrassing, as QCD guides our intuition
toward drastic differences between heavy and light quarks. Heavy
quarks interact by exchanging gluons. On the other hand, the
dynamics of light quarks is dominated by chiral symmetry, which
seems hardly reducible to a local potential.\vspace{-2mm}
\subsub{The role of color}\vspace{-2mm}
One of the main motivations for introducing color was to account for
the antisymmetrization of the quarks in baryons
\cite{Greenberg:1964pe}. In the harmonic oscillator  and its various
developments, the quarks in $N$, $\Delta$, $\Omega^-$, etc., are in
a symmetric overall S-wave, and the spin--isospin part is also
symmetric. An antisymmetric $3\times3\times3\to1$ coupling of color
ensures Fermi statistics.

Then, in this color scheme, a quark in a baryon sees a color
$\bar{3}$ set of two quarks, which is analogous to the antiquark
seen by a quark in an ordinary meson.  This is the beginning of the
diquark idea which will be discussed below.

QCD gives a picture where the quarks interact moderately at short
distances, according to  ``asymptotic freedom'', and more strongly
at large distances, where a linear confinement is suggested by many
studies, though not yet rigorously proved.  The question is whether
a Coulomb plus linear potential mimics QCD well enough so that
reliable predictions can be done. A related question is whether the
interaction among quarks in baryons is of pairwise nature.

Another problem, raised in the late 70s in papers dealing with
``color chemistry'' \cite{Chan:1978nk}, is whether the color
representations used by hadrons are restricted to $3$ (quarks, antidiquarks),
$\bar{3}$ (antiquarks, diquarks) and $1$ (hadrons). Namely is the
octet, which corresponds to gluons, restricted to the crossed
channel, i.e., used only to mediate the interaction, or does it play
a constituent role (glueballs, hybrid mesons and baryons)? Are there
multiquark states containing color-sextet or color-octet clusters?
Experimental evidence for the existence of hadrons with ``hidden
color" in the pre-LEAR area was overruled in high-statistics
experiments in the early phase of LEAR \cite{Walcher:1989qp}.
\subsection{Models of ground-state baryons}
\subsub{Potential models}
The simplest model consists of
\begin{equation}\label{mod:eq:Hpot}
H=\sum_{i=1}^3\frac{\vec{p}_i^2}{2 m_i}+V(\vec{r}_1,\vec{r}_2,\vec{r}_3)-\frac{(\sum_i \vec{p}_i)^2}{2\sum_i m_i}~,
\end{equation}
where $V$ is a suitable translation-invariant interaction, the best
known choice being the harmonic oscillator
\begin{equation}\label{mod:eq:Vho}
V(\vec{r}_1,\vec{r}_2,\vec{r}_3)=\frac{2 K}{3}\sum_{i<j} r_{ij}^2~,
\end{equation}
where $r_{ij}=|\vec{r}_j-\vec{r}_i|$. The ground state is the
minimum of $H$, which can be reached for instance by variational
methods. For equal masses $m_i=m$,  one can introduce the Jacobi coordinates
\begin{equation}\label{mod:eq:Jac}
\vec{\rho}=\vec{r}_2-\vec{r}_1~,
\quad
\vec{\lambda}=\frac{2\vec{r}_3-\vec{r}_1-\vec{r}_2}{\sqrt3}~,
\end{equation}
and minimize approximately (\ref{mod:eq:Hpot}) with
the Gaussian trial wave function
\begin{equation}\label{mod:eq:howf}
\Psi_0(\vec{\rho},\vec{\lambda})=\left(\frac{\alpha^2}{\pi^2}\right)^{3/4}\exp\left[-\frac{\alpha}{2}\left(\vec{\rho}^2+\vec{\lambda}^2\right)\right]~,
\end{equation}
which is the exact solution for (\ref{mod:eq:Vho}) provided
$\alpha=\sqrt{Km}$.

For the spin $S=3/2$ baryons, this symmetric orbital wave function is associated with a symmetric isospin wave function and a symmetric spin state such as $| \uparrow\uparrow\uparrow\rangle$.

For the nucleon, a mixed-symmetric spin doublet (here for $S_z=+1/2$,
\begin{equation}\label{mod:eq:spin-half}
S_{\rho,\lambda}=\left\{\frac{|\uparrow\downarrow\uparrow\rangle-|\downarrow\uparrow\uparrow\rangle}{\sqrt2},\frac{2\uparrow\uparrow\downarrow\rangle-|\downarrow\uparrow\uparrow\rangle-|\uparrow\downarrow\uparrow\rangle}{\sqrt6}\right\},
\end{equation}
is combined to an isospin doublet (here for proton)
\begin{equation}\label{mod:eq:isospin-half}
I_{\rho,\lambda}=\left\{\frac{(udu)-(duu)}{\sqrt2},
\frac{2(uud)-(duu)-(udu)}{\sqrt 6}\right\},
\end{equation}
in a spin--isospin wave function
\begin{equation}\label{mod:eq:spin-isospin-N}
(S_\lambda I_\lambda+ S_\rho I_\rho)/ \sqrt{2}
\end{equation}
which is symmetric under permutations. The extension to unequal
masses is straightforward.

It is amazing that  simple potential models provide a good survey of
ground-state baryons with various flavor content. If the potential
$V$ is taken as being \emph{flavor independent}, as suggested by
QCD, then the Schr{\"o}dinger equation exhibits regularity and
convexity properties \cite{Richard:1992uk,Nussinov:1999sx}.  For
instance,
 \begin{equation}\label{mod:eq:conv}
 M_{QQq}+M_{qqq}< 2 M_{Qqq} \quad\text{if}\quad Q\neq q~.
 \end{equation}\vspace{-2mm}
\subsub{From mesons to baryons}\vspace{-2mm}
In most papers dealing with potential models of baryons, a pairwise
interaction is assumed,
\begin{equation}
\label{mod:eq:paiwise}
V(\vec{r}_1,\vec{r}_2,\vec{r}_3)=\frac{1}{2}\sum_{i<j} v(r_{ij})~,
\end{equation}
for instance $v(r)=\sigma r -a/r +b$. It is then argued
\cite{PhysRev.139.B1006,Stanley:1980fe,Greenberg:1981xn,Richard:1980tg}
that the potential between two quarks in a baryon is half the
quark--antiquark potential in a meson. This result is exact for the
one-gluon-exchange potential, or more generally, any color-octet
exchange, which contains an explicit
$\tilde{\lambda}_i.\tilde{\lambda}_j$ color operator, with
expectation values $-16/3$ for $3\times\bar{3}\to 1$ and  $-8/3$ for
$3\times 3\to \bar{3}$. This ``1/2'' rule also holds if two quarks
are close together and seen by the third one as a localized
$\bar{3}$ source which is equivalent to an antiquark.  More
generally, the $t$-channel color structure of $v$ contains a singlet
and an octet. The singlet cannot contribute to confinement,
otherwise all quarks of the universe would be tightly  bound. The
simplest ansatz is to assume a pure color octet exchange, and this
is why a factor $1/2$ is introduced in Eq.~(\ref{mod:eq:paiwise}).

With this ``1/2'' rule, amazing Hall--Post type of inequalities can
be derived between meson and baryon ground states masses
\cite{Richard:1992uk}. The simplest is for spin-averaged mass values
\be
M_{Q\overline{Q}}/2 \le  M_{QQQ}/3,
\ee
satisfied by, e.g., $\phi(1020)$ and $\Omega^-(1672)$.

\begin{figure}[!!htpc]
\includegraphics[width=.15\textwidth]{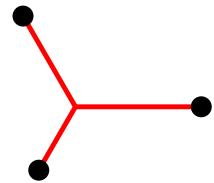}
\caption{\label{qcd:fig:Y} Three-quark confinement in the string limit.}
\end{figure}

However, QCD suggests that the linear potential $v(r)=\sigma\, r$ acting on the
quark--antiquark pair of mesons is not generalized as $\sigma \sum
r_{ij}/2$ in baryons, but by the so-called $Y$-shape potential
\be\label{mod:eq:Yshape}
V(\vec{r}_1,\vec{r}_2,\vec{r}_3)=\sigma \min(d_1+d_2+d_3),
\ee
where $d_i$ is the distance of a junction to the $i^\text{th}$
quark. Adjusting the location of the junction corresponds to the
problem of Fermat and Torriccelli, whose generalization to more than
three terminals is called the minimal Steiner-tree problem. If an
angle of triangle is larger than $120^\circ$,
then the junction coincides with this vertex, otherwise it views each
side under $120^\circ$, as  shown in Fig.~\ref{qcd:fig:Y}.
Unfortunately, $V$ given by the $Y$-shape (\ref{mod:eq:Yshape})
differs little from the result of  the ``1/2'' rule, and one cannot
probe this three-body dynamics from the baryon spectrum. The
difference between the additive model $V\propto \sum
\tilde{\lambda}_i.\tilde{\lambda}_j v(r_{ij})$ and the minimal-path
ansatz (Steiner tree) becomes more dramatic in the multiquark sector
\cite{Vijande:2007ix}.\vspace{-2mm}
\subsub{Hyperfine forces}\vspace{-2mm}
To explain why the $\Delta$ with spin 3/2 is above the nucleon of
spin 1/2, and similarly  $\Sigma^*>\Sigma$, $\Xi^*>\Xi$, etc.,
the spin-independent potential $V$ has to be supplemented by a
spin--spin term, which is usually treated at first order, but
sometimes non-pertubatively, after suitable regularization.
\paragraph{Chromomagnetism}
The most popular model is the one-gluon-exchange
\cite{DeRujula:1975ge}, inspired by the Breit--Fermi term of QED. A
slightly more general formulation involves a \emph{chromomagnetic}
interaction of the form
\begin{equation}\label{mod:eq:chromo}
V_\text{CM} = \sum_{i<j} \frac{\tilde\lambda_i^{c)}.\lambda_j^{(c)} \,\vec{\sigma}_i.\vec{\sigma}_j}{m_i\,m_j} v_{ss}(r_{ij})~,
\end{equation}
where $v_{ss}$ is very short--ranged.  One of the most striking
success of chromomagnetism is the explanation of the
$\Sigma-\Lambda$ splitting.  For both states,
$\sum_{i<j}\vec{\sigma}_i.\vec{\sigma}_j=-3$ since we have an
overall spin $S=1/2$. However, for the $\Lambda$, this strength is
concentrated into the light-quark pair, and thus the downward shift
is more important, due to the $m_i^{-1}\,m_j^{-1}$ dependence of the
operator (\ref{mod:eq:chromo}).

Another success is the prediction of the hyperfine splittings when
the strange quark is replaced by a quark with charm or beauty. While
the $\Sigma-\Lambda$ mass difference remain large, the
$\Sigma^*-\Sigma$ gap is much reduced. This is exactly the pattern
observed for charm and beauty baryons.  See, e.g.,
\cite{Richard:1983mu} for a study on how this effect depends on the
assumed shape of the confining potential $v(r)$. The subtle
interplay between the $m_i^{-1}\,m_j^{-1}$ dependence of the
chromomagnetic operator and the short-range correlation induced by
the central potential has been analyzed for beauty baryons, leading
to successful predictions. See, e.g., \cite{Karliner:2008sv} and
references therein.
\paragraph{Instantons, good diquarks}
However, it has been stressed that chromomagnetism is not the unique
solution. In particular, an \emph{instanton-induced} interaction
\cite{'tHooft:1976fv} also accounts very well for the hyperfine
splittings. See, e.g.,
\cite{Shuryak:1988bf,Loring:2001kx,Semay:2001th}. It can be written
as
\begin{equation}
V_{SS}=-4 \sum_{i<j} g_{ij} \,\mathcal{P}^{[i,j] }\mathcal{P}^{S=0}\,\delta^{(3)}(\vec{r}_j-\vec{r}_j)~,
\end{equation}
with the projection on the spin $S=0$ state  and on the antisymmetric flavor state for each pair.
The dimensionless coupling $g_{ij}$ is stronger for light quarks than for $[ns]$. This explains the $\Sigma-\Lambda$ mass difference, and other splittings within the ground states. Of course, the instanton-induced interaction differs more strikingly  from chromomagnetism in the case of mesons, in particular for pseudoscalar and scalar mesons \cite{Klempt:1995ku}.

An interesting concept has been introduced
\cite{Wilczek:2004im,Jaffe:2004ph}, that of \emph{good diquarks}
with spin $S=0$, which is lower in mass than its vector counter part
with $S=1$. For light quark, the favored pair is in an antisymmetric
isospin state $I=0$. Then the spectrum can be analyzed without
referring to a specific dynamical model for the hyperfine
interaction. However, this concept has been often associated to an
extreme quark--diquark picture of baryon excitations, with many
fewer levels than in the usual three-quark picture. Also, the
concept of good diquark became rather sulfurous when associated to
speculations about multiquark  states which were neither supported
by genuine few-body calculations nor confirmed by the data. We shall
use here the concept of good diquark without endorsing its more
extreme developments.

Note that the diquark model was invented much earlier, and has been
often rediscovered. For a historical survey, see, e.g.,
\cite{Lichtenberg:1996rg}.
\paragraph{Goldstone boson exchange}
In conventional potential models, one starts with a degenerate
ground state near 1100\,MeV, and then  a splitting between the $N$
and the $\Delta$ is introduced. More recently, models have been
developed where one starts from a unique state near 2\,GeV, and then
introduce a Goldstone-boson exchange (GBE) that reads,
\cite{Glozman:1995fu}
\begin{multline}\label{mod:eq:GBE}
V_\text{OGE}=\sum_{i<j}
\frac{g^2}{4\pi}\frac{1}{4 m_im_j}\,\tilde{\lambda}_i^{F}.\tilde{\lambda}_j^{F}\,\vec{\sigma}_i.\vec{\sigma}_j
\,\times\\
\left[\frac{\mu^2\,\exp(-\mu r_{ij})}{r_{ij}} - 4\pi\, \delta^{(3)}(\vec{r}_{ij})\right]~.
\end{multline}
which pushes down both $N$ and $\Delta$ but the former with larger strength.

This interaction is inspired by the one-pion-exchange potential in
nuclear physics. However, in describing the nucleon--nucleon
interaction, the contact term is usually neglected, as hidden by all
uncertainties about the origin of the hard-core interaction at short
distances. Here this is the reverse: the Yukawa tail plays a minor
role, and the splitting of baryons is due to the contact term, which
is regularized in explicit models exploiting this dynamics.

We note in this approach an important flavor-dependence, as the pion
does not couple to heavy quarks. It is not obvious how this
interaction has to be adapted to the meson sector.

The GBE model has been studied by several groups, in particular
\cite{Valcarce:1995dm,Dziembowski:1996cv,Melde:2008yr}.\vspace{-2mm}
\subsub{Improved pictures of ground-state baryons}\vspace{-2mm}
The naive quark model, with its non-relativistic kinematics, frozen
number of constituents, instantaneous interaction, etc., is far from
being fully satisfactory. Several improved pictured have been
proposed.  We briefly review some of them.  However, in a review
devoted to baryon spectroscopy, we cannot set on the same footing
constituent models giving predictions for the whole spectrum of
excited states and sophisticated QCD-inspired studies which are
restricted to the ground state or at most to the first
excitations.\vspace{-3mm}
\paragraph{Quark models with relativistic kinematics}
It is now rather customary to replace the non-relativistic
contribution of constituent mass and kinetic energy,
$m+\vec{p}^2/2$, by the relativistic operator
$({m^2+\vec{p}^2})^{1/2}$. Examples are
\cite{Basdevant:1985ux,Capstick:1986bm}. This is more satisfactory,
but does not solve the problems inherent to the choice of the
dynamics. For instance, with a standard Coulomb-plus-linear
interaction, the  lowest nucleon excitation has negative parity.
\vspace{-3mm}
\paragraph{Relativistic quark models}
This is a more ambitious approach, aiming at a covariant formalism,
even though some approximations are eventually unavoidable in the
calculations. A recent example is \cite{Melde:2008yr} and a
benchmark  is the work by the Bonn group
\cite{Metsch:2003ix,Migura:2006ep,Loring:2001kv,Loring:2001kx,Loring:2001ky},
whose starting point is the Bethe--Salpeter equation. Here, not only
the masses and the static properties can be estimated, but also the
form factors and  quark distributions \cite{Migura:2006en,
Haupt:2006em,VanDyck:2008zz}.\vspace{-3mm}

\paragraph{The MIT bag model}
The MIT bag model stages massless or very light quarks moving freely
inside of cavity of radius $R$ which is adjusted to minimize the bag
energy. A good fit to the ground states of light baryons was
achieved \cite{DeGrand:1975cf}, and this model motivated a variety
of developments. However, the model does not permit an easy estimate
of the excitation spectrum. In particular, the center-of-mass motion
cannot be removed explicitly.\vspace{-3mm}
\paragraph{The bag model for heavy quarks}
The MIT bag model is not suited for heavy quarks. For heavy
$(Q\overline{Q})$ or $(QQQ)$, \cite{Hasenfratz:1977dt} built a bag
to confine the gluon field for any given quark configuration. The
gluon energy is interpreted as the quark potential. Note that in the
case of baryons \cite{Hasenfratz:1980ka}, this model leads to a
$Y$-shape interaction, as discussed above. The case of hadrons with
both heavy and light quarks is less easy. See, e.g.,
\cite{Bernotas:2008bu}, for a recent update.\vspace{-3mm}
\paragraph{The cloudy bag}
A problem with the MIT bag model is the discontinuity of the
axial-vector current across the bag surface. Or in a more empirical
point of view, two nucleons do not interact once their separation
exceeds twice the bag radius. Introducing a pion field around the
nucleon  \cite{Brown:1979ui} or even inside \cite{Thomas:1981vc}
restores a more physical picture.

Starting from a bag of large radius $R\sim1\;$fm, one ends with a
smaller radius $R<1\;$fm for the three-quark domain, and a pion
field extending beyond $1\;$fm. In fact $R$ is not sharply
determined, and the Stony-Brook group got even variants with rather
small radius\footnote{In an ideal scenario, there is a perfect
duality between the three-quark and the pion field picture, named
the ``Cheshire-cat principle''\protect\cite{Nadkarni:1985dm}.}. In
this limit, the details of the quark part become invisible: the
quark core just serves a source of the pion field, and carries the
baryon number, and one recovers the Skyrmion model and other soliton
models.\vspace{-3mm}
\paragraph{Skyrmions and other soliton models}
In this approach, the main emphasis is the coupling of meson to
baryons. Hence the aim is less to perfectly reproduce the spectrum
of high excitations than to account for the low-energy interactions.
For a survey, references, and comparison with experimental data,
see, e.g., \cite{Karliner:1986wq,Weigel:2008zz}. There are many
variants, in particular in the way of treating strangeness and
heavier flavors. For instance, in \cite{Rho:1992yy}, the hyperons
are considered as bound states of a topological soliton and $K$, $D$
or $B$ mesons. \vspace{-3mm}
\paragraph{Chiral perturbation theory and beyond}
There is an old idea by Weinberg and others, QCD is replaced at low
energy by effective Lagrangians which share the same symmetries. The
couplings are treated as free parameters and are used (consistently,
i.e., at the same order in the expansion in powers of the momentum
and quark masses) to calculate other properties. After fruitful
developments in the physics of mesons
\cite{Ecker:1988te,Donoghue:1988ed}, this approach was also applied
to nucleons \cite{Bernard:1995dp} and became widely used. At small
energies, chiral perturbation theory is exact. An extension to
higher energies is possible by the implementation of unitarity
\cite{Oller:2000ma}. Further developments include strangeness, in
particular to describe the $\Lambda_{1/2^-}(1405)$, and exact gauge
invariance for photoproduction \cite{Borasoy:2007ku}. See, e.g.,
\cite{Bernard:2007zu} for a recent survey.\vspace{-3mm}
\paragraph{QCD sum rules}
This beautiful approach to non-perturbative QCD was initiated by
\cite{Shifman:1978bx}, and then developed by several groups.  For a
summary of early applications, see \cite{Reinders:1984sr}. The
extension to baryons is non trivial, since several operators can be
chosen to describe a  given state. After a pioneering paper
\cite{Ioffe:1981kw}, the situation was clarified in
\cite{Chung:1981cc}, and subsequent papers devoted to various flavor
combinations \cite{Dosch:1988vv,Bagan:1993ii,Bagan:1994dy}.
Recently, sum rules were extended to cover octet--decuplet splittings
of heavy baryons \cite{Albuquerque:2009pr}.

\begin{figure}[pt]
\bc
\includegraphics[width=0.45\textwidth]{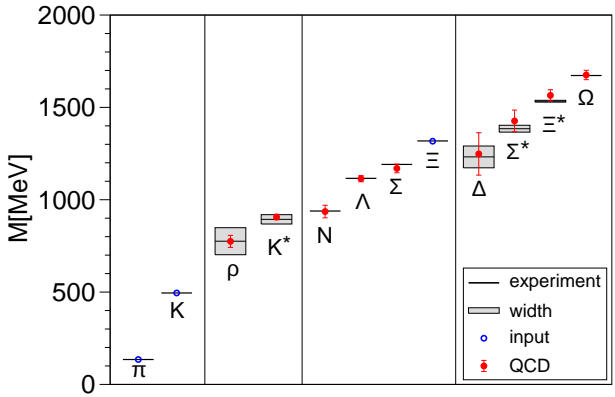}
\vspace{-3mm}\ec \caption{\label{mod:fig:latticemass}The light
hadron spectrum of QCD. Horizontal lines and bands are the
experimental values with their decay widths. The lattice results are
shown by solid circles. Vertical error bars represent the combined
statistical and systematic error estimates. $\pi$, $K$, and $\Xi$
masses input quantities.}
\end{figure}

The idea is to link, via the analytic properties, the perturbative
domain of QCD, where calculations can be done exactly, and the
non-perturbative domain, which can be described in terms of a few
basic constants. These can be adjusted forming a few physical
quantities, which can be used to calculate other
quantities.\vspace{-3mm}
\paragraph{Lattice QCD}
Here,  QCD is reformulated as a field theory in a discretized
phase-space and solved using very astute and powerful techniques
which require, however, expensive computing means. In the domain of
hadron spectroscopy, the best-known applications of lattice QCD are
those dealing with glueballs and hybrid mesons, and also scalar
mesons, but recently the physics of baryons has also been studied.
Figure \ref{mod:fig:latticemass} shows the remarkable achievements
of lattice QCD. Pion masses down to 190\,MeV were used to
extrapolate to the physical point and lattice sizes of up to 6\,fm
\cite{Durr:2008zz}.

Lattice techniques have also been applied to single-charm baryons
\cite{Lewis:2001iz} and even to double-charm baryons
\cite{Flynn:2003vz,Brambilla:2004wf}.

\subsection{\label{sub:se-phegs}Phenomenology of ground-state baryons}
\subsub{Missing states}
Almost all ground-state baryons containing  light or strange quarks
and at most one heavy quark are now  identified. Still missing are
the isospin partners $\Sigma_b^0$ and $\Xi_b^0$ and the spin
excitations ($S=3/2$) of the recently discovered $\Xi_b$ and
$\Omega_b$.

The existence of $\Xi_{cc}^+(3519)$ is uncertain. Its predicted mass
\cite{Fleck:1989mb,Korner:1994nh} is about 100\,MeV larger and
recent calculations give even larger mass values. As compared to a
naive equal-spacing for $p(940)$,  $\Lambda_c^+(2286)$ and
$\Xi_{cc}$, the first correction is that $\Xi_{cc}$ is shifted down
by the heavy--heavy interaction in the chromoelectric sector, see
Eq.~(\ref{mod:eq:conv}). However, both $p$ and $\Lambda_c$ are
shifted down by the favorable chromomagnetic interaction among light
quarks.

As the $(b\bar{c})$ meson has been observed, one should be able to
detect $(bcq)$ baryons with charm and beauty, with two $S=1/2$
states in the ground state, and one $S=3/2$ state. Next will come
the double-beauty sector, and ultimately, baryons with three heavy
quarks.\vspace{-2mm}

\begin{figure}[pt]
\bc
\includegraphics[width=.48\textwidth,height=.40\textwidth]{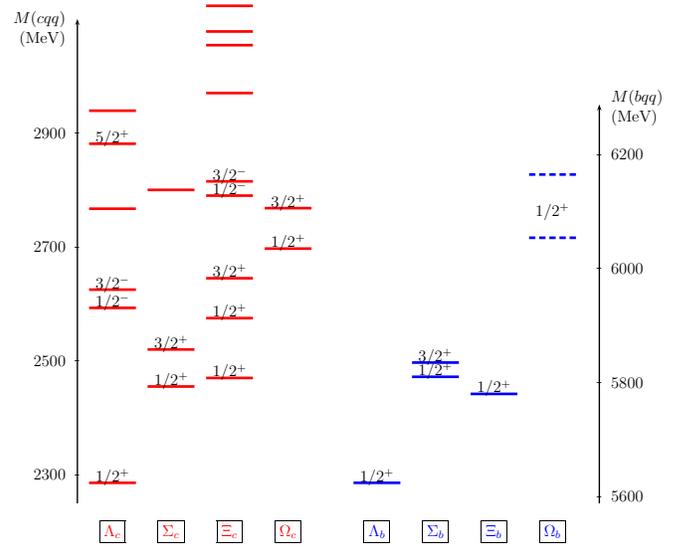}
\ec
\caption{\label{mod:fig:hb} Comparison of the excitation energy
single-charm and single-beauty baryons above the $\Lambda_c$
($\Lambda_b$ mass. The quantum numbers are deduced from the quark
model \cite{Wohl:2008gw}. For the $\Omega_b$ both  {\sc\small D\O}
(top) and {\sc\small CDF} (bottom) results ares shown as dotted
lines. \vspace{-2mm}}
\end{figure}

\subsub{Regularities}\vspace{-2mm}
The masses exhibit a smooth behavior in flavor space, which is
compatible with the expectation based on  potential models
incorporating flavor independence. Moreover, ``heavy quark symmetry
implies that all of the mass splittings are independent of the heavy
quark flavor'', to quote \cite{Isgur:1991wq}. A comparison is made
on Fig.~\ref{mod:fig:hb} of the known single-charm and single-beauty
baryons. The comparison suffers from the small number of beauty
baryons but it is clearly seen that the cost of single-strangeness
excitation $\Xi_Q-\Lambda_Q$ is very similar for $Q=c$ and $Q=b$.

\begin{figure}[pb]
\begin{center}
\includegraphics[width=0.35\textwidth,height=0.45\textwidth,angle=-90]{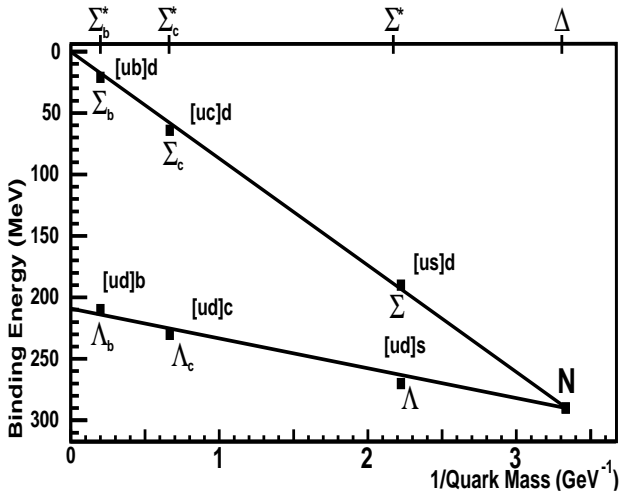}
\end{center}
\caption {\label{gooddiquark}Mass difference between spin-3/2
baryons ($\Delta$, $\Sigma^*$, $\Sigma_c^*$, $\Sigma_b^*$) and
spin-1/2 baryons. In spin-1/2 baryons, the light-light or a
heavy-light quark pair can have spin zero. The spin-0 diquark is
indicated by $[q_1q_2]$. The masses are drawn as a function of the
inverse (constituent) quark mass. \vspace{-1mm}}
\end{figure}

For the double-strangeness excitations, the $\Omega_b(6165)^0$ of
{\sc\small D\O} is problematic. Most models predict $\Omega_b$ with
mass of about 6050\,MeV, 110--120\,MeV lower than the observed mass.
The measurement by {\sc\small CDF}, 6054\,MeV, is in better
agreement with the expectations.\vspace{-2mm}
\subsub{Hyperfine splittings}\vspace{-2mm}
The hyperfine splitting is also varying smoothly from one
configuration to another. Again, this is compatible with the mass
dependence introduced in the chromomagnetic model: an explicit
$m_i^{-1} m_j^{-1}$ in the operator, which is partially cancelled
out by the reinforcement of the short-range correlations when the
masses increase. However, a similar pattern could be reached in
other approaches to hyperfine splitting.  Figure \ref{gooddiquark}
illustrates the regularities of the hyperfine effects in hyperons
when the heavy quark is varied.

\begin{table}[pt]
\caption{\label{Lambda-Sigma-mass-gap}Masses (in MeV) of $\Lambda$
and $\Sigma$ and $\Sigma^*$ baryons quoted from
\cite{Amsler:2008zz}, and mass gaps $\delta M$ between $J^P=1/2^+$
baryons containing `good' diquarks and $J^P$=$3/2^+$ baryons with
all pairs in spin triplet. The quantum numbers of the heavy baryons
are quark model predictions.}
\begin{center}
\renewcommand{\arraystretch}{1.0}
\begin{tabular}{ccccccccccc}
\hline\hline\\[-1.6ex] \multicolumn{1}{c}{$1/2$}& \multicolumn{1}{c}{Mass}
&\quad& $1/2$ & \multicolumn{1}{c}{Mass}
&\quad& $3/2$ & \multicolumn{1}{c}{Mass}  \\[.5ex]
\hline\\[-1.7ex]
\vspace{ .5mm}
 $\Lambda^{0}$   & \hspace{-1mm} $1115.68\pm 0.01$&&
 $\Sigma^{0}$    &  $1192.64\pm{0.04}$\  &&
 $\Sigma^{*0}$   &  $1383.7\pm1.0$ \\
 $\delta M$ &$\left[ud\right]=-271$&&&$\left[us\right]=-191$ &&&  0 MeV\\[2mm]
$\Lambda_c^{0}$ & \hspace{-1mm} $2286.46\pm 0.14$&&
 $\Sigma_c^{0}$  &  $2457.76\pm{0.18}$ &&
 $\Sigma_c^{*0}$ &  $2518.0\pm0.5$\\
 $\delta M$ &$\left[ud\right]=-231$&&&$\left[uc\right]=-60$ &&&  0 MeV\\[2mm]
 $\Lambda_b^{0}$ & \hspace{-1mm} $5619.7\pm 1.7$ &&
 $\Sigma_b^{0}$  &  $5811.5\pm 1.7$ &&
 $\Sigma_b^{*0}$ &  $5832.7\pm 1.9$\\
$\delta M$ &$\left[ud\right]=-213$&&&$\left[ub\right]=-21$ &&&  0 MeV\\[2mm]
 $\Xi_c^{0}$ & \hspace{-1mm} $2471.0\pm 0.4$ &&
 $\Xi ^\prime{}_c^0$  & $2578.0\pm 2.9$  &&
 $\Xi_c^{*0}$ &  $2646.1\pm 1.2$\\
$\delta M$ &$\left[ds\right]=-174$&&&$\left[dc\right]=-70$ &&&  0 MeV\\[1mm]
\hline\hline
\end{tabular}\vspace{-5mm}
\end{center}
 \end{table}

The $\Sigma_Q^*-\Sigma_Q$  is expected to vanish as $M_Q\to\infty$,
with a  $M_Q^{-1}$ in the limit where the change of the wave
function is neglected. In this limit, the combination
$2\Sigma_Q^*+\Sigma_Q-3\Lambda_Q$ is expected to be constant, and
this is rather well confirmed by the data, with about 613, 634 and
618\,MeV for $Q=s$, $c$ and $b$, respectively.

To a good approximation, the hyperfine effect in the pair $q_1q_2$
is found independent of the third quark, this leading to a variety
of sum rules if taken seriously. See, e.g.,
\cite{Lichtenberg:1995kg,Franklin:2008hx}. Within the point of view
of good diquarks, one can, indeed,  measure the downward shift due
to quark pairs in spin-singlet,  starting from the $S=3/2$ baryon
where all pairs are in a spin triplet. As seen in Table
\ref{Lambda-Sigma-mass-gap}, one obtains $[ud]\approx 250$\,MeV, for
$[us]\approx 170$\,MeV, $[uc]\approx 65$\,MeV, and $[ub]\approx
20$\,MeV.\vspace{-2mm}
 \subsub{Isospin splittings}\vspace{-2mm}
This was a subject of many investigations. Before the quark model,
the neutron to proton mass difference has been related by
\cite{Cottingham:1963} to electron--nucleon scattering. In the quark
model, as underlined in \cite{Isgur:1979ed}, there are many
contributions to mass differences within an isospin multiplets, and
the various terms often tend to cancel. There are: the quark-mass
difference $m_d-m_u$; the induced change of chromoelectric energy;
the change in the strength of the chromomagnetic forces; the Coulomb
repulsion; the magnetic interaction; etc. The effects have been
estimated by several groups \cite{Isgur:1979ed, Varga:1998wp} and
extended to heavy quarks \cite{Lichtenberg:1977mv,Franklin:1999bg}.
There is also a contribution to isospin splittings from meson loops,
with pions and baryons in the loops having different masses and
couplings. This effect was emphasized recently for heavy baryons
\cite{Guo:2008ns}.
\subsection{Models of baryon excitations}
While for the ground-state baryons, there is a variety of pictures,
some of them being directly guided by QCD, for the excitation
spectrum, one should still rely on explicit constituent models, and
among them the harmonic oscillator.\vspace{-2mm}
\subsub{Harmonic oscillator}
\paragraph{HO: equal masses}
This is the simplest model, corresponding to (\ref{mod:eq:Hpot})
with all $m_i=m$ and (\ref{mod:eq:Vho}).  Then the relative motion
is described by
\be\label{mod:eq:horel}
\frac{\vec{p}_\rho^2}{m}+K \vec{\rho}^2+\frac{\vec{p}_\lambda^2}{m}+K \vec{\lambda}^2~,
\ee
leading the energy spectrum
\begin{equation}\label{mod:eq:ho-levels}
\sqrt{\frac{K}{m}}\left(6+2 l_\rho + 4 \mathtt{n}_\rho+ 2 l_\lambda +
4 \mathtt{n}_\lambda\right)= \sqrt{\frac{K}{m}}\left(6+
2\mathsf{N}\right),
\end{equation}
in an obvious notation for the orbital momenta
$l_{\rho,\lambda}=0,\, 1,\ldots$ and radial numbers
$\mathtt{n}_{\rho,\lambda}=0,\, 1,\ldots$ attached to each degree of
freedom. The wave functions are also explicitly known.  For the
ground state, it is the Gaussian (\ref{mod:eq:howf}). For
excitations, it also contains a polynomial which reflects the
rotation and permutation properties and ensure the orthogonality.

Note the first radial excitation of the nucleon and $\Delta$, a
symmetric combination of the states with $l_\rho=l_\lambda=0$ and
either $\mathtt{n}_{\rho,\lambda}=(0,1)$ or $(1,0)$ which is
\emph{below} the first negative-parity excitation. This will be
further discussed in connection with alternative models and with the
data.\vspace{-2mm}
\paragraph{HO: unequal masses}
For baryons with one heavy quark, $(qqQ)$, the masses are $(m,m,M)$.
The case of double-charm baryons is deduced by $m\leftrightarrow M$.
The second term in (\ref{mod:eq:horel}) has now a reduced mass $\mu$
with $\mu^{-1}= (2 M^{-1}+ m^{-1})/3$ replacing $m$. Then the energy
levels are modified as
\begin{equation}\label{mod:eq:ho-levels1}
\sqrt{\frac{K}{m}}\left(3+2 l_\rho + 4 \mathtt{n}_\rho\right)
+ \sqrt{\frac{K}{\mu}}\left(3+2 l_\lambda + 4 \mathtt{n}_\lambda\right)
\end{equation}
Hence the $\lambda $ excitation are lower than their $\rho$ analogs
for single-flavor baryons. For baryons with double flavor, the first
excitation are within the heavy-quark sector. The wave function is a
slight generalization of (\ref{mod:eq:howf}), with
$\sqrt{Km}\vec{\rho}^2+\sqrt{K\mu}\vec{\lambda}^2$ in the Gaussian
and the corresponding changes in the normalization.

If the three constituents masses are different, then the Hamiltonian
describing the relative motion is still of the type
\be\label{mod:eq:horel1}
\frac{\vec{p}_x^2}{m_x}+K \vec{x}^2+\frac{\vec{p}_y^2}{m_y}+K \vec{y}^2~,
\ee
with $\vec{x}$ and $\vec{y}$ are combinations of the Jacobi
variables $\vec{\rho}$ and $\vec{\lambda}$ which are obtained,
together with the reduced masses $m_x$ and $m_y$ by the
diagonalization of a $2\times2$ matrix.\vspace{-2mm}
\subsub{Potential models}\vspace{-2mm}
If the potential $V$ is not harmonic, the non-relativistic
Hamiltonian (\ref{mod:eq:Hpot}) can be solved numerically using
powerful techniques developed in nuclear physics, such as Faddeev
equations, hyperspherical expansion, or correlated Gaussians.  While
convergence is easily reached for the energy levels, some additional
effort is usually required to measure the short-range correlations
within the wave function.

Some approximations can be envisaged, as an alternative to the full
three-body calculation.  Some of them are purely technical, for
instance truncating the hyperspherical expansion to the lowest partial
wave. Some others shed some light on the baryon structure.  For
instance, doubly-flavored baryons $(QQq)$ have clear diquark--quark
structure, but the internal diquark dynamics is influenced by the
third quark, an effect which is unfortunately often forgotten%
\footnote{In the case of the harmonic oscillator, exactly 1/3 of
the strength binding $QQ$ is due to the third quark}. $(QQq)$ can
also be treated $\mathrm{H}_2{}^+$ in atomic physics, with $QQ$
moving in a Born--Oppenheimer potential  generated by the light
degrees of freedom \cite{Fleck:1989mb}.

It should be stressed that different models used for the interquark
potential give similar ordering for the first levels. In the HO, the
radial excitation energy is twice the orbital one.  With a linear
confinement, the ratio is smaller, but still the radial excitation
remains above the orbital one, if the potential is local and
flavor-independent \cite{Hogaasen:1982rb}.  Pushing the radial
excitation below the orbital one require drastic changes of the
dynamics, like these of the OBE model.\vspace{-2mm}
\subsub{Relativistic models}\vspace{-2mm}
For relativistic models, the solution can be found by variational
methods, i.e., by expanding the wave function on a basis, usually
chosen as containing Gaussians of different range parameters. The
level order of the first levels is similar to the pattern found in
non-relativistic models.

For high orbital excitations, an interesting result was obtained
\cite{Martin:1985hw}. The levels are well described in the
semi-classical approximation. For low $L$, the lowest state is
symmetric, all quarks sharing  equally the orbital momentum. For
higher $L$, there is a spontaneous breaking of symmetry, and in the
ground-state, two quarks have a relative $l_\rho=0$ while the third
quark  takes $l_\lambda=L$. Hence diquarks are generated
\emph{dynamically} at high $L$, even for a purely linear
interaction. There is no need for short-range forces to form the
diquark.  With relativistic kinematics and linear confinement, both
in the naive 1/2 rule version (Eq.~\ref{mod:eq:paiwise}) or in the
more elaborate $Y$-shape version (Eq.~\ref{mod:eq:Yshape}) a linear
Regge trajectory is obtained, with the same slope as for
mesons.\vspace{-2mm}
 \subsub{Regge phenomenology}\vspace{-2mm}

The Regge theory, first developed in
\cite{Regge:1959mz,Regge:1960zc}, connects the high energy behavior
of the scattering amplitude with singularities in the complex
angular momentum plane of the partial wave amplitudes in the crossed
($t$) channel. It is based on rather general properties of the
S-matrix, on unitarity, analyticity and crossing symmetry. The
simplest singularities are poles (Regge poles). According to the
Chew--Frautschi conjecture \cite{Chew:1961ev,Chew:1962eu}), the
poles fall onto linear trajectories in $M^2,J$ planes. In the Regge
theory, the $t$-channel exchange of a particle with spin $J$ is
replaced by the exchange of a trajectory. Regge-trajectory exchange
is thus a natural generalization of a usual exchange of a particle
with spin $J$ to complex values of $J$. The method established an
important connection between high energy scattering and the spectrum
of hadrons. There is a discussion if Regge trajectories are linear,
parallel, or not \cite{Inopin:1999nf,Tang:2000tb}. No systematic
errors in the mass assignments were, however, included in these
discussions. We will assume linearity and do not see any significant
deviation from linear trajectories.\vspace{-2mm}
 \subsub{Solving QCD}
 \paragraph{QCD sum rules, Lattice QCD}
In QCD sum rules or in lattice QCD, one can reach the ground-state
configuration of any given set of quantum numbers, in particular the
leading Regge trajectory. The difficulty is only to build the
corresponding operators.

The first excitations of the nucleons have received much attention
\cite{Melnitchouk:2002eg}. With the large lattices available, one
could presumably get access to the states on the leading Regge
trajectory, each being the ground state in its $J^P$ sector. This is
probably delicate for the radial excitations, which are derived from
the same operator as the lowest states and for which one should
first remove the leading contribution of the ground state. The
theoretical uncertainty is thus larger. The question is whether,
when the light quark mass vanishes, one observes a change in the
hierarchy of excitation, with the positive-parity  excitation
becoming lower than the negative-parity one. This is still
controversial. The latest results are, however, encouraging:
\cite{Mathur:2003zf} compared the radial and orbital excitations of
the nucleon as a function of the assumed light-quark mass $m_n$, and
found that the former is usually above the latter except for very
small $m_n$, where a crossing is observed, and thus the same
ordering  as the experimental one. This result indicates that the
anomalous ordering is particular to the light quark dynamics. It
remains to be checked by other groups, with attention in particular
to finite size effects \cite{Sasaki:2005ug}. Among the recent
contributions, one may cite
\cite{Mathur:2003zf,Sasaki:2005ap,Basak:2007kj,Bulava:2009jb,Drach:2009dh}.
In this latter article, excitations up to $J=5/2$ have been studied.
\paragraph{AdS/QCD}

\begin{figure}[pb]
\includegraphics[width=0.4\textwidth,height=0.6\textwidth]{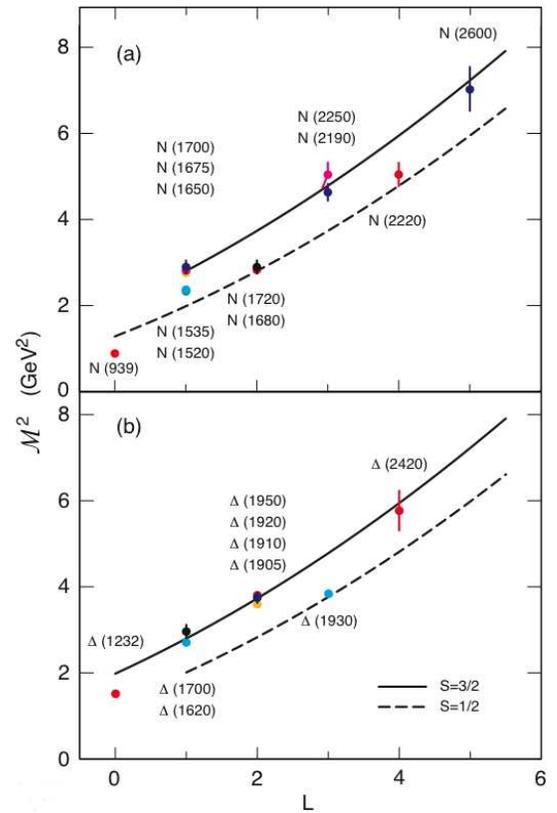}
\caption{\label{TB}Light baryon orbital spectrum for $N^*$ (a) and
$\Delta^*$ (b). The lower dashed curves correspond to baryon states
dual to spin-1/2 modes in the bulk and the upper continuous curve to
states dual to spin-3/2 modes \cite{deTeramond:2005su}. }
\end{figure}

\begin{figure*}[pt]
\centerline{
\includegraphics[width=0.7\textwidth,height=6cm,clip]{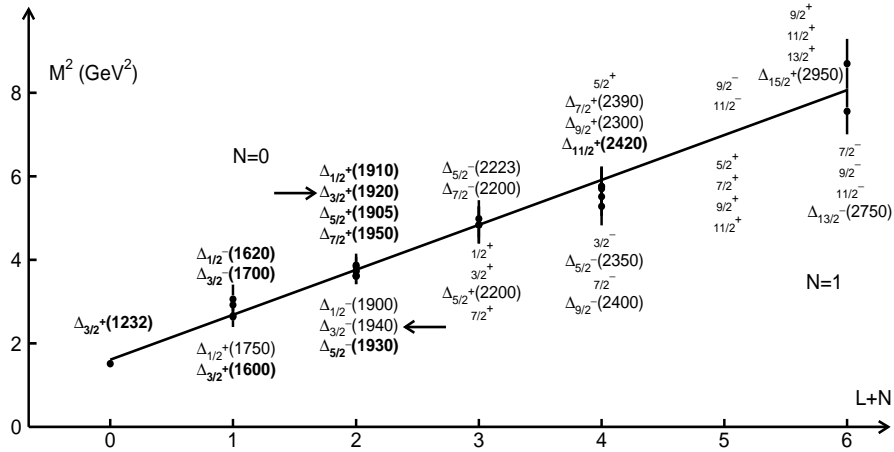}
}
\caption{\label{delta_table} Regge trajectory for $\Delta^*$
resonances as a function of the leading intrinsic orbital angular
momentum $L$ and the radial excitation quantum number $N$
(corresponding to $n_1+n_2$ in quark models) \cite{Klempt:2008rq}.
The line represents a prediction based on AdS/QCD correspondence
(soft wall) \cite{Forkel:2007cm,Forkel:2007tz}. Resonances with
$\mathtt N=0$ and $\mathtt N=1$ are listed above or below the
trajectory. The mass predictions are 1.27, 1.64, 1.92, 2.20, 2.43,
2.64, 2.84\,GeV for $L+\mathtt N=0,1,\cdots 6$, respectively. }
\end{figure*}
A new approach to quantum field theory is presently pursued, the
so-called AdS/CFT correspondence (Anti de Sitter/Conformal Field
Theory), which establishes a duality between string theories defined
on the 5-dimensional AdS space-time and conformal field theories in
physical space-time, see, e.g., \cite{Brodsky:2006uq}. It is assumed
that the effective strong coupling is approximately constant in an
appropriate range of momentum transfer, and that the quark masses
can be neglected. Then QCD becomes a nearly conformal field theory
and the AdS/CFT correspondence can be applied to QCD. The hadron
spectrum and strong interaction dynamics can then be calculated from
a holographic dual string theory defined on five-dimensional AdS
space. For an appropriate choice of the metrics, a semi-classical
approximation to QCD follows which incorporates both color
confinement and conformal short-distance behavior. Confinement is
parameterized by a cut-off in AdS space in the infrared region
(``hard wall'') \cite{Polchinski:2001tt}. Applied to baryon
spectroscopy, AdS/QCD yields a mass relation $M \propto
\mathrm{L}+\mathtt{N}$ \cite{deTeramond:2005su,Brodsky:2008pg},
where  $\mathrm{L}$ and $\mathtt{N}$ are orbital and radial
excitation quantum numbers corresponding to ${\rm
L}=l_{\rho}+l_{\lambda}$ and $\tt {N} =
\tt{n}_{\rho}+\tt{n}_{\lambda}$ in quark models. Spin 1/2 and spin
3/2 baryons require different AdS boundary conditions and lead to
different offset masses. The predictions are shown in Fig.~\ref{TB}.
The lower mass of nucleon resonances with $S=1/2$ can be related to
the effect of ``good'' diquarks \cite{Jaffe:2003sg,Wilczek:2004im}:
diquarks with vanishing spin and isospin are energetically favored
compared to ``bad'' diquarks. Of course, $\Delta$ resonances have
isospin 3/2 and contain no ``good'' diquarks. Problems occur for
$\Delta(1232)$ which is too low in mass and for
$\Delta_{1/2^-}(1620)$ and $\Delta_{3/2^-}(1700)$ which are on the
``wrong'' trajectory. $\Delta_{5/2^-}(1930)$ is treated as spin 1/2
state with $L=3$; in section \ref{se:sub-band}, this state is
combined with $\Delta_{1/2^-}(1900)$ and $\Delta_{3/2^-}(1940)$ to
form a triplet with $L=1, S=3/2, N=1$ quantum numbers in \emph{the
third excitation band}. \cite{deTeramond:2005su} require the
existence of a further to-be-discovered $\Delta$ state with
$J^P=7/2^-$ at 1.9 to 2.0\,GeV.

The use of orbital angular momentum  ${\rm L}$ to classify baryon
resonances has been often criticized, see \cite{Glozman:2009fj}, and
\cite{deTeramond:2009qx,Afonin:2009zk} for a refutation. In
non-relativistic models with anharmonic confinement and
spin-dependent forces, and in relativistic models better suited for
light quarks, each state contains a superposition of several angular
momentum configurations. However quark models with the same
constituent quark rest mass for all excitations are probably not
realistic. An alternative is, e.g., the Nambu picture
\cite{Nambu:1961tp} where the mass of a hadron is distributed along
a string connecting nearly massless quarks. Perhaps the total
angular momentum ${\rm L}$ occurring in recent mass formulas
reflects the length of the inner flux tube linking the quarks.



In \cite{Forkel:2007cm,Forkel:2007tz}, the mass spectrum of light
mesons and baryons was predicted using AdS/QCD in the soft-wall
approximation. The approach relies on deformations of the AdS
metric, governed by one free mass scale proportional to
$\Lambda_{\rm QCD}$ and leads to the same boundary conditions for
$S=1/2$ and $S=3/2$ baryons. Relations between ground state masses
and trajectory slopes
\begin{equation}\label{Delta_ADS}
\begin{split}
&M^2 = 4\lambda^2 (\mathrm{L}+\mathtt{N}+1/2)\qquad\text{for mesons}\\
&M^2 = 4\lambda^2 (\mathrm{L}+\mathtt{N}+3/2)\qquad \text{for
baryons}
\end{split}
\end{equation}
were derived. Using the slope of the $\Delta$ trajectory, baryon
masses were calculated. However, it is argued \cite{Forkel:2007cm}
that hyperfine interactions are not included in AdS/QCD and that the
parameter $\lambda$ in (\ref{Delta_ADS}) should be re-tuned. This
changes the offset (the predicted ground-state mass) and the Regge
slope, and the resulting compromise might show problems for small
and large angular momenta.

The predicted masses for $\Delta$ baryons are plotted as a function
of $\mathrm{L}+\mathtt{N}$ in Fig.~\ref{delta_table} which includes
all resonances (except the one-star $\Delta_{1/2^-}(2150)$ which
would fit well with quantum numbers $L=1,{\tt N}=2$ and 2.2\,GeV
predicted mass). The agreement is excellent and the remaining
problems seen in Fig.~\ref{TB}b disappear.

For nucleon resonances, we need to keep track that some diquarks
have spin zero and are in S-wave. Then Eq.~(\ref{Delta_ADS}) is
rewritten as \cite{Forkel:2008un}
\begin{equation}
M^2 = a\cdot ({L}+{N}+3/2)-\,b\cdot \alpha_D\,\left[{\mathrm
GeV^2}\right]
\label{eq1}
\end{equation}
with $a=1.04$\,GeV$^2$ and $b=1.46$\,GeV$^2$. For the lowest states,
$\alpha_D$ can be interpreted as the fraction of good diquarks and
calculated explicitly from standard quark-model wave functions. In
Table \ref{Gooddiquarks}, the same $\alpha_D$ is assumed along a
trajectory, and its coefficient $b$ is tuned to reproduce the
$\Delta(1232)$--$N(940)$ splitting. Also shown are the quark spin,
the orbital angular momentum  and the radial quantum number
$\tt{N}$. It is remarkable that the masses of all 48 $N$ and
$\Delta$ resonances are very well reproduced using just two
parameters. One parameter is related to confinement and was already
used to describe the $\Delta$ mass spectrum, the second one accounts
for hyperfine effects. It reduces the size of the nucleon by a
fraction which depends on $\alpha_D$.

The precision of the mass calculated by Eq.~(\ref{eq1}) is by far
better than quark model predictions even though the latter have a
significantly larger number of parameters. The mean difference
$\delta M/M$ is 2.5\% for Eq.~(\ref{eq1}), 5.6\% for the
Capstick-Isgur model (with 7 parameters) \cite{Capstick:1986bm}, and
5.1\% (5.4\%) for the two variants of the Bonn model
\cite{Loring:2001kx} (5 parameters). The Skyrme model
\cite{Karliner:1986wq} with 2 parameters predicts only half of the
observed states, with $\delta M/M=9.1$\%. The masses of Table
\ref{all-light-N} were used for the comparison.\vspace{-2mm}
 \begin{table}[!!!htbc]
 \caption{\label{Gooddiquarks}Nucleon and $\Delta$ resonances and
suggested quantum numbers. The predicted masses are calculated using
Eq.~(\ref{eq1}).}
\renewcommand{\arraystretch}{1.20}
\begin{center}
\begin{footnotesize}\begin{tabular}{cccccccr}
\hline\hline
L & N & S & $\alpha_D$ & \multicolumn{3}{c}{Resonance}&Pred.\\
\hline
0 & 0 & 1/2 & 1/2 & $N_{1/2^+}(940)$ &       && 943\\
0 & 1 & 1/2 & 1/2 & $N_{1/2^+}(1440)$&         &&  1396      \\
0 & 2 & 1/2 & 1/2 & $N_{1/2^+}(1710)$&         &&  1735       \\
0 & 3 & 1/2 & 1/2 & $N_{1/2^+}(2100)$&         &&  2017       \\
1 & 0 & 1/2 & 1/4 & $N_{1/2^-}(1535)$,&$N_{3/2^-}(1520)$ &&1516  \\
1 & 1 & 1/2 & 1/4 & $N_{1/2^-}(1905)$,&$N_{3/2^-}(1860)$ &&1833  \\
1 & 2 & 1/2 & 1/4 & $N_{1/2^-}(2090)$,&$N_{3/2^-}(2080)$ &&2102  \\
1 & 0 & 3/2 & 0 & $N_{1/2^-}(1650)$,&$N_{3/2^-}(1700)$,&$N_{5/2^-}(1675)$&1628 \\
2 & 0 & 1/2 & 1/2 & $N_{3/2^+}(1720)$,&$N_{5/2^+}(1680)$&&1735 \\
2 & 0 & 3/2 & 0   & $N_{1/2^+}(1880)$,&$N_{3/2^+}(1900)$,&$\cdots$&1932  \\
2 & 0 & 3/2 & 0   &$\cdots$& $N_{5/2^+}(1870)$,& $N_{7/2^+}(1990)$&1932     \\
3 & 0 & 1/2 & 1/4 & $N_{5/2^-}(2200)$,&$N_{7/2^-}(2190)$ &&2102  \\
3 & 0 & 3/2 & 0 & $N_{9/2^-}(2250)$  &&&  2184 \\
4 & 0 & 1/2 & 1/2 & $N_{9/2^+}(2220)$&          &&      2265  \\
5 & 0 & 1/2 &1/4 & $N_{11/2^-}(2600)$&  &&2557                   \\
6 & 0 & 1/2 & 1/2 & $N_{13/2^+}(2700)$& &&2693\\[0.6ex]              \hline\\[-3ex]
0 & 0 & 3/2 &  0  & $\Delta_{3/2^+}(1232)$ &          &&  1261       \\
0 & 1 & 3/2 &  0  & $\Delta_{3/2^+}(1600)$ &          &&  1628       \\
1 & 0 & 1/2 & 0 & $\Delta_{1/2^-}(1620)$,&$\Delta_{3/2^-}(1700)$&&1628 \\
1 & 1 & 3/2 & 0 & $\Delta_{1/2^-}(1900)$,&$\Delta_{3/2^-}(1940)$,&$\Delta_{5/2^-}(1930)$&1926  \\
1 & 2 & 1/2 & 0 & $\Delta_{1/2^-}(2150)$&&&2184 \\
2 & 0 & 3/2 & 0   &$\Delta_{1/2^+}(1910)$,& $\Delta_{3/2^+}(1920)$, & $\cdots$ & 1926     \\
2 & 0 & 3/2 & 0   &$\cdots$&$\Delta_{5/2^+}(1905)$,& $\Delta_{7/2^+}(1950)$  & 1926     \\
3 & 0 & 1/2 & 0 & &$\Delta_{7/2^-}(2200)$ && 2184\\
3 & 1 & 3/2 & 0 & $\Delta_{5/2^-}(2350)$,&$\Delta_{9/2^-}(2400)$ && 2415\\
4 & 0 & 3/2 & 0 & $\Delta_{7/2^+}(2390)$,&$\Delta_{9/2^+}(2300)$,&$\Delta_{11/2^+}(2420)$& 2415\\
5 & 1 & 3/2 & 0 & $\Delta_{13/2^-}(2750)$ & &&2820\\
6 & 0 & 3/2 & 0 & $\Delta_{15/2^+}(2950)$ & &&2820\\
\hline\hline
\end{tabular}
\renewcommand{\arraystretch}{1.0}
\end{footnotesize}
\vspace{-5mm}
\end{center}
\end{table}
 %
%
 \subsub{Hyperon resonances}\vspace{-2mm}
Little experimental information is added since the review of
\cite{Hey:1982aj}. We just notice that the mass spectrum of strange
baryons is well reproduced by adding a term
\begin{eqnarray}\label{LS_ADS}
M^2_{\Sigma^*(1385)}-M^2_{\Delta(1232)}=0.40\,\left[{\mathrm
GeV^2}\right]
\end{eqnarray}
to Eq. (\ref{eq1}). The SU(3)$_{\rm f}$ singlet states
$\Lambda_{1/2^-}(1405)$, $\Lambda_{3/2^-}(1520)$, and probably
$\Lambda_{7/2^-}(2100)$ have good diquark fractions $\alpha_D=3/2$.

 \subsection{Baryon decays}
Hadron decays are a decisive element of any theory of strong
interactions. The fact that so many resonances -- expected in
symmetric quark models -- are missing in the data could find a
natural explanation if the missing states have weak coupling only to
$N\pi$. Indeed, this is what most models predict.\vspace{-2mm}
 \subsub{Hadron decays on the lattice}\vspace{-2mm}
 \begin{figure}[pb]
\includegraphics[width=0.4\textwidth,height=0.35\textwidth]{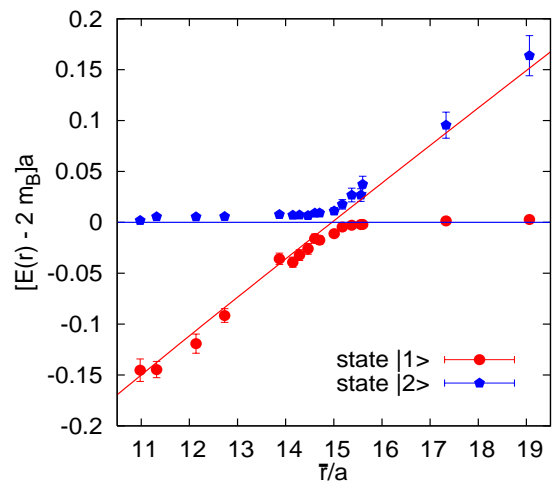}
\caption{\label{fig:potential}Pair creation on a lattice, calculated
for mesons. A sea quark--antiquark pair is created in the vacuum. At
large distances, two-meson states are energetically preferred.  For
static quarks, the levels cross at some distance $R$ (with $a\approx
0.083$\,fm), the string breaking introduces mixing of the energy
levels defined by the potential $V(R)$ and the threshold $2m(B)$
\cite{Michael:2005kw}.}
\end{figure}
 An intuitive understanding of hadron decays can be achieved by
 inspection of the potential energy between two static quarks. The
energy can be described by the superposition of a Coulomb-like
potential and a linearly rising (confinement) potential. At
sufficiently large separations, for $R \approx 1.2$\,fm, the total
energy suffices to produce two (color-neutral) objects: string
breaking occurs. String breaking in mesons can be simulated on a
lattice \cite{Michael:2005kw}. Figure~\ref{fig:potential} displays
the energy levels due to a $q\bar q$ and a two-meson system in an
adiabatic approximation. In a hadronic reaction, the sudden
approximation $-$ where the system follows the straight line $-$ is
more realistic, and mesons can be excited to large energies. Similar
calculations for baryons have not yet been made but the physics
picture should remain the same.\vspace{-2mm}
 \subsub{Models of hadron decays}\vspace{-2mm}
The operators responsible for strong decays of baryon resonances are
unknown. Models need to be constructed with some mechanism in mind;
this can be either elementary meson emission from a baryon, quark
pair creation, string breaking, or flux-tube breaking. In the latter
three cases, a quark pair is created in a process which is often
modeled by assuming $^3P_0$ quantum numbers for the quark pair. A
survey of models, theoretical results and a comparison with data is
given by \cite{Capstick:2000qj}. They conclude that none of the
models does ``what can be termed an excellent job of describing what
is known about baryon strong decays. The main features seem to be
well described, but many of the details are simply incorrect". More
recent widths calculations \cite{Melde:2005hy,Sengl:2007yq} confirm
this statement.\vspace{-2mm}
\subsection{\label{se:sub-band}The band structure of baryon excitations}
The harmonic oscillator provides a frame to classify baryons
resonances. Non-harmonic corrections, relativistic effects, and in
particular spin-dependent forces induce splitting of degenerate
states and mixing of states with the same total spin and parity
$J^P$ but, of course, the number of expected states remains the
same. In this section, the observed baryon resonances are mapped
onto HO quark model states, in an attempt to identify classes of
resonances which are missing. The systematic of observed and missing
resonances may provide hints at the dynamics which lead to the
observed spectrum of baryon resonances.

We focus the discussion on excited states of nucleon and $\Delta$,
and include low-mass $\Lambda$ and $\Sigma$. There is not much known
on the quantum numbers of $\Xi$ and $\Omega$ baryons. An exception
is the recent determination of the $\Xi_{1/2^+}(1690)$ quantum
numbers from $\Lambda_c\to (\Lambda K^0_S)K^+$ decays
\cite{Petersen:2006xd}. A similar classification of baryon
resonances was suggested by \cite{Melde:2008yr}. For low-lying
states, most assignments agree; discrepancies show that present data
do not suffice to identify all states in a unique way.\vspace{-2mm}
 \subsub{First excitation band}\vspace{-2mm}
The first excitation band $(D,L^P_{\textsf{N}})=(70,1^-_1)$ contains
negative-parity resonances. With the SU(3)$_{\rm f}$ decomposition
 \begin{eqnarray}\label{band1}
 70 = {^2}10\ \oplus\ \ {^4}8\ \oplus\ \ {^2}8\ \oplus\ {^2}1,
 \label{70}
 \end{eqnarray}
we expect as non-strange baryons a SU(3)$_{\rm f}$-octet  spin
doublet, a SU(3)$_{\rm f}$-octet spin triplet (a degenerate
quartet), and a SU(3)$_{\rm f}$-decuplet spin doublet. In Table
\ref{Firstband}, the low-mass negative parity states are collected.
The multiplet structure is easily recognized in the data.
Configuration mixing is of course possible for states with the same
$J^P$.

In the hyperon sector, a few expected states have not yet been
observed. A missing state is indicated in Table \ref{Firstband} by
an $x$. Based on eqs. (\ref{Delta_ADS}-\ref{LS_ADS}), we expect all
missing $\Lambda$ and $\Sigma$ states to fall into the 1750 to
1850\,MeV mass range. We have omitted the one-star \Sdthreeabf. The
Crystal Ball Collaboration studied the reaction
$K^-p\to\Lambda\pi^0$ in the c.m.\ energy range 1565 to 1600\,MeV
\cite{Olmsted:2003is}. Their results disagreed strikingly with older
fits which included the \Sdthreeabf\ resonance. Instead, they proved
the absence of any reasonably narrow resonance in this mass range.
 \begin{table}[pt]
\caption{\label{Firstband} The negative parity states of the first
excitation band $(D,L^P_{\textsf{N}})=(70,1^-_1)$. An x stands for a
missing state.}
\begin{center}
\begin{footnotesize}
\renewcommand{\arraystretch}{1.5}
\begin{tabular}{ccccc}
\hline\hline
 ${\bf\ D}; s$     && $J=1/2$ &$J=3/2$ &$J=5/2$ \\
\hline
${\bf 70, 8}; 1/2$&&\Nsoneabf & \Ndthreeabf &  \\
${\bf 70, 8}; 3/2$&& \Nsonebbf & \Ndthreebbf &
\Ndfivea\\
${\bf 70,10}; 1/2$&&\Dsoneabf & \Ddthreeabf \\
\hline
${\bf\ 1, 1}; 1/2$&&\Lsoneabf &\Ldthreeabf   &  \\
${\bf 70, 8}; 1/2$&&\Lsonebbf &\Ldthreebbf  &  \\
${\bf 70, 8}; 3/2$&&\Lsonecbf & $x$ &\Ldfiveabf \\
\hline
${\bf 70, 8}; 1/2$&&\Ssoneabf & \Sdthreebbf&  \\
${\bf 70, 8}; 3/2$&& \Ssonebbf&$x$&\Sdfiveabf\\
${\bf 70,10}; 1/2$&  & $x$ & $x$  & \\
\hline\hline
\end{tabular}
\renewcommand{\arraystretch}{1.0}
\end{footnotesize}
\end{center}
\end{table}

In the $\Lambda$ sector, the  \Lsoneabf\ and \Ldthreeabf\ are
considerably lower in mass than \Lsonebbf\ and \Ldthreebbf. In quark
models, this might be due to favorable hyperfine effects acting on a
pair of light quarks with $l_\rho=0$ and spin 0.  There is also a
copious literature on the effect of coupling to decay channels, or
multiquark components in these states
\cite{Oset:1997it,Choe:1997wz}.

A similar effect can be observed in heavy-flavor baryons. The mass
difference between the $\Lambda_c^+$ ground state and the first
excited states (a doublet) is 325\,MeV, rather low for an orbital
excitation. Like the \Lsoneabf, the two negative-parity states
$\Lambda_c^+(2595)$ and $\Lambda_c^+(2625)$ benefit from the
attractive spin--spin splitting for the light quark pair.

The classification of low-mass negative-parity states in Table
\ref{Firstband} is rather conventional. Nevertheless, we point out
some trivialities. Pairs of states with $J^P=1/2^-$ or $3/2^-$ can
mix (see Eq. \ref{qms} in section \ref{Naming scheme}). The mixing
angle between the two $1/2^-$ states was calculated to be
$-31.7^{\circ}$ \cite{Isgur:1977ef}; for the two $3/2^-$ states, it
is $6^{\circ}$. The probability to find a $S=3/2$ in the
$N_{1/2^-}(1535)$ is $\propto\sin^2\Theta_{1/2^-}=0.28$, the mean
mass separation between the triplet and the doublet is about
150\,MeV. A mixing angle of $30^{\circ}$ does not prevent
identification of the leading component.

In this spirit, we will try to identify leading components also for
higher excitation bands. We are aware of the fact that with
increasing mass, the predicted complexity of the spectrum increases
dramatically, and mixing of states is expected to become a severe
problem. Hence the assignments will become more and more
speculative. The reason why we include a discussion on higher
excitation bands are three-fold: first, there are unexpected
clusters of resonances of different spin-parities (but forming spin
multiplets) spanning a narrow mass interval. Second, the observed
multiplets can be arranged into a small number of
$(D,L^P_{\textsf{N}})$ supermultiplets which sometimes are
completely filled while others remain empty. And third, the observed
multiplets can be characterized by $L$ and $\textsf{N}$, just those
variables which result from AdS/QCD.\vspace{-2mm}
 \subsub{The second excitation band}\vspace{-2mm}
 In the HO model, the second excitation band contains states with
 either two units of angular momentum or one unit of radial excitation,
 with proper antisymmetrization in the case of identical quarks:
\begin{subequations}\label{band2}
  \begin{eqnarray}
  (D,L^P_{\textsf{N}})=&\underline{(56,2^+_2)},& \underline{(70,2^+_2)},\vspace{-4mm}\\
  (D,L^P_{\textsf{N}})=&           & (20,1^+_2),\vspace{-4mm}\\
  (D,L^P_{\textsf{N}})=&\underline{(56,0^+_2)},& \underline{(70,0^+_2)},\vspace{-2mm}
  \end{eqnarray}
\end{subequations}
 with either $(l_{\rho},
l_{\lambda})=(0,2)$ and $(2,0)$ yielding the $(56,2^+_2)$ multiplet,
or with $l_{\rho}, l_{\lambda}=1,1$ coupling to $L=0,1,2$ yielding
$(70,2^+_2)$, $(20,1^+_2)$, and $(70,0^+_2)$. The $(56,0^+_2)$
supermultiplet comprises the first radial excitations with
$(n_{\rho},n_{\lambda})=(0,1)$ or $(1,0)$. Both multiplets with
$L^P=0$ contain nucleons with spin-parity $1/2^+$, while for
decuplet states, $J^P=3/2^+$ for 56-plet members and $J^P=1/2^+$ for
70-plet members.

We begin with $(D,L^P_{\textsf{N}})=(56,0^+_2)$. The most
controversial state is the Roper resonance $N_{1/2^+}(1440)$. In the
HO model, it is degenerate with other $N=2$ states, but in the
experimental spectrum of the nucleon and $\Delta$, it is almost
degenerate, and even slightly below the $N=1$ states with negative
parity. Anharmonic corrections push this state down, and this
perturbative result is confirmed in the hypercentral approximation
\cite{Hogaasen:1982rb}, which is a better approximation to
confinement that is not quadratic. Even in exact treatments of the
three-body problem, but with local, flavor independent potentials of
confining type, the Roper resonance comes always \emph{above} the
first negative-parity states.

The ``wrong" mass of the Roper resonance has initiated a
longstanding debate if it is dynamically generated {\it or} if it is
the nucleon first radial excitation and a quark-model state. We
think it is both. An enlightening discussion of the (im-)possibility
to distinguish meson-meson molecules from four-quark states can be
found in \cite{Jaffe:2007id}. In Table \ref{Scalar}, the
lowest-lying resonances having the same quantum numbers as their
respective ground states and the mass square distance to them are
listed. In colloquia, Nefkens calls them Roper, Loper, Soper, Xoper,
and Doper \cite{Nefkens:prcomm}, to underline that they play similar
roles. If the Roper resonance should be generated by $\Delta\pi$
dynamics without any relation to the quark-model
$(D,L^P_{\textsf{N}})=(56,0^+_2)$ state, $\Sigma_{1/2^+}(1660)$ and
$\Xi_{1/2^+}(1690)$ could be generated by the same mechanism (making
use of $\Sigma_{3/2^+}(1385)\pi$ and $\Xi_{3/2^+}(1530)\pi$). But
there is no analogous mechanism which would lead to
$\Lambda_{1/2^+}(1600)$ and $\Delta_{3/2^+}(1600)$. Understanding
$N_{1/2^+}(1440)$ from the interaction of  mesons and baryons is an
important step in understanding baryons and their interactions;
S-wave thresholds may have an important impact on the precise
location of poles and on the observed branching ratios. The pattern
of states and their approximate mass values seem, however, not or
hardly affected.

\begin{table}[pt]
\caption{\label{Scalar}Members of the
$(D,L^P_{\textsf{N}})=(56,0^+_2)$ and
$(D,L^P_{\textsf{N}})=(70,0^+_2)$ multiplets in the second
excitation band and mass square difference (in GeV$^2$) to the
respective ground state. The expected values for the mass square
differences are 1.08 and 2.16 GeV$^2$, respectively (see Eq.
(\ref{Delta_ADS}) and Table \ref{Gooddiquarks}).\vspace{-3mm} }
\begin{center}
\begin{footnotesize}
\renewcommand{\arraystretch}{1.5}
\begin{tabular}{ccccc}
\hline\hline 56, 8; 1/2&$N_{1/2^+}(1440)$&$\Lambda_{1/2^+}(1600)$&
$\Sigma_{1/2^+}(1660)$ & $\Xi_{1/2^+}(1690)$\vspace{-1mm} \\
\scriptsize$\delta M^2$&\scriptsize 1.19$\pm$0.11&\scriptsize
1.31$\pm$ 0.11& 1.34$\pm$0.11&\scriptsize 1.13$\pm$0.03 \\
56, 10; 3/2&$\Delta_{3/2^+}(1600)$&&x&x\\
\scriptsize$\delta M^2$&\scriptsize$1.04\pm0.15$&\\
\hline 70, 8; 1/2&$ N_{1/2^+}(1710)$&$\Lambda_{1/2^+}(1810)$&
$\Sigma_{1/2^+}(1770)$ & x\vspace{-1mm} \\
\scriptsize$\delta M^2$&\scriptsize 2.04$\pm$0.15&\scriptsize
2.03$\pm$ 0.15&\scriptsize $1.72\pm0.16$&\\
70, 10; 1/2&$\Delta_{1/2^+}(1750)$&&$\Sigma_{1/2^+}(1880)$&x\vspace{-1mm}\\
\scriptsize$\delta M^2$&\scriptsize$1.54\pm0.16$&&\scriptsize2.12$\pm$0.11\\
\hline\hline
\end{tabular}
\renewcommand{\arraystretch}{1.0}
\end{footnotesize}\vspace{-3mm}
\end{center}
\end{table}

Commonly, $N_{1/2^+}(1710)$ and $\Delta_{1/2^+}(1750)$ are
candidates assigned to $(D,L^P_{\textsf{N}})=(70,0^+_2)$, and
$\Sigma_{1/2^+}(1880)$  belongs to it as well. These baryons
represent a new class: the two angular momenta $l_{\rho}$ and
$l_{\lambda}$ are both one and couple to zero. $N_{1/2^+}(1710)$
could also be assigned to the forth excitation band, with 2 units of
radial excitation, but this interpretation is forbidden for
$\Delta_{1/2^+}(1750)$ and unlikely for $\Sigma_{1/2^+}(1880)$. The
former is a 1-star resonance, the latter one has two stars; the PDG
entry $\Sigma_{1/2^+}(1880)$ represents all claims above
$\Sigma_{1/2^+}(1770)$. Supposing their existence, we interpret the
three states as members of the $(D,L^P)=(70,0^+)$ multiplet.

We now turn to $(D,L^P_{\textsf{N}})=(56,2^+_2)$. In the nucleon
spectrum, there should be (at 1.62\,GeV) a spin doublet, in the
$\Delta$ spectrum a spin quartet (at 1.92\,GeV). These are all
readily identified in the spectrum (see Table
\ref{Secondexcitationband}). For the $\Lambda$ and $\Sigma$
excitations, the corresponding states should be at 1.84\,GeV and
2.03\,GeV. All but one state are observed.

The situation is more difficult for
$(D,L^P_{\textsf{N}})=(70,2^+_2)$. We expect a spin doublet
(1.78\,GeV; 1.90\,GeV) and a spin quartet (1.92\,GeV; 2.03\,GeV) of
octet states (mass estimates are for non-strange and strange
baryons). The anchor for $L=2, S=3/2$ states are those having
$J^P=7/2^+$. These are the 2-star \Nfsevenabf\ and the 1-star
\Lfsevenabf. The nucleon quartet can be completed, the $\Lambda$
quartet misses two states, and there is no evidence for a second
$\Sigma$ quartet. Most of the states have 1 or 2 stars, except the
3-star \Lffivebbf.

The interpretation of \Spthreebbf, \Sffivebbf, and \Sfsevenabf\ is
ambiguous; in  Table \ref{Secondexcitationband} these states are
assigned to the decuplet but they may as well be octet states. As
56-plet, they are strange partners of the quartet of $\Delta$
resonances mentioned above which are observed clearly in $\pi N$
scattering. As 70-plet, they would be partners of the more elusive
\Nponecbf, \Npthreebbf, \Nffivebbf, and \Nfsevenabf.

\begin{table}[pt]
\caption{\label{Secondexcitationband}$(D,L^P_{\textsf{N}})=(56,2^+_2)$,
$(D,L^P_{\textsf{N}})=(70,2^+_2)$, and
$(D,L^P_{\textsf{N}})=(20,1^+_2)$ resonances in the second
excitation band.}
\begin{center}
\begin{footnotesize}
\renewcommand{\arraystretch}{1.5}
\hspace{-3mm}\begin{tabular}{ccccc} \hline\hline
$D;\quad s$     & $J=1/2$ &$J=3/2$ &$J=5/2$&$J=7/2$ \\
\hline
56, 8; 1/2&\hspace{-1mm}&\hspace{-1mm}\Npthreeabf\hspace{-1mm}&\hspace{-1mm}\Nffiveabf&  \\
56, 8; 1/2&\hspace{-1mm}&\hspace{-1mm}\Lpthreeabf\hspace{-1mm}&\hspace{-1mm}\Lffiveabf &  \\
56, 8; 1/2&\hspace{-1mm}&\hspace{-1mm}\Spthreeabf\hspace{-1mm}&\hspace{-1mm}\Sffiveabf &  \\
56,10; 3/2&\Dponebbf\hspace{-1mm}&\hspace{-1mm}
\Dpthreecbf\hspace{-1mm}&\hspace{-1mm}\Dffiveabf\hspace{-1mm}&\hspace{-1mm}\Dfsevenabf\\
56,10; 3/2&\hspace{-1mm}x&\hspace{-1mm}
\Spthreebbf\hspace{-1mm}&\hspace{-1mm}\Sffivebbf\hspace{-1mm}&\hspace{-1mm}\Sfsevenabf\\
\hline
70, 8; 3/2&\Nponecbf\hspace{-1mm}&\hspace{-1mm}\Npthreebbf\hspace{-1mm}&\hspace{-1mm}\Nffivebbf\hspace{-1mm}&\hspace{-1mm}\Nfsevenabf\\
70, 8; 3/2&x&x&\hspace{-1mm}\Lffivebbf\hspace{-1mm}&\hspace{-1mm}\Lfsevenabf\\
70, 8; 3/2&x&x&x&\qquad x\quad\hfill ($\Sigma$)\\
70, 8; 1/2&&x&x&\hfill ($N,\Lambda,\Sigma$)\\
70,10; 1/2&&x&x&\hfill ($\Delta,\Sigma$)\\\hline
20, 8; 1/2&x & x&&\hfill ($N,\Lambda,\Sigma$)\\
\hline\hline
\end{tabular}
\renewcommand{\arraystretch}{1.0}
\end{footnotesize}
\end{center}\vspace{-3mm}
\end{table}

In the second excitation band, the 56-plet is nearly complete and
most states are well established. Spatial wave functions can be
constructed which require excitation of one oscillator only. The
70-plet spatial wave functions have components in which a single
oscillator is excited and components with both oscillators being
excited. Several candidates exist, mostly however with 1- or 2-star
status.

In the non-strange sector, four supermultiplets, underlined in Eq.
\ref{band2}, are nearly full while the
$(D,L^P_{\textsf{N}})=(20,1^+_2)$ multiplet is empty. It has an
antisymmetric spatial wave function which is $\vec\rho\times\vec
\lambda\,\psi_0$ in the HO model. Clearly, the wave function has no
component with only one oscillator excited. Assuming (somewhat
deliberately) that in $\pi N$ scattering and in production
experiments, only one of the oscillators is excited, we can
``understand" the absence of this state in the observed spectrum,
provided mixing with nearby states having identical quantum numbers
is small.\vspace{-2mm}
 \subsub{The third excitation band}\vspace{-2mm}
In the third  band, the number of expected states increases
significantly. In the harmonic oscillator basis, the following
multiplets are predicted:
\begin{subequations}\label{band3}
  \begin{eqnarray}
(D,L^P_{\textsf{N}})&=&\underline{(56,1^-_3)},\ 2\times (70,1^-_3), (20,1^-_3),\vspace{-4mm} \\
(D,L^P_{\textsf{N}})&=&(70,2^-_3),                             \vspace{-4mm}\\
(D,L^P_{\textsf{N}})&=&(56,3^-_3),\ \underline{(70,3^-_3)},\
(20,3^-_3),\vspace{-2mm}
  \end{eqnarray}
\end{subequations}
Thus, 45 $N^*$ and $\Delta^*$ resonances are expected while only 12
resonances are found in the 1800 to 2300 MeV mass range. Most of
them are decorated with 1 or 2 stars, and some of them will be
assigned to the fifth band. All candidates belong just to the two
underlined multiplets. The breakdown into states of defined spin and
parity is given in Table \ref{Band3-missing}.

We first look for nucleon resonances with mass below 2.3\,GeV and
large angular momenta. These are \Ngsevenabf\  and \Ngnineabf. Based
on the Regge trajectory of Fig.~\ref{delta_table}, we assign
$L$\,=\,3 to both of them. We propose the assignments of Table
\ref{Thirdband} as $(D,L^P_{\textsf{N}})$\,=\,$(70,3^-_3)$ states:
\Ngnineabf\ is a 4-star ``stretched" state with $L=3, S=3/2$; these
often leave a more significant trace in the data then states which
would fall onto a daughter Regge trajectory. Likewise, we propose
\Ngsevenabf\ to have $L=3, S=1/2$ with spin and orbital angular
momenta aligned. The two states \Ndfivebbf\ and \Ngsevenabf\ could
also be members of the spin quartet.  The $N_{5/2^-}(2070)$ is
observed, jointly with \Nsoneabf\ and \Npthreeabf, to have strong
coupling to $N\eta$. The pattern is used in \cite{Bartholomy:2007zz}
to argue that the state has $S=1/2$. The two resonances \Nsonedbf\
and \Ndthreedbf\ are tentatively interpreted as second radial
excitations and are assigned to
$(D,L^P_{\textsf{N}})$\,=\,$(70,1^-_5)$.

There is a striking sequence of negative-parity $\Delta$ states in
the 1900-2000\,MeV region, the \Dsonebbf , \Ddthreebbf, and
\Ddfiveabf\ resonances. They could belong to two different doublets
with $L=1$ and $L=3$; the partner of \Ddfiveabf\ would then be
\Dgsevenabf. In view of the absence of a large $\vec L\cdot\vec S$
splitting in other cases, the mass separation seems rather large,
and we do not follow this path. A future discovery of a $7/2^-$
state below 2\,GeV - as predicted by Glozman (pr. comm.) - would
lead to a different interpretation.

We assign the three states to a triplet in the
$(D,L^P_{\textsf{N}})$\,=\,$(56,1^-_3)$ multi\-plet. The triplet is
separated in mass square from the doublet \Dsonea , \Ddthreea\ by
0.94 GeV$^2$ (which is similar to the $N(1440)$--$N(940)$ mass
square difference). If this is true, there must be a spin doublet
nucleon pair of resonances with $J=1/2^-$ and $J=3/2^-$ below
1.9\,GeV (to allow for a mass shift by a finite good-diquark
fraction). This pair indeed exists, even though with debatable
confidence. The states are listed in Table \ref{Thirdband}. The
56-multiplet is full.

\begin{table}[pt]
\caption{\label{Band3-missing}Number of expected states in the third
excitation band and observed states in the 1.8 to 2.4\,GeV mass
range ($N$ and $\Delta$).
 }\renewcommand{\arraystretch}{1.2}
\begin{footnotesize}
\begin{tabular}{cccccc}
\hline\hline \hspace{-1mm}&\hspace{-1mm} $N_{1/2^-}$
\hspace{-1mm}&\hspace{-1mm} $N_{3/2^-}$ \hspace{-1mm}&\hspace{-1mm}
$N_{5/2^-}$ \hspace{-1mm}&\hspace{-1mm} $N_{7/2^-}$\hspace{-1mm}&
\hspace{-1mm}$N_{9/2^-}$\hspace{-1mm}\\
exptd\hspace{-7mm}&   7   &    9   &    8   &      5  &    1       \\
obsvd\hspace{-7mm}&   2   &    2   &    1   &      1  &    1        \\
\hline\hline \hspace{-1mm}&\hspace{-1mm} $\Delta_{1/2^-}$
\hspace{-1mm}&\hspace{-1mm} $\Delta_{3/2^-}$
\hspace{-1mm}&\hspace{-1mm} $\Delta_{5/2^-}$
\hspace{-1mm}&\hspace{-1mm} $\Delta_{7/2^-}$\hspace{-1mm}&
\hspace{-1mm}$\Delta_{9/2^-}$\hspace{-1mm}\\
exptd\hspace{-7mm}&   3   &    5   &    4   &      2  &    1       \\
obsvd\hspace{-7mm}&   2   &    1   &    2   &      1  &    1
\\ \hline\hline\end{tabular}
\end{footnotesize}
\renewcommand{\arraystretch}{1.0}
\end{table}
\begin{table}[pt]
\caption{\label{Thirdband}The negative parity states of the third
excitation band $(D,L^P_{\textsf{N}})=(56,1^-_3)$ and
$(D,L^P_{\textsf{N}})=(70,3^-_3)$. }
\begin{center}
\begin{footnotesize}
\renewcommand{\arraystretch}{1.5}
\begin{tabular}{ccccc}
\hline\hline
 ${\bf\ D}; s$    & &$J=1/2$ & $J=3/2$ &$J=5/2$  \\
\hline
${\bf 56, 8}; 1/2$& & \Nsonecbf          & \Ndthreecbf     &\\
${\bf 56,10}; 3/2$&&\Dsonebbf & \Ddthreebbf & \Ddfiveabf  \\
\hline\hline
 ${\bf\ D}; s$     &$J=3/2$ & $J=5/2$ &$J=7/2$ &$J=9/2$ \\
\hline
${\bf 70, 8}; 1/2$&&$N_{5/2^-}(2070)$& \Ngsevenabf\  &  \\
${\bf 70, 8}; 3/2$&x&\Ndfivebbf            &  $x$    &\Ngnineabf \\
${\bf 70,10}; 1/2$&& $x$ & \Dgsevenabf & \\
\hline\hline
\end{tabular}
\renewcommand{\arraystretch}{1.0}
\end{footnotesize}
\end{center}
\end{table}

The assignment of the three states \Dsonebbf , \Ddthreebbf, and
\Ddfiveabf\ assumes that they are of the same kind. For quark
models, they are found at a rather low mass, $M\approx 2200$\,MeV is
expected. \cite{Gonzalez:2008pv} suggest to explain at least
\Ddfiveabf\ as $\rho\Delta$ bound state while the other two are
predicted to have a large $\rho\Delta$ component.

Does this finding imply that we can neglect \Ddfiveabf\ for our
discussion of quark model states? We do not believe so. Chiral
dynamics is an important tool to understand properties of baryon
(and meson) resonances. But it addresses the same objects. The
famous \Nsoneabf\ can be understood as dynamically generated
resonance. But it is a quark model state as well. Resonances are not
independent of their decays, they can often be constructed from
their decays, but this does not imply that they are supernumerous
from the quark model point of view.

\Dgnineabf\ has a mass which makes it unlikely to have (dominantly)
$L=5$ intrinsic orbital angular momentum. With $L=3$, it needs a
quark spin $S=3/2$. Using quark model arguments only, \Dgsevenabf\
and \Dgnineabf\ could both be
$(D,L^P_{\textsf{N}})$\,=\,$(56,3^-_3)$ multiplet members. However,
there is a 200\,MeV mass difference between the two states and, in
view of Fig.~\ref{delta_table}, we assign \Dgsevenabf\ to the
$(D,L^P_{\textsf{N}})$\,=\,$(70,3^-_3)$ and \Dgnineabf\ to
$(D,L^P_{\textsf{N}})=(56,3^-_5)$. We thus propose that
odd-angular-momentum $\Delta$ states are in a 56-plet if and only if
there is a simultaneous excitation of the radial quantum number. The
\Ddfivebbf\ resonance could be a spin partner of either \Dgsevenabf\
or \Dgnineabf , or the entry may comprise two resonances. The
\Dsonecbf\ is the third state with these quantum numbers. It might
be a second radial excitation and belong to
$(D,L^P_{\textsf{N}})$\,=\,$(70,1^-_5)$.\vspace{-2mm}
 \subsub{Further excitation bands}\vspace{-2mm}
In the forth band, the number of states is exploding while data are
scarce. Expected is a large number of multiplets (\ref{band4}),
\begin{subequations}\label{band4}
  \begin{eqnarray}
 (D,L^P_{\textsf{N}})&=&2\times (56,0^+_4),\  2\times
 (70,0^+_4),\vspace{-4mm}\\
 (D,L^P_{\textsf{N}})&=&(20,1^+_4),\ (70,1^+_4),\vspace{-4mm}\\
 (D,L^P_{\textsf{N}})&=&2\times (56,2^+_4),\ 3\times (70,2^+_4),\
 (20,2^+_4),\vspace{-4mm}\\
 (D,L^P_{\textsf{N}})&=&(70,3^+_4),\ (20,3^+_4),\vspace{-4mm}\\
 (D,L^P_{\textsf{N}})&=&\underline{(56,4^+_4)},\ 2\times (70,4^+_4),\vspace{-2mm}
  \end{eqnarray}
\end{subequations}
while only few of them are found (Table \ref{Band4-missing}). The
large number of expected states is one of the unsolved issues in
baryon spectroscopy. It is known as the problem of the {\it missing
resonances}. Equation~(\ref{band4}) gives the decomposition of
expected states into multiplets. While 93 $N$ and $\Delta$
resonances are expected, 4 are found. All four observed states,
\Nhnineabf, \Dfsevenbbf, \Dhnineabf, and \Dhelevenabf, when
interpreted as $L=4$ $S=1/2$ nucleon and $S=3/2$ $\Delta$
resonances, belong to the $(D,L^P_{\textsf{N}})=(56,4^+_4)$
supermultiplet, in which only two states, a $N_{7/2^+}$ and a
$\Delta_{5/2^+}$, are missing.

The spectrum continues with \Ddfivebbf\ and \Dgnineabf\ ($L=3,N=1$),
\Nielevenabf\ ($L=5,N=0$) in the $5^{\rm th}$, with \Nkthirteenabf\
and \Dkfifteenabf\ ($L=6,N=0$) in the $6^{\rm th}$, and
\Dithirteenabf\ ($L=5,N=1$) in the $7^{\rm th}$ band. The number of
expected states increases dramatically. We conjecture that at high
masses, beyond 2.3\,GeV, all observed nucleons have $\vec J=\vec L+
\vec S$ have spin 1/2 and all $\Delta$ resonances, spin 3/2.

\begin{table}[pt]
\caption{\label{Band4-missing}Number of expected states in the forth
excitation band and observed states in the 2.1 to 2.5\,GeV mass
range ($N$ and $\Delta$).
 }\renewcommand{\arraystretch}{1.2}
\begin{footnotesize}
\begin{tabular}{ccccccc}
\hline\hline \hspace{-1mm}&\hspace{-1mm} $N_{1/2^+}$
\hspace{-1mm}&\hspace{-1mm} $N_{3/2^+}$ \hspace{-1mm}&\hspace{-1mm}
$N_{5/2^+}$ \hspace{-1mm}&\hspace{-1mm} $N_{7/2^+}$\hspace{-1mm}&
\hspace{-1mm}$N_{9/2^+}$\hspace{-1mm}&
\hspace{-1mm}$N_{11/2^+}$\hspace{-1mm}\\
exptd\hspace{-7mm}&   10  &    14   &    16   &   12  &    7&2       \\
obsvd\hspace{-7mm}&   0   &    0   &    0   &   0 &  1  &    0        \\
\hline\hline \hspace{-1mm}&\hspace{-1mm} $\Delta_{1/2^+}$
\hspace{-1mm}&\hspace{-1mm} $\Delta_{3/2^+}$
\hspace{-1mm}&\hspace{-1mm} $\Delta_{5/2^+}$
\hspace{-1mm}&\hspace{-1mm} $\Delta_{7/2^+}$\hspace{-1mm}&
\hspace{-1mm}$\Delta_{9/2^+}$\hspace{-1mm}&
\hspace{-1mm}$\Delta_{11/2^+}$\hspace{-1mm}\\
exptd\hspace{-7mm}&   5   &    8   &    8   &    7    &    3&1       \\
obsvd\hspace{-7mm}&   0   &    0   &    1   &      1  &    1 &
\\ \hline\hline\end{tabular}
\end{footnotesize}
\renewcommand{\arraystretch}{1.0}
\end{table}
\vspace{-2mm}
 \subsub{Dynamical conclusions}\vspace{-2mm}
In the low-mass region, in the first excitation shell, the quark
model gives a perfect match of the number of expected and observed
states. Starting from $\mathtt{N}=2$, only states are realized in
which the $\rho$ and the $\lambda$ oscillator are excited coherently
(e.g., with a wave function $\propto\rho^2+\lambda^2$) while states
with both oscillators excited simultaneously (e.g., with a wave
function $\propto\rho\times\lambda$) have not been observed. If
mixing were important, their absence in the spectrum would pose a
severe problem for any quark model.

Positive-parity nucleon resonances with $L=2, S=3/2$ will have
$J^P=7/2^+$; indeed, a two-star \Nfsevenabf\ exists. Above, there is
a \Nhnineabf\ but no $11/2^+$ partner which should exist if
\Nhnineabf\ had $L=4, S=3/2$. Instead it likely has $L=4, S=1/2$.
Likewise, \Nkthirteenabf\ exists but no $15/2^+$ nucleon, and we
assign $L=6, S=1/2$. The four states $N(940)$, \Nffivea, \Nhnineabf,
and \Nkthirteenabf\ belong to the leading nucleon Regge trajectory.

Negative-parity nucleon resonances with the largest total angular
momenta (in a given mass interval) are \Ndfivea, \Ngnineabf,
\Nielevenabf, where the former two resonances obviously have $L=1$
and $L=3$ and $S=3/2$, and the latter one $L=5,S=1/2$. We conclude
that for up to $L=3$, nucleon resonances can have spin $S=1/2$ or
have spin $S=3/2$ while for high masses, the observed nucleon
resonances have spin $S=1/2$.

High-spin positive parity $\Delta$ resonances are readily identified
as \Dpthreea, \Dfsevenabf, \Dhelevenabf, \Dkfifteenabf\ with
$L=0,2,4,6$ and $S=3/2$ as leading contributions (and possibly some
small higher-$L$ components). The observed positive-parity $\Delta$
resonances can all be assigned to spin $S=3/2$ multiplets. The
\Dponeabf\ resonance is the only positive-parity $I=3/2$ resonances
which belongs to a 70-plet.

The negative-parity sector is a bit more complicated. $\Delta$
resonances with $L=1$, $S=3/2$ are forbidden for $n_{\lambda}=0$,
and resonances have either $S=1/2$, $n_{\lambda}=0$ (and belong to a
70-plet) or $S=3/2$, $n_{\lambda}=1$ (and belong to a 56-plet). For
$L=3$, $\Delta$ resonances still have either $S=1/2$,
$n_{\lambda}=0$ and belong to $70,3_3^-$, or they have $S=3/2$,
$n_{\lambda}=1$ ($56,1_3^-$) even though HO wave functions do not
forbid either $S=3/2$, $n_{\lambda}=0$ ($56,3_3^-$) or $S=1/2$,
$n_{\lambda}=1$ ($70,1_3^-$). For $L=5$, only $S=3/2$ and
$n_{\lambda}=1$ is observed.

In summary, most observed $\Delta$ resonances fall into 56-plets, in
70-plets $\Delta$ resonances are seen only up to the third shell.
Nucleon resonances above the third shell are in a 56-plet when they
have positive, in a 70-plet for negative parity. There are no states
which would need to be assigned to a 20-plet. In other words, the
experimentally known resonances above the third shell can be
described by a diquark in S-wave (with the $\rho$-oscillator in the
ground state) and the $\lambda$ oscillator carrying the full
excitation.

This rule leads to a selection of allowed multiplets which are
summarized in Table \ref{Martin}.\vspace{-3mm}

\begin{table}[pt]
\caption{\label{Martin}Observed multiplets at large angular
momenta}\bc
\renewcommand{\arraystretch}{1.5}
\begin{footnotesize}\begin{tabular}{cccc}
\hline\hline $N^*$ with  $P=+$: && \multirow{2}{*}{spin $S=1/2$} &
\multirow{2}{*}{$l_{\lambda}=L$; $n_{\rho}=0$}\\
$N^*$ with $P=-$: &&& \\\hline $\Delta^*$ with  $P=+$:
&&\multirow{2}{*}{spin $S=3/2$;} & $l_{\lambda}=L$; $n_{\rho}=0$\\
$\Delta^*$ with $P=-$:  && & $l_{\lambda}=L$;
$n_{\rho}=1$\\\hline\hline
\end{tabular}
\end{footnotesize}
\renewcommand{\arraystretch}{1.0}\vspace{-3mm}\ec
\end{table}

\subsection{Exotic baryons}
The search for exotic mesons, spin-parity exotics and crypto-exotic
states, has been a continuous stimulation of the field. Examples are
the $\pi_1(1400)$ and $\pi_1(1600)$ mesons with $J^{PC}=1^{-+}$
(quantum numbers which cannot come from $q\bar q$), the flavor
exotic states $Z^{\pm}(4050)$, $Z^{\pm}(4248)$, and $Z^{\pm}(4430)$
\cite{Abe:2007wga,Mizuk:2008me} (decaying into a pion and a $c\bar
c$ resonance), or mesons like $f_0(980)$, $a_0(980)$, $f_0(1500)$,
$X(3872)$ \cite{Abe:2007wga} which have attracted a large number of
theoretical papers trying to understand their nature either as
quarkonium states or as crypto-exotic states, as glueballs, as
weakly or tightly bound tetraquarks or as molecular states (among
other more exotic interpretations). The existence of exotic mesons
as additional states in meson spectroscopy is not beyond doubt: see,
e.g.,  \cite{Klempt:2007cp} for a critical and \cite{Crede:2008vw}
for a more optimistic view.

Intruders into the world of baryons would be identified
unambiguously when they have quantum numbers which differ from those
of $qqq$ baryons. There are no spin-parity exotic quantum numbers in
baryon spectroscopy, but flavor exotic states (containing an
antiquark in the flavor wave function) might exist. Most discussion
is directed to the question if crypto-exotic baryons exist.

Examples of baryons which may deserve an interpretation beyond the
quark model are $N_{1/2^+}(1440)$ which is found at an unexpectedly
low mass, $N_{1/2^-}(1535)$, a resonance which is observed at the
expected mass but with an unusual large decay branching ratio to
$N\eta$, and the $\Lambda_{1/2^-}(1405)$ and $\Lambda_{3/2^-}(1520)$
resonances with their low mass and unusual splitting. A consistent
\cite{Liu:2005pm,Zou:2007mk} -- even though controversial
\cite{Sibirtsev:2006ia}, \cite{Liu:2006ym} -- picture for these
possibly crypto-exotic baryons ascribes the mass pattern to a large
$qqqq\bar q$ fraction in the baryonic wave functions.\vspace{-2mm}
 \subsub{\label{pentaq}Pentaquarks}\vspace{-2mm}
The question of the existence of multiquark hadrons has been raised
at the beginning of the quark model, and is regularly revisited,
either due to fleeting experimental evidence or to theoretical
speculations. In the late 60's some analyses suggested a possible
resonance with baryon number $B=1$ and strangeness $S=-1$, opposite
to that of the $\Lambda$ or $\Sigma$ hyperons.

In 1976, a stable dihyperon $H$ was proposed \cite{Jaffe:1976yi},
whose tentative binding was due to coherence in the chromomagnetic
interaction. In 1987, Gignoux et al., and, independently, Lipkin
\cite{Gignoux:1987cn,Lipkin:1987sk} showed that the same mechanism
leads to a stable $(Q\bar{q}^4)$ below the threshold for spontaneous
dissociation into $(Q\bar q)+ (\bar{q}^3)$. This calculation, and
Jaffe's for his $H=(u^2d^2s^2)$ gave 300\,MeV of binding if the
light quark are treated in the SU(3)$_f$ limit (and $Q$ infinitely heavy
for the pentaquark) and if the short-range correlation $\langle
\delta^{(3)}(r_{ij})\rangle$ is borrowed from ordinary baryons.
However, relaxing these strong assumptions always goes in the
direction of less and less binding, and even instability. The $H$
was searched for in dozens of experiments \cite{Ashery:1995ha}. The
1987-vintage pentaquark was searched for by the experiment E791 at
Fermilab, \cite{Aitala:1997ja}, but the results are not conclusive.

Some years ago, a lighter pentaquark was found in photoproduction,
called $\Theta^+(1540)$ \cite{Nakano:2003qx}, inspired by the
beautiful theoretical speculation in a chiral soliton model
predicting an  (anti-) decuplet of narrow baryons
\cite{Diakonov:1997mm}, following, in turn,  a number of earlier
papers. The $\Theta^+(1540)$ was confirmed in a series of
low-statistics experiments. The decuplet was enriched by the doubly
charged $\Phi(1860)$ \cite{Alt:2003vb}; the missing members were
identified with $N_{1/2^+}(1710)$ and $\Sigma_{1/2^+}(1890)$. A
narrow peak in the $p D^{*-}$ and $\bar pD^{*+}$ distributions
signaled a baryon with an intrinsic $\bar c$-quark,
$\Theta_c^0(3100)$ \cite{Aktas:2004qf}.

These observations initiated a large number of further experimental
and theoretical studies which were reviewed by \cite{Dzierba:2004db}
and \cite{Hicks:2005gp}. Recent experiments had partly a very
significant increase in statistics but no narrow pentaquark state
was confirmed. It it exists, the $\Theta^+(1540)$ must be very
narrow: from the absence of a signal in the reaction $K^+d\to
K^0pp$, an upper limit of about 1\,MeV can be derived
\cite{Cahn:2003wq,Sibirtsev:2004bg,Workman:2004im}. The list of
experiments and upper limits for pentaquark production can be found
in PDG \cite{Wohl:2008st} from where we quote the final conclusion:
{\it The whole story - the discoveries themselves, the tidal wave of
papers by theorists and phenomenologists that followed, and the
eventual ``undiscovery"  - is a curious episode in the history of
science.} The evidence for a pentaquark interpretation
\cite{Kuznetsov:2008hj} of a narrow peak in the $n\eta$ invariant
mass spectrum at 1680\,MeV is weak; the peak is observed in
photoproduction of $\eta$-mesons off neutrons in a deuteron
\cite{Kuznetsov:2007gr,Jaegle:2008ux,Fantini:2008zz} but the data
are not really in conflict with standard properties of
$N_{1/2^-}(1535)$ and $N_{1/2^-}(1650)$ and interference between
them \cite{Anisovich:2008wd,Doring:2009qr}.\vspace{-2mm}
 \subsub{Dynamically generated resonances}\vspace{-2mm}
A number of baryon resonances has been suggested to be due to the
dynamics of the meson--baryon interaction. Before entering a
discussion of individual cases, we specify different views of the
meaning of ``dynamically generated resonances''. The $\Delta(1232)$
resonance, e.g., can be considered  as $\pi N$ resonance
\cite{Chew:1955zz}, and this remains the most efficient tool to
describe $\pi$-nucleus scattering, as a propagation of $\Delta$-hole
excitations. Alternatively, the $\Delta(1232)$ is easily described
in the quark model, mainly as a state of three light  quarks,
$(qqq)$, with spins aligned, but its higher Fock states certainly
accounts for  an overlap with  $\pi N$. Some quark models are
supplemented by explicit accounts for hadron--hadron components, as
e.g., \cite{Vijande:2008zn} for mesons with charm and strangeness.
Years ago, a model-independent analyses of the effect of hadronic
loops was proposed by \cite{Tornqvist:1984fy,Tornqvist:1985fi}.

When a resonance is close to the threshold for an important decay
mode, in particular for decays into two hadrons in S-wave, the
molecular component can become large. For the $\Delta(1232)$, this
is mostly a matter of taste whether it is first described as a quark
state acquiring hadron--hadron components, or built first from the
interaction of its decay products, i.e., generated dynamically.

The problem becomes of course much more delicate when dynamical
resonances are predicted atop the quark-model states, or when the
light quark baryons are disregarded altogether and replaced by a
systematics of meson--baryon excitations. A convincing formalism is
available: an effective field theory in terms of hadrons, with the
symmetries of QCD, and coupling adjusted by fitting the low-energy
strong-interaction data \cite{Gasser:1983yg,Gasser:1984gg}. However,
it is not obvious which spectrum would emerge.

Dynamically generated states can possibly be identified by a study
of their behavior as a function of the number of colors
\cite{Lutz:2001yb}. \cite{Hanhart:2007cm} points out that the
analytic structure of the meson-baryon scattering matrix at
important thresholds is different for (tightly-bound) $qqq$ states
and (weakly-bound) molecular states, and this provides a means to
identify the nature of a resonance. Chiral dynamics with unitarity
constraints and explicit resonance fields have provided a very good
picture of meson--nucleon scattering. When such a formalism is
implemented, additional resonances (genuine quark-model states) are
sometimes no longer required to fit the data, see, e.g.,
\cite{Meissner:1999vr,Doring:2004kt}.

We now turn to a discussion of some specific cases. \vspace{-3mm}

\paragraph{The Roper resonance}
The Roper resonance $N_{1/2^+}(1440)$ is the lowest-mass nucleon
resonance and has the quantum numbers of the nucleon. Its most
natural explanation as first radial excitation is incompatible with
quark models in which the radial excitation requires two
harmonic-oscillator quanta while the negative parity states like
$N_{1/2^-}(1535)$ require one quantum only. Even including
anharmonicity, the mass of the first radial excitation should always
be above the first orbital-angular-momentum excitation. Within the
constituent quark model with one-gluon-exchange
\cite{Capstick:1986bm} or instanton induced forces
\cite{Loring:2001kx}, the Roper $N_{1/2^+}(1440)$ should have a mass
80\,MeV above the $N_{1/2^-}(1535)$ mass, and not $\approx100$\,MeV
below it.

Models using Goldstone-boson exchange interactions
\cite{Glozman:1995fu} improve on the Roper mass but this success is
counterbalanced by two shortcomings:  the interaction is (1)
inappropriate to calculate the full hadronic spectrum, and (2)
restricted to light baryons. Only the lowest-mass excitations were
calculated with a comparatively large number of adjustable
parameters.

\begin{figure}[pt]
\begin{center}
\includegraphics[width=\columnwidth]{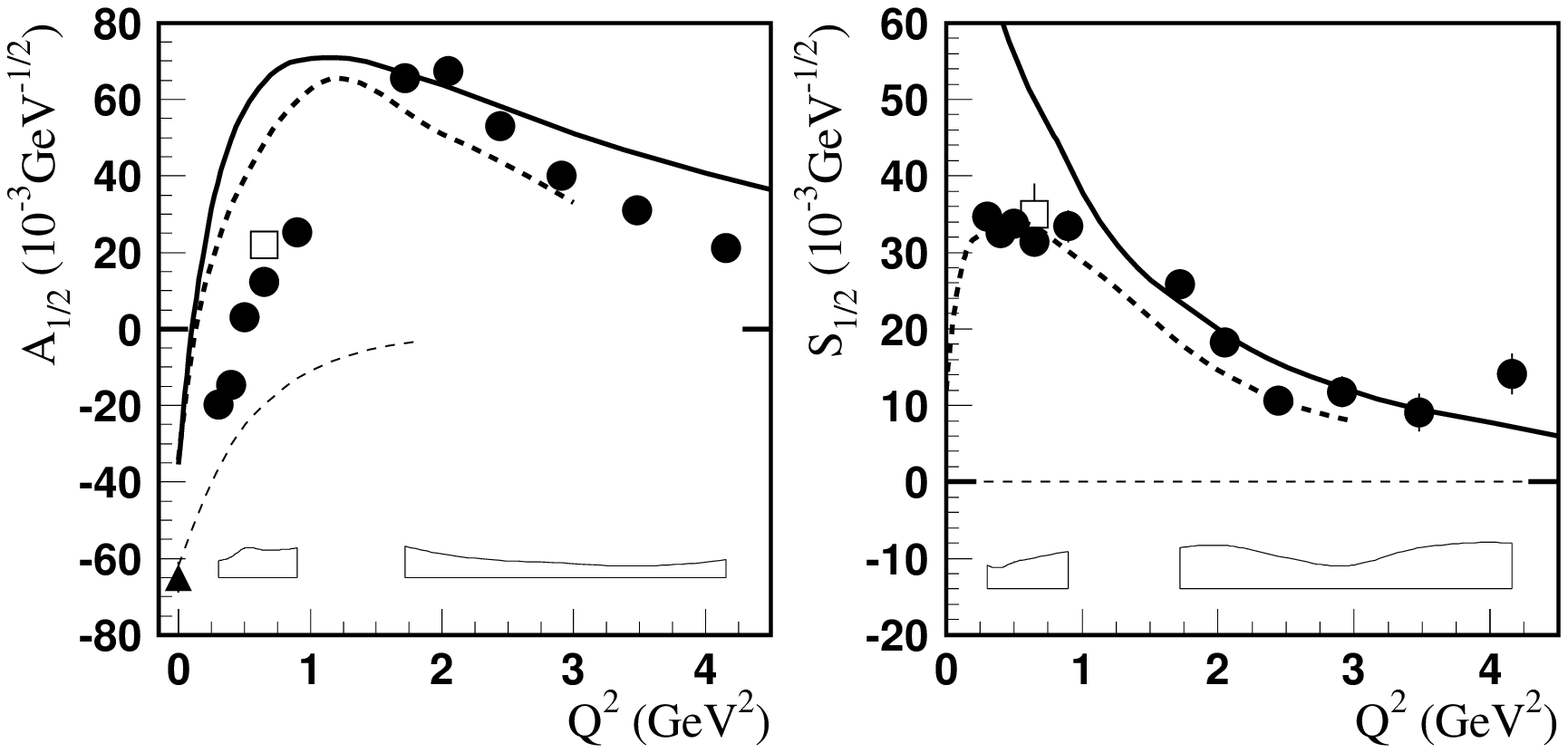}
\end{center}\vspace{-3mm} \caption{\small Helicity amplitudes for the
$\gamma^* p \rightarrow~N(1440)P_{11}$ transition. The full circles
are recent results from CLAS \cite{Aznauryan:2009mx}, open boxes are
results of an earlier analysis which included 2$\pi$
electroproduction data \cite{Aznauryan:2005tp}. The bands show the
model uncertainties. The full triangle at $Q^2=0$ is the PDG
estimate \cite{Amsler:2008zz}. The thick curves correspond to quark
models assuming that $N(1440)P_{11}$ is a first radial excitation of
the $3q$ ground state: \cite{Capstick:1994ne} (dashed),
\cite{Aznauryan:2007ja} (solid). The thin dashed curves are obtained
assuming that $N(1440)P_{11}$ is a gluonic baryon excitation (q$^3$G
hybrid state) \cite{Li:1991yba}.
\label{P11}}\vspace{-2mm}
\end{figure}

The Roper resonance has a surprisingly large width, and the
transition photo-coupling amplitude has even the wrong sign
\cite{Capstick:1994ne}. Some calculations on a lattice support the
idea that the Roper is not the radial excitation of the nucleon
\cite{Burch:2006cc,Borasoy:2006fk} (but others come to the contrary
conclusion \cite{Mathur:2003zf,Mahbub:2009aa}). These difficulties,
to explain the properties of the Roper resonance, encouraged
attempts to interpret the data dynamically, without introducing a
resonance. In a coupled-channel meson exchange model based on an
effective chiral-symmetric Lagrangian by \cite{Krehl:1999km}, no
genuine $qqq$-resonance was needed to fit $\pi N$ phase shifts and
inelasticity, in agreement with \cite{Schneider:2006bd}. Thus,
$N_{1/2^+}(1440)$ is often interpreted as an intruder into the world
of $qqq$ baryons. Yet, the sign change in the helicity amplitude
(Fig.~\ref{P11}) as a function of $Q^2$
\cite{Aznauryan:2008pe,Aznauryan:2009mx} does not support this
interpretation; it rather suggest a node in the wave function and
thus a radially excited state. The result does of course not rule
out a $qqqq\bar q$ ($N\pi$) component in the wave function as
suggested by \cite{Li:2006nm,JuliaDiaz:2006av}.

There has been the claim that the Roper resonance region might house
two resonances \cite{Morsch:2000xi}, one at 1390\,MeV with a small
elastic width and large coupling to $N\pi\pi$, and a second one at
higher mass -- around 1460\,MeV -- with a large elastic width and
small $N\pi\pi$ coupling. This idea was tested in
\cite{Sarantsev:2007bk} analyzing the over-constrained set of
reactions $\pi^-p\to N\pi$, $\pi^-p\to n\pi^0\pi^0$, $\gamma p\to
N\pi$, $\gamma p\to p\pi^0\pi^0$. A second pole was rejected unless
its width was sufficiently narrow to allow the resonance to have its
full phase motion in between the masses at which data are available.
We note in passing that in EBAC, no photo-produced Roper resonance
was found  in fits to the total cross section \cite{Kamano:2009im}.
But of course, such fits are much less sensitive to the underlying
dynamics than event-based likelihood fits performed by
\cite{Sarantsev:2007bk}.

We mention here a few further $N_{1/2^+}$ states: a narrow $N(1680)$
which might have been observed in $n\eta$ photoproduction was
already discussed as $N_{1/2^+}(1680)$ in the section on
pentaquarks. A $N_{1/2^+}(1880)$ was recently reported by
\cite{Castelijns:2007qt} from photoproduction and has been observed
by \cite{Manley:1984jz} in the reaction $\pi^-p\to p\pi^+\pi^-$. The
latter observation is listed in the PDG under $N_{1/2^+}(2100)$. The
$N_{1/2^+}(1710)$ resonance, questioned in the most recent analysis
of $\pi N$ elastic scattering \cite{Arndt:2006bf}, was required in
fits to $\pi N\to N\eta$ and $\pi N\to \Lambda K$
\cite{Ceci:2005vf,Ceci:2006ra}. \vspace{-3mm}

\paragraph{$N_{1/2^-}(1535)$ 3-quark resonance or $N\eta$-$\Sigma K$
coupled channel effect?} This resonance is observed at a mass
expected in quark models but its large decay branching ratio to
$N\eta$ invited speculations that it might be created dynamically.
An effective chiral Lagrangian, relying on an expansion in
increasing powers of derivatives of the meson fields and quark
masses, has been successful in understanding many $N_{1/2^-}(1535)$
properties (and of the meson-baryon system at low energies)
\cite{Kaiser:1995eg}. More recent studies -- with more data but
similar conclusions -- are presented in
\cite{Hyodo:2008xr,Geng:2008cv,Doring:2008sv}. \cite{Doring:2009uc}
studied the pole structure of $N_{1/2^-}(1535)$ and
$N_{1/2^-}(1650)$. If a dynamically generated $N_{1/2^-}(1535)$ is
introduced and an additional pole (as quark model state), the latter
pole moves far into the complex plane and provides an almost energy
independent background while the dynamically generated
$N_{1/2^-}(1535)$ pole appears as a stable object. However, the
dynamical generation of $N_{1/2^-}(1535)$ is tied to its strong
couplings to $K\Lambda$ and $K\Sigma$. If theses couplings are
reduced by about 40\% or 50\%, the dynamically generated resonance
disappears.

Experimentally, response functions, photo-couplings, and $\eta N$
coupling strengths as functions of the invariant squared momentum
transfer (measured for $Q^2$=0.13--3.3\,GeV$^2$) were deduced from a
measurement of cross sections for the reaction $ep \to e'\eta p$ for
total center of mass energies $W$=1.5--2.3\,GeV
\cite{Denizli:2007tq}.
 \begin{figure}[pt]
\bc
\includegraphics[width=.35\textwidth]{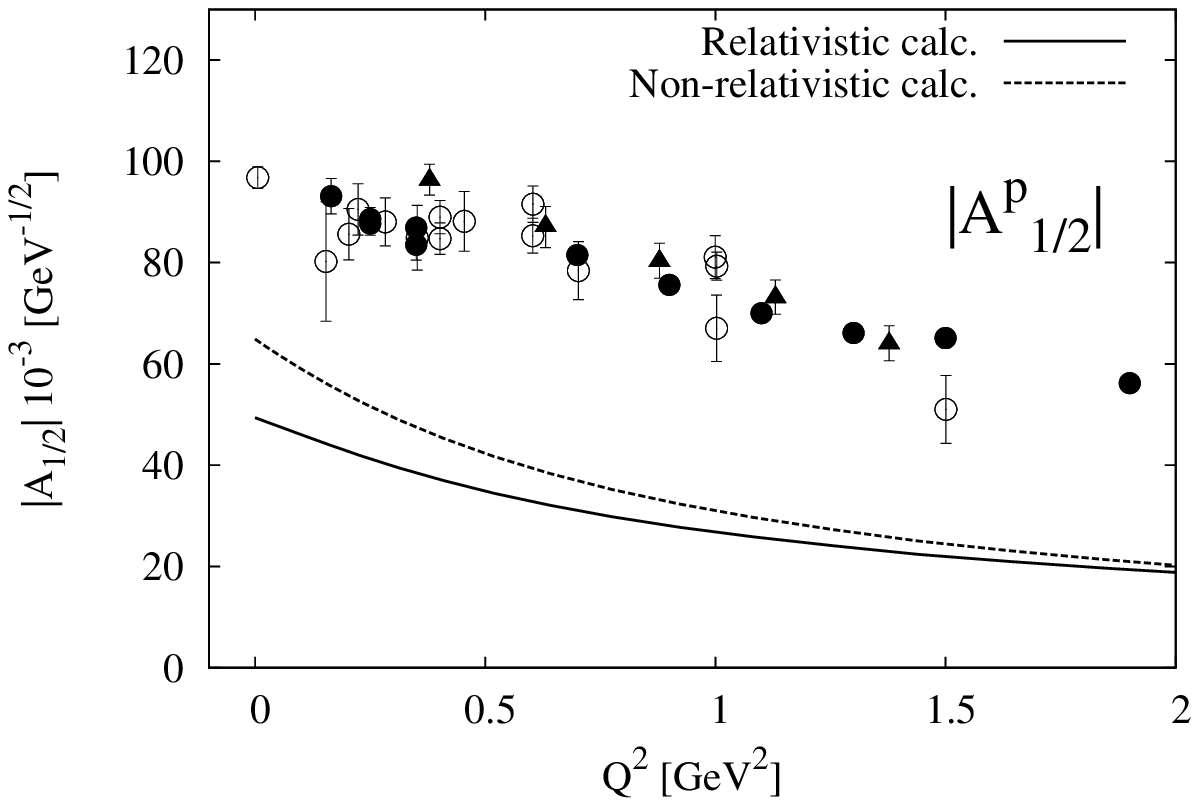}\\
\includegraphics[width=.35\textwidth]{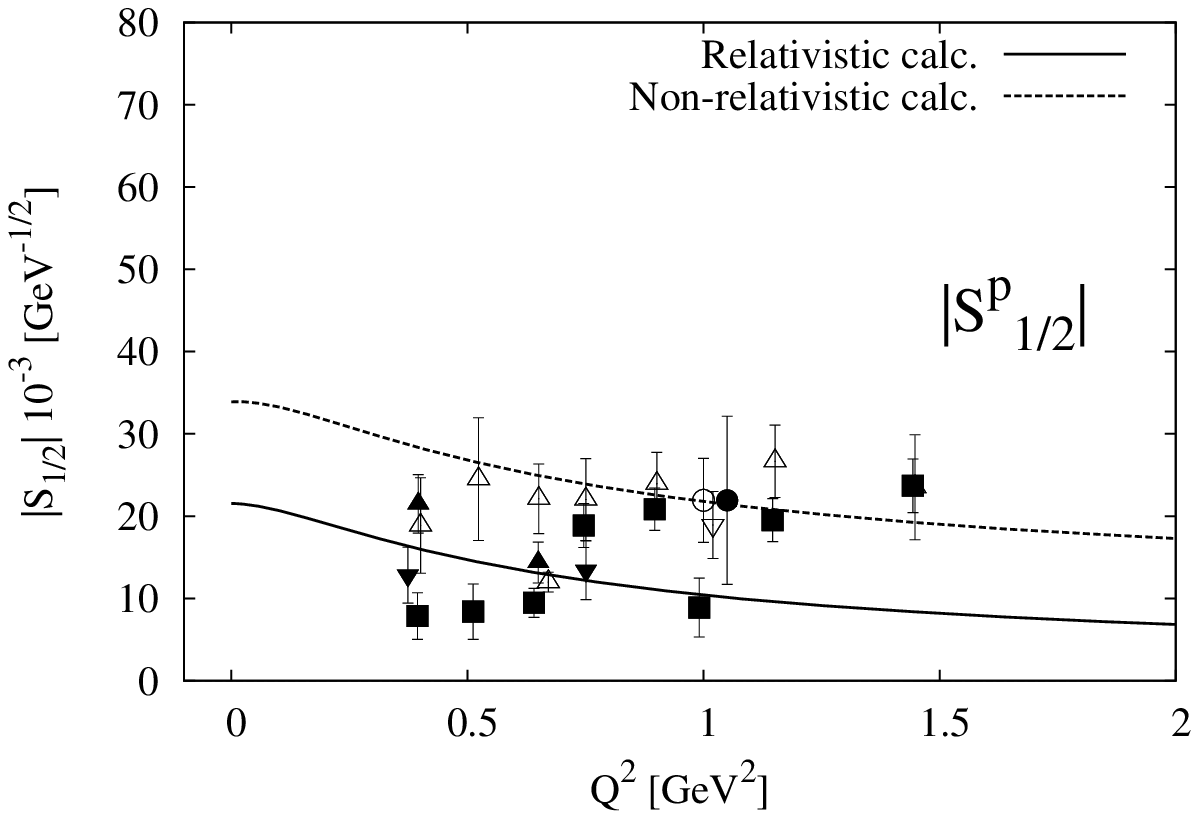}
\vspace{-3mm}\ec \caption {\label{s11} Transverse (top) and
longitudinal (bottom) helicity amplitudes for the $\gamma^*p\to
N_{1/2^-}(1535)$ transition. The data points are from {\sc\small
CLAS} \cite{Denizli:2007tq}. The value at the photon point is from
\cite{Amsler:2008zz}. The lines represent calculations by
\cite{Jido:2007sm}.\vspace{-2mm}}
\end{figure}
 The helicity amplitudes were calculated within a coupled channel
chiral unitary approach assuming that $N_{1/2^-}(1535)$ is
dynamically generated from the strong interaction of mesons and
baryons \cite{Jido:2007sm}. The $Q^2$ dependence is reproduced, the
absolute height not (a quantity which is difficult to determine
reliably from the data). The ratios obtained between the $S_{1/2}$
and $A_{1/2}$ for the two charge states of the $N_{1/2^-}(1535)$
agree qualitatively with experiment. They are not inconsistent with
this resonance being dynamically generated. However, there are
indications -- e.g.,  the harder $Q^2$ dependence in the data
compared to the prediction -- that a genuine quark-state component
could improve the agreement between experiment and the
model.\vspace{-3mm}

\paragraph{$\Lambda_{1/2^-}(1405)$}
 One of the first
historical examples is $\Lambda_{1/2^-}(1405)$ which was suggested
to be a $\overline K N$ quasi-bound state
\cite{Dalitz:1959dn,Dalitz:1960du}. This approach has been often
revisited, since the $\Lambda_{1/2^-}(1405)$ is one of the
resonances having a mass which is difficult to reproduce in quark
models. It falls just below the $N\overline K$ threshold; hence the
attractive interaction between $N$ and $\overline K$ and the
coupling to the $\Sigma\pi$ channel could lead to a threshold
enhancement or attract the pole of a not-too-far $qqq$ resonance
\cite{Dalitz:1967fp}. In models exploiting chiral symmetry and
imposing unitarity, $\Lambda_{1/2^-}(1405)$ can be generated
dynamically from the interaction of mesons and baryons in coupled
channels.

This is a unique (or rare) example where the predictions of chiral
dynamics and the quark model are at variance. Quark models predict
one $1/2^-$ resonance at 1400\,MeV, $\Lambda_{1/2^-}(1405)$. A
detailed study within a chiral unitary model revealed that the
$N\overline K$--$\Sigma\pi$ coupled channel effects is considerably
more complex. \cite{Jido:2003cb} suggest that
$\Lambda_{1/2^-}(1405)$ may contain two resonances; one - mainly
SU$(3)_{\rm f}$ singlet - at 1360\,MeV with a larger width and a
stronger coupling to $\pi\Sigma$, the other one at 1426\,MeV, which
is mostly SU$(3)_{\rm f}$ octet and couples more strongly to the
$N\overline K$. The lower mass state is mostly observed in the
$\pi^- p \to K^0 \pi \Sigma$ reaction while the reaction $K^- p \to
\pi^0 \pi^0 \Sigma^0$ produces a relatively narrow ($\Gamma =
38$\,MeV) peak at 1420\,MeV \cite{Magas:2005vu,Oller:2006jw}.
However, it is not yet clear how much SU(3) breaking invalidates
these conclusions; possibly, the second pole could even dissolve in
the background \cite{Borasoy:2006sr}.

We propose to test these ideas by a measurement of
$J/\psi\to\Lambda_{1/2^-}\overline\Lambda_{1/2^-}$ where
$\Lambda_{1/2^-}$ stands for the conventional
$\Lambda_{1/2^-}(1405)$ resonance or the two-resonance structure of
\cite{Jido:2003cb} and to measure the frequency with which the
following decay sequences occur:
\begin{subequations}\label{Lambda-1-8}
  \begin{eqnarray}
J/\psi \to & (\Lambda_{1/2^-}\to \pi \Sigma)& (\overline\Lambda_{1/2^-}\to\pi \overline\Sigma)\\
J/\psi \to & (\Lambda_{1/2^-}\to \bar K N)& (\overline\Lambda_{1/2^-}\to\pi \overline\Sigma)\\
J/\psi \to & (\Lambda_{1/2^-}\to \bar K N)& (\overline\Lambda_{1/2^-}\to
K \overline N)\,.
   \end{eqnarray}
\end{subequations}
In $J/\psi$ decays SU$(3)_{\rm f}$ singlet and octet states can be
produced pairwise, but simultaneous production of one octet and one
singlet state is suppressed. If there were two states, there should
be correlations between $\Lambda_{1/2^-}(1405)$ and
$\overline\Lambda_{1/2^-}$ decays; for a single-state resonances,
the decays are uncorrelated. We anticipate that the latter
prediction is correct. Assuming a two-pole structure of
$\Lambda_{1/2^-}(1405)$, the correlation in the
$\Lambda_{1/2^-}(1405)\to\Sigma\pi$ (+c.c.) decay modes is
calculated in \cite{Li:2004bg}. We note in passing that
\cite{Wohl:2008gw} compares light and heavy baryons and concludes
that $\Lambda_{1/2^-}(1405)$ is a 3-quark resonance.\vspace{-2mm}
 \subsub{Baryonic hybrids}\vspace{-2mm}
Baryons with properties incompatible with quark model predictions
can be suspected to be baryonic hybrids. This fate is shared by a
number of states, the Roper resonance $N_{1/2^+}(1440)$ being one
example.  Likewise, $\Lambda_{1/2^+}(1600)$
\cite{Kisslinger:2003hk}, $\Sigma_{1/2^+}(1600)$ and
$\Xi_{1/2^+}(1660)$ have low masses and could be hybrids as well.
The mass gap between $\Lambda_{1/2^-}(1405)$ and
$\Lambda_{3/2^-}(1520)$ is larger than expected in quark models but
can be reproduced assuming them to be of hybrid nature
\cite{Kittel:2005jm} where a possible hybrid nature is also
suggested for $\Lambda_c(2593)$ and $\Lambda_c(2676)$.

First bag-model predictions suggested that some hybrids could have
masses just below 2\,GeV \cite{Barnes:1982fj,Golowich:1982kx} making
a hybrid interpretation of $N_{1/2^+}(1440)$ unlikely. Also in a
non-relativistic flux-tube model, the lowest hybrid-baryon mass was
estimated to be $1870\pm100$\,MeV
\cite{Barnes:1995hc,Capstick:2002wm}. Within a relativistic quark
model, \cite{Gerasyuta:2002hg} arrived at hybrid masses suggesting
that $N_{1/2^+}(1710)$ and $\Delta_{3/2^+}(1600)$ could be hybrid
baryons.  QCD sum rules predict, however, a hybrid mass of 1500\,MeV
and $N_{1/2^+}(1440)$ remains a hybrid candidate
\cite{Kisslinger:1995yw}.

The most convincing experimental evidence providing an
interpretation of the Roper resonance is derived from recent
measurements of nucleon resonance transition form factors.
Figure~\ref{P11} shows the transverse and longitudinal
electro-coupling amplitudes $A_{1/2}$ and $S_{1/2}$ of the
transition to the $N_{1/2^+}(1440)$ resonance. At the photon point
$A_{1/2}$ is negative. The amplitude rises steeply with $Q^2$ and a
sign change occurs near $Q^2=0.5$~GeV$^2$. At $Q^2=2\,$GeV$^2$ the
amplitude has about the same magnitude but opposite sign as at
$Q^2=0$. Then it falls off slowly. The longitudinal amplitude
$S_{1/2}$ is large at low $Q^2$ and drops off smoothly with
increasing $Q^2$. The bold curves represent various quark model
calculations, the thin dashed line is for a gluonic excitation
\cite{Li:1991yba}. The hybrid hypothesis misses the sign change in
$A_{1/2}$;  $S_{1/2}$ is predicted to vanish identically. In
contrast, most quark models qualitatively reproduce the experimental
findings: the Roper $N_{1/2^+}(1440)$ resonance is the first radial
excitation of the nucleon.\vspace{-2mm}
 \subsub{Parity doublets, chiral multiplets}\vspace{-2mm}
The existence of parity doublets in the baryon spectrum has been
noticed as early as 1968 in \cite{Minami:1968nr}, and arguments in
favor of their existence were given even before (see
\cite{Afonin:2007mj} for a review). Parity doublets are expected in
a world of chiral symmetry. The large mass difference between the
nucleon and its chiral partner with $J=1/2$ but negative parity,
$N_{1/2^-}(1535)$, evidences that chiral symmetry is broken
spontaneously.  Glozman deserves the credit to have consistently
pointed out - in at least 20 papers on arXiv, we quote here
\cite{Glozman:1999tk,Cohen:2001gb,Cohen:2002st} - that at high
masses, mesons and baryons often occur in nearly mass-degenerate
pairs of states with given spin but opposite parity: parity doublets
are observed and possibly even parity quartets in which all (four)
nucleon and $\Delta$ states with identical $J^P$ are degenerate in
mass. \cite{Bicudo:2009cr} argue that the mass splittings between
these parity partners decrease with increasing baryon mass, and that
the decreasing mass difference can be used to probe the running
quark mass in the mid-infrared power-law regime.

Table \ref{chiral} summarizes the experimental status of multiplets
for $J^P=1/2^{\pm},\cdots 9/2^{\pm}$. In spite of an intense
discussion in the literature, reviewed, e.g., by \cite{Jaffe:2006jy}
and \cite{Glozman:2007ek}, there is no consensus whether parity
doubling emerges from the spin-orbital dynamics of the 3-quark
system, if it reflects a deep symmetry in QCD, or if they do not
exist at all in nature. With the present status of the data, this
question will likely remain unsettled. New data and new analyses are
needed.

\begin{table}[pt]\renewcommand{\arraystretch}{1.5}
\caption{\label{chiral}Parity doublets and chiral multiplets of
$N^*$ and $\Delta^*$ resonances of high mass. List and star rating
are taken from Table \ref{all-light-N}. \vspace*{-3mm}}
\bc
\renewcommand{\arraystretch}{1.3}
\begin{footnotesize}
\begin{tabular}{ccccc} \hline\hline\\[-3ex]
$J$=$\frac{1}{2}$&N$_{1/2^+}(1710)$&\hspace{-0.2mm}N$_{1/2^-}(1650)$&\hspace{-0.2mm}$\Delta_{1/2^+}(1750)$&\hspace{-0.2mm}$\Delta_{1/2^-}(1620)$\vspace{-1.5mm}\\
&***&****&&****\vspace{-1mm}\\
$J$=$\frac{3}{2}$&N$_{3/2^+}(1720)$&\hspace{-0.2mm}N$_{3/2^-}(1700)$&\hspace{-0.2mm}$\Delta_{3/2^+}(1600)$&\hspace{-0.2mm}$\Delta_{3/2^-}(1700)$\vspace{-1.5mm}\\
&****&***&***&****\vspace{-1mm}\\
$J$=$\frac{5}{2}$&N$_{5/2^+}(1680)$&\hspace{-0.2mm}N$_{5/2^-}(1675)$&\multicolumn{2}{c}{no chiral partners}\vspace{-1.5mm}\\
&****&****&&\vspace{-1mm}\\
$J$=$\frac{1}{2}$&N$_{1/2^+}(1880)$&\hspace{-0.2mm}N$_{1/2^-}(1905)$&\hspace{-0.2mm}$\Delta_{1/2^+}(1910)$&\hspace{-0.2mm}$\Delta_{1/2^-}(1900)$\vspace{-1.5mm}\\
&**&*&****&**\vspace{-1mm}\\
$J$=$\frac{3}{2}$&N$_{3/2^+}(1900)$&\hspace{-0.2mm}N$_{3/2^-}(1860)$&\hspace{-0.2mm}$\Delta_{3/2^+}(1920)$&\hspace{-0.2mm}$\Delta_{3/2^-}(1940)$\vspace{-1.5mm}\\
&*&**&***&**\vspace{-1mm}\\
$J$=$\frac{5}{2}$&N$_{5/2^+}(1870)$&\hspace{-0.2mm}no ch. partner\hspace{-0.2mm}&\hspace{-0.2mm}$\Delta_{5/2^+}(1905)$&\hspace{-0.2mm}$\Delta_{5/2^-}(1930)$\vspace{-1.5mm}\\
&**&&****&**\vspace{-1mm}\\
$J$=$\frac{7}{2}$&N$_{7/2^+}(1990)$&\hspace{-0.2mm}no ch. partner\hspace{-0.2mm}&\hspace{-0.2mm}$\Delta_{7/2^+}(1950)$&\hspace{-0.2mm}no ch. partner\vspace{-1.5mm}\\
&**&&****&\vspace{-1mm}\\
$J$=$\frac{7}{2}$&\hspace{-0.2mm}no ch. partner\hspace{-0.2mm}&\hspace{-0.2mm}N$_{7/2^-}(2190)$&\hspace{-0.2mm}no ch. partner&\hspace{-0.2mm}$\Delta_{7/2^-}(2200)$\vspace{-1.5mm}\\
&&****&&*\vspace{-1mm}\\
$J$=$\frac{9}{2}$&N$_{9/2^+}(2220)$&\hspace{-0.2mm}N$_{9/2^-}(2250)$&\hspace{-0.2mm}$\Delta_{9/2^+}(2300)$&\hspace{-0.2mm}$\Delta_{9/2^-}(2400)$\vspace{-1.5mm}\\
&****&****&**&**\\
\hline\hline
\end{tabular}
\end{footnotesize} \vspace*{-2mm}
\renewcommand{\arraystretch}{1.0}
\ec
\end{table}
In the harmonic oscillator approximation, a three-quark system is
characterized by successive shells of positive and negative parity.
Formally, this corresponds to masses being proportional to
$\mathrm{L}+2\mathtt{N}$. Parity doubling is not expected. In
AdS/QCD parity doubling arises naturally due to the $\mathrm
L+\mathtt{N}$ dependence of the nucleonic mass levels. Within their
collective model of baryons by \cite{Bijker:1994yr,Bijker:1996tr},
parity doubling is explained by the ``geometric structure" of
excitations \cite{Iachello:1989mw}. In Regge phenomenology, the
separation of states scales with $\delta M^2$\,=\,const, or
$M_1-M_2$\,=\,const/($M_1+M_2$). Experimentally, the masses of
states with positive and negative parity often show mass-degeneracy,
but not in all cases.  Clearly, a definition is needed when two
masses are called mass degenerate (within experimental errors) or
not. In Table \ref{chiral}, we have not accepted as parity partners
having a mass spacing in the order of the normal shell separation.
Based on quantitative tests, \cite{Klempt:2002tt} and
\cite{Shifman:2007xn} remain skeptical if the observed mass pattern
are related to a fundamental symmetry of QCD; it could as well be
due a dynamical symmetry like absence of spin-orbit forces.

\markboth{\sl Baryon spectroscopy} {\sl Summary and prospects}
\section{\label{se:sum}Summary and prospects}
The recent years have seen a remarkable boost in our knowledge of
baryons with heavy flavors, with the number of known baryons with
$b$-quarks increasing from 1 to 7 in the last 4 years, and that of
charmed states from 16 to 34. However, many points remain to be
clarified: in most cases, the quantum numbers of  heavy-flavor
baryons are deduced from quark-model expectation, and a direct
measurement would be desirable. One exception is the
$\Lambda_c(2880)$,   determined experimentally to be $J^P=5/2^+$
exploiting the decay angular distribution in the sequential
$\Lambda_c(2880)^+\to(\Sigma_c(2455)\pi)^+$ decay
(Fig.~\ref{angular_2880}), but the mass spectrum suggests rather
spin $1/2$ or $3/2$ and negative parity. The heaviest baryon known
so far, $\Omega_b$, may have a mass of 6.165\,GeV which seems almost
100\,MeV too high by comparison with the strangeness-excitation
energy in the sector of charmed baryons, but this result has been
challenged recently.

The double-charmed baryon, $\Xi_{cc}^+$ has been seen in only one
experiment, and the measured mass seems a little too low as compared
to model prediction. It is surprising that the mechanism of double
$c\bar{c}$ production, which is responsible, e.g., for the
observation of $J/\psi+\eta_c$ in $e^+e^-$ collisions does not
produce more often $cc+\bar{c}\bar{c}$, whose hadronization would
lead to double-charm baryons. Triple-charm (or $(ccb)$, $(cbb)$ or
$(bbb)$) spectroscopy will be to baryons what heavy quarkonium is
for mesons: a laboratory for high-precision QCD studies. It is
expected, for instance, that the analog of the Roper resonance for
these baryons would be stable, and lie below the negative-parity
excitations.

The experimental prospects for heavy baryon spectroscopy are bright
provided the chances are used. Remember that discussions and even
workshops are regularly held to use the production potential of
heavy-ion collisions for the spectroscopy of exotic  and
heavy-flavored hadrons, but the corresponding upgrade of detectors,
triggers and analysis programs has not yet started.

Doubled charmed baryons will probably be produced abundantly at LHC
and even $(ccc)$ states are not beyond the possibility. See, for
instance, \cite{Berezhnoi:1998aa,GomshiNobary:2007xk} for estimates
of the production rates. The upgrade of {\sc\small BELLE} will
improve the statistics in $B$ decays and of background $e^+e^-$
annihilation events very substantially; most information we have at
present stems from the predecessors {\sc\small BaBaR}, the present
{\sc\small BELLE} and from {\sc\small CLEO}. {\sc\small PANDA}
offers a further unique possibility to study the physics of heavy
flavors.

Light baryon spectroscopy has come again into the focus of a large
community. The quark model still provides the most convincing
picture. Even its simplest version, the  harmonic oscillator,
accounts for the number of expected states at low masses, and the
description is improved by using a better central potential and
spin-dependent forces.  In its relativistic variants,
electromagnetic properties such helicity amplitudes of
photoproduction, magnetic moments, and form factors can be
calculated as well. However, at higher excitations, the quark model
leads to the problem of ``missing resonances''. Here we recall that
the masses of ground state baryons do not arise from the motion of
relativistic quarks but rather from chiral symmetry breaking.
Possibly, chiral symmetry breaking is also the primary source for
the masses of excited baryons, where chiral symmetry could be broken
in an extended volume.

The question of dynamically generated resonances will require
further clarification. The states predicted by quark--gluon dynamics
need long-range corrections with higher Fock configurations, which
are dominated by the meson--baryon interaction. On the other hand,
resonances can be described starting from a purely hadronic picture,
with the recent improvements provided by effective theories and
chiral dynamics, but in this approach, short-range corrections lead
back to interacting  quarks. The situation is perhaps similar to
that of atoms in a magnetic field, for which both the weak-field and
a strong-field limits are relatively simple. For intermediate
fields, the truncated weak-field and strong-field expansions give
different predictions, that a superficial observer could
misinterpret as a  doubling of the atomic levels. For baryon
resonances, quark-model wave functions and meson--baryon states have
clearly a sizable overlap, hence their superposition should be
handled with care.

Experimentally, intense efforts are undertaken to carry out
photoproduction experiments with linearly and circularly polarized
photons and protons polarized along the direction of the incoming
photon beam, or transversely. The reaction $\gamma p\to\Lambda K^+$
offers the best chance to perform a complete experiment, in which
the full photoproduction amplitude can be reconstructed in an
energy-independent partial-wave analysis. Important steps have been
marked by experiments like {\sc\small CBELSA}, {\sc\small CLAS},
{\sc\small GRAAL}, {\sc\small LEPS}, and different experiments at
MAMI; several groups are attacking the difficult task of extracting
from the data resonant and non-resonant contributions in
energy-dependent partial-wave analyses. The confirmation of a few
states ($N_{3/2^+}(1900)$, $\Delta_{3/2^+}(1920)$,
$\Delta_{3/2^-}(1940)$) which had been observed in the old analyses
of H\"ohler and of Cutkosky and which were missing in the recent
analysis of the GWU group substantiates the hope that
photoproduction of multi-particle final states is a well-suited
method for uncovering new baryon resonances.

The known baryon resonances show a few  very surprising results.
First, the apparent absence (or smallness) of forces beyond
confinement and hyperfine interactions leads to clear spin
multiplets and thus allows one to assign intrinsic orbital and spin
angular momenta to a given baryon resonance. The four nearly
mass-degenerate states $\Delta_{1/2^+}(1910)$,
$\Delta_{3/2^+}(1920)$, $\Delta_{5/2^+}(1905)$, and
$\Delta_{7/2^+}(1950)$ form a quartet of resonances. It is counting
the number of states and not relying on a model which determines the
total quark spin to $S=3/2$ and the orbital angular momentum to
$L=2$. Mixing with other states is not excluded, but giving mixing
angles is (so far) a model-dependent statement. On this basis, all
nucleon and $\Delta$ resonances can be assigned to a few SU(6)
multiplets while other multiplets remain completely empty. At large
masses, all known resonances are compatible with nucleon excitations
having a total quark spin $S=1/2$ and $\Delta$ excitations having
$S=3/2$. At low energies, including the second excitation shell, the
full richness offered by the 3-particle problem seems to be
realized, except for one multiplet with an antisymmetric orbital
wave function in which the angular momenta of the two oscillators
with $l_{\rho}=1$ and $l_{\lambda}=1$ couple to a total angular
momentum $L=1$. Based on the systematics of baryon masses, we expect
a spin doublet $N_{1/2^+}$ and $N_{3/2^+}$ at a mass of about
$1.75-1.85$\,GeV. Since both oscillators are excited, direct
production of these states may be suppressed. But the two states
could mix with the two known states $N_{1/2^+}(1710)$ and
$N_{3/2^+}(1720)$, and we expect a pattern which is difficult to
resolve. Indeed, inconsistencies in the properties of the two
resonances as produced in photoproduction and in $\pi N$ elastic
scattering may be a first hint for these elusive resonances. In the
intermediate region, in the third shell, some multiplets are rather
completely filled while others remain empty. There is no obvious
systematic behavior which states are observed and which ones not. It
is an open question if these states are not realized because of an
unknown dynamical selection rule or if they just have escaped
experimental verification. We note that in most cases there is, for
isospin $I$ and strangeness S, only one resonance is found
experimentally with a set of quantum numbers $L$, $\tt N$, $S$, and
$J$ while quark model predict an increasing (with mass) number of
states all having the same quantum numbers.

The masses of nucleon and $\Delta$ resonances exhibit intriguing
spin-parity doublets, pairs of states with $J^P=J^{\pm}$, and even
evidence for four mass-degenerate  nucleon and two $\Delta$
resonances, all having the same $J$. The absence of strong
spin-orbit forces leads to a degeneracy of states with given $L$ and
$S$ but coupling to different $J$. Thus, the spectrum reveals a high
level of symmetries. Different interpretations have been offered to
explain the symmetries, restoration of chiral symmetry in the
high-mass region (Table \ref{chiral}), and AdS/QCD (Table
\ref{Gooddiquarks}). The two interpretations predict different mass
values for the lowest-mass $\Delta_{7/2^-}$ state. In AdS/QCD this
state should have intrinsic $L=3, S=1/2$ and 2.12\,GeV mass. When
chiral symmetry is restored, it should be found at 1.95\,GeV. A
search for the lowest-mass $\Delta_{7/2^-}$ resonance is thus
urgently requested.

Photo-induced reactions seems to favor production of
low-angular-momentum states while pion-induced reactions (at least
$\pi N$ elastic scattering) is rich in high-angular-momentum states.
To get a complete picture, hadron-induced reactions will be needed
for a full understanding of the baryon resonance spectrum.
\cite{Bugg:2007vq} has underlined that relatively simple experiments
with no charged-particle tracking and with no magnetic field but a
good electromagnetic calorimeter and a polarized target would give
decisive new information on the hadronic mass spectrum, for both
mesons and baryons, provided a good pion beam -- which in the
sixties of last century used to be the most natural thing in the
world -- would be available. A perfect laboratory for such
experiments would be JPARC at KEK.

The chances for breakthroughs in the spectroscopy of light and heavy
baryons are there and need to be pursued. The additional degrees of
freedom in baryons -- compared to the much simpler mesons -- offer
the possibility to test how strong QCD responds in such a complex
environment: which of the multitude of configurations are realized
and what are the effective agents and forces leading to the highly
degenerate pattern of energy levels. A related question is whether
iterating the binding mechanisms seen at work for baryons lead to
exotic hadrons, in particular multiquark states.

\acknowledgements{
This review was initiated by the stimulating environment of the
Sonderforschungsbereich SFB/TR16. E.K. is indebted to all his
colleagues in SFB/TR16. Special thanks go to Ulrike Thoma and to her
contributions in an early stage of this review. Illuminating
discussions with Bernard Metsch and Herbert Petry are gratefully
acknowledged as well as the cooperation with Alexey Anisovich,
Victor Nikonov and Andrey Sarantsev within the Bonn--Gatchina Partial
Wave Analysis Group.
We also benefitted from comments on an early version of this review, in particular by Eulogio Oset and Ulf Mei{\ss}ner.}
%

\markboth{\sl Baryon spectroscopy } {\sl Bibliography}

\begin{thebibliography}{479}
\expandafter\ifx\csname natexlab\endcsname\relax\def\natexlab#1{#1}\fi
\expandafter\ifx\csname bibnamefont\endcsname\relax
  \def\bibnamefont#1{#1}\fi
\expandafter\ifx\csname bibfnamefont\endcsname\relax
  \def\bibfnamefont#1{#1}\fi
\expandafter\ifx\csname citenamefont\endcsname\relax
  \def\citenamefont#1{#1}\fi
\expandafter\ifx\csname url\endcsname\relax
  \def\url#1{\texttt{#1}}\fi
\expandafter\ifx\csname urlprefix\endcsname\relax\def\urlprefix{URL }\fi
\providecommand{\bibinfo}[2]{#2}
\providecommand{\eprint}[2][]{\url{#2}}

\bibitem[{Aaltonen \emph{et~al.}(2007{\natexlab{a}})\citenamefont{Aaltonen}
  \emph{et~al.}}]{Aaltonen:2007rw}
\bibinfo{author}{\bibnamefont{Aaltonen}, \bibfnamefont{T.}}, \emph{et~al.},
  \bibinfo{year}{2007}{\natexlab{a}}, \bibinfo{journal}{Phys. Rev. Lett.}
  \textbf{\bibinfo{volume}{99}}, \bibinfo{pages}{202001}.

\bibitem[{Aaltonen \emph{et~al.}(2007{\natexlab{b}})\citenamefont{Aaltonen}
  \emph{et~al.}}]{Aaltonen:2007un}
\bibinfo{author}{\bibnamefont{Aaltonen}, \bibfnamefont{T.}}, \emph{et~al.},
  \bibinfo{year}{2007}{\natexlab{b}}, \bibinfo{journal}{Phys. Rev. Lett.}
  \textbf{\bibinfo{volume}{99}}, \bibinfo{pages}{052002}.

\bibitem[{Aaltonen \emph{et~al.}(2009)\citenamefont{Aaltonen}
  \emph{et~al.}}]{Aaltonen:2009ny}
\bibinfo{author}{\bibnamefont{Aaltonen}, \bibfnamefont{T.}}, \emph{et~al.},
  \bibinfo{year}{2009}, \eprint{0905.3123}.

\bibitem[{Abazov \emph{et~al.}(2006)\citenamefont{Abazov}
  \emph{et~al.}}]{Abazov:2005pn}
\bibinfo{author}{\bibnamefont{Abazov}, \bibfnamefont{V.~M.}}, \emph{et~al.},
  \bibinfo{year}{2006}, \bibinfo{journal}{Nucl. Instrum. Meth.}
  \textbf{\bibinfo{volume}{A565}}, \bibinfo{pages}{463}.

\bibitem[{Abazov \emph{et~al.}(2007)\citenamefont{Abazov}
  \emph{et~al.}}]{Abazov:2007ub}
\bibinfo{author}{\bibnamefont{Abazov}, \bibfnamefont{V.~M.}}, \emph{et~al.},
  \bibinfo{year}{2007}, \bibinfo{journal}{Phys. Rev. Lett.}
  \textbf{\bibinfo{volume}{99}}, \bibinfo{pages}{052001}.

\bibitem[{Abazov \emph{et~al.}(2008)\citenamefont{Abazov}
  \emph{et~al.}}]{Abazov:2008qm}
\bibinfo{author}{\bibnamefont{Abazov}, \bibfnamefont{V.~M.}}, \emph{et~al.},
  \bibinfo{year}{2008}, \bibinfo{journal}{Phys. Rev. Lett.}
  \textbf{\bibinfo{volume}{101}}, \bibinfo{pages}{232002}.

\bibitem[{Abe \emph{et~al.}(2007)\citenamefont{Abe} \emph{et~al.}}]{Abe:2006rz}
\bibinfo{author}{\bibnamefont{Abe}, \bibfnamefont{K.}}, \emph{et~al.},
  \bibinfo{year}{2007}, \bibinfo{journal}{Phys. Rev. Lett.}
  \textbf{\bibinfo{volume}{98}}, \bibinfo{pages}{262001}.

\bibitem[{Abe \emph{et~al.}(2008)\citenamefont{Abe}
  \emph{et~al.}}]{Abe:2007wga}
\bibinfo{author}{\bibnamefont{Abe}, \bibfnamefont{K.}}, \emph{et~al.},
  \bibinfo{year}{2008}, \bibinfo{journal}{Phys. Rev. Lett.}
  \textbf{\bibinfo{volume}{100}}, \bibinfo{pages}{142001}.

\bibitem[{Ablikim \emph{et~al.}(2006)\citenamefont{Ablikim}
  \emph{et~al.}}]{Ablikim:2004ug}
\bibinfo{author}{\bibnamefont{Ablikim}, \bibfnamefont{M.}}, \emph{et~al.},
  \bibinfo{year}{2006}, \bibinfo{journal}{Phys. Rev. Lett.}
  \textbf{\bibinfo{volume}{97}}, \bibinfo{pages}{062001}.

\bibitem[{Abrams \emph{et~al.}(1974)\citenamefont{Abrams}
  \emph{et~al.}}]{Abrams:1974yy}
\bibinfo{author}{\bibnamefont{Abrams}, \bibfnamefont{G.~S.}}, \emph{et~al.},
  \bibinfo{year}{1974}, \bibinfo{journal}{Phys. Rev. Lett.}
  \textbf{\bibinfo{volume}{33}}, \bibinfo{pages}{1453}.

\bibitem[{Acosta \emph{et~al.}(2005)\citenamefont{Acosta}
  \emph{et~al.}}]{Acosta:2004yw}
\bibinfo{author}{\bibnamefont{Acosta}, \bibfnamefont{D.~E.}}, \emph{et~al.},
  \bibinfo{year}{2005}, \bibinfo{journal}{Phys. Rev.}
  \textbf{\bibinfo{volume}{D71}}, \bibinfo{pages}{032001}.

\bibitem[{\citenamefont{Afonin}(2007)}]{Afonin:2007mj}
\bibinfo{author}{\bibnamefont{Afonin}, \bibfnamefont{S.~S.}},
  \bibinfo{year}{2007}, \bibinfo{journal}{Int. J. Mod. Phys.}
  \textbf{\bibinfo{volume}{A22}}, \bibinfo{pages}{4537}.

\bibitem[{\citenamefont{Afonin}(2009)}]{Afonin:2009zk}
\bibinfo{author}{\bibnamefont{Afonin}, \bibfnamefont{S.~S.}},
  \bibinfo{year}{2009}, \eprint{0904.0737}.

\bibitem[{Ahrens \emph{et~al.}(2000)\citenamefont{Ahrens}
  \emph{et~al.}}]{Ahrens:2000bc}
\bibinfo{author}{\bibnamefont{Ahrens}, \bibfnamefont{J.}}, \emph{et~al.},
  \bibinfo{year}{2000}, \bibinfo{journal}{Phys. Rev. Lett.}
  \textbf{\bibinfo{volume}{84}}, \bibinfo{pages}{5950}.

\bibitem[{Ahrens \emph{et~al.}(2001)\citenamefont{Ahrens}
  \emph{et~al.}}]{Ahrens:2001qt}
\bibinfo{author}{\bibnamefont{Ahrens}, \bibfnamefont{J.}}, \emph{et~al.},
  \bibinfo{year}{2001}, \bibinfo{journal}{Phys. Rev. Lett.}
  \textbf{\bibinfo{volume}{87}}, \bibinfo{pages}{022003}.

\bibitem[{Ahrens \emph{et~al.}(2003)\citenamefont{Ahrens}
  \emph{et~al.}}]{Ahrens:2006yx}
\bibinfo{author}{\bibnamefont{Ahrens}, \bibfnamefont{J.}}, \emph{et~al.},
  \bibinfo{year}{2003}, \bibinfo{journal}{Phys. Rev. Lett.}
  \textbf{\bibinfo{volume}{97}}.

\bibitem[{Ahrens \emph{et~al.}(2005)\citenamefont{Ahrens}
  \emph{et~al.}}]{Ahrens:2005ia}
\bibinfo{author}{\bibnamefont{Ahrens}, \bibfnamefont{J.}}, \emph{et~al.},
  \bibinfo{year}{2005}, \bibinfo{journal}{Phys. Lett.}
  \textbf{\bibinfo{volume}{B624}}, \bibinfo{pages}{173}.

\bibitem[{Ahrens \emph{et~al.}(2006)\citenamefont{Ahrens}
  \emph{et~al.}}]{Ahrens:2006gp}
\bibinfo{author}{\bibnamefont{Ahrens}, \bibfnamefont{J.}}, \emph{et~al.},
  \bibinfo{year}{2006}, \bibinfo{journal}{Phys. Rev.}
  \textbf{\bibinfo{volume}{C74}}, \bibinfo{pages}{045204}.

\bibitem[{Ahrens \emph{et~al.}(2007)\citenamefont{Ahrens}
  \emph{et~al.}}]{Ahrens:2007zzj}
\bibinfo{author}{\bibnamefont{Ahrens}, \bibfnamefont{J.}}, \emph{et~al.},
  \bibinfo{year}{2007}, \bibinfo{journal}{Eur. Phys. J.}
  \textbf{\bibinfo{volume}{A34}}, \bibinfo{pages}{11}.

\bibitem[{Aitala \emph{et~al.}(1998)\citenamefont{Aitala}
  \emph{et~al.}}]{Aitala:1997ja}
\bibinfo{author}{\bibnamefont{Aitala}, \bibfnamefont{E.~M.}}, \emph{et~al.},
  \bibinfo{year}{1998}, \bibinfo{journal}{Phys. Rev. Lett.}
  \textbf{\bibinfo{volume}{81}}, \bibinfo{pages}{44}.

\bibitem[{Ajaka \emph{et~al.}(2006)\citenamefont{Ajaka}
  \emph{et~al.}}]{Ajaka:2006bn}
\bibinfo{author}{\bibnamefont{Ajaka}, \bibfnamefont{J.}}, \emph{et~al.},
  \bibinfo{year}{2006}, \bibinfo{journal}{Phys. Rev. Lett.}
  \textbf{\bibinfo{volume}{96}}, \bibinfo{pages}{132003}.

\bibitem[{Ajaka \emph{et~al.}(2008)\citenamefont{Ajaka}
  \emph{et~al.}}]{Ajaka:2008zz}
\bibinfo{author}{\bibnamefont{Ajaka}, \bibfnamefont{J.}}, \emph{et~al.},
  \bibinfo{year}{2008}, \bibinfo{journal}{Phys. Rev. Lett.}
  \textbf{\bibinfo{volume}{100}}, \bibinfo{pages}{052003}.

\bibitem[{Aker \emph{et~al.}(1992)\citenamefont{Aker}
  \emph{et~al.}}]{Aker:1992ny}
\bibinfo{author}{\bibnamefont{Aker}, \bibfnamefont{E.}}, \emph{et~al.},
  \bibinfo{year}{1992}, \bibinfo{journal}{Nucl. Instrum. Meth.}
  \textbf{\bibinfo{volume}{A321}}, \bibinfo{pages}{69}.

\bibitem[{Aktas \emph{et~al.}(2004)\citenamefont{Aktas}
  \emph{et~al.}}]{Aktas:2004qf}
\bibinfo{author}{\bibnamefont{Aktas}, \bibfnamefont{A.}}, \emph{et~al.},
  \bibinfo{year}{2004}, \bibinfo{journal}{Phys. Lett.}
  \textbf{\bibinfo{volume}{B588}}, \bibinfo{pages}{17}.

\bibitem[{Albrecht \emph{et~al.}(1989)\citenamefont{Albrecht}
  \emph{et~al.}}]{Albrecht:1988vy}
\bibinfo{author}{\bibnamefont{Albrecht}, \bibfnamefont{H.}}, \emph{et~al.},
  \bibinfo{year}{1989}, \bibinfo{journal}{Nucl. Instrum. Meth.}
  \textbf{\bibinfo{volume}{A275}}, \bibinfo{pages}{1}.

\bibitem[{Albrecht \emph{et~al.}(1993)\citenamefont{Albrecht}
  \emph{et~al.}}]{Albrecht:1993pt}
\bibinfo{author}{\bibnamefont{Albrecht}, \bibfnamefont{H.}}, \emph{et~al.},
  \bibinfo{year}{1993}, \bibinfo{journal}{Phys. Lett.}
  \textbf{\bibinfo{volume}{B317}}, \bibinfo{pages}{227}.

\bibitem[{Albrecht \emph{et~al.}(1997)\citenamefont{Albrecht}
  \emph{et~al.}}]{Albrecht:1997qa}
\bibinfo{author}{\bibnamefont{Albrecht}, \bibfnamefont{H.}}, \emph{et~al.},
  \bibinfo{year}{1997}, \bibinfo{journal}{Phys. Lett.}
  \textbf{\bibinfo{volume}{B402}}, \bibinfo{pages}{207}.

\bibitem[{Albrow \emph{et~al.}(1970)\citenamefont{Albrow}
  \emph{et~al.}}]{Albrow:1971xa}
\bibinfo{author}{\bibnamefont{Albrow}, \bibfnamefont{M.~G.}}, \emph{et~al.},
  \bibinfo{year}{1970}, \bibinfo{journal}{Nucl. Phys.}
  \textbf{\bibinfo{volume}{B25}}, \bibinfo{pages}{9}.

\bibitem[{Albrow \emph{et~al.}(1972)\citenamefont{Albrow}
  \emph{et~al.}}]{Albrow:1972ky}
\bibinfo{author}{\bibnamefont{Albrow}, \bibfnamefont{M.~G.}}, \emph{et~al.},
  \bibinfo{year}{1972}, \bibinfo{journal}{Nucl. Phys.}
  \textbf{\bibinfo{volume}{B37}}, \bibinfo{pages}{594}.

\bibitem[{\citenamefont{Albuquerque}
  \emph{et~al.}(2009)\citenamefont{Albuquerque, Narison, and
  Nielsen}}]{Albuquerque:2009pr}
\bibinfo{author}{\bibnamefont{Albuquerque}, \bibfnamefont{R.~A.}},
  \bibinfo{author}{\bibfnamefont{S.}~\bibnamefont{Narison}}, and
  \bibinfo{author}{\bibfnamefont{M.}~\bibnamefont{Nielsen}},
  \bibinfo{year}{2009}, \eprint{0904.3717}.

\bibitem[{Alekseev \emph{et~al.}(1991)\citenamefont{Alekseev}
  \emph{et~al.}}]{Alekseev:1989cy}
\bibinfo{author}{\bibnamefont{Alekseev}, \bibfnamefont{I.~G.}}, \emph{et~al.},
  \bibinfo{year}{1991}, \bibinfo{journal}{Nucl. Phys.}
  \textbf{\bibinfo{volume}{B348}}, \bibinfo{pages}{257}.

\bibitem[{Alekseev \emph{et~al.}(1995)\citenamefont{Alekseev}
  \emph{et~al.}}]{Alekseev:1995nw}
\bibinfo{author}{\bibnamefont{Alekseev}, \bibfnamefont{I.~G.}}, \emph{et~al.},
  \bibinfo{year}{1995}, \bibinfo{journal}{Phys. Lett.}
  \textbf{\bibinfo{volume}{B351}}, \bibinfo{pages}{585}.

\bibitem[{Alekseev \emph{et~al.}(1997)\citenamefont{Alekseev}
  \emph{et~al.}}]{Alekseev:1996gs}
\bibinfo{author}{\bibnamefont{Alekseev}, \bibfnamefont{I.~G.}}, \emph{et~al.},
  \bibinfo{year}{1997}, \bibinfo{journal}{Phys. Rev.}
  \textbf{\bibinfo{volume}{C55}}, \bibinfo{pages}{2049}.

\bibitem[{Alekseev \emph{et~al.}(2000)\citenamefont{Alekseev}
  \emph{et~al.}}]{Alekseev:2000nk}
\bibinfo{author}{\bibnamefont{Alekseev}, \bibfnamefont{I.~G.}}, \emph{et~al.},
  \bibinfo{year}{2000}, \bibinfo{journal}{Phys. Lett.}
  \textbf{\bibinfo{volume}{B485}}, \bibinfo{pages}{32}.

\bibitem[{Alekseev \emph{et~al.}(2006)\citenamefont{Alekseev}
  \emph{et~al.}}]{Alekseev:2005zr}
\bibinfo{author}{\bibnamefont{Alekseev}, \bibfnamefont{I.~G.}}, \emph{et~al.},
  \bibinfo{year}{2006}, \bibinfo{journal}{Eur. Phys. J.}
  \textbf{\bibinfo{volume}{C45}}, \bibinfo{pages}{383}.

\bibitem[{Alexander \emph{et~al.}(1999)\citenamefont{Alexander}
  \emph{et~al.}}]{Alexander:1999ud}
\bibinfo{author}{\bibnamefont{Alexander}, \bibfnamefont{J.~P.}}, \emph{et~al.},
  \bibinfo{year}{1999}, \bibinfo{journal}{Phys. Rev. Lett.}
  \textbf{\bibinfo{volume}{83}}, \bibinfo{pages}{3390}.

\bibitem[{Alt \emph{et~al.}(2004)\citenamefont{Alt} \emph{et~al.}}]{Alt:2003vb}
\bibinfo{author}{\bibnamefont{Alt}, \bibfnamefont{C.}}, \emph{et~al.},
  \bibinfo{year}{2004}, \bibinfo{journal}{Phys. Rev. Lett.}
  \textbf{\bibinfo{volume}{92}}, \bibinfo{pages}{042003}.

\bibitem[{Ambrozewicz \emph{et~al.}(2007)\citenamefont{Ambrozewicz}
  \emph{et~al.}}]{Ambrozewicz:2006zj}
\bibinfo{author}{\bibnamefont{Ambrozewicz}, \bibfnamefont{P.}}, \emph{et~al.},
  \bibinfo{year}{2007}, \bibinfo{journal}{Phys. Rev.}
  \textbf{\bibinfo{volume}{C75}}, \bibinfo{pages}{045203}.

\bibitem[{Ammar \emph{et~al.}(2001)\citenamefont{Ammar}
  \emph{et~al.}}]{Ammar:2000uh}
\bibinfo{author}{\bibnamefont{Ammar}, \bibfnamefont{R.}}, \emph{et~al.},
  \bibinfo{year}{2001}, \bibinfo{journal}{Phys. Rev. Lett.}
  \textbf{\bibinfo{volume}{86}}, \bibinfo{pages}{1167}.

\bibitem[{Amsler \emph{et~al.}(2008)\citenamefont{Amsler}
  \emph{et~al.}}]{Amsler:2008zz}
\bibinfo{author}{\bibnamefont{Amsler}, \bibfnamefont{C.}}, \emph{et~al.},
  \bibinfo{year}{2008}, \bibinfo{journal}{Phys. Lett.}
  \textbf{\bibinfo{volume}{B667}}, \bibinfo{pages}{1}.

\bibitem[{Anciant \emph{et~al.}(2000)\citenamefont{Anciant}
  \emph{et~al.}}]{Anciant:2000az}
\bibinfo{author}{\bibnamefont{Anciant}, \bibfnamefont{E.}}, \emph{et~al.},
  \bibinfo{year}{2000}, \bibinfo{journal}{Phys. Rev. Lett.}
  \textbf{\bibinfo{volume}{85}}, \bibinfo{pages}{4682}.

\bibitem[{Anderson \emph{et~al.}(1969)\citenamefont{Anderson}
  \emph{et~al.}}]{Anderson:1969jw}
\bibinfo{author}{\bibnamefont{Anderson}, \bibfnamefont{R.~L.}}, \emph{et~al.},
  \bibinfo{year}{1969}, \bibinfo{journal}{Phys. Rev. Lett.}
  \textbf{\bibinfo{volume}{23}}, \bibinfo{pages}{721}.

\bibitem[{Anderson \emph{et~al.}(1976)\citenamefont{Anderson}
  \emph{et~al.}}]{Anderson:1976ph}
\bibinfo{author}{\bibnamefont{Anderson}, \bibfnamefont{R.~L.}}, \emph{et~al.},
  \bibinfo{year}{1976}, \bibinfo{journal}{Phys. Rev.}
  \textbf{\bibinfo{volume}{D14}}, \bibinfo{pages}{679}.

\bibitem[{\citenamefont{Anisovich} \emph{et~al.}(2008)\citenamefont{Anisovich,
  Anisovich, Matveev, Nikonov, Nyiri, and Sarantsev}}]{Gatchina:2008aaa}
\bibinfo{author}{\bibnamefont{Anisovich}, \bibfnamefont{A.}},
  \bibinfo{author}{\bibfnamefont{V.}~\bibnamefont{Anisovich}},
  \bibinfo{author}{\bibfnamefont{M.}~\bibnamefont{Matveev}},
  \bibinfo{author}{\bibfnamefont{V.}~\bibnamefont{Nikonov}},
  \bibinfo{author}{\bibfnamefont{J.}~\bibnamefont{Nyiri}}, and
  \bibinfo{author}{\bibfnamefont{A.}~\bibnamefont{Sarantsev}},
  \bibinfo{year}{2008}, \bibinfo{journal}{World Scientific, Singapore} ,
  \bibinfo{pages}{580}.


\bibitem[{\citenamefont{Anisovich} \emph{et~al.}(2005)\citenamefont{Anisovich,
  Klempt, Sarantsev, and Thoma}}]{Anisovich:2004zz}
\bibinfo{author}{\bibnamefont{Anisovich}, \bibfnamefont{A.}},
  \bibinfo{author}{\bibfnamefont{E.}~\bibnamefont{Klempt}},
  \bibinfo{author}{\bibfnamefont{A.}~\bibnamefont{Sarantsev}}, and
  \bibinfo{author}{\bibfnamefont{U.}~\bibnamefont{Thoma}},
  \bibinfo{year}{2005}, \bibinfo{journal}{Eur. Phys. J.}
  \textbf{\bibinfo{volume}{A24}}, \bibinfo{pages}{111}.


\bibitem[{Anisovich \emph{et~al.}(2005)\citenamefont{Anisovich}
  \emph{et~al.}}]{Anisovich:2005tf}
\bibinfo{author}{\bibnamefont{Anisovich}, \bibfnamefont{A.~V.}}, \emph{et~al.},
  \bibinfo{year}{2005}, \bibinfo{journal}{Eur. Phys. J.}
  \textbf{\bibinfo{volume}{A25}}, \bibinfo{pages}{427}.

\bibitem[{Anisovich \emph{et~al.}(2007{\natexlab{a}})\citenamefont{Anisovich}
  \emph{et~al.}}]{Anisovich:2007zz}
\bibinfo{author}{\bibnamefont{Anisovich}, \bibfnamefont{A.~V.}}, \emph{et~al.},
  \bibinfo{year}{2007{\natexlab{a}}}, \bibinfo{journal}{Eur. Phys. J.}
  \textbf{\bibinfo{volume}{A34}}, \bibinfo{pages}{129}.

\bibitem[{Anisovich \emph{et~al.}(2007{\natexlab{b}})\citenamefont{Anisovich}
  \emph{et~al.}}]{Anisovich:2007bq}
\bibinfo{author}{\bibnamefont{Anisovich}, \bibfnamefont{A.~V.}}, \emph{et~al.},
  \bibinfo{year}{2007{\natexlab{b}}}, \bibinfo{journal}{Eur. Phys. J.}
  \textbf{\bibinfo{volume}{A34}}, \bibinfo{pages}{243}.

\bibitem[{Anisovich \emph{et~al.}(2009)\citenamefont{Anisovich}
  \emph{et~al.}}]{Anisovich:2008wd}
\bibinfo{author}{\bibnamefont{Anisovich}, \bibfnamefont{A.~V.}}, \emph{et~al.},
  \bibinfo{year}{2009}, \bibinfo{journal}{Eur. Phys. J.}
  \textbf{\bibinfo{volume}{A41}}, \bibinfo{pages}{13}.

\bibitem[{\citenamefont{Anisovich} \emph{et~al.}(2009)\citenamefont{Anisovich,
  Klempt, Nikonov, Sarantsev, and Thoma}}]{Anisovich:2009zy}
\bibinfo{author}{\bibnamefont{Anisovich}, \bibfnamefont{A.}},
   \bibinfo{year}{2009}, \bibinfo{journal}{submitted to Eur. Phys. J.,}
   \eprint{arXiv:0911.5277 [hep-ph]}.

\bibitem[{\citenamefont{Arndt} \emph{et~al.}(2006)\citenamefont{Arndt, Briscoe,
  Strakovsky, and Workman}}]{Arndt:2006bf}
\bibinfo{author}{\bibnamefont{Arndt}, \bibfnamefont{R.~A.}},
  \bibinfo{author}{\bibfnamefont{W.~J.} \bibnamefont{Briscoe}},
  \bibinfo{author}{\bibfnamefont{I.~I.} \bibnamefont{Strakovsky}}, and
  \bibinfo{author}{\bibfnamefont{R.~L.} \bibnamefont{Workman}},
  \bibinfo{year}{2006}, \bibinfo{journal}{Phys. Rev.}
  \textbf{\bibinfo{volume}{C74}}, \bibinfo{pages}{045205}.

\bibitem[{\citenamefont{Artru} \emph{et~al.}(2009)\citenamefont{Artru, Elchikh,
  Richard, Soffer, and Teryaev}}]{Artru:2008cp}
\bibinfo{author}{\bibnamefont{Artru}, \bibfnamefont{X.}},
  \bibinfo{author}{\bibfnamefont{M.}~\bibnamefont{Elchikh}},
  \bibinfo{author}{\bibfnamefont{J.-M.} \bibnamefont{Richard}},
  \bibinfo{author}{\bibfnamefont{J.}~\bibnamefont{Soffer}}, and
  \bibinfo{author}{\bibfnamefont{O.~V.} \bibnamefont{Teryaev}},
  \bibinfo{year}{2009}, \bibinfo{journal}{Phys. Rept.}
  \textbf{\bibinfo{volume}{470}}, \bibinfo{pages}{1}.

\bibitem[{Artuso \emph{et~al.}(2001)\citenamefont{Artuso}
  \emph{et~al.}}]{Artuso:2000xy}
\bibinfo{author}{\bibnamefont{Artuso}, \bibfnamefont{M.}}, \emph{et~al.},
  \bibinfo{year}{2001}, \bibinfo{journal}{Phys. Rev. Lett.}
  \textbf{\bibinfo{volume}{86}}, \bibinfo{pages}{4479}.

\bibitem[{\citenamefont{Ashery}(1996)}]{Ashery:1995ha}
\bibinfo{author}{\bibnamefont{Ashery}, \bibfnamefont{D.}},
  \bibinfo{year}{1996}, \bibinfo{journal}{Hyperfine Interact.}
  \textbf{\bibinfo{volume}{103}}, \bibinfo{pages}{253}.

\bibitem[{Assafiri \emph{et~al.}(2003)\citenamefont{Assafiri}
  \emph{et~al.}}]{Assafiri:2003mv}
\bibinfo{author}{\bibnamefont{Assafiri}, \bibfnamefont{Y.}}, \emph{et~al.},
  \bibinfo{year}{2003}, \bibinfo{journal}{Phys. Rev. Lett.}
  \textbf{\bibinfo{volume}{90}}, \bibinfo{pages}{222001}.

\bibitem[{Aston \emph{et~al.}(1988)\citenamefont{Aston}
  \emph{et~al.}}]{Aston:1988yn}
\bibinfo{author}{\bibnamefont{Aston}, \bibfnamefont{D.}}, \emph{et~al.},
  \bibinfo{year}{1988}, \bibinfo{journal}{Phys. Lett.}
  \textbf{\bibinfo{volume}{B215}}, \bibinfo{pages}{799}.

\bibitem[{Aston \emph{et~al.}(1990)\citenamefont{Aston}
  \emph{et~al.}}]{Aston:1990ys}
\bibinfo{author}{\bibnamefont{Aston}, \bibfnamefont{D.}}, \emph{et~al.},
  \bibinfo{year}{1990}, \bibinfo{note}{invited talk given at 15th APS Div. of
  Particles and Fields General Mtg., Houston, TX, Jan 3-6, 1990}.

\bibitem[{Athar \emph{et~al.}(2005)\citenamefont{Athar}
  \emph{et~al.}}]{Athar:2004ni}
\bibinfo{author}{\bibnamefont{Athar}, \bibfnamefont{S.~B.}}, \emph{et~al.},
  \bibinfo{year}{2005}, \bibinfo{journal}{Phys. Rev.}
  \textbf{\bibinfo{volume}{D71}}, \bibinfo{pages}{051101}.

\bibitem[{Aubert \emph{et~al.}(1974)\citenamefont{Aubert}
  \emph{et~al.}}]{Aubert:1974js}
\bibinfo{author}{\bibnamefont{Aubert}, \bibfnamefont{J.~J.}}, \emph{et~al.},
  \bibinfo{year}{1974}, \bibinfo{journal}{Phys. Rev. Lett.}
  \textbf{\bibinfo{volume}{33}}, \bibinfo{pages}{1404}.

\bibitem[{Aubert \emph{et~al.}(2002)\citenamefont{Aubert}
  \emph{et~al.}}]{Aubert:2001tu}
\bibinfo{author}{\bibnamefont{Aubert}, \bibfnamefont{B.}}, \emph{et~al.},
  \bibinfo{year}{2002}, \bibinfo{journal}{Nucl. Instrum. Meth.}
  \textbf{\bibinfo{volume}{A479}}, \bibinfo{pages}{1}.

\bibitem[{Aubert \emph{et~al.}(2005)\citenamefont{Aubert}
  \emph{et~al.}}]{Aubert:2005gt}
\bibinfo{author}{\bibnamefont{Aubert}, \bibfnamefont{B.}}, \emph{et~al.},
  \bibinfo{year}{2005}, \bibinfo{journal}{Phys. Rev.}
  \textbf{\bibinfo{volume}{D72}}, \bibinfo{pages}{052006}.

\bibitem[{Aubert \emph{et~al.}(2006{\natexlab{a}})\citenamefont{Aubert}
  \emph{et~al.}}]{Aubert:2006uw}
\bibinfo{author}{\bibnamefont{Aubert}, \bibfnamefont{B.}}, \emph{et~al.},
  \bibinfo{year}{2006}{\natexlab{a}}, \bibinfo{note}{$33^{\rm rd}$
  International Conference on High Energy Physics (ICHEP 06), Moscow, Russia,
  26 Jul - 2 Aug 2006}, \eprint{hep-ex/0607042}.

\bibitem[{Aubert \emph{et~al.}(2006{\natexlab{b}})\citenamefont{Aubert}
  \emph{et~al.}}]{Aubert:2006je}
\bibinfo{author}{\bibnamefont{Aubert}, \bibfnamefont{B.}}, \emph{et~al.},
  \bibinfo{year}{2006}{\natexlab{b}}, \bibinfo{journal}{Phys. Rev. Lett.}
  \textbf{\bibinfo{volume}{97}}, \bibinfo{pages}{232001}.

\bibitem[{Aubert \emph{et~al.}(2006{\natexlab{c}})\citenamefont{Aubert}
  \emph{et~al.}}]{Aubert:2006qw}
\bibinfo{author}{\bibnamefont{Aubert}, \bibfnamefont{B.}}, \emph{et~al.},
  \bibinfo{year}{2006}{\natexlab{c}}, \bibinfo{journal}{Phys. Rev.}
  \textbf{\bibinfo{volume}{D74}}, \bibinfo{pages}{011103}.

\bibitem[{Aubert \emph{et~al.}(2006{\natexlab{d}})\citenamefont{Aubert}
  \emph{et~al.}}]{Aubert:2006rv}
\bibinfo{author}{\bibnamefont{Aubert}, \bibfnamefont{B.}}, \emph{et~al.},
  \bibinfo{year}{2006}{\natexlab{d}}, \bibinfo{note}{$33^{\rm rd}$
  International Conference on High Energy Physics (ICHEP 06), Moscow, Russia,
  26 Jul - 2 Aug 2006}, \eprint{hep-ex/0607086}.

\bibitem[{Aubert \emph{et~al.}(2007)\citenamefont{Aubert}
  \emph{et~al.}}]{Aubert:2006sp}
\bibinfo{author}{\bibnamefont{Aubert}, \bibfnamefont{B.}}, \emph{et~al.},
  \bibinfo{year}{2007}, \bibinfo{journal}{Phys. Rev. Lett.}
  \textbf{\bibinfo{volume}{98}}, \bibinfo{pages}{012001}.

\bibitem[{Aubert \emph{et~al.}(2008{\natexlab{a}})\citenamefont{Aubert}
  \emph{et~al.}}]{Aubert:2007dt}
\bibinfo{author}{\bibnamefont{Aubert}, \bibfnamefont{B.}}, \emph{et~al.},
  \bibinfo{year}{2008}{\natexlab{a}}, \bibinfo{journal}{Phys. Rev.}
  \textbf{\bibinfo{volume}{D77}}, \bibinfo{pages}{012002}.

\bibitem[{Aubert \emph{et~al.}(2008{\natexlab{b}})\citenamefont{Aubert}
  \emph{et~al.}}]{Aubert:2008if}
\bibinfo{author}{\bibnamefont{Aubert}, \bibfnamefont{B.}}, \emph{et~al.},
  \bibinfo{year}{2008{\natexlab{b}}}, \bibinfo{journal}{Phys. Rev.}
  \textbf{\bibinfo{volume}{D78}}, \bibinfo{pages}{112003}.

\bibitem[{Augustin \emph{et~al.}(1974)\citenamefont{Augustin}
  \emph{et~al.}}]{Augustin:1974xw}
\bibinfo{author}{\bibnamefont{Augustin}, \bibfnamefont{J.~E.}}, \emph{et~al.},
  \bibinfo{year}{1974}, \bibinfo{journal}{Phys. Rev. Lett.}
  \textbf{\bibinfo{volume}{33}}, \bibinfo{pages}{1406}.

\bibitem[{Avakian \emph{et~al.}(2004)\citenamefont{Avakian}
  \emph{et~al.}}]{Avakian:2003pk}
\bibinfo{author}{\bibnamefont{Avakian}, \bibfnamefont{H.}}, \emph{et~al.},
  \bibinfo{year}{2004}, \bibinfo{journal}{Phys. Rev.}
  \textbf{\bibinfo{volume}{D69}}, \bibinfo{pages}{112004}.

\bibitem[{Avery \emph{et~al.}(1989)\citenamefont{Avery}
  \emph{et~al.}}]{Avery:1988uh}
\bibinfo{author}{\bibnamefont{Avery}, \bibfnamefont{P.}}, \emph{et~al.},
  \bibinfo{year}{1989}, \bibinfo{journal}{Phys. Rev. Lett.}
  \textbf{\bibinfo{volume}{62}}, \bibinfo{pages}{863}.

\bibitem[{Aznauryan \emph{et~al.}(2005)\citenamefont{Aznauryan}
  \emph{et~al.}}]{Aznauryan:2005tp}
 \bibinfo{author}{\bibnamefont{Aznauryan}, \bibfnamefont{I.~G.}},
 \bibinfo{author}{\bibnamefont{V.~D.}, \bibfnamefont{Burkert}},
 \bibinfo{author}{\bibnamefont{G.~V.}, \bibfnamefont{Fedotov}},
 \bibinfo{author}{\bibnamefont{B.~S.}, \bibfnamefont{Ishkhanov}},
 \bibinfo{author}{\bibnamefont{V.~I.}, \bibfnamefont{Mokeev}},
  \bibinfo{year}{2005},
  \bibinfo{journal}{Phys. Rev.} \textbf{\bibinfo{volume}{C72}},
  \bibinfo{pages}{045201}.

\bibitem[{\citenamefont{Aznauryan}(2007)}]{Aznauryan:2007ja}
\bibinfo{author}{\bibnamefont{Aznauryan}, \bibfnamefont{I.~G.}},
  \bibinfo{year}{2007}, \bibinfo{journal}{Phys. Rev.}
  \textbf{\bibinfo{volume}{C76}}, \bibinfo{pages}{025212}.

\bibitem[{Aznauryan \emph{et~al.}(2008)\citenamefont{Aznauryan}
  \emph{et~al.}}]{Aznauryan:2008pe}
\bibinfo{author}{\bibnamefont{Aznauryan}, \bibfnamefont{I.~G.}}, \emph{et~al.},
  \bibinfo{year}{2008},
  \bibinfo{journal}{Phys. Rev.} \textbf{\bibinfo{volume}{C78}},
  \bibinfo{pages}{045209}.

\bibitem[{\citenamefont{Aznauryan} \emph{et~al.}(2008)\citenamefont{Aznauryan,
  Burkert, and Lee}}]{Aznauryan:2008us}
\bibinfo{author}{\bibnamefont{Aznauryan}, \bibfnamefont{I.~G.}},
  \bibinfo{author}{\bibfnamefont{V.~D.} \bibnamefont{Burkert}}, and
  \bibinfo{author}{\bibfnamefont{T.~S.~H.} \bibnamefont{Lee}},
  \bibinfo{year}{2008}, \eprint{0810.0997}.

\bibitem[{\citenamefont{Aznauryan}(2009)\citenamefont{Aznauryan}
  \emph{et~al.}}]{Aznauryan:2009mx}
\bibinfo{author}{\bibnamefont{Aznauryan}, \bibfnamefont{I.~G.}},
  \bibinfo{year}{2007}, \bibinfo{journal}{Phys. Rev.}
  \textbf{\bibinfo{volume}{C80}}, \bibinfo{pages}{055203}.

\bibitem[{\citenamefont{Bagan} \emph{et~al.}(1993)\citenamefont{Bagan, Chabab,
  Dosch, and Narison}}]{Bagan:1993ii}
\bibinfo{author}{\bibnamefont{Bagan}, \bibfnamefont{E.}},
  \bibinfo{author}{\bibfnamefont{M.}~\bibnamefont{Chabab}},
  \bibinfo{author}{\bibfnamefont{H.~G.} \bibnamefont{Dosch}}, and
  \bibinfo{author}{\bibfnamefont{S.}~\bibnamefont{Narison}},
  \bibinfo{year}{1993}, \bibinfo{journal}{Phys. Lett.}
  \textbf{\bibinfo{volume}{B301}}, \bibinfo{pages}{243}.

\bibitem[{\citenamefont{Bagan} \emph{et~al.}(1994)\citenamefont{Bagan, Dosch,
  Gosdzinsky, Narison, and Richard}}]{Bagan:1994dy}
\bibinfo{author}{\bibnamefont{Bagan}, \bibfnamefont{E.}},
  \bibinfo{author}{\bibfnamefont{H.~G.} \bibnamefont{Dosch}},
  \bibinfo{author}{\bibfnamefont{P.}~\bibnamefont{Gosdzinsky}},
  \bibinfo{author}{\bibfnamefont{S.}~\bibnamefont{Narison}}, and
  \bibinfo{author}{\bibfnamefont{J.~M.} \bibnamefont{Richard}},
  \bibinfo{year}{1994}, \bibinfo{journal}{Z. Phys.}
  \textbf{\bibinfo{volume}{C64}}, \bibinfo{pages}{57}.

\bibitem[{Ballam \emph{et~al.}(1972)\citenamefont{Ballam}
  \emph{et~al.}}]{Ballam:1971yd}
\bibinfo{author}{\bibnamefont{Ballam}, \bibfnamefont{J.}}, \emph{et~al.},
  \bibinfo{year}{1972}, \bibinfo{journal}{Phys. Rev.}
  \textbf{\bibinfo{volume}{D5}}, \bibinfo{pages}{545}.

\bibitem[{Ballam \emph{et~al.}(1973)\citenamefont{Ballam}
  \emph{et~al.}}]{Ballam:1972eq}
\bibinfo{author}{\bibnamefont{Ballam}, \bibfnamefont{J.}}, \emph{et~al.},
  \bibinfo{year}{1973}, \bibinfo{journal}{Phys. Rev.}
  \textbf{\bibinfo{volume}{D7}}, \bibinfo{pages}{3150}.

\bibitem[{Barber \emph{et~al.}(1982)\citenamefont{Barber}
  \emph{et~al.}}]{Barber:1981fj}
\bibinfo{author}{\bibnamefont{Barber}, \bibfnamefont{D.~P.}}, \emph{et~al.},
  \bibinfo{year}{1982}, \bibinfo{journal}{Zeit. Phys.}
  \textbf{\bibinfo{volume}{C12}}, \bibinfo{pages}{1}.

\bibitem[{Barber \emph{et~al.}(1984)\citenamefont{Barber}
  \emph{et~al.}}]{Barber:1985fr}
\bibinfo{author}{\bibnamefont{Barber}, \bibfnamefont{D.~P.}}, \emph{et~al.},
  \bibinfo{year}{1984}, \bibinfo{journal}{Z. Phys.}
  \textbf{\bibinfo{volume}{C26}}, \bibinfo{pages}{343}.

\bibitem[{Bari \emph{et~al.}(1991{\natexlab{a}})\citenamefont{Bari}
  \emph{et~al.}}]{Bari:1991ty}
\bibinfo{author}{\bibnamefont{Bari}, \bibfnamefont{G.}}, \emph{et~al.},
  \bibinfo{year}{1991}{\natexlab{a}}, \bibinfo{journal}{Nuovo Cim.}
  \textbf{\bibinfo{volume}{A104}}, \bibinfo{pages}{1787}.

\bibitem[{Bari \emph{et~al.}(1991{\natexlab{b}})\citenamefont{Bari}
  \emph{et~al.}}]{Bari:1991in}
\bibinfo{author}{\bibnamefont{Bari}, \bibfnamefont{G.}}, \emph{et~al.},
  \bibinfo{year}{1991}{\natexlab{b}}, \bibinfo{journal}{Nuovo Cim.}
  \textbf{\bibinfo{volume}{A104}}, \bibinfo{pages}{571}.

\bibitem[{\citenamefont{Barker} \emph{et~al.}(1975)\citenamefont{Barker,
  Donnachie, and Storrow}}]{Barker:1975bp}
\bibinfo{author}{\bibnamefont{Barker}, \bibfnamefont{I.~S.}},
  \bibinfo{author}{\bibfnamefont{A.}~\bibnamefont{Donnachie}}, and
  \bibinfo{author}{\bibfnamefont{J.~K.} \bibnamefont{Storrow}},
  \bibinfo{year}{1975}, \bibinfo{journal}{Nucl. Phys.}
  \textbf{\bibinfo{volume}{B95}}, \bibinfo{pages}{347}.

\bibitem[{Barmin \emph{et~al.}(2003)\citenamefont{Barmin}
  \emph{et~al.}}]{Barmin:2003vv}
\bibinfo{author}{\bibnamefont{Barmin}, \bibfnamefont{V.~V.}}, \emph{et~al.},
  \bibinfo{year}{2003}, \bibinfo{journal}{Phys. Atom. Nucl.}
  \textbf{\bibinfo{volume}{66}}, \bibinfo{pages}{1715}.

\bibitem[{\citenamefont{Barnes and Close}(1983)}]{Barnes:1982fj}
\bibinfo{author}{\bibnamefont{Barnes}, \bibfnamefont{T.}}, and
  \bibinfo{author}{\bibfnamefont{F.~E.} \bibnamefont{Close}},
  \bibinfo{year}{1983}, \bibinfo{journal}{Phys. Lett.}
  \textbf{\bibinfo{volume}{B123}}, \bibinfo{pages}{89}.

\bibitem[{\citenamefont{Barnes} \emph{et~al.}(1995)\citenamefont{Barnes, Close,
  and Swanson}}]{Barnes:1995hc}
\bibinfo{author}{\bibnamefont{Barnes}, \bibfnamefont{T.}},
  \bibinfo{author}{\bibfnamefont{F.~E.} \bibnamefont{Close}}, and
  \bibinfo{author}{\bibfnamefont{E.~S.} \bibnamefont{Swanson}},
  \bibinfo{year}{1995}, \bibinfo{journal}{Phys. Rev.}
  \textbf{\bibinfo{volume}{D52}}, \bibinfo{pages}{5242}.

\bibitem[{Barnes \emph{et~al.}(1964)\citenamefont{Barnes}
  \emph{et~al.}}]{Barnes:1964pd}
\bibinfo{author}{\bibnamefont{Barnes}, \bibfnamefont{V.~E.}}, \emph{et~al.},
  \bibinfo{year}{1964}, \bibinfo{journal}{Phys. Rev. Lett.}
  \textbf{\bibinfo{volume}{12}}, \bibinfo{pages}{204}.

\bibitem[{Barrow \emph{et~al.}(2001)\citenamefont{Barrow}
  \emph{et~al.}}]{Barrow:2001ds}
\bibinfo{author}{\bibnamefont{Barrow}, \bibfnamefont{S.~P.}}, \emph{et~al.},
  \bibinfo{year}{2001}, \bibinfo{journal}{Phys. Rev.}
  \textbf{\bibinfo{volume}{C64}}, \bibinfo{pages}{044601}.

\bibitem[{Bartalini \emph{et~al.}(2002)\citenamefont{Bartalini}
  \emph{et~al.}}]{Bartalini:2002cj}
\bibinfo{author}{\bibnamefont{Bartalini}, \bibfnamefont{O.}}, \emph{et~al.},
  \bibinfo{year}{2002}, \bibinfo{journal}{Phys. Lett.}
  \textbf{\bibinfo{volume}{B544}}, \bibinfo{pages}{113}.

\bibitem[{Bartalini \emph{et~al.}(2005)\citenamefont{Bartalini}
  \emph{et~al.}}]{Bartalini:2005wx}
\bibinfo{author}{\bibnamefont{Bartalini}, \bibfnamefont{O.}}, \emph{et~al.},
  \bibinfo{year}{2005}, \bibinfo{journal}{Eur. Phys. J.}
  \textbf{\bibinfo{volume}{A26}}, \bibinfo{pages}{399}.

\bibitem[{Bartalini \emph{et~al.}(2007)\citenamefont{Bartalini}
  \emph{et~al.}}]{Bartalini:2007fg}
\bibinfo{author}{\bibnamefont{Bartalini}, \bibfnamefont{O.}}, \emph{et~al.},
  \bibinfo{year}{2007}, \bibinfo{journal}{Eur. Phys. J.}
  \textbf{\bibinfo{volume}{A33}}, \bibinfo{pages}{169}.

\bibitem[{Barth \emph{et~al.}(2003{\natexlab{a}})\citenamefont{Barth}
  \emph{et~al.}}]{Barth:2003kv}
\bibinfo{author}{\bibnamefont{Barth}, \bibfnamefont{J.}}, \emph{et~al.},
  \bibinfo{year}{2003}{\natexlab{a}}, \bibinfo{journal}{Eur. Phys. J.}
  \textbf{\bibinfo{volume}{A18}}, \bibinfo{pages}{117}.

\bibitem[{Barth \emph{et~al.}(2003{\natexlab{b}})\citenamefont{Barth}
  \emph{et~al.}}]{Barth:2003bq}
\bibinfo{author}{\bibnamefont{Barth}, \bibfnamefont{J.}}, \emph{et~al.},
  \bibinfo{year}{2003}{\natexlab{b}}, \bibinfo{journal}{Eur. Phys. J.}
  \textbf{\bibinfo{volume}{A17}}, \bibinfo{pages}{269}.

\bibitem[{Barth \emph{et~al.}(2003{\natexlab{c}})\citenamefont{Barth}
  \emph{et~al.}}]{Barth:2003es}
\bibinfo{author}{\bibnamefont{Barth}, \bibfnamefont{J.}}, \emph{et~al.},
  \bibinfo{year}{2003}{\natexlab{c}}, \bibinfo{journal}{Phys. Lett.}
  \textbf{\bibinfo{volume}{B572}}, \bibinfo{pages}{127}.

\bibitem[{Bartholomy \emph{et~al.}(2005)\citenamefont{Bartholomy}
  \emph{et~al.}}]{Bartholomy:2004uz}
\bibinfo{author}{\bibnamefont{Bartholomy}, \bibfnamefont{O.}}, \emph{et~al.},
  \bibinfo{year}{2005}, \bibinfo{journal}{Phys. Rev. Lett.}
  \textbf{\bibinfo{volume}{94}}, \bibinfo{pages}{012003}.

\bibitem[{Bartholomy \emph{et~al.}(2007)\citenamefont{Bartholomy}
  \emph{et~al.}}]{Bartholomy:2007zz}
\bibinfo{author}{\bibnamefont{Bartholomy}, \bibfnamefont{O.}}, \emph{et~al.},
  \bibinfo{year}{2007}, \bibinfo{journal}{Eur. Phys. J.}
  \textbf{\bibinfo{volume}{A33}}, \bibinfo{pages}{133}.

\bibitem[{Basak \emph{et~al.}(2007)\citenamefont{Basak}
  \emph{et~al.}}]{Basak:2007kj}
\bibinfo{author}{\bibnamefont{Basak}, \bibfnamefont{S.}}, \emph{et~al.},
  \bibinfo{year}{2007}, \bibinfo{journal}{Phys. Rev.}
  \textbf{\bibinfo{volume}{D76}}, \bibinfo{pages}{074504}.

\bibitem[{\citenamefont{Basdevant and Boukraa}(1986)}]{Basdevant:1985ux}
\bibinfo{author}{\bibnamefont{Basdevant}, \bibfnamefont{J.~L.}}, and
  \bibinfo{author}{\bibfnamefont{S.}~\bibnamefont{Boukraa}},
  \bibinfo{year}{1986}, \bibinfo{journal}{Z. Phys.}
  \textbf{\bibinfo{volume}{C30}}, \bibinfo{pages}{103}.

\bibitem[{Battaglieri \emph{et~al.}(2001)\citenamefont{Battaglieri}
  \emph{et~al.}}]{Battaglieri:2001xv}
\bibinfo{author}{\bibnamefont{Battaglieri}, \bibfnamefont{M.}}, \emph{et~al.},
  \bibinfo{year}{2001}, \bibinfo{journal}{Phys. Rev. Lett.}
  \textbf{\bibinfo{volume}{87}}, \bibinfo{pages}{172002}.

\bibitem[{Battaglieri \emph{et~al.}(2003)\citenamefont{Battaglieri}
  \emph{et~al.}}]{Battaglieri:2002pr}
\bibinfo{author}{\bibnamefont{Battaglieri}, \bibfnamefont{M.}}, \emph{et~al.},
  \bibinfo{year}{2003}, \bibinfo{journal}{Phys. Rev. Lett.}
  \textbf{\bibinfo{volume}{90}}, \bibinfo{pages}{022002}.

\bibitem[{Battaglieri \emph{et~al.}(2008)\citenamefont{Battaglieri}
  \emph{et~al.}}]{Battaglieri:2008ps}
\bibinfo{author}{\bibnamefont{Battaglieri}, \bibfnamefont{M.}}, \emph{et~al.},
  \bibinfo{year}{2008}, \bibinfo{journal}{Phys. Rev. Lett.}
  \textbf{\bibinfo{volume}{102}}, \bibinfo{pages}{102001}.

\bibitem[{Battaglieri \emph{et~al.}(2009)\citenamefont{Battaglieri}
  \emph{et~al.}}]{Battaglieri:2009fq}
\bibinfo{author}{\bibnamefont{Battaglieri}, \bibfnamefont{M.}}, \emph{et~al.},
  \bibinfo{year}{2009}, \bibinfo{journal}{Phys. Rev.}
  \textbf{\bibinfo{volume}{D 80}}, \bibinfo{pages}{072005}.

\bibitem[{\citenamefont{Beck}(2006)}]{Beck:2006prc}
\bibinfo{author}{\bibnamefont{Beck}, \bibfnamefont{R.}}, \bibinfo{year}{2006},
  \bibinfo{journal}{Eur. Phys. J.} \textbf{\bibinfo{volume}{A28, s01}},
  \bibinfo{pages}{173}.

\bibitem[{\citenamefont{Berezhnoi} \emph{et~al.}(1998)\citenamefont{Berezhnoi,
  Kiselev, Likhoded, and Onishchenko}}]{Berezhnoi:1998aa}
\bibinfo{author}{\bibnamefont{Berezhnoi}, \bibfnamefont{A.~V.}},
  \bibinfo{author}{\bibfnamefont{V.~V.} \bibnamefont{Kiselev}},
  \bibinfo{author}{\bibfnamefont{A.~K.} \bibnamefont{Likhoded}}, and
  \bibinfo{author}{\bibfnamefont{A.~I.} \bibnamefont{Onishchenko}},
  \bibinfo{year}{1998}, \bibinfo{journal}{Phys. Rev.}
  \textbf{\bibinfo{volume}{D57}}, \bibinfo{pages}{4385}.

\bibitem[{\citenamefont{Bernard} \emph{et~al.}(1995)\citenamefont{Bernard,
  Kaiser, and Mei{\ss}ner}}]{Bernard:1995dp}
\bibinfo{author}{\bibnamefont{Bernard}, \bibfnamefont{V.}},
  \bibinfo{author}{\bibfnamefont{N.}~\bibnamefont{Kaiser}}, and
  \bibinfo{author}{\bibfnamefont{U.-G.} \bibnamefont{Mei{\ss}ner}},
  \bibinfo{year}{1995}, \bibinfo{journal}{Int. J. Mod. Phys.}
  \textbf{\bibinfo{volume}{E4}}, \bibinfo{pages}{193}.

\bibitem[{\citenamefont{Bernard}(2008)}]{Bernard:2007zu}
\bibinfo{author}{\bibnamefont{Bernard}, \bibfnamefont{V.}},
  \bibinfo{year}{2008}, \bibinfo{journal}{Prog. Part. Nucl. Phys.}
  \textbf{\bibinfo{volume}{60}}, \bibinfo{pages}{82}.

\bibitem[{\citenamefont{Bernotas and Simonis}(2008)}]{Bernotas:2008bu}
\bibinfo{author}{\bibnamefont{Bernotas}, \bibfnamefont{A.}}, and
  \bibinfo{author}{\bibfnamefont{V.}~\bibnamefont{Simonis}},
  \bibinfo{year}{2008}, \eprint{0808.1220}.

\bibitem[{Biagi \emph{et~al.}(1983)\citenamefont{Biagi}
  \emph{et~al.}}]{Biagi:1983en}
\bibinfo{author}{\bibnamefont{Biagi}, \bibfnamefont{S.~F.}}, \emph{et~al.},
  \bibinfo{year}{1983}, \bibinfo{journal}{Phys. Lett.}
  \textbf{\bibinfo{volume}{B122}}, \bibinfo{pages}{455}.

\bibitem[{Biagi \emph{et~al.}(1985)\citenamefont{Biagi}
  \emph{et~al.}}]{Biagi:1984mu}
\bibinfo{author}{\bibnamefont{Biagi}, \bibfnamefont{S.~F.}}, \emph{et~al.},
  \bibinfo{year}{1985}, \bibinfo{journal}{Z. Phys.}
  \textbf{\bibinfo{volume}{C28}}, \bibinfo{pages}{175}.

\bibitem[{\citenamefont{Bicudo}
\emph{et~al.}(2009)\citenamefont{Bicudo, Cardoso, Cauteren, and
Llanes-Estrada}}]{Bicudo:2009cr}
\bibinfo{author}{\bibnamefont{Bicudo}, \bibfnamefont{P.}},
  \bibinfo{author}{\bibfnamefont{M.}~\bibnamefont{Cardoso}},
  \bibinfo{author}{\bibfnamefont{T.}~\bibnamefont{Van Cauteren }}, and
  \bibinfo{author}{\bibfnamefont{F.~J.}~\bibnamefont{Llanes-Estrada}},
  \bibinfo{year}{2009}, \bibinfo{journal}{Phys. Rev. Lett.}
  \textbf{\bibinfo{volume}{103}}, \bibinfo{pages}{092003}.

\bibitem[{\citenamefont{Bienlein and Bloom}(1981)}]{Bienlein:1981rn}
\bibinfo{author}{\bibnamefont{Bienlein}, \bibfnamefont{J.~K.}}, and
  \bibinfo{author}{\bibfnamefont{E.~D.} \bibnamefont{Bloom}},
  \bibinfo{year}{1981}, \bibinfo{note}{pRC 81/09}.

\bibitem[{\citenamefont{Bijker} \emph{et~al.}(1994)\citenamefont{Bijker,
  Iachello, and Leviatan}}]{Bijker:1994yr}
\bibinfo{author}{\bibnamefont{Bijker}, \bibfnamefont{R.}},
  \bibinfo{author}{\bibfnamefont{F.}~\bibnamefont{Iachello}}, and
  \bibinfo{author}{\bibfnamefont{A.}~\bibnamefont{Leviatan}},
  \bibinfo{year}{1994}, \bibinfo{journal}{Ann. Phys.}
  \textbf{\bibinfo{volume}{236}}, \bibinfo{pages}{69}.

\bibitem[{\citenamefont{Bijker} \emph{et~al.}(1997)\citenamefont{Bijker,
  Iachello, and Leviatan}}]{Bijker:1996tr}
\bibinfo{author}{\bibnamefont{Bijker}, \bibfnamefont{R.}},
  \bibinfo{author}{\bibfnamefont{F.}~\bibnamefont{Iachello}}, and
  \bibinfo{author}{\bibfnamefont{A.}~\bibnamefont{Leviatan}},
  \bibinfo{year}{1997}, \bibinfo{journal}{Phys. Rev.}
  \textbf{\bibinfo{volume}{D55}}, \bibinfo{pages}{2862}.

\bibitem[{\citenamefont{Bijker} \emph{et~al.}(2000)\citenamefont{Bijker,
  Iachello, and Leviatan}}]{Bijker:2000gq}
\bibinfo{author}{\bibnamefont{Bijker}, \bibfnamefont{R.}},
  \bibinfo{author}{\bibfnamefont{F.}~\bibnamefont{Iachello}}, and
  \bibinfo{author}{\bibfnamefont{A.}~\bibnamefont{Leviatan}},
  \bibinfo{year}{2000}, \bibinfo{journal}{Annals Phys.}
  \textbf{\bibinfo{volume}{284}}, \bibinfo{pages}{89}.

\bibitem[{Biselli \emph{et~al.}(2008)\citenamefont{Biselli}
  \emph{et~al.}}]{Biselli:2008ug}
\bibinfo{author}{\bibnamefont{Biselli}, \bibfnamefont{A.~S.}},
  \emph{et~al.}, \bibinfo{year}{2008},
  \bibinfo{journal}{Phys. Rev.} \textbf{\bibinfo{volume}{C78}},
  \bibinfo{pages}{045204}.

\bibitem[{Blomqvist \emph{et~al.}(1998)\citenamefont{Blomqvist}
  \emph{et~al.}}]{Blomqvist:1998xn}
\bibinfo{author}{\bibnamefont{Blomqvist}, \bibfnamefont{K.~I.}}, \emph{et~al.},
  \bibinfo{year}{1998}, \bibinfo{journal}{Nucl. Instrum. Meth.}
  \textbf{\bibinfo{volume}{A403}}, \bibinfo{pages}{263}.

\bibitem[{Bock \emph{et~al.}(1998)\citenamefont{Bock}
  \emph{et~al.}}]{Bock:1998rk}
\bibinfo{author}{\bibnamefont{Bock}, \bibfnamefont{A.}}, \emph{et~al.},
  \bibinfo{year}{1998}, \bibinfo{journal}{Phys. Rev. Lett.}
  \textbf{\bibinfo{volume}{81}}, \bibinfo{pages}{534}.

\bibitem[{\citenamefont{Borasoy} \emph{et~al.}(2006)\citenamefont{Borasoy,
  Bruns, Mei{\ss}ner, and Lewis}}]{Borasoy:2006fk}
\bibinfo{author}{\bibnamefont{Borasoy}, \bibfnamefont{B.}},
  \bibinfo{author}{\bibfnamefont{P.~C.} \bibnamefont{Bruns}},
  \bibinfo{author}{\bibfnamefont{U.~G.} \bibnamefont{Mei{\ss}ner}}, and
  \bibinfo{author}{\bibfnamefont{R.}~\bibnamefont{Lewis}},
  \bibinfo{year}{2006}, \bibinfo{journal}{Phys. Lett.}
  \textbf{\bibinfo{volume}{B641}}, \bibinfo{pages}{294}.

\bibitem[{\citenamefont{Borasoy} \emph{et~al.}(2006)\citenamefont{Borasoy,
  Mei{\ss}ner, and Nissler}}]{Borasoy:2006sr}
\bibinfo{author}{\bibnamefont{Borasoy}, \bibfnamefont{B.}},
  \bibinfo{author}{\bibfnamefont{U.~G.} \bibnamefont{Mei{\ss}ner}}, and
  \bibinfo{author}{\bibfnamefont{R.}~\bibnamefont{Nissler}},
  \bibinfo{year}{2006}, \bibinfo{journal}{Phys. Rev.}
  \textbf{\bibinfo{volume}{C74}}, \bibinfo{pages}{055201}.

\bibitem[{\citenamefont{Borasoy} \emph{et~al.}(2007)\citenamefont{Borasoy, Bruns,
  Mei{\ss}ner, and Nissler}}]{Borasoy:2007ku}
\bibinfo{author}{\bibnamefont{Borasoy}, \bibfnamefont{B.}},
  \bibinfo{author}{\bibfnamefont{U.~G.} \bibnamefont{Mei{\ss}ner}}, and
  \bibinfo{author}{\bibfnamefont{R.}~\bibnamefont{Nissler}},
  \bibinfo{year}{2007}, \bibinfo{journal}{Eur. Phys. J.}
  \textbf{\bibinfo{volume}{A34}}, \bibinfo{pages}{161}.


\bibitem[{Bradford \emph{et~al.}(2006)\citenamefont{Bradford}
  \emph{et~al.}}]{Bradford:2005pt}
\bibinfo{author}{\bibnamefont{Bradford}, \bibfnamefont{R.}}, \emph{et~al.},
  \bibinfo{year}{2006}, \bibinfo{journal}{Phys. Rev.}
  \textbf{\bibinfo{volume}{C73}}, \bibinfo{pages}{035202}.

\bibitem[{Bradford \emph{et~al.}(2007)\citenamefont{Bradford}
  \emph{et~al.}}]{Bradford:2006ba}
\bibinfo{author}{\bibnamefont{Bradford}, \bibfnamefont{R.}}, \emph{et~al.},
  \bibinfo{year}{2007}, \bibinfo{journal}{Phys. Rev.}
  \textbf{\bibinfo{volume}{C75}}, \bibinfo{pages}{035205}.

\bibitem[{Brambilla \emph{et~al.}(2004)\citenamefont{Brambilla}
  \emph{et~al.}}]{Brambilla:2004wf}
\bibinfo{author}{\bibnamefont{Brambilla}, \bibfnamefont{N.}}, \emph{et~al.},
  \bibinfo{year}{2004}, \bibinfo{note}{published as CERN Yellow Report,
  CERN-2005-005, Geneva: CERN, 2005. 487 p.}, \eprint{hep-ph/0412158}.

\bibitem[{\citenamefont{Briscoe} \emph{et~al.}(2005)\citenamefont{Briscoe,
  Arndt, Strakovsky, and Workman}}]{Briscoe:2005ek}
\bibinfo{author}{\bibnamefont{Briscoe}, \bibfnamefont{W.}},
  \bibinfo{author}{\bibfnamefont{R.}~\bibnamefont{Arndt}},
  \bibinfo{author}{\bibfnamefont{I.}~\bibnamefont{Strakovsky}}, and
  \bibinfo{author}{\bibfnamefont{R.}~\bibnamefont{Workman}},
  \bibinfo{year}{2005}, \bibinfo{note}{international Workshop on the Physics of
  Excited Baryons (NSTAR 05), Tallahassee, Florida, 10-15 Oct 2005}.

\bibitem[{\citenamefont{Brodsky}(2007)}]{Brodsky:2006uq}
\bibinfo{author}{\bibnamefont{Brodsky}, \bibfnamefont{S.~J.}},
  \bibinfo{year}{2007}, \bibinfo{journal}{Eur. Phys. J.}
  \textbf{\bibinfo{volume}{A31}}, \bibinfo{pages}{638}.

\bibitem[{\citenamefont{Brodsky and de~Teramond}(2008)}]{Brodsky:2008pg}
\bibinfo{author}{\bibnamefont{Brodsky}, \bibfnamefont{S.~J.}}, and
  \bibinfo{author}{\bibfnamefont{G.~F.} \bibnamefont{de~Teramond}},
  \bibinfo{year}{2008}, \eprint{arXiv:0802.0514 [hep-ph]}.

\bibitem[{\citenamefont{Brown and Rho}(1979)}]{Brown:1979ui}
\bibinfo{author}{\bibnamefont{Brown}, \bibfnamefont{G.~E.}}, and
  \bibinfo{author}{\bibfnamefont{M.}~\bibnamefont{Rho}}, \bibinfo{year}{1979},
  \bibinfo{journal}{Phys. Lett.} \textbf{\bibinfo{volume}{B82}},
  \bibinfo{pages}{177}.

\bibitem[{Brown \emph{et~al.}(1978)\citenamefont{Brown}
  \emph{et~al.}}]{Brown:1978hw}
\bibinfo{author}{\bibnamefont{Brown}, \bibfnamefont{R.~M.}}, \emph{et~al.},
  \bibinfo{year}{1978}, \bibinfo{journal}{Nucl. Phys.}
  \textbf{\bibinfo{volume}{B144}}, \bibinfo{pages}{287}.

\bibitem[{\citenamefont{Bugg}(2007)}]{Bugg:2007vq}
\bibinfo{author}{\bibnamefont{Bugg}, \bibfnamefont{D.~V.}},
  \bibinfo{year}{2007}, \eprint{0709.1256}.

\bibitem[{Bulava \emph{et~al.}(2009)\citenamefont{Bulava}
  \emph{et~al.}}]{Bulava:2009jb}
\bibinfo{author}{\bibnamefont{Bulava}, \bibfnamefont{J.~M.}}, \emph{et~al.},
  \bibinfo{year}{2009}, \bibinfo{journal}{Phys. Rev.}
  \textbf{\bibinfo{volume}{D79}}, \bibinfo{pages}{034505}.

\bibitem[{Burch \emph{et~al.}(2006)\citenamefont{Burch}
  \emph{et~al.}}]{Burch:2006cc}
\bibinfo{author}{\bibnamefont{Burch}, \bibfnamefont{T.}}, \emph{et~al.},
  \bibinfo{year}{2006}, \bibinfo{journal}{Phys. Rev.}
  \textbf{\bibinfo{volume}{D74}}, \bibinfo{pages}{014504}.

\bibitem[{\citenamefont{Butterworth and Wing}(2005)}]{Butterworth:2005aq}
\bibinfo{author}{\bibnamefont{Butterworth}, \bibfnamefont{J.~M.}}, and
  \bibinfo{author}{\bibfnamefont{M.}~\bibnamefont{Wing}}, \bibinfo{year}{2005},
  \bibinfo{journal}{Rept. Prog. Phys.} \textbf{\bibinfo{volume}{68}},
  \bibinfo{pages}{2773}.

\bibitem[{\citenamefont{Cahn}(2004)}]{Cahn:2003wq}
\bibinfo{author}{\bibnamefont{Cahn}, \bibfnamefont{R.~N.}}, and
  \bibinfo{author}{\bibfnamefont{G.~H.} \bibnamefont{Trilling}},
  \bibinfo{year}{2004}, \bibinfo{journal}{Phys. Rev.}
  \textbf{\bibinfo{volume}{D69}}, \bibinfo{pages}{011501}.

\bibitem[{\citenamefont{Cano and Gonzalez}(1998)}]{Cano:1998wz}
\bibinfo{author}{\bibnamefont{Cano}, \bibfnamefont{F.}}, and
  \bibinfo{author}{\bibfnamefont{P.} \bibnamefont{Gonzalez}},
  \bibinfo{year}{1998}, \bibinfo{journal}{Phys. Lett.}
  \textbf{\bibinfo{volume}{B431}}, \bibinfo{pages}{270}.

\bibitem[{\citenamefont{Cano} \emph{et~al.}(1996)\citenamefont{Cano, Gonzalez,
  Noguera, and Desplanques}}]{Cano:1996ep}
\bibinfo{author}{\bibnamefont{Cano}, \bibfnamefont{F.}},
  \bibinfo{author}{\bibfnamefont{P.}~\bibnamefont{Gonzalez}},
  \bibinfo{author}{\bibfnamefont{S.}~\bibnamefont{Noguera}}, and
  \bibinfo{author}{\bibfnamefont{B.}~\bibnamefont{Desplanques}},
  \bibinfo{year}{1996}, \bibinfo{journal}{Nucl. Phys.}
  \textbf{\bibinfo{volume}{A603}}, \bibinfo{pages}{257}.

\bibitem[{\citenamefont{Capstick and Isgur}(1986)}]{Capstick:1986bm}
\bibinfo{author}{\bibnamefont{Capstick}, \bibfnamefont{S.}}, and
  \bibinfo{author}{\bibfnamefont{N.}~\bibnamefont{Isgur}},
  \bibinfo{year}{1986}, \bibinfo{journal}{Phys. Rev.}
  \textbf{\bibinfo{volume}{D34}}, \bibinfo{pages}{2809}.

\bibitem[{\citenamefont{Capstick and Keister}(1995)}]{Capstick:1994ne}
\bibinfo{author}{\bibnamefont{Capstick}, \bibfnamefont{S.}}, and
  \bibinfo{author}{\bibfnamefont{B.~D.} \bibnamefont{Keister}},
  \bibinfo{year}{1995}, \bibinfo{journal}{Phys. Rev.}
  \textbf{\bibinfo{volume}{D51}}, \bibinfo{pages}{3598}.

\bibitem[{\citenamefont{Capstick and Page}(2002)}]{Capstick:2002wm}
\bibinfo{author}{\bibnamefont{Capstick}, \bibfnamefont{S.}}, and
  \bibinfo{author}{\bibfnamefont{P.~R.} \bibnamefont{Page}},
  \bibinfo{year}{2002}, \bibinfo{journal}{Phys. Rev.}
  \textbf{\bibinfo{volume}{C66}}, \bibinfo{pages}{065204}.

\bibitem[{\citenamefont{Capstick and Roberts}(2000)}]{Capstick:2000qj}
\bibinfo{author}{\bibnamefont{Capstick}, \bibfnamefont{S.}}, and
  \bibinfo{author}{\bibfnamefont{W.}~\bibnamefont{Roberts}},
  \bibinfo{year}{2000}, \bibinfo{journal}{Prog. Part. Nucl. Phys.}
  \textbf{\bibinfo{volume}{45}}, \bibinfo{pages}{S241}.

\bibitem[{\citenamefont{Cardarelli}
  \emph{et~al.}(1997)\citenamefont{Cardarelli, Pace, Salme, and
  Simula}}]{Cardarelli:1996vn}
\bibinfo{author}{\bibnamefont{Cardarelli}, \bibfnamefont{F.}},
  \bibinfo{author}{\bibfnamefont{E.}~\bibnamefont{Pace}},
  \bibinfo{author}{\bibfnamefont{G.}~\bibnamefont{Salme}}, and
  \bibinfo{author}{\bibfnamefont{S.}~\bibnamefont{Simula}},
  \bibinfo{year}{1997}, \bibinfo{journal}{Phys. Lett.}
  \textbf{\bibinfo{volume}{B397}}, \bibinfo{pages}{13}.

\bibitem[{Carman \emph{et~al.}(2003)\citenamefont{Carman}
  \emph{et~al.}}]{Carman:2002se}
\bibinfo{author}{\bibnamefont{Carman}, \bibfnamefont{D.~S.}}, \emph{et~al.},
   \bibinfo{year}{2003},
  \bibinfo{journal}{Phys. Rev. Lett.} \textbf{\bibinfo{volume}{90}},
  \bibinfo{pages}{131804}.

\bibitem[{Carman \emph{et~al.}(2009)\citenamefont{Carman}
  \emph{et~al.}}]{Carman:2009fi}
\bibinfo{author}{\bibnamefont{Carman}, \bibfnamefont{D.~S.}}, \emph{et~al.},
  \bibinfo{year}{2009},
  \bibinfo{journal}{Phys. Rev.} \textbf{\bibinfo{volume}{C79}},
  \bibinfo{pages}{065205}.

\bibitem[{Castelijns \emph{et~al.}(2008)\citenamefont{Castelijns}
  \emph{et~al.}}]{Castelijns:2007qt}
\bibinfo{author}{\bibnamefont{Castelijns}, \bibfnamefont{R.}}, \emph{et~al.},
  \bibinfo{year}{2008}, \bibinfo{journal}{Eur. Phys. J.}
  \textbf{\bibinfo{volume}{A35}}, \bibinfo{pages}{39}.

\bibitem[{Cazzoli \emph{et~al.}(1975)\citenamefont{Cazzoli}
  \emph{et~al.}}]{Cazzoli:1975et}
\bibinfo{author}{\bibnamefont{Cazzoli}, \bibfnamefont{E.~G.}}, \emph{et~al.},
  \bibinfo{year}{1975}, \bibinfo{journal}{Phys. Rev. Lett.}
  \textbf{\bibinfo{volume}{34}}, \bibinfo{pages}{1125}.

\bibitem[{\citenamefont{Ceci}
  \emph{et~al.}(2006{\natexlab{a}})\citenamefont{Ceci, Svarc, and
  Zauner}}]{Ceci:2006ra}
\bibinfo{author}{\bibnamefont{Ceci}, \bibfnamefont{S.}},
  \bibinfo{author}{\bibfnamefont{A.}~\bibnamefont{Svarc}}, and
  \bibinfo{author}{\bibfnamefont{B.}~\bibnamefont{Zauner}},
  \bibinfo{year}{2006}{\natexlab{a}}, \bibinfo{journal}{Phys. Rev. Lett.}
  \textbf{\bibinfo{volume}{97}}, \bibinfo{pages}{062002}.

\bibitem[{\citenamefont{Ceci}
  \emph{et~al.}(2006{\natexlab{b}})\citenamefont{Ceci, Svarc, and
  Zauner}}]{Ceci:2005vf}
\bibinfo{author}{\bibnamefont{Ceci}, \bibfnamefont{S.}},
  \bibinfo{author}{\bibfnamefont{A.}~\bibnamefont{Svarc}}, and
  \bibinfo{author}{\bibfnamefont{B.}~\bibnamefont{Zauner}},
  \bibinfo{year}{2006}{\natexlab{b}}, \bibinfo{journal}{Few Body Sys.}
  \textbf{\bibinfo{volume}{39}}, \bibinfo{pages}{27}.

\bibitem[{\citenamefont{Ceci} \emph{et~al.}(2008)\citenamefont{Ceci, Svarc,
  Zauner, Manley, and Capstick}}]{Ceci:2006jj}
\bibinfo{author}{\bibnamefont{Ceci}, \bibfnamefont{S.}},
  \bibinfo{author}{\bibfnamefont{A.}~\bibnamefont{Svarc}},
  \bibinfo{author}{\bibfnamefont{B.}~\bibnamefont{Zauner}},
  \bibinfo{author}{\bibfnamefont{M.}~\bibnamefont{Manley}}, and
  \bibinfo{author}{\bibfnamefont{S.}~\bibnamefont{Capstick}},
  \bibinfo{year}{2008}, \bibinfo{journal}{Phys. Lett.}
  \textbf{\bibinfo{volume}{B659}}, \bibinfo{pages}{228}.

\bibitem[{Chan \emph{et~al.}(1978)\citenamefont{Chan}
  \emph{et~al.}}]{Chan:1978nk}
\bibinfo{author}{\bibnamefont{Chan}, \bibfnamefont{H.-M.}}, \emph{et~al.},
  \bibinfo{year}{1978}, \bibinfo{journal}{Phys. Lett.}
  \textbf{\bibinfo{volume}{B76}}, \bibinfo{pages}{634}.

\bibitem[{Chekanov \emph{et~al.}(2004)\citenamefont{Chekanov}
  \emph{et~al.}}]{Chekanov:2004kn}
\bibinfo{author}{\bibnamefont{Chekanov}, \bibfnamefont{S.}}, \emph{et~al.},
  \bibinfo{year}{2004}, \bibinfo{journal}{Phys. Lett.}
  \textbf{\bibinfo{volume}{B591}}, \bibinfo{pages}{7}.

\bibitem[{\citenamefont{Cheng and Chua}(2007)}]{Cheng:2006dk}
\bibinfo{author}{\bibnamefont{Cheng}, \bibfnamefont{H.-Y.}}, and
  \bibinfo{author}{\bibfnamefont{C.-K.} \bibnamefont{Chua}},
  \bibinfo{year}{2007}, \bibinfo{journal}{Phys. Rev.}
  \textbf{\bibinfo{volume}{D75}}, \bibinfo{pages}{014006}.

\bibitem[{\citenamefont{Cheng} \emph{et~al.}(2009)\citenamefont{Cheng,
  Wang, and Zhang}}]{Cheng:2009tm}
\bibinfo{author}{\bibnamefont{Cheng}, \bibfnamefont{B.}},
  \bibinfo{author}{\bibfnamefont{D.-X.}~\bibnamefont{Wang}}, and
  \bibinfo{author}{\bibfnamefont{A.}~\bibnamefont{Zhang}},
  \bibinfo{year}{2009}, \eprint{arXiv:0906.3934}.

\bibitem[{\citenamefont{Chew and Frautschi}(1961)}]{Chew:1961ev}
\bibinfo{author}{\bibnamefont{Chew}, \bibfnamefont{G.~F.}}, and
  \bibinfo{author}{\bibfnamefont{S.~C.} \bibnamefont{Frautschi}},
  \bibinfo{year}{1961}, \bibinfo{journal}{Phys. Rev. Lett.}
  \textbf{\bibinfo{volume}{7}}, \bibinfo{pages}{394}.

\bibitem[{\citenamefont{Chew and Frautschi}(1962)}]{Chew:1962eu}
\bibinfo{author}{\bibnamefont{Chew}, \bibfnamefont{G.~F.}}, and
  \bibinfo{author}{\bibfnamefont{S.~C.} \bibnamefont{Frautschi}},
  \bibinfo{year}{1962}, \bibinfo{journal}{Phys. Rev. Lett.}
  \textbf{\bibinfo{volume}{8}}, \bibinfo{pages}{41}.

\bibitem[{\citenamefont{Chew and Low}(1956)}]{Chew:1955zz}
\bibinfo{author}{\bibnamefont{Chew}, \bibfnamefont{G.~F.}}, and
  \bibinfo{author}{\bibfnamefont{M.~L.} \bibnamefont{Goldhaber}},
  \bibinfo{author}{\bibfnamefont{F.~E.} \bibnamefont{Low}},
  \bibinfo{author}{\bibfnamefont{S.~C.} \bibnamefont{Frautschi}},
  \bibinfo{year}{1956}, \bibinfo{journal}{Phys. Rev.}
  \textbf{\bibinfo{volume}{101}}, \bibinfo{pages}{1570}.

\bibitem[{\citenamefont{{\protect{Chew, Goldberger, Low,
  Nambu}}}(1957)}]{Chew:1957tf}
\bibinfo{author}{\bibnamefont{{\protect{Chew, Goldberger, Low, Nambu}}}},
  \bibinfo{year}{1957}, \bibinfo{journal}{Phys. Rev.}
  \textbf{\bibinfo{volume}{106}}, \bibinfo{pages}{1345}.

\bibitem[{\citenamefont{Chiang and Tabakin}(1997)}]{Chiang:1996em}
\bibinfo{author}{\bibnamefont{Chiang}, \bibfnamefont{W.-T.}}, and
  \bibinfo{author}{\bibfnamefont{F.}~\bibnamefont{Tabakin}},
  \bibinfo{year}{1997}, \bibinfo{journal}{Phys. Rev.}
  \textbf{\bibinfo{volume}{C55}}, \bibinfo{pages}{2054}.

\bibitem[{\citenamefont{Chiang} \emph{et~al.}(2003)\citenamefont{Chiang, Yang,
  Tiator, Vanderhaeghen, and Drechsel}}]{Chiang:2002vq}
\bibinfo{author}{\bibnamefont{Chiang}, \bibfnamefont{W.-T.}},
  \bibinfo{author}{\bibfnamefont{S.~N.} \bibnamefont{Yang}},
  \bibinfo{author}{\bibfnamefont{L.}~\bibnamefont{Tiator}},
  \bibinfo{author}{\bibfnamefont{M.}~\bibnamefont{Vanderhaeghen}}, and
  \bibinfo{author}{\bibfnamefont{D.}~\bibnamefont{Drechsel}},
  \bibinfo{year}{2003}, \bibinfo{journal}{Phys. Rev.}
  \textbf{\bibinfo{volume}{C68}}, \bibinfo{pages}{045202}.

\bibitem[{Chistov \emph{et~al.}(2006)\citenamefont{Chistov}
  \emph{et~al.}}]{Chistov:2006zj}
\bibinfo{author}{\bibnamefont{Chistov}, \bibfnamefont{R.}}, \emph{et~al.},
  \bibinfo{year}{2006}, \bibinfo{journal}{Phys. Rev. Lett.}
  \textbf{\bibinfo{volume}{97}}, \bibinfo{pages}{162001}.

\bibitem[{\citenamefont{Choe}(1998)}]{Choe:1997wz}
\bibinfo{author}{\bibnamefont{Choe}, \bibfnamefont{S.}}, \bibinfo{year}{1998},
  \bibinfo{journal}{Eur. Phys. J.} \textbf{\bibinfo{volume}{A3}},
  \bibinfo{pages}{65}.

\bibitem[{\citenamefont{Chung} \emph{et~al.}(1982)\citenamefont{Chung, Dosch,
  Kremer, and Schall}}]{Chung:1981cc}
\bibinfo{author}{\bibnamefont{Chung}, \bibfnamefont{Y.}},
  \bibinfo{author}{\bibfnamefont{H.~G.} \bibnamefont{Dosch}},
  \bibinfo{author}{\bibfnamefont{M.}~\bibnamefont{Kremer}}, and
  \bibinfo{author}{\bibfnamefont{D.}~\bibnamefont{Schall}},
  \bibinfo{year}{1982}, \bibinfo{journal}{Nucl. Phys.}
  \textbf{\bibinfo{volume}{B197}}, \bibinfo{pages}{55}.

\bibitem[{\citenamefont{Cohen and Glozman}(2002{\natexlab{a}})}]{Cohen:2001gb}
\bibinfo{author}{\bibnamefont{Cohen}, \bibfnamefont{T.~D.}}, and
  \bibinfo{author}{\bibfnamefont{L.~Y.} \bibnamefont{Glozman}},
  \bibinfo{year}{2002}{\natexlab{a}}, \bibinfo{journal}{Phys. Rev.}
  \textbf{\bibinfo{volume}{D65}}, \bibinfo{pages}{016006}.

\bibitem[{\citenamefont{Cohen and Glozman}(2002{\natexlab{b}})}]{Cohen:2002st}
\bibinfo{author}{\bibnamefont{Cohen}, \bibfnamefont{T.~D.}}, and
  \bibinfo{author}{\bibfnamefont{L.~Y.} \bibnamefont{Glozman}},
  \bibinfo{year}{2002}{\natexlab{b}}, \bibinfo{journal}{Int. J. Mod. Phys.}
  \textbf{\bibinfo{volume}{A17}}, \bibinfo{pages}{1327}.

\bibitem[{\citenamefont{Copley} \emph{et~al.}(1979)\citenamefont{Copley, Isgur,
  and Karl}}]{Copley:1979wj}
\bibinfo{author}{\bibnamefont{Copley}, \bibfnamefont{L.~A.}},
  \bibinfo{author}{\bibfnamefont{N.}~\bibnamefont{Isgur}}, and
  \bibinfo{author}{\bibfnamefont{G.}~\bibnamefont{Karl}}, \bibinfo{year}{1979},
  \bibinfo{journal}{Phys. Rev.} \textbf{\bibinfo{volume}{D20}},
  \bibinfo{pages}{768}, \bibinfo{note}{erratum-ibid.D23:817,1981}.

\bibitem[{\citenamefont{Corthals}
  \emph{et~al.}(2007{\natexlab{a}})\citenamefont{Corthals, Ireland,
  Van~Cauteren, and Ryckebusch}}]{Corthals:2006nz}
\bibinfo{author}{\bibnamefont{Corthals}, \bibfnamefont{T.}},
  \bibinfo{author}{\bibfnamefont{D.~G.} \bibnamefont{Ireland}},
  \bibinfo{author}{\bibfnamefont{T.}~\bibnamefont{Van~Cauteren}}, and
  \bibinfo{author}{\bibfnamefont{J.}~\bibnamefont{Ryckebusch}},
  \bibinfo{year}{2007}{\natexlab{a}}, \bibinfo{journal}{Phys. Rev.}
  \textbf{\bibinfo{volume}{C75}}, \bibinfo{pages}{045204}.

\bibitem[{\citenamefont{Corthals} \emph{et~al.}(2006)\citenamefont{Corthals,
  Ryckebusch, and Van~Cauteren}}]{Corthals:2005ce}
\bibinfo{author}{\bibnamefont{Corthals}, \bibfnamefont{T.}},
  \bibinfo{author}{\bibfnamefont{J.}~\bibnamefont{Ryckebusch}}, and
  \bibinfo{author}{\bibfnamefont{T.}~\bibnamefont{Van~Cauteren}},
  \bibinfo{year}{2006}, \bibinfo{journal}{Phys. Rev.}
  \textbf{\bibinfo{volume}{C73}}, \bibinfo{pages}{045207}.

\bibitem[{\citenamefont{Corthals}
  \emph{et~al.}(2007{\natexlab{b}})\citenamefont{Corthals, Van~Cauteren,
  Van~Craeyveld, Ryckebusch, and Ireland}}]{Corthals:2007kc}
\bibinfo{author}{\bibnamefont{Corthals}, \bibfnamefont{T.}},
  \bibinfo{author}{\bibfnamefont{T.}~\bibnamefont{Van~Cauteren}},
  \bibinfo{author}{\bibfnamefont{P.}~\bibnamefont{Van~Craeyveld}},
  \bibinfo{author}{\bibfnamefont{J.}~\bibnamefont{Ryckebusch}}, and
  \bibinfo{author}{\bibfnamefont{D.~G.} \bibnamefont{Ireland}},
  \bibinfo{year}{2007}{\natexlab{b}}, \bibinfo{journal}{Phys. Lett.}
  \textbf{\bibinfo{volume}{B656}}, \bibinfo{pages}{186}.

\bibitem[{\citenamefont{Cottingham}(1963)}]{Cottingham:1963}
\bibinfo{author}{\bibnamefont{Cottingham}, \bibfnamefont{W.~N.}},
  \bibinfo{year}{1963}, \bibinfo{journal}{Ann. Phys. (N.Y.)}
  \textbf{\bibinfo{volume}{25}}, \bibinfo{pages}{424}.

\bibitem[{Cox \emph{et~al.}(1969)\citenamefont{Cox} \emph{et~al.}}]{Cox:1968yg}
\bibinfo{author}{\bibnamefont{Cox}, \bibfnamefont{C.~R.}}, \emph{et~al.},
  \bibinfo{year}{1969}, \bibinfo{journal}{Phys. Rev.}
  \textbf{\bibinfo{volume}{184}}, \bibinfo{pages}{1453}.

\bibitem[{Craig \emph{et~al.}(2003)\citenamefont{Craig}
  \emph{et~al.}}]{Craig:2003yd}
\bibinfo{author}{\bibnamefont{Craig}, \bibfnamefont{K.}}, \emph{et~al.},
  \bibinfo{year}{2003}, \bibinfo{journal}{Phys. Rev. Lett.}
  \textbf{\bibinfo{volume}{91}}, \bibinfo{pages}{102301}.

\bibitem[{\citenamefont{Crede and Meyer}(2009)}]{Crede:2008vw}
\bibinfo{author}{\bibnamefont{Crede}, \bibfnamefont{V.}}, and
  \bibinfo{author}{\bibfnamefont{C.~A.} \bibnamefont{Meyer}},
  \bibinfo{year}{2009}, \bibinfo{journal}{Prog. Part. Nucl. Phys. 63}
  \textbf{\bibinfo{volume}{2009}}, \bibinfo{pages}{74}.

\bibitem[{Crede \emph{et~al.}(2005)\citenamefont{Crede}
  \emph{et~al.}}]{Crede:2003ax}
\bibinfo{author}{\bibnamefont{Crede}, \bibfnamefont{V.}}, \emph{et~al.},
  \bibinfo{year}{2005}, \bibinfo{journal}{Phys. Rev. Lett.}
  \textbf{\bibinfo{volume}{94}}, \bibinfo{pages}{012004}.

\bibitem[{Crede \emph{et~al.}(2009)\citenamefont{Crede}
  \emph{et~al.}}]{Crede:2009zz}
\bibinfo{author}{\bibnamefont{Crede}, \bibfnamefont{V.}}, \emph{et~al.},
  \bibinfo{year}{2009}, \bibinfo{journal}{Phys. Rev.}
  \textbf{\bibinfo{volume}{C 80}}, \bibinfo{pages}{055202}.

\bibitem[{Csorna \emph{et~al.}(2001)\citenamefont{Csorna}
  \emph{et~al.}}]{Csorna:2000hw}
\bibinfo{author}{\bibnamefont{Csorna}, \bibfnamefont{S.~E.}}, \emph{et~al.},
  \bibinfo{year}{2001}, \bibinfo{journal}{Phys. Rev. Lett.}
  \textbf{\bibinfo{volume}{86}}, \bibinfo{pages}{4243}.

\bibitem[{\citenamefont{Cutkosky} \emph{et~al.}(1979)\citenamefont{Cutkosky,
  Forsyth, Hendrick, and Kelly}}]{Cutkosky:1979fy}
\bibinfo{author}{\bibnamefont{Cutkosky}, \bibfnamefont{R.~E.}},
  \bibinfo{author}{\bibfnamefont{C.~P.} \bibnamefont{Forsyth}},
  \bibinfo{author}{\bibfnamefont{R.~E.} \bibnamefont{Hendrick}}, and
  \bibinfo{author}{\bibfnamefont{R.~L.} \bibnamefont{Kelly}},
  \bibinfo{year}{1979}, \bibinfo{journal}{Phys. Rev.}
  \textbf{\bibinfo{volume}{D20}}, \bibinfo{pages}{2804, 2839}.

\bibitem[{\citenamefont{Cutkosky} \emph{et~al.}(1980)\citenamefont{Cutkosky,
  Forsyth, Babcock, Kelly, and Hendrick}}]{Cutkosky:1980rh}
\bibinfo{author}{\bibnamefont{Cutkosky}, \bibfnamefont{R.~E.}},
  \bibinfo{author}{\bibfnamefont{C.~P.} \bibnamefont{Forsyth}},
  \bibinfo{author}{\bibfnamefont{J.~B.} \bibnamefont{Babcock}},
  \bibinfo{author}{\bibfnamefont{R.~L.} \bibnamefont{Kelly}}, and
  \bibinfo{author}{\bibfnamefont{R.~E.} \bibnamefont{Hendrick}},
  \bibinfo{year}{1980}, \bibinfo{note}{presented at 4th Int. Conf. on Baryon
  Resonances, Toronto, Canada, Jul 14-16, 1980}.

\bibitem[{\citenamefont{Dalitz and Tuan}(1959)}]{Dalitz:1959dn}
\bibinfo{author}{\bibnamefont{Dalitz}, \bibfnamefont{R.~H.}}, and
  \bibinfo{author}{\bibfnamefont{S.~F.} \bibnamefont{Tuan}},
  \bibinfo{year}{1959}, \bibinfo{journal}{Phys. Rev. Lett.}
  \textbf{\bibinfo{volume}{2}}, \bibinfo{pages}{425}.

\bibitem[{\citenamefont{Dalitz and Tuan}(1960)}]{Dalitz:1960du}
\bibinfo{author}{\bibnamefont{Dalitz}, \bibfnamefont{R.~H.}}, and
  \bibinfo{author}{\bibfnamefont{S.~F.} \bibnamefont{Tuan}},
  \bibinfo{year}{1960}, \bibinfo{journal}{Annals Phys.}
  \textbf{\bibinfo{volume}{10}}, \bibinfo{pages}{307}.

\bibitem[{\citenamefont{Dalitz} \emph{et~al.}(1967)\citenamefont{Dalitz, Wong,
  and Rajasekaran}}]{Dalitz:1967fp}
\bibinfo{author}{\bibnamefont{Dalitz}, \bibfnamefont{R.~H.}},
  \bibinfo{author}{\bibfnamefont{T.~C.} \bibnamefont{Wong}}, and
  \bibinfo{author}{\bibfnamefont{G.}~\bibnamefont{Rajasekaran}},
  \bibinfo{year}{1967}, \bibinfo{journal}{Phys. Rev.}
  \textbf{\bibinfo{volume}{153}}, \bibinfo{pages}{1617}.

\bibitem[{Dalton \emph{et~al.}(2008)\citenamefont{Dalton}
  \emph{et~al.}}]{Dalton:2008ff}
\bibinfo{author}{\bibnamefont{Dalton}, \bibfnamefont{M.}}, \emph{et~al.},
  \bibinfo{year}{2008}, \eprint{0804.3509}.

\bibitem[{\citenamefont{Danilov and Mizuk}(2008)}]{Danilov:2008zz}
\bibinfo{author}{\bibnamefont{Danilov}, \bibfnamefont{M.~V.}}, and
  \bibinfo{author}{\bibfnamefont{R.~V.} \bibnamefont{Mizuk}},
  \bibinfo{year}{2008}, \bibinfo{journal}{Phys. Atom. Nucl.}
  \textbf{\bibinfo{volume}{71}}, \bibinfo{pages}{605}.

\bibitem[{Dannhausen \emph{et~al.}(2001)\citenamefont{Dannhausen}
  \emph{et~al.}}]{Dannhausen:2001yz}
\bibinfo{author}{\bibnamefont{Dannhausen}, \bibfnamefont{H.~W.}},
  \emph{et~al.}, \bibinfo{year}{2001}, \bibinfo{journal}{Eur. Phys. J.}
  \textbf{\bibinfo{volume}{A11}}, \bibinfo{pages}{441}.

\bibitem[{De~Masi \emph{et~al.}(2008)\citenamefont{De~Masi}
  \emph{et~al.}}]{DeMasi:2007id}
\bibinfo{author}{\bibnamefont{De~Masi}, \bibfnamefont{R.}}, \emph{et~al.},
  \bibinfo{year}{2008}, \bibinfo{journal}{Phys. Rev.}
  \textbf{\bibinfo{volume}{C77}}, \bibinfo{pages}{042201}.

\bibitem[{\citenamefont{De~Rujula} \emph{et~al.}(1975)\citenamefont{De~R{\'u}jula,
  Georgi, and Glashow}}]{DeRujula:1975ge}
\bibinfo{author}{\bibnamefont{De~R{\'u}jula}, \bibfnamefont{A.}},
  \bibinfo{author}{\bibfnamefont{H.}~\bibnamefont{Georgi}}, and
  \bibinfo{author}{\bibfnamefont{S.~L.} \bibnamefont{Glashow}},
  \bibinfo{year}{1975}, \bibinfo{journal}{Phys. Rev.}
  \textbf{\bibinfo{volume}{D12}}, \bibinfo{pages}{147}.

\bibitem[{\citenamefont{DeGrand} \emph{et~al.}(1975)\citenamefont{DeGrand,
  Jaffe, Johnson, and Kiskis}}]{DeGrand:1975cf}
\bibinfo{author}{\bibnamefont{DeGrand}, \bibfnamefont{T.~A.}},
  \bibinfo{author}{\bibfnamefont{R.~L.} \bibnamefont{Jaffe}},
  \bibinfo{author}{\bibfnamefont{K.}~\bibnamefont{Johnson}}, and
  \bibinfo{author}{\bibfnamefont{J.~E.} \bibnamefont{Kiskis}},
  \bibinfo{year}{1975}, \bibinfo{journal}{Phys. Rev.}
  \textbf{\bibinfo{volume}{D12}}, \bibinfo{pages}{2060}.

\bibitem[{Denizli \emph{et~al.}(2007)\citenamefont{Denizli}
  \emph{et~al.}}]{Denizli:2007tq}
\bibinfo{author}{\bibnamefont{Denizli}, \bibfnamefont{H.}}, \emph{et~al.},
  \bibinfo{year}{2007}, \bibinfo{journal}{Phys. Rev.}
  \textbf{\bibinfo{volume}{C76}}, \bibinfo{pages}{015204}.

\bibitem[{DiSalvo \emph{et~al.}(2009)\citenamefont{DiSalvo}
  \emph{et~al.}}]{DiSalvo:2009zz}
\bibinfo{author}{\bibnamefont{DiSalvo}, \bibfnamefont{R.}}, \emph{et~al.},
  \bibinfo{year}{2009}, \bibinfo{journal}{Eur. Rev. J.}
  \textbf{\bibinfo{volume}{A42}}, \bibinfo{pages}{151}.

\bibitem[{\citenamefont{Diakonov} \emph{et~al.}(1997)\citenamefont{Diakonov,
  Petrov, and Polyakov}}]{Diakonov:1997mm}
\bibinfo{author}{\bibnamefont{Diakonov}, \bibfnamefont{D.}},
  \bibinfo{author}{\bibfnamefont{V.}~\bibnamefont{Petrov}}, and
  \bibinfo{author}{\bibfnamefont{M.~V.} \bibnamefont{Polyakov}},
  \bibinfo{year}{1997}, \bibinfo{journal}{Z. Phys.}
  \textbf{\bibinfo{volume}{A359}}, \bibinfo{pages}{305}.

\bibitem[{\citenamefont{Donoghue} \emph{et~al.}(1989)\citenamefont{Donoghue,
  Ramirez, and Valencia}}]{Donoghue:1988ed}
\bibinfo{author}{\bibnamefont{Donoghue}, \bibfnamefont{J.~F.}},
  \bibinfo{author}{\bibfnamefont{C.}~\bibnamefont{Ramirez}}, and
  \bibinfo{author}{\bibfnamefont{G.}~\bibnamefont{Valencia}},
  \bibinfo{year}{1989}, \bibinfo{journal}{Phys. Rev.}
  \textbf{\bibinfo{volume}{D39}}, \bibinfo{pages}{1947}.

\bibitem[{\citenamefont{\protect{D\"oring}}(2004)}]{Doring:2004kt}
\bibinfo{author}{\bibnamefont{\protect{D\"oring}}, \bibfnamefont{M.}}, and
  \bibinfo{author}{\bibfnamefont{E.} \bibnamefont{Oset}}, and
  \bibinfo{author}{\bibfnamefont{M.~J.} \bibnamefont{Vicente Vacas}},
  \bibinfo{year}{2004}, \bibinfo{journal}{Phys. Rev.}
  \textbf{\bibinfo{volume}{C70}}, \bibinfo{pages}{045203}.

\bibitem[{\citenamefont{\protect{D\"oring}} \emph{et~al.}(2008)\citenamefont{\protect D\"oring, Oset,
  and Zou}}]{Doring:2008sv}
\bibinfo{author}{\bibnamefont{\protect{D\"oring}}, \bibfnamefont{M.}},
  \bibinfo{author}{\bibfnamefont{E.}~\bibnamefont{Oset}}, and
  \bibinfo{author}{\bibfnamefont{B.~S.} \bibnamefont{Zou}},
  \bibinfo{year}{2008}, \bibinfo{journal}{Phys. Rev.}
  \textbf{\bibinfo{volume}{C78}}, \bibinfo{pages}{025207}.

\bibitem[{\citenamefont{\protect{D\"oring}} \emph{et~al.}(2009)\citenamefont{\protect D\"oring,
  and Nakayama}}]{Doring:2009qr}
\bibinfo{author}{\bibnamefont{\protect{D\"oring}}, \bibfnamefont{M.}},
  \bibinfo{author}{\bibfnamefont{K.} \bibnamefont{Nakayama}},
  \bibinfo{year}{2009}, \eprint{arXiv:0909.3538 [nucl-th]}.

\bibitem[{\citenamefont{\protect{D\"oring}} \emph{et~al.}(2009)\citenamefont{\protect D\"oring,
  and Nakayama}}]{Doring:2009uc}
\bibinfo{author}{\bibnamefont{\protect{D\"oring}}, \bibfnamefont{M.}},
  \bibinfo{author}{\bibfnamefont{K.} \bibnamefont{Nakayama}},
  \bibinfo{year}{2009}, \eprint{arXiv:0906.2949 [nucl-th]}.

\bibitem[{\citenamefont{Dosch} \emph{et~al.}(1989)\citenamefont{Dosch, Jamin,
  and Narison}}]{Dosch:1988vv}
\bibinfo{author}{\bibnamefont{Dosch}, \bibfnamefont{H.~G.}},
  \bibinfo{author}{\bibfnamefont{M.}~\bibnamefont{Jamin}}, and
  \bibinfo{author}{\bibfnamefont{S.}~\bibnamefont{Narison}},
  \bibinfo{year}{1989}, \bibinfo{journal}{Phys. Lett.}
  \textbf{\bibinfo{volume}{B220}}, \bibinfo{pages}{251}.

\bibitem[{Drach \emph{et~al.}(2008)\citenamefont{Drach}
  \emph{et~al.}}]{Drach:2009dh}
\bibinfo{author}{\bibnamefont{Drach}, \bibfnamefont{V.}}, \emph{et~al.},
  \bibinfo{year}{2008}, \bibinfo{journal}{PoS}
  \textbf{\bibinfo{volume}{LAT2008}}, \bibinfo{pages}{123}.

\bibitem[{\citenamefont{Drechsel} \emph{et~al.}(2007)\citenamefont{Drechsel,
  Kamalov, and Tiator}}]{Drechsel:2007if}
\bibinfo{author}{\bibnamefont{Drechsel}, \bibfnamefont{D.}},
  \bibinfo{author}{\bibfnamefont{S.~S.} \bibnamefont{Kamalov}}, and
  \bibinfo{author}{\bibfnamefont{L.}~\bibnamefont{Tiator}},
  \bibinfo{year}{2007}, \bibinfo{journal}{Eur. Phys. J.}
  \textbf{\bibinfo{volume}{A34}}, \bibinfo{pages}{69}.

\bibitem[{\citenamefont{Drechsel and Walcher}(2008)}]{Drechsel:2007sq}
\bibinfo{author}{\bibnamefont{Drechsel}, \bibfnamefont{D.}}, and
  \bibinfo{author}{\bibfnamefont{T.}~\bibnamefont{Walcher}},
  \bibinfo{year}{2008}, \bibinfo{journal}{Rev. Mod. Phys.}
  \textbf{\bibinfo{volume}{80}}, \bibinfo{pages}{731}.

\bibitem[{\citenamefont{Drell and Hearn}(1966)}]{Drell:1966jv}
\bibinfo{author}{\bibnamefont{Drell}, \bibfnamefont{S.~D.}}, and
  \bibinfo{author}{\bibfnamefont{A.~C.} \bibnamefont{Hearn}},
  \bibinfo{year}{1966}, \bibinfo{journal}{Phys. Rev. Lett.}
  \textbf{\bibinfo{volume}{16}}, \bibinfo{pages}{908}.

\bibitem[{Dugger \emph{et~al.}(2002)\citenamefont{Dugger}
  \emph{et~al.}}]{Dugger:2002ft}
\bibinfo{author}{\bibnamefont{Dugger}, \bibfnamefont{M.}}, \emph{et~al.},
  \bibinfo{year}{2002}, \bibinfo{journal}{Phys. Rev. Lett.}
  \textbf{\bibinfo{volume}{89}}, \bibinfo{pages}{222002}.

\bibitem[{Dugger \emph{et~al.}(2006)\citenamefont{Dugger}
  \emph{et~al.}}]{Dugger:2005my}
\bibinfo{author}{\bibnamefont{Dugger}, \bibfnamefont{M.}}, \emph{et~al.},
  \bibinfo{year}{2006}, \bibinfo{journal}{Phys. Rev. Lett.}
  \textbf{\bibinfo{volume}{96}}, \bibinfo{pages}{062001}.

\bibitem[{Dugger \emph{et~al.}(2007)\citenamefont{Dugger}
  \emph{et~al.}}]{Dugger:2007bt}
\bibinfo{author}{\bibnamefont{Dugger}, \bibfnamefont{M.}}, \emph{et~al.},
  \bibinfo{year}{2007}, \bibinfo{journal}{Phys. Rev.}
  \textbf{\bibinfo{volume}{C76}}, \bibinfo{pages}{025211}.

\bibitem[{Dugger \emph{et~al.}(2009)\citenamefont{Dugger}
  \emph{et~al.}}]{Dugger:2009pn}
\bibinfo{author}{\bibnamefont{Dugger}, \bibfnamefont{M.}}, \emph{et~al.},
  \bibinfo{year}{2009},
  \bibinfo{journal}{Phys. Rev.} \textbf{\bibinfo{volume}{C79}},
  \bibinfo{pages}{065206}.

\bibitem[{\citenamefont{Durand} \emph{et~al.}(2008)\citenamefont{Durand,
  Julia-Diaz, Lee, Saghai, and Sato}}]{Durand:2008es}
\bibinfo{author}{\bibnamefont{Durand}, \bibfnamefont{J.}},
  \bibinfo{author}{\bibfnamefont{B.}~\bibnamefont{Julia-Diaz}},
  \bibinfo{author}{\bibfnamefont{T.~S.~H.} \bibnamefont{Lee}},
  \bibinfo{author}{\bibfnamefont{B.}~\bibnamefont{Saghai}}, and
  \bibinfo{author}{\bibfnamefont{T.}~\bibnamefont{Sato}}, \bibinfo{year}{2008},
  \bibinfo{journal}{Phys. Rev.} \textbf{\bibinfo{volume}{C78}},
  \bibinfo{pages}{025204}.

\bibitem[{{\protect{D\"urr}} \emph{et~al.}(2008)\citenamefont{{\protect D\"urr}}
  \emph{et~al.}}]{Durr:2008zz}
\bibinfo{author}{\bibnamefont{\protect{D\"urr}}, \bibfnamefont{S.}},
  \emph{et~al.}, \bibinfo{year}{2008}, \bibinfo{journal}{Science}
  \textbf{\bibinfo{volume}{322}}, \bibinfo{pages}{1224}.

\bibitem[{Dutz \emph{et~al.}(2003)\citenamefont{Dutz}
  \emph{et~al.}}]{Dutz:2003mm}
\bibinfo{author}{\bibnamefont{Dutz}, \bibfnamefont{H.}}, \emph{et~al.},
  \bibinfo{year}{2003}, \bibinfo{journal}{Phys. Rev. Lett.}
  \textbf{\bibinfo{volume}{91}}, \bibinfo{pages}{192001}.

\bibitem[{\citenamefont{Dziembowski}
  \emph{et~al.}(1996)\citenamefont{Dziembowski, Fabre de~la Ripelle, and
  Miller}}]{Dziembowski:1996cv}
\bibinfo{author}{\bibnamefont{Dziembowski}, \bibfnamefont{Z.}},
  \bibinfo{author}{\bibfnamefont{M.}~\bibnamefont{Fabre de~la Ripelle}}, and
  \bibinfo{author}{\bibfnamefont{G.~A.} \bibnamefont{Miller}},
  \bibinfo{year}{1996}, \bibinfo{journal}{Phys. Rev.}
  \textbf{\bibinfo{volume}{C53}}, \bibinfo{pages}{2038}.

\bibitem[{\citenamefont{Dzierba} \emph{et~al.}(2005)\citenamefont{Dzierba,
  Meyer, and Szczepaniak}}]{Dzierba:2004db}
\bibinfo{author}{\bibnamefont{Dzierba}, \bibfnamefont{A.~R.}},
  \bibinfo{author}{\bibfnamefont{C.~A.} \bibnamefont{Meyer}}, and
  \bibinfo{author}{\bibfnamefont{A.~P.} \bibnamefont{Szczepaniak}},
  \bibinfo{year}{2005}, \bibinfo{journal}{J. Phys. Conf. Ser.}
  \textbf{\bibinfo{volume}{9}}, \bibinfo{pages}{192}.

\bibitem[{\citenamefont{Ebert} \emph{et~al.}(2008)\citenamefont{Ebert, Faustov,
  and Galkin}}]{Ebert:2007nw}
\bibinfo{author}{\bibnamefont{Ebert}, \bibfnamefont{D.}},
  \bibinfo{author}{\bibfnamefont{R.~N.} \bibnamefont{Faustov}}, and
  \bibinfo{author}{\bibfnamefont{V.~O.} \bibnamefont{Galkin}},
  \bibinfo{year}{2008}, \bibinfo{journal}{Phys. Lett.}
  \textbf{\bibinfo{volume}{B659}}, \bibinfo{pages}{612}.

\bibitem[{\citenamefont{Ecker} \emph{et~al.}(1989)\citenamefont{Ecker, Gasser,
  Pich, and de~Rafael}}]{Ecker:1988te}
\bibinfo{author}{\bibnamefont{Ecker}, \bibfnamefont{G.}},
  \bibinfo{author}{\bibfnamefont{J.}~\bibnamefont{Gasser}},
  \bibinfo{author}{\bibfnamefont{A.}~\bibnamefont{Pich}}, and
  \bibinfo{author}{\bibfnamefont{E.}~\bibnamefont{de~Rafael}},
  \bibinfo{year}{1989}, \bibinfo{journal}{Nucl. Phys.}
  \textbf{\bibinfo{volume}{B321}}, \bibinfo{pages}{311}.

\bibitem[{Edwards \emph{et~al.}(1995)\citenamefont{Edwards}
  \emph{et~al.}}]{Edwards:1994ar}
\bibinfo{author}{\bibnamefont{Edwards}, \bibfnamefont{K.~W.}}, \emph{et~al.},
  \bibinfo{year}{1995}, \bibinfo{journal}{Phys. Rev. Lett.}
  \textbf{\bibinfo{volume}{74}}, \bibinfo{pages}{3331}.

\bibitem[{Egiyan \emph{et~al.}(2006)\citenamefont{Egiyan}
  \emph{et~al.}}]{Egiyan:2006ks}
\bibinfo{author}{\bibnamefont{Egiyan}, \bibfnamefont{H.}}, \emph{et~al.},
  \bibinfo{year}{2006}, \bibinfo{journal}{Phys. Rev.}
  \textbf{\bibinfo{volume}{C73}}, \bibinfo{pages}{025204}.

\bibitem[{Elsner \emph{et~al.}(2006)\citenamefont{Elsner}
  \emph{et~al.}}]{Elsner:2005cz}
\bibinfo{author}{\bibnamefont{Elsner}, \bibfnamefont{D.}}, \emph{et~al.},
  \bibinfo{year}{2006}, \bibinfo{journal}{Eur. Phys. J.}
  \textbf{\bibinfo{volume}{A27}}, \bibinfo{pages}{91}.

\bibitem[{Elsner \emph{et~al.}(2007)\citenamefont{Elsner}
  \emph{et~al.}}]{Elsner:2007hm}
\bibinfo{author}{\bibnamefont{Elsner}, \bibfnamefont{D.}}, \emph{et~al.},
  \bibinfo{year}{2007}, \bibinfo{journal}{Eur. Phys. J.}
  \textbf{\bibinfo{volume}{A33}}, \bibinfo{pages}{147}.

\bibitem[{Elsner \emph{et~al.}(2009)\citenamefont{Elsner}
  \emph{et~al.}}]{Elsner:2008sn}
\bibinfo{author}{\bibnamefont{Elsner}, \bibfnamefont{D.}}, \emph{et~al.},
  \bibinfo{year}{2009}, \bibinfo{journal}{Eur. Phys. J.}
  \textbf{\bibinfo{volume}{A39}}, \bibinfo{pages}{373}.

\bibitem[{Engelfried \emph{et~al.}(1998)\citenamefont{Engelfried}
  \emph{et~al.}}]{Engelfried:1997rp}
\bibinfo{author}{\bibnamefont{Engelfried}, \bibfnamefont{J.}}, \emph{et~al.},
  \bibinfo{year}{1998}, \bibinfo{journal}{Nucl. Instrum. Meth.}
  \textbf{\bibinfo{volume}{A409}}, \bibinfo{pages}{439}.

\bibitem[{Erbe \emph{et~al.}(1968)\citenamefont{Erbe}
  \emph{et~al.}}]{Erbe:1968ke}
\bibinfo{author}{\bibnamefont{Erbe}, \bibfnamefont{R.}}, \emph{et~al.},
  \bibinfo{year}{1968}, \bibinfo{journal}{Phys. Rev.}
  \textbf{\bibinfo{volume}{175}}, \bibinfo{pages}{1669}.

\bibitem[{Fantini \emph{et~al.}(2008)\citenamefont{Fantini}
  \emph{et~al.}}]{Fantini:2008zz}
\bibinfo{author}{\bibnamefont{Fantini}, \bibfnamefont{A.}}, \emph{et~al.},
  \bibinfo{year}{2008}, \bibinfo{journal}{Phys. Rev.}
  \textbf{\bibinfo{volume}{C78}}, \bibinfo{pages}{015203}.

\bibitem[{Fedotov \emph{et~al.}(2009)\citenamefont{Fedotov}
  \emph{et~al.}}]{Fedotov:2008gy}
\bibinfo{author}{\bibnamefont{Fedotov}, \bibfnamefont{G.}}, \emph{et~al.},
  \bibinfo{year}{2009}, \bibinfo{journal}{Phys. Rev.}
  \textbf{\bibinfo{volume}{C79}}, \bibinfo{pages}{015204}.

\bibitem[{\citenamefont{Feuster and Mosel}(1998)}]{Feuster:1997pq}
\bibinfo{author}{\bibnamefont{Feuster}, \bibfnamefont{T.}}, and
  \bibinfo{author}{\bibfnamefont{U.} \bibnamefont{Mosel}},
  \bibinfo{year}{1998}, \bibinfo{journal}{Phys. Rev.}
  \textbf{\bibinfo{volume}{C58}}, \bibinfo{pages}{457}.

\bibitem[{\citenamefont{Feuster and Mosel}(1999)}]{Feuster:1998cj}
\bibinfo{author}{\bibnamefont{Feuster}, \bibfnamefont{T.}}, and
  \bibinfo{author}{\bibfnamefont{U.} \bibnamefont{Mosel}},
  \bibinfo{year}{1998}, \bibinfo{journal}{Phys. Rev.}
  \textbf{\bibinfo{volume}{C59}}, \bibinfo{pages}{460}.

\bibitem[{\citenamefont{Fleck and Richard}(1989)}]{Fleck:1989mb}
\bibinfo{author}{\bibnamefont{Fleck}, \bibfnamefont{S.}}, and
  \bibinfo{author}{\bibfnamefont{J.~M.} \bibnamefont{Richard}},
  \bibinfo{year}{1989}, \bibinfo{journal}{Prog. Theor. Phys.}
  \textbf{\bibinfo{volume}{82}}, \bibinfo{pages}{760}.

\bibitem[{\citenamefont{Fleck and Richard}(1990)}]{Fleck:1990ma}
\bibinfo{author}{\bibnamefont{Fleck}, \bibfnamefont{S.}}, and
  \bibinfo{author}{\bibfnamefont{J.~M.} \bibnamefont{Richard}},
  \bibinfo{year}{1990}, \bibinfo{journal}{Part. World}
  \textbf{\bibinfo{volume}{1}}, \bibinfo{pages}{67}.

\bibitem[{\citenamefont{Flynn} \emph{et~al.}(2003)\citenamefont{Flynn, Mescia,
  and Tariq}}]{Flynn:2003vz}
\bibinfo{author}{\bibnamefont{Flynn}, \bibfnamefont{J.~M.}},
  \bibinfo{author}{\bibfnamefont{F.}~\bibnamefont{Mescia}}, and
  \bibinfo{author}{\bibfnamefont{A.~S.~B.} \bibnamefont{Tariq}}, \bibinfo{year}{2003},
  \bibinfo{journal}{JHEP} \textbf{\bibinfo{volume}{07}}, \bibinfo{pages}{066}.

\bibitem[{\citenamefont{Forkel}
  \emph{et~al.}(2007{\natexlab{a}})\citenamefont{Forkel, Beyer, and
  Frederico}}]{Forkel:2007tz}
\bibinfo{author}{\bibnamefont{Forkel}, \bibfnamefont{H.}},
  \bibinfo{author}{\bibfnamefont{M.}~\bibnamefont{Beyer}}, and
  \bibinfo{author}{\bibfnamefont{T.}~\bibnamefont{Frederico}},
  \bibinfo{year}{2007}{\natexlab{a}}, \bibinfo{journal}{Int. J. Mod. Phys.}
  \textbf{\bibinfo{volume}{E16}}, \bibinfo{pages}{2794}.

\bibitem[{\citenamefont{Forkel}
  \emph{et~al.}(2007{\natexlab{b}})\citenamefont{Forkel, Beyer, and
  Frederico}}]{Forkel:2007cm}
\bibinfo{author}{\bibnamefont{Forkel}, \bibfnamefont{H.}},
  \bibinfo{author}{\bibfnamefont{M.}~\bibnamefont{Beyer}}, and
  \bibinfo{author}{\bibfnamefont{T.}~\bibnamefont{Frederico}},
  \bibinfo{year}{2007}{\natexlab{b}}, \bibinfo{journal}{JHEP}
  \textbf{\bibinfo{volume}{07}}, \bibinfo{pages}{077}.

\bibitem[{\citenamefont{Forkel and Klempt}(2009)}]{Forkel:2008un}
\bibinfo{author}{\bibnamefont{Forkel}, \bibfnamefont{H.}}, and
  \bibinfo{author}{\bibfnamefont{E.}~\bibnamefont{Klempt}},
  \bibinfo{year}{2009}, \bibinfo{journal}{Phys. Lett.}
  \textbf{\bibinfo{volume}{B679}}, \bibinfo{pages}{77}.

\bibitem[{Frabetti \emph{et~al.}(1994)\citenamefont{Frabetti}
  \emph{et~al.}}]{Frabetti:1993hg}
\bibinfo{author}{\bibnamefont{Frabetti}, \bibfnamefont{P.~L.}}, \emph{et~al.},
  \bibinfo{year}{1994}, \bibinfo{journal}{Phys. Rev. Lett.}
  \textbf{\bibinfo{volume}{72}}, \bibinfo{pages}{961}.

\bibitem[{Frabetti \emph{et~al.}(1996)\citenamefont{Frabetti}
  \emph{et~al.}}]{Frabetti:1995sb}
\bibinfo{author}{\bibnamefont{Frabetti}, \bibfnamefont{P.~L.}}, \emph{et~al.},
  \bibinfo{year}{1996}, \bibinfo{journal}{Phys. Lett.}
  \textbf{\bibinfo{volume}{B365}}, \bibinfo{pages}{461}.

\bibitem[{\citenamefont{Franklin}(1999)}]{Franklin:1999bg}
\bibinfo{author}{\bibnamefont{Franklin}, \bibfnamefont{J.}},
  \bibinfo{year}{1999}, \bibinfo{journal}{Phys. Rev.}
  \textbf{\bibinfo{volume}{D59}}, \bibinfo{pages}{117502}.

\bibitem[{\citenamefont{Franklin}(2008)}]{Franklin:2008hx}
\bibinfo{author}{\bibnamefont{Franklin}, \bibfnamefont{J.}},
  \bibinfo{year}{2008}, \eprint{0811.2143}.

\bibitem[{Frolov \emph{et~al.}(1999)\citenamefont{Frolov}
  \emph{et~al.}}]{Frolov:1998pw}
\bibinfo{author}{\bibnamefont{Frolov}, \bibfnamefont{V.~V.}}, \emph{et~al.},
  \bibinfo{year}{1999}, \bibinfo{journal}{Phys. Rev. Lett.}
  \textbf{\bibinfo{volume}{82}}, \bibinfo{pages}{45}.

\bibitem[{Gabler \emph{et~al.}(1994)\citenamefont{Gabler}
  \emph{et~al.}}]{Gabler:1994ay}
\bibinfo{author}{\bibnamefont{Gabler}, \bibfnamefont{A.~R.}}, \emph{et~al.},
  \bibinfo{year}{1994}, \bibinfo{journal}{Nucl. Instrum. Meth.}
  \textbf{\bibinfo{volume}{A346}}, \bibinfo{pages}{168}.

\bibitem[{\citenamefont{Gaillard} \emph{et~al.}(1975)\citenamefont{Gaillard,
  Lee, and Rosner}}]{Gaillard:1974mw}
\bibinfo{author}{\bibnamefont{Gaillard}, \bibfnamefont{M.~K.}},
  \bibinfo{author}{\bibfnamefont{B.~W.} \bibnamefont{Lee}}, and
  \bibinfo{author}{\bibfnamefont{J.~L.} \bibnamefont{Rosner}},
  \bibinfo{year}{1975}, \bibinfo{journal}{Rev. Mod. Phys.}
  \textbf{\bibinfo{volume}{47}}, \bibinfo{pages}{277}.

\bibitem[{Gaiser \emph{et~al.}(1986)\citenamefont{Gaiser}
  \emph{et~al.}}]{Gaiser:1985ix}
\bibinfo{author}{\bibnamefont{Gaiser}, \bibfnamefont{J.}}, \emph{et~al.},
  \bibinfo{year}{1986}, \bibinfo{journal}{Phys. Rev.}
  \textbf{\bibinfo{volume}{D34}}, \bibinfo{pages}{711}.

\bibitem[{\citenamefont{Garcia-Recio}
  \emph{et~al.}(2003)\citenamefont{Garcia-Recio, Nieves, Ruiz~Arriola, and
  Vicente~Vacas}}]{GarciaRecio:2002td}
\bibinfo{author}{\bibnamefont{Garcia-Recio}, \bibfnamefont{C.}},
  \bibinfo{author}{\bibfnamefont{J.}~\bibnamefont{Nieves}},
  \bibinfo{author}{\bibfnamefont{E.}~\bibnamefont{Ruiz~Arriola}}, and
  \bibinfo{author}{\bibfnamefont{M.~J.} \bibnamefont{Vicente~Vacas}},
  \bibinfo{year}{2003}, \bibinfo{journal}{Phys. Rev.}
  \textbf{\bibinfo{volume}{D67}}, \bibinfo{pages}{076009}.

\bibitem[{\citenamefont{Garcilazo} \emph{et~al.}(2007)\citenamefont{Garcilazo,
  Vijande, and Valcarce}}]{Garcilazo:2007eh}
\bibinfo{author}{\bibnamefont{Garcilazo}, \bibfnamefont{H.}},
  \bibinfo{author}{\bibfnamefont{J.}~\bibnamefont{Vijande}}, and
  \bibinfo{author}{\bibfnamefont{A.}~\bibnamefont{Valcarce}},
  \bibinfo{year}{2007}, \bibinfo{journal}{J. Phys.}
  \textbf{\bibinfo{volume}{G34}}, \bibinfo{pages}{961}.

\bibitem[{\citenamefont{Gasser and Leutwyler}(1984)}]{Gasser:1983yg}
\bibinfo{author}{\bibnamefont{Gasser}, \bibfnamefont{J.}}, and
  \bibinfo{author}{\bibfnamefont{H.}~\bibnamefont{Leutwyler}},
  \bibinfo{year}{1984}, \bibinfo{journal}{Ann. Phys.}
  \textbf{\bibinfo{volume}{158}}, \bibinfo{pages}{142}.

\bibitem[{\citenamefont{Gasser and Leutwyler}(1985)}]{Gasser:1984gg}
\bibinfo{author}{\bibnamefont{Gasser}, \bibfnamefont{J.}}, and
  \bibinfo{author}{\bibfnamefont{H.}~\bibnamefont{Leutwyler}},
  \bibinfo{year}{1985}, \bibinfo{journal}{Nucl. Phys.}
  \textbf{\bibinfo{volume}{B250}}, \bibinfo{pages}{465}.

\bibitem[{\citenamefont{Gell-Mann}(1962)}]{GellMann:1962xb}
\bibinfo{author}{\bibnamefont{Gell-Mann}, \bibfnamefont{M.}},
  \bibinfo{year}{1962}, \bibinfo{journal}{Phys. Rev.}
  \textbf{\bibinfo{volume}{125}}, \bibinfo{pages}{1067}.

\bibitem[{\citenamefont{Gell-Mann and Ne'eman}(1964)}]{Gell-Mann:101798}
\bibinfo{author}{\bibnamefont{Gell-Mann}, \bibfnamefont{M.}}, and
  \bibinfo{author}{\bibfnamefont{Y.}~\bibnamefont{Ne'eman}},
  \bibinfo{year}{1964}, \emph{\bibinfo{title}{The eightfold way}}, Frontiers in
  Physics (\bibinfo{publisher}{Benjamin}, \bibinfo{address}{New York, NY}).

\bibitem[{\citenamefont{Geng} \emph{et~al.}(2009)\citenamefont{Geng, Oset, Zou,
  and Doring}}]{Geng:2008cv}
\bibinfo{author}{\bibnamefont{Geng}, \bibfnamefont{L.~S.}},
  \bibinfo{author}{\bibfnamefont{E.}~\bibnamefont{Oset}},
  \bibinfo{author}{\bibfnamefont{B.~S.} \bibnamefont{Zou}}, and
  \bibinfo{author}{\bibfnamefont{M.}~\bibnamefont{Doring}},
  \bibinfo{year}{2009}, \bibinfo{journal}{Phys. Rev.}
  \textbf{\bibinfo{volume}{C79}}, \bibinfo{pages}{025203}.

\bibitem[{\citenamefont{Gerasimov}(1966)}]{Gerasimov:1965et}
\bibinfo{author}{\bibnamefont{Gerasimov}, \bibfnamefont{S.~B.}},
  \bibinfo{year}{1966}, \bibinfo{journal}{Sov. J. Nucl. Phys.}
  \textbf{\bibinfo{volume}{2}}, \bibinfo{pages}{430}, \bibinfo{note}{[Yad.\
  Fiz.\ {\bf 2} (1966) 598].}

\bibitem[{\citenamefont{Gerasyuta and Kochkin}(2002)}]{Gerasyuta:2002hg}
\bibinfo{author}{\bibnamefont{Gerasyuta}, \bibfnamefont{S.~M.}}, and
  \bibinfo{author}{\bibfnamefont{V.~I.} \bibnamefont{Kochkin}},
  \bibinfo{year}{2002}, \bibinfo{journal}{Phys. Rev.}
  \textbf{\bibinfo{volume}{D66}}, \bibinfo{pages}{116001}.

\bibitem[{\citenamefont{Gignoux} \emph{et~al.}(1987)\citenamefont{Gignoux,
  Silvestre-Brac, and Richard}}]{Gignoux:1987cn}
\bibinfo{author}{\bibnamefont{Gignoux}, \bibfnamefont{C.}},
  \bibinfo{author}{\bibfnamefont{B.}~\bibnamefont{Silvestre-Brac}}, and
  \bibinfo{author}{\bibfnamefont{J.~M.} \bibnamefont{Richard}},
  \bibinfo{year}{1987}, \bibinfo{journal}{Phys. Lett.}
  \textbf{\bibinfo{volume}{B193}}, \bibinfo{pages}{323}.

\bibitem[{Glander \emph{et~al.}(2004)\citenamefont{Glander}
  \emph{et~al.}}]{Glander:2003jw}
\bibinfo{author}{\bibnamefont{Glander}, \bibfnamefont{K.~H.}}, \emph{et~al.},
  \bibinfo{year}{2004}, \bibinfo{journal}{Eur. Phys. J.}
  \textbf{\bibinfo{volume}{A19}}, \bibinfo{pages}{251}.

\bibitem[{\citenamefont{Glozman}(2000)}]{Glozman:1999tk}
\bibinfo{author}{\bibnamefont{Glozman}, \bibfnamefont{L.~Y.}},
  \bibinfo{year}{2000}, \bibinfo{journal}{Phys. Lett.}
  \textbf{\bibinfo{volume}{B475}}, \bibinfo{pages}{329}.

\bibitem[{\citenamefont{Glozman}(2007)}]{Glozman:2007ek}
\bibinfo{author}{\bibnamefont{Glozman}, \bibfnamefont{L.~Y.}},
  \bibinfo{year}{2007}, \bibinfo{journal}{Phys. Rept.}
  \textbf{\bibinfo{volume}{444}}, \bibinfo{pages}{1}.

\bibitem[{\citenamefont{Glozman}(2009)}]{Glozman:2009fj}
\bibinfo{author}{\bibnamefont{Glozman}, \bibfnamefont{L.~Y.}},
  \bibinfo{year}{2009}, \eprint{0903.3923}.

\bibitem[{\citenamefont{Glozman and Riska}(1996)}]{Glozman:1995fu}
\bibinfo{author}{\bibnamefont{Glozman}, \bibfnamefont{L.~Y.}}, and
  \bibinfo{author}{\bibfnamefont{D.~O.} \bibnamefont{Riska}},
  \bibinfo{year}{1996}, \bibinfo{journal}{Phys. Rept.}
  \textbf{\bibinfo{volume}{268}}, \bibinfo{pages}{263}.

\bibitem[{\citenamefont{Godfrey and Isgur}(1985)}]{Godfrey:1985xj}
\bibinfo{author}{\bibnamefont{Godfrey}, \bibfnamefont{S.}}, and
  \bibinfo{author}{\bibfnamefont{N.}~\bibnamefont{Isgur}},
  \bibinfo{year}{1985}, \bibinfo{journal}{Phys. Rev.}
  \textbf{\bibinfo{volume}{D32}}, \bibinfo{pages}{189}.

\bibitem[{Goldhaber \emph{et~al.}(1976)\citenamefont{Goldhaber}
  \emph{et~al.}}]{Goldhaber:1976xn}
\bibinfo{author}{\bibnamefont{Goldhaber}, \bibfnamefont{G.}}, \emph{et~al.},
  \bibinfo{year}{1976}, \bibinfo{journal}{Phys. Rev. Lett.}
  \textbf{\bibinfo{volume}{37}}, \bibinfo{pages}{255}.

\bibitem[{\citenamefont{Goloskokov}(2008)}]{Goloskokov:2007fd}
\bibinfo{author}{\bibnamefont{Goloskokov}, \bibfnamefont{S.~V.}},
  \bibinfo{year}{2008}, \bibinfo{journal}{Eur. Phys. J. ST}
  \textbf{\bibinfo{volume}{162}}, \bibinfo{pages}{25}.

\bibitem[{\citenamefont{Golowich} \emph{et~al.}(1983)\citenamefont{Golowich,
  Haqq, and Karl}}]{Golowich:1982kx}
\bibinfo{author}{\bibnamefont{Golowich}, \bibfnamefont{E.}},
  \bibinfo{author}{\bibfnamefont{E.}~\bibnamefont{Haqq}}, and
  \bibinfo{author}{\bibfnamefont{G.}~\bibnamefont{Karl}}, \bibinfo{year}{1983},
  \bibinfo{journal}{Phys. Rev.} \textbf{\bibinfo{volume}{D28}},
  \bibinfo{pages}{160}.

\bibitem[{\citenamefont{Gomez~Tejedor and Oset}(1996)}]{GomezTejedor:1995pe}
\bibinfo{author}{\bibnamefont{Gomez~Tejedor}, \bibfnamefont{J.~A.}}, and
  \bibinfo{author}{\bibfnamefont{E.}~\bibnamefont{Oset}}, \bibinfo{year}{1996},
  \bibinfo{journal}{Nucl. Phys.} \textbf{\bibinfo{volume}{A600}},
  \bibinfo{pages}{413}.

\bibitem[{\citenamefont{Gomshi~Nobary and
  Sepahvand}(2007)}]{GomshiNobary:2007xk}
\bibinfo{author}{\bibnamefont{Gomshi~Nobary}, \bibfnamefont{M.~A.}}, and
  \bibinfo{author}{\bibfnamefont{R.}~\bibnamefont{Sepahvand}},
  \bibinfo{year}{2007}, \bibinfo{journal}{Phys. Rev.}
  \textbf{\bibinfo{volume}{D76}}, \bibinfo{pages}{114006}.

\bibitem[{\citenamefont{Gonzalez} \emph{et~al.}(2009)\citenamefont{Gonzalez,
  Oset, and Vijande}}]{Gonzalez:2008pv}
\bibinfo{author}{\bibnamefont{Gonzalez}, \bibfnamefont{P.}},
  \bibinfo{author}{\bibfnamefont{E.}~\bibnamefont{Oset}}, and
  \bibinfo{author}{\bibfnamefont{J.}~\bibnamefont{Vijande}},
  \bibinfo{year}{2009}, \bibinfo{journal}{Phys. Rev.}
  \textbf{\bibinfo{volume}{C79}}, \bibinfo{pages}{025209}.

\bibitem[{\citenamefont{Gopal}(1980)}]{Gopal:1980ur}
\bibinfo{author}{\bibnamefont{Gopal}, \bibfnamefont{G.~P.}},
  \bibinfo{year}{1980}, \bibinfo{note}{invited talk given at 4th Int. Conf. on
  Baryon Resonances, Toronto, Canada, Jul 14-16, 1980}.

\bibitem[{\citenamefont{Greenberg}(1964)}]{Greenberg:1964pe}
\bibinfo{author}{\bibnamefont{Greenberg}, \bibfnamefont{O.~W.}},
  \bibinfo{year}{1964}, \bibinfo{journal}{Phys. Rev. Lett.}
  \textbf{\bibinfo{volume}{13}}, \bibinfo{pages}{598}.

\bibitem[{\citenamefont{Greenberg and Lipkin}(1981)}]{Greenberg:1981xn}
\bibinfo{author}{\bibnamefont{Greenberg}, \bibfnamefont{O.~W.}}, and
  \bibinfo{author}{\bibfnamefont{H.~J.} \bibnamefont{Lipkin}},
  \bibinfo{year}{1981}, \bibinfo{journal}{Nucl. Phys.}
  \textbf{\bibinfo{volume}{A370}}, \bibinfo{pages}{349}.

\bibitem[{\citenamefont{Guberina} \emph{et~al.}(2000)\citenamefont{Guberina,
  Melic, and Stefancic}}]{Guberina:2000de}
\bibinfo{author}{\bibnamefont{Guberina}, \bibfnamefont{B.}},
  \bibinfo{author}{\bibfnamefont{B.}~\bibnamefont{Melic}}, and
  \bibinfo{author}{\bibfnamefont{H.}~\bibnamefont{Stefancic}},
  \bibinfo{year}{2000}, \bibinfo{journal}{Phys. Lett.}
  \textbf{\bibinfo{volume}{B484}}, \bibinfo{pages}{43}.

\bibitem[{\citenamefont{Guberina} \emph{et~al.}(1986)\citenamefont{Guberina,
  Ruckl, and Trampetic}}]{Guberina:1986gd}
\bibinfo{author}{\bibnamefont{Guberina}, \bibfnamefont{B.}},
  \bibinfo{author}{\bibfnamefont{R.}~\bibnamefont{Ruckl}}, and
  \bibinfo{author}{\bibfnamefont{J.}~\bibnamefont{Trampetic}},
  \bibinfo{year}{1986}, \bibinfo{journal}{Z. Phys.}
  \textbf{\bibinfo{volume}{C33}}, \bibinfo{pages}{297}.

\bibitem[{\citenamefont{Guo} \emph{et~al.}(2008)\citenamefont{Guo, Hanhart, and
  Mei{\ss}ner}}]{Guo:2008ns}
\bibinfo{author}{\bibnamefont{Guo}, \bibfnamefont{F.-K.}},
  \bibinfo{author}{\bibfnamefont{C.}~\bibnamefont{Hanhart}}, and
  \bibinfo{author}{\bibfnamefont{U.-G.} \bibnamefont{Mei{\ss}ner}},
  \bibinfo{year}{2008}, \bibinfo{journal}{JHEP} \textbf{\bibinfo{volume}{09}},
  \bibinfo{pages}{136}.

\bibitem[{Guo \emph{et~al.}(2007)\citenamefont{Guo} \emph{et~al.}}]{Guo:2007dw}
\bibinfo{author}{\bibnamefont{Guo}, \bibfnamefont{L.}}, \emph{et~al.},
  \bibinfo{year}{2007}, \bibinfo{journal}{Phys. Rev.}
  \textbf{\bibinfo{volume}{C76}}, \bibinfo{pages}{025208}.

\bibitem[{Gutz \emph{et~al.}(2008)\citenamefont{Gutz}
  \emph{et~al.}}]{Gutz:2008zz}
\bibinfo{author}{\bibnamefont{Gutz}, \bibfnamefont{E.}}, \emph{et~al.},
  \bibinfo{year}{2008}, \bibinfo{journal}{Eur. Phys. J.}
  \textbf{\bibinfo{volume}{A35}}, \bibinfo{pages}{291}.

\bibitem[{Hadjidakis \emph{et~al.}(2005)\citenamefont{Hadjidakis}
  \emph{et~al.}}]{Hadjidakis:2004zm}
\bibinfo{author}{\bibnamefont{Hadjidakis}, \bibfnamefont{C.}}, \emph{et~al.},
  \bibinfo{year}{2005}, \bibinfo{journal}{Phys. Lett.}
  \textbf{\bibinfo{volume}{B605}}, \bibinfo{pages}{256}.

\bibitem[{\citenamefont{Han and Nambu}(1965)}]{PhysRev.139.B1006}
\bibinfo{author}{\bibnamefont{Han}, \bibfnamefont{M.~Y.}}, and
  \bibinfo{author}{\bibfnamefont{Y.}~\bibnamefont{Nambu}},
  \bibinfo{year}{1965}, \bibinfo{journal}{Phys. Rev.}
  \textbf{\bibinfo{volume}{139}}(\bibinfo{number}{4B}), \bibinfo{pages}{B1006}.

\bibitem[{\citenamefont{Hanhart}(2008)}]{Hanhart:2007cm}
\bibinfo{author}{\bibnamefont{Hanhart}, \bibfnamefont{C.}},
  \bibinfo{year}{2008}, \bibinfo{journal}{Eur. Phys. J.}
  \textbf{\bibinfo{volume}{A35}}, \bibinfo{pages}{271}.

\bibitem[{\citenamefont{Hasenfratz}
  \emph{et~al.}(1980)\citenamefont{Hasenfratz, Horgan, Kuti, and
  Richard}}]{Hasenfratz:1980ka}
\bibinfo{author}{\bibnamefont{Hasenfratz}, \bibfnamefont{P.}},
  \bibinfo{author}{\bibfnamefont{R.~R.} \bibnamefont{Horgan}},
  \bibinfo{author}{\bibfnamefont{J.}~\bibnamefont{Kuti}}, and
  \bibinfo{author}{\bibfnamefont{J.~M.} \bibnamefont{Richard}},
  \bibinfo{year}{1980}, \bibinfo{journal}{Phys. Lett.}
  \textbf{\bibinfo{volume}{B94}}, \bibinfo{pages}{401}.

\bibitem[{\citenamefont{Hasenfratz and Kuti}(1978)}]{Hasenfratz:1977dt}
\bibinfo{author}{\bibnamefont{Hasenfratz}, \bibfnamefont{P.}}, and
  \bibinfo{author}{\bibfnamefont{J.}~\bibnamefont{Kuti}}, \bibinfo{year}{1978},
  \bibinfo{journal}{Phys. Rept.} \textbf{\bibinfo{volume}{40}},
  \bibinfo{pages}{75}.

\bibitem[{\citenamefont{Haupt} \emph{et~al.}(2006)\citenamefont{Haupt, Metsch,
  and Petry}}]{Haupt:2006em}
\bibinfo{author}{\bibnamefont{Haupt}, \bibfnamefont{C.}},
  \bibinfo{author}{\bibfnamefont{B.}~\bibnamefont{Metsch}}, and
  \bibinfo{author}{\bibfnamefont{H.-R.} \bibnamefont{Petry}},
  \bibinfo{year}{2006}, \bibinfo{journal}{Eur. Phys. J.}
  \textbf{\bibinfo{volume}{A28}}, \bibinfo{pages}{213}.

\bibitem[{\citenamefont{He} \emph{et~al.}(2007)\citenamefont{He, Li, Liu, and
  Zeng}}]{He:2006is}
\bibinfo{author}{\bibnamefont{He}, \bibfnamefont{X.-G.}},
  \bibinfo{author}{\bibfnamefont{X.-Q.} \bibnamefont{Li}},
  \bibinfo{author}{\bibfnamefont{X.}~\bibnamefont{Liu}}, and
  \bibinfo{author}{\bibfnamefont{X.-Q.} \bibnamefont{Zeng}},
  \bibinfo{year}{2007}, \bibinfo{journal}{Eur. Phys. J.}
  \textbf{\bibinfo{volume}{C51}}, \bibinfo{pages}{883}.

\bibitem[{\citenamefont{Helbing}(2006)}]{Helbing:2006zp}
\bibinfo{author}{\bibnamefont{Helbing}, \bibfnamefont{K.}},
  \bibinfo{year}{2006}, \bibinfo{journal}{Prog. Part. Nucl. Phys.}
  \textbf{\bibinfo{volume}{57}}, \bibinfo{pages}{405}.

\bibitem[{\citenamefont{Hey and Kelly}(1983)}]{Hey:1982aj}
\bibinfo{author}{\bibnamefont{Hey}, \bibfnamefont{A.~J.~G.}}, and
  \bibinfo{author}{\bibfnamefont{R.~L.} \bibnamefont{Kelly}},
  \bibinfo{year}{1983}, \bibinfo{journal}{Phys. Rept.}
  \textbf{\bibinfo{volume}{96}}, \bibinfo{pages}{71}.

\bibitem[{\citenamefont{Hicks}(2005)}]{Hicks:2005gp}
\bibinfo{author}{\bibnamefont{Hicks}, \bibfnamefont{K.~H.}},
  \bibinfo{year}{2005}, \bibinfo{journal}{Prog. Part. Nucl. Phys.}
  \textbf{\bibinfo{volume}{55}}, \bibinfo{pages}{647}.

\bibitem[{Hicks \emph{et~al.}(2007)\citenamefont{Hicks}
  \emph{et~al.}}]{Hicks:2007zz}
\bibinfo{author}{\bibnamefont{Hicks}, \bibfnamefont{K.}}, \emph{et~al.},
  \bibinfo{year}{2007}, \bibinfo{journal}{Phys. Rev.}
  \textbf{\bibinfo{volume}{C76}}, \bibinfo{pages}{042201}.

\bibitem[{Hicks \emph{et~al.}(2009)\citenamefont{Hicks}
  \emph{et~al.}}]{Hicks:2008yn}
\bibinfo{author}{\bibnamefont{Hicks}, \bibfnamefont{K.}}, \emph{et~al.},
  \bibinfo{year}{2007}, \bibinfo{journal}{Phys. Rev. Lett.}
  \textbf{\bibinfo{volume}{102}}, \bibinfo{pages}{012501}.

\bibitem[{Hleiqawi \emph{et~al.}(2007)\citenamefont{Hleiqawi}
  \emph{et~al.}}]{Hleiqawi:2007ad}
\bibinfo{author}{\bibnamefont{Hleiqawi}, \bibfnamefont{I.}}, \emph{et~al.},
  \bibinfo{year}{2007}, \bibinfo{journal}{Phys. Rev.}
  \textbf{\bibinfo{volume}{C75}}, \bibinfo{pages}{042201}.

\bibitem[{\citenamefont{Hogaasen and Richard}(1983)}]{Hogaasen:1982rb}
\bibinfo{author}{\bibnamefont{Hogaasen}, \bibfnamefont{H.}}, and
  \bibinfo{author}{\bibfnamefont{J.~M.} \bibnamefont{Richard}},
  \bibinfo{year}{1983}, \bibinfo{journal}{Phys. Lett.}
  \textbf{\bibinfo{volume}{B124}}, \bibinfo{pages}{520}.

\bibitem[{\citenamefont{{\protect H\"ohler}}(2004)}]{Hohler:2004pr}
\bibinfo{author}{\bibnamefont{{\protect H\"ohler}}, \bibfnamefont{G.}},
  \bibinfo{year}{2004}, \bibinfo{journal}{Determination of pole parameters in
  $\pi N$ scattering, report, unpublished\hspace{-1mm}} .

\bibitem[{\citenamefont{{\protect H\"ohler}}
  \emph{et~al.}(1979)\citenamefont{{\protect H\"ohler}, Kaiser, Koch, and
  Pietarinen}}]{Hohler:1979yr}
\bibinfo{author}{\bibnamefont{{\protect H\"ohler}}, \bibfnamefont{G.}},
  \bibinfo{author}{\bibfnamefont{F.}~\bibnamefont{Kaiser}},
  \bibinfo{author}{\bibfnamefont{R.}~\bibnamefont{Koch}}, and
  \bibinfo{author}{\bibfnamefont{E.}~\bibnamefont{Pietarinen}},
  \bibinfo{year}{1979}, \bibinfo{journal}{Physics Data}
  \textbf{\bibinfo{volume}{12-1}}.

\bibitem[{\citenamefont{{\protect{H\"ohler}} and
  Workman}(2008)}]{Hohler:2004gt}
\bibinfo{author}{\bibnamefont{{\protect{H\"ohler}}}, \bibfnamefont{G.}}, and
  \bibinfo{author}{\bibfnamefont{R.~L.} \bibnamefont{Workman}},
  \bibinfo{year}{2008}, \bibinfo{note}{{N and Delta Resonances}, in
  \cite{Yao:2006px}}.

\bibitem[{\citenamefont{'t~Hooft}(1976)}]{'tHooft:1976fv}
\bibinfo{author}{\bibnamefont{'t~Hooft}, \bibfnamefont{G.}},
  \bibinfo{year}{1976}, \bibinfo{journal}{Phys. Rev.}
  \textbf{\bibinfo{volume}{D14}}, \bibinfo{pages}{3432}.

\bibitem[{Horn \emph{et~al.}(2008{\natexlab{a}})\citenamefont{Horn}
  \emph{et~al.}}]{Horn:2007pp}
\bibinfo{author}{\bibnamefont{Horn}, \bibfnamefont{I.}}, \emph{et~al.},
  \bibinfo{year}{2008}{\natexlab{a}}, \bibinfo{journal}{Phys. Rev. Lett.}
  \textbf{\bibinfo{volume}{101}}, \bibinfo{pages}{202002}.

\bibitem[{Horn \emph{et~al.}(2008{\natexlab{b}})\citenamefont{Horn}
  \emph{et~al.}}]{Horn:2008qv}
\bibinfo{author}{\bibnamefont{Horn}, \bibfnamefont{I.}}, \emph{et~al.},
  \bibinfo{year}{2008}{\natexlab{b}}, \bibinfo{journal}{Eur. Phys. J.}
  \textbf{\bibinfo{volume}{A38}}, \bibinfo{pages}{173}.

\bibitem[{\citenamefont{Hyodo} \emph{et~al.}(2008)\citenamefont{Hyodo, Jido,
  and Hosaka}}]{Hyodo:2008xr}
\bibinfo{author}{\bibnamefont{Hyodo}, \bibfnamefont{T.}},
  \bibinfo{author}{\bibfnamefont{D.}~\bibnamefont{Jido}}, and
  \bibinfo{author}{\bibfnamefont{A.}~\bibnamefont{Hosaka}},
  \bibinfo{year}{2008}, \bibinfo{journal}{Phys. Rev.}
  \textbf{\bibinfo{volume}{C78}}, \bibinfo{pages}{025203}.

\bibitem[{\citenamefont{Hyslop} \emph{et~al.}(1992)\citenamefont{Hyslop, Arndt,
  Roper, and Workman}}]{Hyslop:1992cs}
\bibinfo{author}{\bibnamefont{Hyslop}, \bibfnamefont{J.~S.}},
  \bibinfo{author}{\bibfnamefont{R.~A.} \bibnamefont{Arndt}},
  \bibinfo{author}{\bibfnamefont{L.~D.} \bibnamefont{Roper}}, and
  \bibinfo{author}{\bibfnamefont{R.~L.} \bibnamefont{Workman}},
  \bibinfo{year}{1992}, \bibinfo{journal}{Phys. Rev.}
  \textbf{\bibinfo{volume}{D46}}, \bibinfo{pages}{961}.

\bibitem[{\citenamefont{Iachello}(1989)}]{Iachello:1989mw}
\bibinfo{author}{\bibnamefont{Iachello}, \bibfnamefont{F.}},
  \bibinfo{year}{1989}, \bibinfo{journal}{Phys. Rev. Lett.}
  \textbf{\bibinfo{volume}{62}}, \bibinfo{pages}{2440}.

\bibitem[{\citenamefont{Iijima and Prebys}(2000)}]{Iijima:2000cq}
\bibinfo{author}{\bibnamefont{Iijima}, \bibfnamefont{T.}}, and
  \bibinfo{author}{\bibfnamefont{E.}~\bibnamefont{Prebys}},
  \bibinfo{year}{2000}, \bibinfo{journal}{Nucl. Instrum. Meth.}
  \textbf{\bibinfo{volume}{A446}}, \bibinfo{pages}{75}.

\bibitem[{\citenamefont{Inopin and Sharov}(2001)}]{Inopin:1999nf}
\bibinfo{author}{\bibnamefont{Inopin}, \bibfnamefont{A.}}, and
  \bibinfo{author}{\bibfnamefont{G.~S.} \bibnamefont{Sharov}},
  \bibinfo{year}{2001}, \bibinfo{journal}{Phys. Rev.}
  \textbf{\bibinfo{volume}{D63}}, \bibinfo{pages}{054023}.

\bibitem[{\citenamefont{Ioffe}(1981)}]{Ioffe:1981kw}
\bibinfo{author}{\bibnamefont{Ioffe}, \bibfnamefont{B.~L.}},
  \bibinfo{year}{1981}, \bibinfo{journal}{Nucl. Phys.}
  \textbf{\bibinfo{volume}{B188}}, \bibinfo{pages}{317},
  \bibinfo{note}{erratum, B191 (1981) 591}.

\bibitem[{Iori \emph{et~al.}(2007)\citenamefont{Iori}
  \emph{et~al.}}]{Iori:2007pw}
\bibinfo{author}{\bibnamefont{Iori}, \bibfnamefont{M.}}, \emph{et~al.},
  \bibinfo{year}{2007}, \eprint{FERMILAB-PUB-07-011-E, hep-ex/0701021}.

\bibitem[{\citenamefont{Isgur}(1980)}]{Isgur:1979ed}
\bibinfo{author}{\bibnamefont{Isgur}, \bibfnamefont{N.}}, \bibinfo{year}{1980},
  \bibinfo{journal}{Phys. Rev.} \textbf{\bibinfo{volume}{D21}},
  \bibinfo{pages}{779}, \bibinfo{note}{erratum: D23 (1981) 817}.

\bibitem[{\citenamefont{Isgur}(2000)}]{Isgur:2000ad}
\bibinfo{author}{\bibnamefont{Isgur}, \bibfnamefont{N.}}, \bibinfo{year}{2000},
  \bibinfo{note}{why N*'s are important. N*2000 Summary. Published in: Newport
  News 2000, Excited nucleons and hadronic structure, p.\ 403-422.},
  \eprint{nucl-th/0007008}.

\bibitem[{\citenamefont{Isgur and Karl}(1977)}]{Isgur:1977ef}
\bibinfo{author}{\bibnamefont{Isgur}, \bibfnamefont{N.}}, and
  \bibinfo{author}{\bibfnamefont{G.}~\bibnamefont{Karl}}, \bibinfo{year}{1977},
  \bibinfo{journal}{Phys. Lett.} \textbf{\bibinfo{volume}{B72}},
  \bibinfo{pages}{109}.

\bibitem[{\citenamefont{Isgur and Wise}(1991)}]{Isgur:1991wq}
\bibinfo{author}{\bibnamefont{Isgur}, \bibfnamefont{N.}}, and
  \bibinfo{author}{\bibfnamefont{M.~B.} \bibnamefont{Wise}},
  \bibinfo{year}{1991}, \bibinfo{journal}{Phys. Rev. Lett.}
  \textbf{\bibinfo{volume}{66}}, \bibinfo{pages}{1130}.

\bibitem[{Jaegle \emph{et~al.}(2008)\citenamefont{Jaegle}
  \emph{et~al.}}]{Jaegle:2008ux}
\bibinfo{author}{\bibnamefont{Jaegle}, \bibfnamefont{I.}}, \emph{et~al.},
  \bibinfo{year}{2008}, \bibinfo{journal}{Phys. Rev. Lett.}
  \textbf{\bibinfo{volume}{100}}, \bibinfo{pages}{252002}.

\bibitem[{\citenamefont{Jaffe}(1977)}]{Jaffe:1976yi}
\bibinfo{author}{\bibnamefont{Jaffe}, \bibfnamefont{R.~L.}},
  \bibinfo{year}{1977}, \bibinfo{journal}{Phys. Rev. Lett.}
  \textbf{\bibinfo{volume}{38}}, \bibinfo{pages}{195}, \bibinfo{note}{erratum,
  ibid.38, 617 (1977)}.

\bibitem[{\citenamefont{Jaffe}(2005)}]{Jaffe:2004ph}
\bibinfo{author}{\bibnamefont{Jaffe}, \bibfnamefont{R.~L.}},
  \bibinfo{year}{2005}, \bibinfo{journal}{Phys. Rept.}
  \textbf{\bibinfo{volume}{409}}, \bibinfo{pages}{1}.

\bibitem[{\citenamefont{Jaffe}(2007)}]{Jaffe:2007id}
\bibinfo{author}{\bibnamefont{Jaffe}, \bibfnamefont{R.~L.}},
  \bibinfo{year}{2007}, \bibinfo{journal}{AIP Conf. Proc.}
  \textbf{\bibinfo{volume}{964}}, \bibinfo{pages}{1}.

\bibitem[{\citenamefont{Jaffe} \emph{et~al.}(2006)\citenamefont{Jaffe, Pirjol,
  and Scardicchio}}]{Jaffe:2006jy}
\bibinfo{author}{\bibnamefont{Jaffe}, \bibfnamefont{R.~L.}},
  \bibinfo{author}{\bibfnamefont{D.}~\bibnamefont{Pirjol}}, and
  \bibinfo{author}{\bibfnamefont{A.}~\bibnamefont{Scardicchio}},
  \bibinfo{year}{2006}, \bibinfo{journal}{Phys. Rept.}
  \textbf{\bibinfo{volume}{435}}, \bibinfo{pages}{157}.

\bibitem[{\citenamefont{Jaffe and Wilczek}(2003)}]{Jaffe:2003sg}
\bibinfo{author}{\bibnamefont{Jaffe}, \bibfnamefont{R.~L.}}, and
  \bibinfo{author}{\bibfnamefont{F.}~\bibnamefont{Wilczek}},
  \bibinfo{year}{2003}, \bibinfo{journal}{Phys. Rev. Lett.}
  \textbf{\bibinfo{volume}{91}}, \bibinfo{pages}{232003}.

\bibitem[{Jessop \emph{et~al.}(1999)\citenamefont{Jessop}
  \emph{et~al.}}]{Jessop:1998wt}
\bibinfo{author}{\bibnamefont{Jessop}, \bibfnamefont{C.~P.}}, \emph{et~al.},
  \bibinfo{year}{1999}, \bibinfo{journal}{Phys. Rev. Lett.}
  \textbf{\bibinfo{volume}{82}}, \bibinfo{pages}{492}.

\bibitem[{\citenamefont{Jido} \emph{et~al.}(2008)\citenamefont{Jido, Doering,
  and Oset}}]{Jido:2007sm}
\bibinfo{author}{\bibnamefont{Jido}, \bibfnamefont{D.}},
  \bibinfo{author}{\bibfnamefont{M.}~\bibnamefont{Doering}}, and
  \bibinfo{author}{\bibfnamefont{E.}~\bibnamefont{Oset}}, \bibinfo{year}{2008},
  \bibinfo{journal}{Phys. Rev.} \textbf{\bibinfo{volume}{C77}},
  \bibinfo{pages}{065207}.

\bibitem[{\citenamefont{Jido} \emph{et~al.}(2003)\citenamefont{Jido, Oller,
  Oset, Ramos, and Mei{\ss}ner}}]{Jido:2003cb}
\bibinfo{author}{\bibnamefont{Jido}, \bibfnamefont{D.}},
  \bibinfo{author}{\bibfnamefont{J.~A.} \bibnamefont{Oller}},
  \bibinfo{author}{\bibfnamefont{E.}~\bibnamefont{Oset}},
  \bibinfo{author}{\bibfnamefont{A.}~\bibnamefont{Ramos}}, and
  \bibinfo{author}{\bibfnamefont{U.~G.} \bibnamefont{Mei{\ss}ner}},
  \bibinfo{year}{2003}, \bibinfo{journal}{Nucl. Phys.}
  \textbf{\bibinfo{volume}{A725}}, \bibinfo{pages}{181}.

\bibitem[{Joo \emph{et~al.}(2002)\citenamefont{Joo} \emph{et~al.}}]{Joo:2001tw}
\bibinfo{author}{\bibnamefont{Joo}, \bibfnamefont{K.}}, \emph{et~al.},
 \bibinfo{year}{2002},
  \bibinfo{journal}{Phys. Rev. Lett.} \textbf{\bibinfo{volume}{88}},
  \bibinfo{pages}{122001}.

\bibitem[{Joo \emph{et~al.}(2003)\citenamefont{Joo} \emph{et~al.}}]{Joo:2003uc}
\bibinfo{author}{\bibnamefont{Joo}, \bibfnamefont{K.}}, \emph{et~al.}, \bibinfo{year}{2003},
  \bibinfo{journal}{Phys. Rev.} \textbf{\bibinfo{volume}{C68}},
  \bibinfo{pages}{032201}.

\bibitem[{Joo \emph{et~al.}(2005)\citenamefont{Joo} \emph{et~al.}}]{Joo:2005gs}
\bibinfo{author}{\bibnamefont{Joo}, \bibfnamefont{K.}}, \emph{et~al.},
  \bibinfo{year}{2005}, \bibinfo{journal}{Phys. Rev.}
  \textbf{\bibinfo{volume}{C72}}, \bibinfo{pages}{058202}.

\bibitem[{\citenamefont{Julia-Diaz}
  \emph{et~al.}(2007)\citenamefont{Julia-Diaz, Lee, Matsuyama, and
  Sato}}]{JuliaDiaz:2007kz}
\bibinfo{author}{\bibnamefont{Julia-Diaz}, \bibfnamefont{B.}},
  \bibinfo{author}{\bibfnamefont{T.~S.~H.} \bibnamefont{Lee}},
  \bibinfo{author}{\bibfnamefont{A.}~\bibnamefont{Matsuyama}}, and
  \bibinfo{author}{\bibfnamefont{T.}~\bibnamefont{Sato}}, \bibinfo{year}{2007},
  \bibinfo{journal}{Phys. Rev.} \textbf{\bibinfo{volume}{C76}},
  \bibinfo{pages}{065201}.

\bibitem[{\citenamefont{Julia-Diaz}
  \emph{et~al.}(2008)\citenamefont{Julia-Diaz, Lee, Matsuyama, Sato, and
  Smith}}]{JuliaDiaz:2007fa}
\bibinfo{author}{\bibnamefont{Julia-Diaz}, \bibfnamefont{B.}},
  \bibinfo{author}{\bibfnamefont{T.~S.~H.} \bibnamefont{Lee}},
  \bibinfo{author}{\bibfnamefont{A.}~\bibnamefont{Matsuyama}},
  \bibinfo{author}{\bibfnamefont{T.}~\bibnamefont{Sato}}, and
  \bibinfo{author}{\bibfnamefont{L.~C.} \bibnamefont{Smith}},
  \bibinfo{year}{2008}, \bibinfo{journal}{Phys. Rev.}
  \textbf{\bibinfo{volume}{C77}}, \bibinfo{pages}{045205}.

\bibitem[{\citenamefont{Julia-Diaz and Riska}(2006)}]{JuliaDiaz:2006av}
\bibinfo{author}{\bibnamefont{Julia-Diaz}, \bibfnamefont{B.}}, and
  \bibinfo{author}{\bibfnamefont{D.~O.} \bibnamefont{Riska}},
  \bibinfo{year}{2006}, \bibinfo{journal}{Nucl. Phys.}
  \textbf{\bibinfo{volume}{A780}}, \bibinfo{pages}{175}.

\bibitem[{\citenamefont{Julia-Diaz}
  \emph{et~al.}(2004)\citenamefont{Julia-Diaz, Riska, and
  Coester}}]{JuliaDiaz:2003gq}
\bibinfo{author}{\bibnamefont{Julia-Diaz}, \bibfnamefont{B.}},
  \bibinfo{author}{\bibfnamefont{D.~O.} \bibnamefont{Riska}}, and
  \bibinfo{author}{\bibfnamefont{F.}~\bibnamefont{Coester}},
  \bibinfo{year}{2004}, \bibinfo{journal}{Phys. Rev.}
  \textbf{\bibinfo{volume}{C69}}, \bibinfo{pages}{035212}.

\bibitem[{Julia-Diaz \emph{et~al.}(2009)\citenamefont{Julia-Diaz}
  \emph{et~al.}}]{JuliaDiaz:2009ww}
\bibinfo{author}{\bibnamefont{Julia-Diaz}, \bibfnamefont{B.}}, \emph{et~al.},
  \bibinfo{year}{2009}, \eprint{0904.1918}.

\bibitem[{Junkersfeld \emph{et~al.}(2007)\citenamefont{Junkersfeld}
  \emph{et~al.}}]{Junkersfeld:2007yr}
\bibinfo{author}{\bibnamefont{Junkersfeld}, \bibfnamefont{J.}}, \emph{et~al.},
  \bibinfo{year}{2007}, \bibinfo{journal}{Eur. Phys. J.}
  \textbf{\bibinfo{volume}{A31}}, \bibinfo{pages}{365}.

\bibitem[{\citenamefont{Kaiser} \emph{et~al.}(1995)\citenamefont{Kaiser,
  Siegel, and Weise}}]{Kaiser:1995eg}
\bibinfo{author}{\bibnamefont{Kaiser}, \bibfnamefont{N.}},
  \bibinfo{author}{\bibfnamefont{P.~B.} \bibnamefont{Siegel}}, and
  \bibinfo{author}{\bibfnamefont{W.}~\bibnamefont{Weise}},
  \bibinfo{year}{1995}, \bibinfo{journal}{Nucl. Phys.}
  \textbf{\bibinfo{volume}{A594}}, \bibinfo{pages}{325}.

\bibitem[{Kalleicher \emph{et~al.}(1997)\citenamefont{Kalleicher}
  \emph{et~al.}}]{Kalleicher:1997qf}
\bibinfo{author}{\bibnamefont{Kalleicher}, \bibfnamefont{F.}}, \emph{et~al.},
  \bibinfo{year}{1997}, \bibinfo{journal}{Z. Phys.}
  \textbf{\bibinfo{volume}{A359}}, \bibinfo{pages}{201}.

\bibitem[{\citenamefont{Kamano} \emph{et~al.}(2009)\citenamefont{Kamano,
  Julia-Diaz, Lee, Matsuyama, and Sato}}]{Kamano:2008gr}
\bibinfo{author}{\bibnamefont{Kamano}, \bibfnamefont{H.}},
  \bibinfo{author}{\bibfnamefont{B.}~\bibnamefont{Julia-Diaz}},
  \bibinfo{author}{\bibfnamefont{T.~S.~H.} \bibnamefont{Lee}},
  \bibinfo{author}{\bibfnamefont{A.}~\bibnamefont{Matsuyama}}, and
  \bibinfo{author}{\bibfnamefont{T.}~\bibnamefont{Sato}}, \bibinfo{year}{2009},
  \eprint{arXiv:0909.1129 [nucl-th]}.

\bibitem[{\citenamefont{Kamano} \emph{et~al.}(2009)\citenamefont{Kamano,
  Julia-Diaz, Lee, Matsuyama, and Sato}}]{Kamano:2009im}
\bibinfo{author}{\bibnamefont{Kamano}, \bibfnamefont{H.}},
  \bibinfo{author}{\bibfnamefont{B.}~\bibnamefont{Julia-Diaz}},
  \bibinfo{author}{\bibfnamefont{T.~S.~H.} \bibnamefont{Lee}},
  \bibinfo{author}{\bibfnamefont{A.}~\bibnamefont{Matsuyama}}, and
  \bibinfo{author}{\bibfnamefont{T.}~\bibnamefont{Sato}}, \bibinfo{year}{2009},
  \bibinfo{journal}{Phys. Rev.} \textbf{\bibinfo{volume}{C79}},
  \bibinfo{pages}{025206}.

\bibitem[{\citenamefont{Karliner}(1986)\citenamefont{Karliner
  and Mattis}}]{Karliner:1986wq}
\bibinfo{author}{\bibnamefont{Karliner}, \bibfnamefont{M.}}, and
  \bibinfo{author}{\bibfnamefont{M.~P.} \bibnamefont{Mattis}},
  \bibinfo{year}{1986}, \bibinfo{journal}{Phys. Rev.}
  \textbf{\bibinfo{volume}{D34}}, \bibinfo{pages}{1991}.

\bibitem[{\citenamefont{Karliner} \emph{et~al.}(2008)\citenamefont{Karliner,
  Keren-Zur, Lipkin, and Rosner}}]{Karliner:2008sv}
\bibinfo{author}{\bibnamefont{Karliner}, \bibfnamefont{M.}},
  \bibinfo{author}{\bibfnamefont{B.}~\bibnamefont{Keren-Zur}},
  \bibinfo{author}{\bibfnamefont{H.~J.} \bibnamefont{Lipkin}}, and
  \bibinfo{author}{\bibfnamefont{J.~L.} \bibnamefont{Rosner}},
  \bibinfo{year}{2008}, \eprint{0804.1575}.

\bibitem[{\citenamefont{Kashevarov}(2009)}]{Kashevarov:2009ww}
\bibinfo{author}{\bibnamefont{Kashevarov}, \bibfnamefont{V.~L.}},
  \bibinfo{year}{2009}, \eprint{0901.3888}.

\bibitem[{Kernel \emph{et~al.}(1989{\natexlab{a}})\citenamefont{Kernel}
  \emph{et~al.}}]{Kernel:1988sua}
\bibinfo{author}{\bibnamefont{Kernel}, \bibfnamefont{G.}}, \emph{et~al.},
  \bibinfo{year}{1989}{\natexlab{a}}, \bibinfo{journal}{Phys. Lett.}
  \textbf{\bibinfo{volume}{B216}}, \bibinfo{pages}{244}.

\bibitem[{Kernel \emph{et~al.}(1989{\natexlab{b}})\citenamefont{Kernel}
  \emph{et~al.}}]{Kernel:1989yf}
\bibinfo{author}{\bibnamefont{Kernel}, \bibfnamefont{G.}}, \emph{et~al.},
  \bibinfo{year}{1989}{\natexlab{b}}, \bibinfo{journal}{Phys. Lett.}
  \textbf{\bibinfo{volume}{B225}}, \bibinfo{pages}{198}.

\bibitem[{Kernel \emph{et~al.}(1990)\citenamefont{Kernel}
  \emph{et~al.}}]{Kernel:1990qx}
\bibinfo{author}{\bibnamefont{Kernel}, \bibfnamefont{G.}}, \emph{et~al.},
  \bibinfo{year}{1990}, \bibinfo{journal}{Z. Phys.}
  \textbf{\bibinfo{volume}{C48}}, \bibinfo{pages}{201}.

\bibitem[{\citenamefont{Kirchbach} \emph{et~al.}(2001)\citenamefont{Kirchbach,
  Moshinsky, and Smirnov}}]{Kirchbach:2001de}
\bibinfo{author}{\bibnamefont{Kirchbach}, \bibfnamefont{M.}},
  \bibinfo{author}{\bibfnamefont{M.}~\bibnamefont{Moshinsky}}, and
  \bibinfo{author}{\bibfnamefont{Y.~F.} \bibnamefont{Smirnov}},
  \bibinfo{year}{2001}, \bibinfo{journal}{Phys. Rev.}
  \textbf{\bibinfo{volume}{D64}}, \bibinfo{pages}{114005}.

\bibitem[{\citenamefont{Kisslinger}(2004)}]{Kisslinger:2003hk}
\bibinfo{author}{\bibnamefont{Kisslinger}, \bibfnamefont{L.~S.}},
  \bibinfo{year}{2004}, \bibinfo{journal}{Phys. Rev.}
  \textbf{\bibinfo{volume}{D69}}, \bibinfo{pages}{054015}.

\bibitem[{\citenamefont{Kisslinger and Li}(1995)}]{Kisslinger:1995yw}
\bibinfo{author}{\bibnamefont{Kisslinger}, \bibfnamefont{L.~S.}}, and
  \bibinfo{author}{\bibfnamefont{Z.~P.} \bibnamefont{Li}},
  \bibinfo{year}{1995}, \bibinfo{journal}{Phys. Rev.}
  \textbf{\bibinfo{volume}{D51}}, \bibinfo{pages}{5986}.

\bibitem[{\citenamefont{Kittel and Farrar}(2005)}]{Kittel:2005jm}
\bibinfo{author}{\bibnamefont{Kittel}, \bibfnamefont{O.}}, and
  \bibinfo{author}{\bibfnamefont{G.~R.} \bibnamefont{Farrar}},
  \bibinfo{year}{2005}, \bibinfo{note}{bONN-TH-2005-03 (unpublished)},
  \eprint{hep-ph/0508150}.

\bibitem[{Klein \emph{et~al.}(2008)\citenamefont{Klein}
  \emph{et~al.}}]{Klein:2008gs}
\bibinfo{author}{\bibnamefont{Klein}, \bibfnamefont{F.}}, \emph{et~al.}, \bibinfo{year}{2008},
  \bibinfo{journal}{Phys. Rev.} \textbf{\bibinfo{volume}{D78}},
  \bibinfo{pages}{117101}.

\bibitem[{\citenamefont{Klempt}(2003)}]{Klempt:2002tt}
\bibinfo{author}{\bibnamefont{Klempt}, \bibfnamefont{E.}},
  \bibinfo{year}{2003}, \bibinfo{journal}{Phys. Lett.}
  \textbf{\bibinfo{volume}{B559}}, \bibinfo{pages}{144}.

\bibitem[{\citenamefont{Klempt}(2008)}]{Klempt:2008rq}
\bibinfo{author}{\bibnamefont{Klempt}, \bibfnamefont{E.}},
  \bibinfo{year}{2008}, \bibinfo{journal}{Eur. Phys. J.}
  \textbf{\bibinfo{volume}{A38}}, \bibinfo{pages}{187}.

\bibitem[{\citenamefont{Klempt} \emph{et~al.}(2006)\citenamefont{Klempt,
  Anisovich, Nikonov, Sarantsev, and Thoma}}]{Klempt:2006sa}
\bibinfo{author}{\bibnamefont{Klempt}, \bibfnamefont{E.}},
  \bibinfo{author}{\bibfnamefont{A.~V.} \bibnamefont{Anisovich}},
  \bibinfo{author}{\bibfnamefont{V.~A.} \bibnamefont{Nikonov}},
  \bibinfo{author}{\bibfnamefont{A.~V.} \bibnamefont{Sarantsev}}, and
  \bibinfo{author}{\bibfnamefont{U.}~\bibnamefont{Thoma}},
  \bibinfo{year}{2006}, \bibinfo{journal}{Eur. Phys. J.}
  \textbf{\bibinfo{volume}{A29}}, \bibinfo{pages}{307}.

\bibitem[{\citenamefont{Klempt} \emph{et~al.}(1995)\citenamefont{Klempt,
  Metsch, Munz, and Petry}}]{Klempt:1995ku}
\bibinfo{author}{\bibnamefont{Klempt}, \bibfnamefont{E.}},
  \bibinfo{author}{\bibfnamefont{B.~C.} \bibnamefont{Metsch}},
  \bibinfo{author}{\bibfnamefont{C.~R.} \bibnamefont{Munz}}, and
  \bibinfo{author}{\bibfnamefont{H.~R.} \bibnamefont{Petry}},
  \bibinfo{year}{1995}, \bibinfo{journal}{Phys. Lett.}
  \textbf{\bibinfo{volume}{B361}}, \bibinfo{pages}{160}.

\bibitem[{\citenamefont{Klempt and Zaitsev}(2007)}]{Klempt:2007cp}
\bibinfo{author}{\bibnamefont{Klempt}, \bibfnamefont{E.}}, and
  \bibinfo{author}{\bibfnamefont{A.}~\bibnamefont{Zaitsev}},
  \bibinfo{year}{2007}, \bibinfo{journal}{Phys. Rept.}
  \textbf{\bibinfo{volume}{454}}, \bibinfo{pages}{1}.

\bibitem[{Knapp \emph{et~al.}(1976)\citenamefont{Knapp}
  \emph{et~al.}}]{Knapp:1976qw}
\bibinfo{author}{\bibnamefont{Knapp}, \bibfnamefont{B.}}, \emph{et~al.},
  \bibinfo{year}{1976}, \bibinfo{journal}{Phys. Rev. Lett.}
  \textbf{\bibinfo{volume}{37}}, \bibinfo{pages}{882}.

\bibitem[{\citenamefont{{\protect{Kn\"ochlein}}}
  \emph{et~al.}(1995)\citenamefont{{\protect{Kn\"ochlein}}, Drechsel, and
  Tiator}}]{Knochlein:1995qz}
\bibinfo{author}{\bibnamefont{{\protect{Kn\"ochlein}}}, \bibfnamefont{G.}},
  \bibinfo{author}{\bibfnamefont{D.}~\bibnamefont{Drechsel}}, and
  \bibinfo{author}{\bibfnamefont{L.}~\bibnamefont{Tiator}},
  \bibinfo{year}{1995}, \bibinfo{journal}{Z. Phys.}
  \textbf{\bibinfo{volume}{A352}}, \bibinfo{pages}{327}.

\bibitem[{Kohri \emph{et~al.}(2009)\citenamefont{Kohri}
  \emph{et~al.}}]{Kohri:2009xe}
\bibinfo{author}{\bibnamefont{Kohri}, \bibfnamefont{H.}}, \emph{et~al.},
\bibinfo{year}{2009}, \eprint{0906.0197}.

\bibitem[{\citenamefont{Kokkedee}(1969)}]{Kokkedee:101899}
\bibinfo{author}{\bibnamefont{Kokkedee}, \bibfnamefont{J.~J.~J.}},
  \bibinfo{year}{1969}, \emph{\bibinfo{title}{The quark model}}, Frontiers in
  Physics (\bibinfo{publisher}{Benjamin}, \bibinfo{address}{New York, NY}),
  \bibinfo{note}{based on a series of lectures given at CERN in Autumn 1967}.

\bibitem[{\citenamefont{Koniuk and Isgur}(1980{\natexlab{a}})}]{Koniuk:1979vy}
\bibinfo{author}{\bibnamefont{Koniuk}, \bibfnamefont{R.}}, and
  \bibinfo{author}{\bibfnamefont{N.}~\bibnamefont{Isgur}},
  \bibinfo{year}{1980}{\natexlab{a}}, \bibinfo{journal}{Phys. Rev.}
  \textbf{\bibinfo{volume}{D21}}, \bibinfo{pages}{1868}.

\bibitem[{\citenamefont{Koniuk and Isgur}(1980{\natexlab{b}})}]{Koniuk:1979vw}
\bibinfo{author}{\bibnamefont{Koniuk}, \bibfnamefont{R.}}, and
  \bibinfo{author}{\bibfnamefont{N.}~\bibnamefont{Isgur}},
  \bibinfo{year}{1980}{\natexlab{b}}, \bibinfo{journal}{Phys. Rev. Lett.}
  \textbf{\bibinfo{volume}{44}}, \bibinfo{pages}{845}.

\bibitem[{\citenamefont{Kopp}(1996)}]{Kopp:1996kg}
\bibinfo{author}{\bibnamefont{Kopp}, \bibfnamefont{S.~E.}},
  \bibinfo{year}{1996}, \bibinfo{journal}{Nucl. Instrum. Meth.}
  \textbf{\bibinfo{volume}{A384}}, \bibinfo{pages}{61}.

\bibitem[{\citenamefont{K{\protect\"o}rner}
  \emph{et~al.}(1994)\citenamefont{K{\protect\"o}rner, Kramer, and
  Pirjol}}]{Korner:1994nh}
\bibinfo{author}{\bibnamefont{K{\protect\"o}rner}, \bibfnamefont{J.~G.}},
  \bibinfo{author}{\bibfnamefont{M.}~\bibnamefont{Kramer}}, and
  \bibinfo{author}{\bibfnamefont{D.}~\bibnamefont{Pirjol}},
  \bibinfo{year}{1994}, \bibinfo{journal}{Prog. Part. Nucl. Phys.}
  \textbf{\bibinfo{volume}{33}}, \bibinfo{pages}{787}.

\bibitem[{Kozlenko \emph{et~al.}(2003)\citenamefont{Kozlenko}
  \emph{et~al.}}]{Kozlenko:2003hu}
\bibinfo{author}{\bibnamefont{Kozlenko}, \bibfnamefont{N.~G.}}, \emph{et~al.},
  \bibinfo{year}{2003}, \bibinfo{journal}{Phys. Atom. Nucl.}
  \textbf{\bibinfo{volume}{66}}, \bibinfo{pages}{110}.

\bibitem[{Krambrich \emph{et~al.}(2009)\citenamefont{Krambrich}
  \emph{et~al.}}]{Krambrich:2009te}
\bibinfo{author}{\bibnamefont{Krambrich}, \bibfnamefont{D.}}, \emph{et~al.},
  \bibinfo{year}{2009}, \bibinfo{journal}{Phys. Rev. Lett.}
  \textbf{\bibinfo{volume}{103}}, \bibinfo{pages}{052002}.

\bibitem[{\citenamefont{Krehl} \emph{et~al.}(2000)\citenamefont{Krehl, Hanhart,
  Krewald, and Speth}}]{Krehl:1999km}
\bibinfo{author}{\bibnamefont{Krehl}, \bibfnamefont{O.}},
  \bibinfo{author}{\bibfnamefont{C.}~\bibnamefont{Hanhart}},
  \bibinfo{author}{\bibfnamefont{S.}~\bibnamefont{Krewald}}, and
  \bibinfo{author}{\bibfnamefont{J.}~\bibnamefont{Speth}},
  \bibinfo{year}{2000}, \bibinfo{journal}{Phys. Rev.}
  \textbf{\bibinfo{volume}{C62}}, \bibinfo{pages}{025207}.

\bibitem[{\citenamefont{Krusche and Schadmand}(2003)}]{Krusche:2003ik}
\bibinfo{author}{\bibnamefont{Krusche}, \bibfnamefont{B.}}, and
  \bibinfo{author}{\bibfnamefont{S.}~\bibnamefont{Schadmand}},
  \bibinfo{year}{2003}, \bibinfo{journal}{Prog. Part. Nucl. Phys.}
  \textbf{\bibinfo{volume}{51}}, \bibinfo{pages}{399}.

\bibitem[{\citenamefont{Kuznetsov}(2007)\citenamefont{Kuznetsov
  \emph{et~al.}}}]{Kuznetsov:2007gr}
\bibinfo{author}{\bibnamefont{Kuznetsov}, \bibfnamefont{V.}}, \emph{et~al.},
  \bibinfo{year}{2007}, \bibinfo{journal}{Phys. Lett.}
  \textbf{\bibinfo{volume}{B647}}, \bibinfo{pages}{23}.

\bibitem[{\citenamefont{Kuznetsov}(2008)\citenamefont{Kuznetsov
  \emph{et~al.}}}]{Kuznetsov:2008hj}
\bibinfo{author}{\bibnamefont{Kuznetsov}, \bibfnamefont{A.}}, \emph{et~al.},
  \bibinfo{year}{2008}, \bibinfo{journal}{Acta Phys.\ Polon.}
  \textbf{\bibinfo{volume}{B 39}}, \bibinfo{pages}{1949}.

\bibitem[{LaCourse and Olsson(1988)\citenamefont{LaCourse and Olsson}}]{LaCourse:1988cu}
\bibinfo{author}{\bibnamefont{LaCourse}, \bibfnamefont{D.}}, and
  \bibinfo{author}{\bibfnamefont{M.~G.} \bibnamefont{Olsson}},
  \bibinfo{year}{2008}, \bibinfo{journal}{Phys. Rev.}
  \textbf{\bibinfo{volume}{D 39}}, \bibinfo{pages}{2751}.

\bibitem[{{\protect{Langg\"artner}} \emph{et~al.}(2001)\citenamefont{Langgartner}
  \emph{et~al.}}]{Langgartner:2001sg}
\bibinfo{author}{\bibnamefont{\protect{Langg\"artner}}, \bibfnamefont{W.}}, \emph{et~al.},
  \bibinfo{year}{2001}, \bibinfo{journal}{Phys. Rev. Lett.}
  \textbf{\bibinfo{volume}{87}}, \bibinfo{pages}{052001}.

\bibitem[{Lawall \emph{et~al.}(2005)\citenamefont{Lawall}
  \emph{et~al.}}]{Lawall:2005np}
\bibinfo{author}{\bibnamefont{Lawall}, \bibfnamefont{R.}}, \emph{et~al.},
  \bibinfo{year}{2005}, \bibinfo{journal}{Eur. Phys. J.}
  \textbf{\bibinfo{volume}{A24}}, \bibinfo{pages}{275}.

\bibitem[{\citenamefont{Lee}(2007)}]{Lee:2009zzo}
\bibinfo{author}{\bibnamefont{Lee}, \bibfnamefont{T.~S.~H.}},
  \bibinfo{year}{2007}, \bibinfo{journal}{Int. J. Mod. Phys.}
  \textbf{\bibinfo{volume}{E18}}, \bibinfo{pages}{1215}.

\bibitem[{\citenamefont{Lesniak} \emph{et~al.}(2008)\citenamefont{Lesniak}}]{Lesiak:2008wz}
\bibinfo{author}{\bibnamefont{Lesiak}, \bibfnamefont{T.}},
    \bibinfo{year}{2008}, \bibinfo{journal}{Phys. Lett.}
  \textbf{\bibinfo{volume}{B665}}, \bibinfo{pages}{9}.

\bibitem[{\citenamefont{Lewis} \emph{et~al.}(2001)\citenamefont{Lewis, Mathur,
  and Woloshyn}}]{Lewis:2001iz}
\bibinfo{author}{\bibnamefont{Lewis}, \bibfnamefont{R.}},
  \bibinfo{author}{\bibfnamefont{N.}~\bibnamefont{Mathur}}, and
  \bibinfo{author}{\bibfnamefont{R.~M.} \bibnamefont{Woloshyn}},
  \bibinfo{year}{2001}, \bibinfo{journal}{Phys. Rev.}
  \textbf{\bibinfo{volume}{D64}}, \bibinfo{pages}{094509}.


\bibitem[{\citenamefont{Li and Oset}(2004)}]{Li:2004bg}
\bibinfo{author}{\bibnamefont{Li}, \bibfnamefont{C.}}, and
  \bibinfo{author}{\bibfnamefont{E.} \bibnamefont{Oset}},
  \bibinfo{year}{2004}, \bibinfo{journal}{Phys. Rev.}
  \textbf{\bibinfo{volume}{C70}}, \bibinfo{pages}{065202}.

\bibitem[{\citenamefont{Li and Riska}(2006)}]{Li:2006nm}
\bibinfo{author}{\bibnamefont{Li}, \bibfnamefont{Q.~B.}}, and
  \bibinfo{author}{\bibfnamefont{D.~O.} \bibnamefont{Riska}},
  \bibinfo{year}{2006}, \bibinfo{journal}{Phys. Rev.}
  \textbf{\bibinfo{volume}{C74}}, \bibinfo{pages}{015202}.

\bibitem[{\citenamefont{Li} \emph{et~al.}(2009)\citenamefont{Li, Shen, and
  Collaboration}}]{LI:2009iw}
\bibinfo{author}{\bibnamefont{Li}, \bibfnamefont{S.}},
  \bibinfo{author}{\bibfnamefont{X.}~\bibnamefont{Shen}}, and
  \bibinfo{author}{\bibfnamefont{f.~t.~B.} \bibnamefont{Collaboration}},
  \bibinfo{year}{2009}, \eprint{0905.1562}.

\bibitem[{\citenamefont{Li} \emph{et~al.}(1992)\citenamefont{Li, Burkert, and
  Li}}]{Li:1991yba}
\bibinfo{author}{\bibnamefont{Li}, \bibfnamefont{Z.-P.}},
  \bibinfo{author}{\bibfnamefont{V.}~\bibnamefont{Burkert}}, and
  \bibinfo{author}{\bibfnamefont{Z.-J.} \bibnamefont{Li}},
  \bibinfo{year}{1992}, \bibinfo{journal}{Phys. Rev.}
  \textbf{\bibinfo{volume}{D46}}, \bibinfo{pages}{70}.

\bibitem[{\citenamefont{Lichtenberg}(1996)}]{Lichtenberg:1996rg}
\bibinfo{author}{\bibnamefont{Lichtenberg}, \bibfnamefont{D.}},
  \bibinfo{year}{1996}, \bibinfo{note}{given at 3rd International Workshop on
  Diquarks and other Models of Compositeness (DIQUARKS III), Turin, Italy,
  28-30 Oct 1996}.

\bibitem[{\citenamefont{Lichtenberg}(1977)}]{Lichtenberg:1977mv}
\bibinfo{author}{\bibnamefont{Lichtenberg}, \bibfnamefont{D.~B.}},
  \bibinfo{year}{1977}, \bibinfo{journal}{Phys. Rev.}
  \textbf{\bibinfo{volume}{D16}}, \bibinfo{pages}{231}.

\bibitem[{\citenamefont{Lichtenberg}
  \emph{et~al.}(1996)\citenamefont{Lichtenberg, Roncaglia, and
  Predazzi}}]{Lichtenberg:1995kg}
\bibinfo{author}{\bibnamefont{Lichtenberg}, \bibfnamefont{D.~B.}},
  \bibinfo{author}{\bibfnamefont{R.}~\bibnamefont{Roncaglia}}, and
  \bibinfo{author}{\bibfnamefont{E.}~\bibnamefont{Predazzi}},
  \bibinfo{year}{1996}, \bibinfo{journal}{Phys. Rev.}
  \textbf{\bibinfo{volume}{D53}}, \bibinfo{pages}{6678}.

\bibitem[{\citenamefont{Lipkin}(1987)}]{Lipkin:1987sk}
\bibinfo{author}{\bibnamefont{Lipkin}, \bibfnamefont{H.~J.}},
  \bibinfo{year}{1987}, \bibinfo{journal}{Phys. Lett.}
  \textbf{\bibinfo{volume}{B195}}, \bibinfo{pages}{484}.

\bibitem[{\citenamefont{Liu and Zou}(2006)}]{Liu:2005pm}
\bibinfo{author}{\bibnamefont{Liu}, \bibfnamefont{B.~C.}}, and
  \bibinfo{author}{\bibfnamefont{B.~S.} \bibnamefont{Zou}},
  \bibinfo{year}{2006}, \bibinfo{journal}{Phys. Rev. Lett.}
  \textbf{\bibinfo{volume}{96}}, \bibinfo{pages}{042002}.

\bibitem[{\citenamefont{Liu and Zou}(2007)}]{Liu:2006ym}
\bibinfo{author}{\bibnamefont{Liu}, \bibfnamefont{B.~C.}}, and
  \bibinfo{author}{\bibfnamefont{B.~S.} \bibnamefont{Zou}},
  \bibinfo{year}{2007}, \bibinfo{journal}{Phys. Rev. Lett.}
  \textbf{\bibinfo{volume}{98}}, \bibinfo{pages}{039102}.

\bibitem[{\citenamefont{Liu} \emph{et~al.}(2008)\citenamefont{Liu, Ke, Qiao,
  Wei, and Li}}]{Liu:2007twb}
\bibinfo{author}{\bibnamefont{Liu}, \bibfnamefont{X.}},
  \bibinfo{author}{\bibfnamefont{H.-W.} \bibnamefont{Ke}},
  \bibinfo{author}{\bibfnamefont{Q.-P.} \bibnamefont{Qiao}},
  \bibinfo{author}{\bibfnamefont{Z.-T.} \bibnamefont{Wei}}, and
  \bibinfo{author}{\bibfnamefont{X.-Q.} \bibnamefont{Li}},
  \bibinfo{year}{2008}, \bibinfo{journal}{Phys. Rev.}
  \textbf{\bibinfo{volume}{D77}}, \bibinfo{pages}{035014}.

\bibitem[{Lleres \emph{et~al.}(2007)\citenamefont{Lleres}
  \emph{et~al.}}]{Lleres:2007tx}
\bibinfo{author}{\bibnamefont{Lleres}, \bibfnamefont{A.}}, \emph{et~al.},
  \bibinfo{year}{2007}, \bibinfo{journal}{Eur. Phys. J.}
  \textbf{\bibinfo{volume}{A31}}, \bibinfo{pages}{79}.

\bibitem[{Lleres \emph{et~al.}(2009)\citenamefont{Lleres}
  \emph{et~al.}}]{Lleres:2008em}
\bibinfo{author}{\bibnamefont{Lleres}, \bibfnamefont{A.}}, \emph{et~al.},
  \bibinfo{year}{2009}, \bibinfo{journal}{Eur. Phys. J.}
  \textbf{\bibinfo{volume}{A39}}, \bibinfo{pages}{149}.

\bibitem[{Lowe \emph{et~al.}(1991)\citenamefont{Lowe}
  \emph{et~al.}}]{Lowe:1991gd}
\bibinfo{author}{\bibnamefont{Lowe}, \bibfnamefont{J.}}, \emph{et~al.},
  \bibinfo{year}{1991}, \bibinfo{journal}{Phys. Rev.}
  \textbf{\bibinfo{volume}{C44}}, \bibinfo{pages}{956}.

\bibitem[{\citenamefont{Lutz and Kolomeitsev}(2002)}]{Lutz:2001yb}
\bibinfo{author}{\bibnamefont{Lutz}, \bibfnamefont{M.~F.~M.}}, and
  \bibinfo{author}{\bibfnamefont{E.~E.} \bibnamefont{Kolomeitsev}},
  \bibinfo{year}{2002}, \bibinfo{journal}{Nucl. Phys.}
  \textbf{\bibinfo{volume}{A700}}, \bibinfo{pages}{193}.

\bibitem[{MacCormick \emph{et~al.}(1996)\citenamefont{MacCormick}
  \emph{et~al.}}]{MacCormick:1996jz}
\bibinfo{author}{\bibnamefont{MacCormick}, \bibfnamefont{M.}}, \emph{et~al.},
  \bibinfo{year}{1996}, \bibinfo{journal}{Phys. Rev.}
  \textbf{\bibinfo{volume}{C53}}, \bibinfo{pages}{41}.

\bibitem[{\citenamefont{Magas} \emph{et~al.}(2005)\citenamefont{Magas, Oset,
  and Ramos}}]{Magas:2005vu}
\bibinfo{author}{\bibnamefont{Magas}, \bibfnamefont{V.~K.}},
  \bibinfo{author}{\bibfnamefont{E.}~\bibnamefont{Oset}}, and
  \bibinfo{author}{\bibfnamefont{A.}~\bibnamefont{Ramos}},
  \bibinfo{year}{2005}, \bibinfo{journal}{Phys. Rev. Lett.}
  \textbf{\bibinfo{volume}{95}}, \bibinfo{pages}{052301}.

\bibitem[{Mahbub \emph{et~al.}(2009)\citenamefont{Mahbub}
  \emph{et~al.}}]{Mahbub:2009aa}
\bibinfo{author}{\bibnamefont{Mahbub}, \bibfnamefont{M.~S.}}, \emph{et~al.},
  \bibinfo{year}{2009}, \bibinfo{journal}{Phys. Lett.}
  \textbf{\bibinfo{volume}{679}}, \bibinfo{pages}{418}.

\bibitem[{\citenamefont{Manley} \emph{et~al.}(1984)\citenamefont{Manley, Arndt,
  Goradia, and Teplitz}}]{Manley:1984jz}
\bibinfo{author}{\bibnamefont{Manley}, \bibfnamefont{D.~M.}},
  \bibinfo{author}{\bibfnamefont{R.~A.} \bibnamefont{Arndt}},
  \bibinfo{author}{\bibfnamefont{Y.}~\bibnamefont{Goradia}}, and
  \bibinfo{author}{\bibfnamefont{V.~L.} \bibnamefont{Teplitz}},
  \bibinfo{year}{1984}, \bibinfo{journal}{Phys. Rev.}
  \textbf{\bibinfo{volume}{D30}}, \bibinfo{pages}{904}.

\bibitem[{\citenamefont{Manley and Saleski}(1992)}]{Manley:1992yb}
\bibinfo{author}{\bibnamefont{Manley}, \bibfnamefont{D.~M.}}, and
  \bibinfo{author}{\bibfnamefont{E.~M.} \bibnamefont{Saleski}},
  \bibinfo{year}{1992}, \bibinfo{journal}{Phys. Rev.}
  \textbf{\bibinfo{volume}{D45}}, \bibinfo{pages}{4002}.

\bibitem[{Manley \emph{et~al.}(2002)\citenamefont{Manley}
  \emph{et~al.}}]{Manley:2002ue}
\bibinfo{author}{\bibnamefont{Manley}, \bibfnamefont{D.~M.}}, \emph{et~al.},
  \bibinfo{year}{2002}, \bibinfo{journal}{Phys. Rev. Lett.}
  \textbf{\bibinfo{volume}{88}}, \bibinfo{pages}{012002}.

\bibitem[{Manweiler \emph{et~al.}(2008)\citenamefont{Manweiler}
  \emph{et~al.}}]{Manweiler:2008zz}
\bibinfo{author}{\bibnamefont{Manweiler}, \bibfnamefont{R.}}, \emph{et~al.},
  \bibinfo{year}{2008}, \bibinfo{journal}{Phys. Rev.}
  \textbf{\bibinfo{volume}{C77}}, \bibinfo{pages}{015205}.

\bibitem[{\citenamefont{Mart and Sulaksono}(2006)}]{Mart:2006dk}
\bibinfo{author}{\bibnamefont{Mart}, \bibfnamefont{T.}}, and
  \bibinfo{author}{\bibfnamefont{A.}~\bibnamefont{Sulaksono}},
  \bibinfo{year}{2006}, \bibinfo{journal}{Phys. Rev.}
  \textbf{\bibinfo{volume}{C74}}, \bibinfo{pages}{055203}.

\bibitem[{\citenamefont{Martin}(1986)}]{Martin:1985hw}
\bibinfo{author}{\bibnamefont{Martin}, \bibfnamefont{A.}},
  \bibinfo{year}{1986}, \bibinfo{journal}{Z. Phys.}
  \textbf{\bibinfo{volume}{C32}}, \bibinfo{pages}{359}.

\bibitem[{Martin \emph{et~al.}(1975)\citenamefont{Martin}
  \emph{et~al.}}]{Martin:1974rt}
\bibinfo{author}{\bibnamefont{Martin}, \bibfnamefont{J.~F.}}, \emph{et~al.},
  \bibinfo{year}{1975}, \bibinfo{journal}{Nucl. Phys.}
  \textbf{\bibinfo{volume}{B89}}, \bibinfo{pages}{253}.

\bibitem[{Mathur \emph{et~al.}(2005)\citenamefont{Mathur}
  \emph{et~al.}}]{Mathur:2003zf}
\bibinfo{author}{\bibnamefont{Mathur}, \bibfnamefont{N.}}, \emph{et~al.},
  \bibinfo{year}{2005}, \bibinfo{journal}{Phys. Lett.}
  \textbf{\bibinfo{volume}{B605}}, \bibinfo{pages}{137}.

\bibitem[{\citenamefont{Matsuyama} \emph{et~al.}(2007)\citenamefont{Matsuyama,
  Sato, and Lee}}]{Matsuyama:2006rp}
\bibinfo{author}{\bibnamefont{Matsuyama}, \bibfnamefont{A.}},
  \bibinfo{author}{\bibfnamefont{T.}~\bibnamefont{Sato}}, and
  \bibinfo{author}{\bibfnamefont{T.~S.~H.} \bibnamefont{Lee}},
  \bibinfo{year}{2007}, \bibinfo{journal}{Phys. Rept.}
  \textbf{\bibinfo{volume}{439}}, \bibinfo{pages}{193}.

\bibitem[{Mattson \emph{et~al.}(2002)\citenamefont{Mattson}
  \emph{et~al.}}]{Mattson:2002vu}
\bibinfo{author}{\bibnamefont{Mattson}, \bibfnamefont{M.}}, \emph{et~al.},
  \bibinfo{year}{2002}, \bibinfo{journal}{Phys. Rev. Lett.}
  \textbf{\bibinfo{volume}{89}}, \bibinfo{pages}{112001}.

\bibitem[{\citenamefont{Meadows}(1980)}]{Meadows:1980vr}
\bibinfo{author}{\bibnamefont{Meadows}, \bibfnamefont{B.~T.}},
  \bibinfo{year}{1980}, \bibinfo{note}{in Proceedings Baryon 1980, Toronto
  (1980), p. 283-300}.

\bibitem[{Mecking \emph{et~al.}(2003)\citenamefont{Mecking}
  \emph{et~al.}}]{Mecking:2003zu}
\bibinfo{author}{\bibnamefont{Mecking}, \bibfnamefont{B.~A.}}, \emph{et~al.},
  \bibinfo{year}{2003}, \bibinfo{journal}{Nucl. Instrum. Meth.}
  \textbf{\bibinfo{volume}{A503}}, \bibinfo{pages}{513}.

\bibitem[{\citenamefont{Mei{\ss}ner}(2000)}]{Meissner:1999vr}
\bibinfo{author}{\bibnamefont{Mei{\ss}ner}, \bibfnamefont{Ulf-G.}}, and
  \bibinfo{author}{\bibfnamefont{J.~A.} \bibnamefont{Oller}},
  \bibinfo{year}{2000}, \bibinfo{journal}{Nucl. Phys.}
  \textbf{\bibinfo{volume}{A673}}, \bibinfo{pages}{311}.

\bibitem[{\citenamefont{Melde} \emph{et~al.}(2008)\citenamefont{Melde, Plessas,
  and Sengl}}]{Melde:2008yr}
\bibinfo{author}{\bibnamefont{Melde}, \bibfnamefont{T.}},
  \bibinfo{author}{\bibfnamefont{W.}~\bibnamefont{Plessas}}, and
  \bibinfo{author}{\bibfnamefont{B.}~\bibnamefont{Sengl}},
  \bibinfo{year}{2008}, \bibinfo{journal}{Phys. Rev.}
  \textbf{\bibinfo{volume}{D77}}, \bibinfo{pages}{114002}.

\bibitem[{\citenamefont{Melde} \emph{et~al.}(2005)\citenamefont{Melde, Plessas,
  and Wagenbrunn}}]{Melde:2005hy}
\bibinfo{author}{\bibnamefont{Melde}, \bibfnamefont{T.}},
  \bibinfo{author}{\bibfnamefont{W.}~\bibnamefont{Plessas}}, and
  \bibinfo{author}{\bibfnamefont{R.~F.} \bibnamefont{Wagenbrunn}},
  \bibinfo{year}{2005}, \bibinfo{journal}{Phys. Rev.}
  \textbf{\bibinfo{volume}{C72}}, \bibinfo{pages}{015207}.

\bibitem[{Melnitchouk \emph{et~al.}(2003)\citenamefont{Melnitchouk}
  \emph{et~al.}}]{Melnitchouk:2002eg}
\bibinfo{author}{\bibnamefont{Melnitchouk}, \bibfnamefont{W.}}, \emph{et~al.},
  \bibinfo{year}{2003}, \bibinfo{journal}{Phys. Rev.}
  \textbf{\bibinfo{volume}{D67}}, \bibinfo{pages}{114506}.

\bibitem[{Merkel \emph{et~al.}(2007)\citenamefont{Merkel}
  \emph{et~al.}}]{Merkel:2007ig}
\bibinfo{author}{\bibnamefont{Merkel}, \bibfnamefont{H.}}, \emph{et~al.},
  \bibinfo{year}{2007}, \bibinfo{journal}{Phys. Rev. Lett.}
  \textbf{\bibinfo{volume}{99}}, \bibinfo{pages}{132301}.

\bibitem[{\citenamefont{Metsch} \emph{et~al.}(2003)\citenamefont{Metsch,
  Loring, Merten, and Petry}}]{Metsch:2003ix}
\bibinfo{author}{\bibnamefont{Metsch}, \bibfnamefont{B.}},
  \bibinfo{author}{\bibfnamefont{U.}~\bibnamefont{Loring}},
  \bibinfo{author}{\bibfnamefont{D.}~\bibnamefont{Merten}}, and
  \bibinfo{author}{\bibfnamefont{H.}~\bibnamefont{Petry}},
  \bibinfo{year}{2003}, \bibinfo{journal}{Eur. Phys. J.}
  \textbf{\bibinfo{volume}{A18}}, \bibinfo{pages}{189}.

\bibitem[{Mibe \emph{et~al.}(2005)\citenamefont{Mibe}
  \emph{et~al.}}]{Mibe:2005er}
\bibinfo{author}{\bibnamefont{Mibe}, \bibfnamefont{T.}}, \emph{et~al.},
  \bibinfo{year}{2005}, \bibinfo{journal}{Phys. Rev. Lett.}
  \textbf{\bibinfo{volume}{95}}, \bibinfo{pages}{182001}.

\bibitem[{\citenamefont{Michael}(2006)}]{Michael:2005kw}
\bibinfo{author}{\bibnamefont{Michael}, \bibfnamefont{C.}},
  \bibinfo{year}{2006}, \bibinfo{journal}{PoS}
  \textbf{\bibinfo{volume}{LAT2005}}, \bibinfo{pages}{008}.

\bibitem[{\citenamefont{Migura}
  \emph{et~al.}(2006{\natexlab{a}})\citenamefont{Migura, Merten, Metsch, and
  Petry}}]{Migura:2006ep}
\bibinfo{author}{\bibnamefont{Migura}, \bibfnamefont{S.}},
  \bibinfo{author}{\bibfnamefont{D.}~\bibnamefont{Merten}},
  \bibinfo{author}{\bibfnamefont{B.}~\bibnamefont{Metsch}}, and
  \bibinfo{author}{\bibfnamefont{H.-R.} \bibnamefont{Petry}},
  \bibinfo{year}{2006}{\natexlab{a}}, \bibinfo{journal}{Eur. Phys. J.}
  \textbf{\bibinfo{volume}{A28}}, \bibinfo{pages}{41}.

\bibitem[{\citenamefont{Migura}
  \emph{et~al.}(2006{\natexlab{b}})\citenamefont{Migura, Merten, Metsch, and
  Petry}}]{Migura:2006en}
\bibinfo{author}{\bibnamefont{Migura}, \bibfnamefont{S.}},
  \bibinfo{author}{\bibfnamefont{D.}~\bibnamefont{Merten}},
  \bibinfo{author}{\bibfnamefont{B.}~\bibnamefont{Metsch}}, and
  \bibinfo{author}{\bibfnamefont{H.-R.} \bibnamefont{Petry}},
  \bibinfo{year}{2006}{\natexlab{b}}, \bibinfo{journal}{Eur. Phys. J.}
  \textbf{\bibinfo{volume}{A28}}, \bibinfo{pages}{55}.

\bibitem[{\citenamefont{Minami}(1968)}]{Minami:1968nr}
\bibinfo{author}{\bibnamefont{Minami}, \bibfnamefont{S.}},
  \bibinfo{year}{1968}, \bibinfo{journal}{Prog. Theor. Phys.}
  \textbf{\bibinfo{volume}{40}}, \bibinfo{pages}{1068}.

\bibitem[{Mizuk \emph{et~al.}(2005)\citenamefont{Mizuk}
  \emph{et~al.}}]{Mizuk:2004yu}
\bibinfo{author}{\bibnamefont{Mizuk}, \bibfnamefont{R.}}, \emph{et~al.},
  \bibinfo{year}{2005}, \bibinfo{journal}{Phys. Rev. Lett.}
  \textbf{\bibinfo{volume}{94}}, \bibinfo{pages}{122002}.

\bibitem[{Mizuk \emph{et~al.}(2008)\citenamefont{Mizuk}
  \emph{et~al.}}]{Mizuk:2008me}
\bibinfo{author}{\bibnamefont{Mizuk}, \bibfnamefont{R.}}, \emph{et~al.},
  \bibinfo{year}{2008}, \bibinfo{journal}{Phys. Rev.}
  \textbf{\bibinfo{volume}{D78}}, \bibinfo{pages}{072004}.

\bibitem[{Mokeev \emph{et~al.}(2008)\citenamefont{Mokeev}
  \emph{et~al.}}]{Mokeev:2008iw}
\bibinfo{author}{\bibnamefont{Mokeev}, \bibfnamefont{V.~I.}}, \emph{et~al.},
  \bibinfo{year}{2008}, \eprint{0809.4158}.

\bibitem[{Mokhtari \emph{et~al.}(1985)\citenamefont{Mokhtari}
  \emph{et~al.}}]{Mokhtari:1985cb}
\bibinfo{author}{\bibnamefont{Mokhtari}, \bibfnamefont{A.}}, \emph{et~al.},
  \bibinfo{year}{1985}, \bibinfo{journal}{Phys. Rev. Lett.}
  \textbf{\bibinfo{volume}{55}}, \bibinfo{pages}{359}.

\bibitem[{Mokhtari \emph{et~al.}(1987)\citenamefont{Mokhtari}
  \emph{et~al.}}]{Mokhtari:1987iy}
\bibinfo{author}{\bibnamefont{Mokhtari}, \bibfnamefont{A.}}, \emph{et~al.},
  \bibinfo{year}{1987}, \bibinfo{journal}{Phys. Rev.}
  \textbf{\bibinfo{volume}{D35}}, \bibinfo{pages}{810}.

\bibitem[{Morand \emph{et~al.}(2005)\citenamefont{Morand}
  \emph{et~al.}}]{Morand:2005ex}
\bibinfo{author}{\bibnamefont{Morand}, \bibfnamefont{L.}}, \emph{et~al.},
  \bibinfo{year}{2005}, \bibinfo{journal}{Eur. Phys. J.}
  \textbf{\bibinfo{volume}{A24}}, \bibinfo{pages}{445}.

\bibitem[{Morrow \emph{et~al.}(2009)\citenamefont{Morrow}
  \emph{et~al.}}]{Morrow:2008ek}
\bibinfo{author}{\bibnamefont{Morrow}, \bibfnamefont{S.~A.}}, \emph{et~al.},
  \bibinfo{year}{2009}, \bibinfo{journal}{Eur. Phys. J.}
  \textbf{\bibinfo{volume}{A39}}, \bibinfo{pages}{5}.

\bibitem[{\citenamefont{Morsch and Zupranski}(2000)}]{Morsch:2000xi}
\bibinfo{author}{\bibnamefont{Morsch}, \bibfnamefont{H.~P.}}, and
  \bibinfo{author}{\bibfnamefont{P.}~\bibnamefont{Zupranski}},
  \bibinfo{year}{2000}, \bibinfo{journal}{Phys. Rev.}
  \textbf{\bibinfo{volume}{C61}}, \bibinfo{pages}{024002}.

\bibitem[{\citenamefont{Nacher} \emph{et~al.}(2001)\citenamefont{Nacher, Oset,
  Vicente, and Roca}}]{Nacher:2000eq}
\bibinfo{author}{\bibnamefont{Nacher}, \bibfnamefont{J.~C.}},
  \bibinfo{author}{\bibfnamefont{E.}~\bibnamefont{Oset}},
  \bibinfo{author}{\bibfnamefont{M.~J.} \bibnamefont{Vicente}}, and
  \bibinfo{author}{\bibfnamefont{L.}~\bibnamefont{Roca}}, \bibinfo{year}{2001},
  \bibinfo{journal}{Nucl. Phys.} \textbf{\bibinfo{volume}{A695}},
  \bibinfo{pages}{295}.

\bibitem[{\citenamefont{Nadkarni and Nielsen}(1986)}]{Nadkarni:1985dm}
\bibinfo{author}{\bibnamefont{Nadkarni}, \bibfnamefont{S.}}, and
  \bibinfo{author}{\bibfnamefont{H.~B.} \bibnamefont{Nielsen}},
  \bibinfo{year}{1986}, \bibinfo{journal}{Nucl. Phys.}
  \textbf{\bibinfo{volume}{B263}}, \bibinfo{pages}{1}.

\bibitem[{Nakabayashi \emph{et~al.}(2006)\citenamefont{Nakabayashi}
  \emph{et~al.}}]{Nakabayashi:2006ut}
\bibinfo{author}{\bibnamefont{Nakabayashi}, \bibfnamefont{T.}}, \emph{et~al.},
  \bibinfo{year}{2006}, \bibinfo{journal}{Phys. Rev.}
  \textbf{\bibinfo{volume}{C74}}, \bibinfo{pages}{035202}.

\bibitem[{Nakano \emph{et~al.}(2003)\citenamefont{Nakano}
  \emph{et~al.}}]{Nakano:2003qx}
\bibinfo{author}{\bibnamefont{Nakano}, \bibfnamefont{T.}}, \emph{et~al.},
  \bibinfo{year}{2003}, \bibinfo{journal}{Phys. Rev. Lett.}
  \textbf{\bibinfo{volume}{91}}, \bibinfo{pages}{012002}.

\bibitem[{Nakano \emph{et~al.}(2009)\citenamefont{Nakano}
  \emph{et~al.}}]{Nakano:2008ee}
\bibinfo{author}{\bibnamefont{Nakano}, \bibfnamefont{T.}}, \emph{et~al.},
  \bibinfo{year}{2009}, \bibinfo{journal}{Phys. Rev.}
  \textbf{\bibinfo{volume}{C 78}}, \bibinfo{pages}{025210}.

\bibitem[{\citenamefont{Nakayama} \emph{et~al.}(2008)\citenamefont{Nakayama,
  Oh, and Haberzettl}}]{Nakayama:2008tg}
\bibinfo{author}{\bibnamefont{Nakayama}, \bibfnamefont{K.}},
  \bibinfo{author}{\bibfnamefont{Y.}~\bibnamefont{Oh}}, and
  \bibinfo{author}{\bibfnamefont{H.}~\bibnamefont{Haberzettl}},
  \bibinfo{year}{2008}, \eprint{0803.3169}.

\bibitem[{\citenamefont{Nambu and Jona-Lasinio}(1961)}]{Nambu:1961tp}
\bibinfo{author}{\bibnamefont{Nambu}, \bibfnamefont{Y.}}, and
  \bibinfo{author}{\bibfnamefont{G.}~\bibnamefont{Jona-Lasinio}},
  \bibinfo{year}{1961}, \bibinfo{journal}{Phys. Rev.}
  \textbf{\bibinfo{volume}{122}}, \bibinfo{pages}{345}.

\bibitem[{Nanova \emph{et~al.}(2008)\citenamefont{Nanova}
  \emph{et~al.}}]{Nanova:2008kr}
\bibinfo{author}{\bibnamefont{Nanova}, \bibfnamefont{M.}}, \emph{et~al.},
  \bibinfo{year}{2008}, \bibinfo{journal}{Eur. Phys. J.}
  \textbf{\bibinfo{volume}{A35}}, \bibinfo{pages}{333}.

\bibitem[{\citenamefont{Narison}(2004)}]{Narison:994162}
\bibinfo{author}{\bibnamefont{Narison}, \bibfnamefont{S.}},
  \bibinfo{year}{2004}, \emph{\bibinfo{title}{QCD as a theory of hadrons: from
  partons to confinement; electronic version}} (\bibinfo{publisher}{Cambridge
  Univ. Press}, \bibinfo{address}{Cambridge}).

\bibitem[{Nasseripour \emph{et~al.}(2008)\citenamefont{Nasseripour}
  \emph{et~al.}}]{Nasseripour:2008fz}
\bibinfo{author}{\bibnamefont{Nasseripour}, \bibfnamefont{R.}}, \emph{et~al.},
  \bibinfo{year}{2008}, \bibinfo{journal}{Phys. Rev.}
  \textbf{\bibinfo{volume}{C77}}, \bibinfo{pages}{065208}.

\bibitem[{Naumann \emph{et~al.}(2003)\citenamefont{Naumann}
  \emph{et~al.}}]{Naumann:2003vf}
\bibinfo{author}{\bibnamefont{Naumann}, \bibfnamefont{J.}}, \emph{et~al.},
  \bibinfo{year}{2003}, \bibinfo{journal}{Nucl. Instrum. Meth.}
  \textbf{\bibinfo{volume}{A498}}, \bibinfo{pages}{211}.

\bibitem[{\citenamefont{Nefkens}(2001)}]{Nefkens:prcomm}
\bibinfo{author}{\bibnamefont{Nefkens}, \bibfnamefont{B.}},
  \bibinfo{year}{2001}, \bibinfo{note}{18th Students' Workshop on
  Electromagnetic Interactions in Bosen/Saar, Sept. 2-7}.

\bibitem[{\citenamefont{Nefkens} \emph{et~al.}(2002)\citenamefont{Nefkens,
  Prakhov, and Starostin}}]{Nefkens:2002rz}
\bibinfo{author}{\bibnamefont{Nefkens}, \bibfnamefont{B.~M.~K.}},
  \bibinfo{author}{\bibfnamefont{S.}~\bibnamefont{Prakhov}}, and
  \bibinfo{author}{\bibfnamefont{A.}~\bibnamefont{Starostin}}, \bibinfo{year}{2002},
  \bibinfo{note}{international Workshop on Chiral Fluctuations in Hadronic
  Matter, Orsay, France, Sep. 26-28, 2001}, \eprint{nucl-ex/0202007}.

\bibitem[{Niiyama \emph{et~al.}(2008)\citenamefont{Niiyama}
  \emph{et~al.}}]{Niiyama:2008rt}
\bibinfo{author}{\bibnamefont{Niiyama}, \bibfnamefont{M.}}, \emph{et~al.},
  \bibinfo{year}{2008}, \bibinfo{journal}{Phys. Rev.}
  \textbf{\bibinfo{volume}{C78}}, \bibinfo{pages}{035202}.

\bibitem[{\citenamefont{Nikonov} \emph{et~al.}(2008)\citenamefont{Nikonov,
  Anisovich, Klempt, Sarantsev, and Thoma}}]{Nikonov:2007br}
\bibinfo{author}{\bibnamefont{Nikonov}, \bibfnamefont{V.~A.}},
  \bibinfo{author}{\bibfnamefont{A.~V.} \bibnamefont{Anisovich}},
  \bibinfo{author}{\bibfnamefont{E.}~\bibnamefont{Klempt}},
  \bibinfo{author}{\bibfnamefont{A.~V.} \bibnamefont{Sarantsev}}, and
  \bibinfo{author}{\bibfnamefont{U.}~\bibnamefont{Thoma}},
  \bibinfo{year}{2008}, \bibinfo{journal}{Phys. Lett.}
  \textbf{\bibinfo{volume}{B662}}, \bibinfo{pages}{245}.

\bibitem[{\citenamefont{Nussinov and Lampert}(2002)}]{Nussinov:1999sx}
\bibinfo{author}{\bibnamefont{Nussinov}, \bibfnamefont{S.}}, and
  \bibinfo{author}{\bibfnamefont{M.~A.} \bibnamefont{Lampert}},
  \bibinfo{year}{2002}, \bibinfo{journal}{Phys. Rept.}
  \textbf{\bibinfo{volume}{362}}, \bibinfo{pages}{193}.

\bibitem[{Ocherashvili \emph{et~al.}(2005)\citenamefont{Ocherashvili}
  \emph{et~al.}}]{Ocherashvili:2004hi}
\bibinfo{author}{\bibnamefont{Ocherashvili}, \bibfnamefont{A.}}, \emph{et~al.},
  \bibinfo{year}{2005}, \bibinfo{journal}{Phys. Lett.}
  \textbf{\bibinfo{volume}{B628}}, \bibinfo{pages}{18}.

\bibitem[{\citenamefont{Oller} \emph{et~al.}(2000)\citenamefont{Oller, Oset,
  and Ramos}}]{Oller:2000ma}
\bibinfo{author}{\bibnamefont{Oller}, \bibfnamefont{J.~A.}},
  \bibinfo{author}{\bibfnamefont{E.}~\bibnamefont{Oset}}, and
  \bibinfo{author}{\bibfnamefont{A.}~\bibnamefont{Ramos}},
  \bibinfo{year}{2000}, \bibinfo{journal}{Prog. Part. Nucl. Phys.}
  \textbf{\bibinfo{volume}{45}}, \bibinfo{pages}{157}.

\bibitem[{\citenamefont{Oller}(2006)\citenamefont{Oller}}]{Oller:2006jw}
\bibinfo{author}{\bibnamefont{Oller}, \bibfnamefont{J.~A.}},
  \bibinfo{year}{2006}, \bibinfo{journal}{Eur. Phys. J.}
  \textbf{\bibinfo{volume}{A28}}, \bibinfo{pages}{63}.

\bibitem[{Olmsted \emph{et~al.}(2004)\citenamefont{Olmsted}
  \emph{et~al.}}]{Olmsted:2003is}
\bibinfo{author}{\bibnamefont{Olmsted}, \bibfnamefont{J.}}, \emph{et~al.},
  \bibinfo{year}{2004}, \bibinfo{journal}{Phys. Lett.}
  \textbf{\bibinfo{volume}{B588}}, \bibinfo{pages}{29}.

\bibitem[{\citenamefont{Oset and Ramos}(1998)}]{Oset:1997it}
\bibinfo{author}{\bibnamefont{Oset}, \bibfnamefont{E.}}, and
  \bibinfo{author}{\bibfnamefont{A.}~\bibnamefont{Ramos}},
  \bibinfo{year}{1998}, \bibinfo{journal}{Nucl. Phys.}
  \textbf{\bibinfo{volume}{A635}}, \bibinfo{pages}{99}.

\bibitem[{Osipenko \emph{et~al.}(2008)\citenamefont{Osipenko}
  \emph{et~al.}}]{Osipenko:2008rv}
\bibinfo{author}{\bibnamefont{Osipenko}, \bibfnamefont{M.}},
\emph{et~al.}, \bibinfo{year}{2008}, \eprint{0809.1153}.

\bibitem[{van Pee \emph{et~al.}(2007)\citenamefont{van Pee}
  \emph{et~al.}}]{vanPee:2007tw}
\bibinfo{author}{\bibnamefont{van Pee}, \bibfnamefont{H.}}, \emph{et~al.},
  \bibinfo{year}{2007}, \bibinfo{journal}{Eur. Phys. J.}
  \textbf{\bibinfo{volume}{A31}}, \bibinfo{pages}{61}.

\bibitem[{\citenamefont{Penner and Mosel}(2002{\natexlab{a}})}]{Penner:2002ma}
\bibinfo{author}{\bibnamefont{Penner}, \bibfnamefont{G.}}, and
  \bibinfo{author}{\bibfnamefont{U.}~\bibnamefont{Mosel}},
  \bibinfo{year}{2002}{\natexlab{a}}, \bibinfo{journal}{Phys. Rev.}
  \textbf{\bibinfo{volume}{C66}}, \bibinfo{pages}{055211}.

\bibitem[{\citenamefont{Penner and Mosel}(2002{\natexlab{b}})}]{Penner:2002md}
\bibinfo{author}{\bibnamefont{Penner}, \bibfnamefont{G.}}, and
  \bibinfo{author}{\bibfnamefont{U.}~\bibnamefont{Mosel}},
  \bibinfo{year}{2002}{\natexlab{b}}, \bibinfo{journal}{Phys. Rev.}
  \textbf{\bibinfo{volume}{C66}}, \bibinfo{pages}{055212}.

\bibitem[{\citenamefont{Petersen}(2006)}]{Petersen:2006xd}
\bibinfo{author}{\bibnamefont{Petersen}, \bibfnamefont{B.~A.}},
  \bibinfo{year}{2006}, \bibinfo{note}{$33^{\rm rd}$ International Conference
  on High Energy Physics (ICHEP 06), Moscow, Russia, 26 Jul - 2 Aug 2006.
  Published in *Moscow 2006, ICHEP* 919-922, hep-ex/0610049}.

\bibitem[{Pocanic \emph{et~al.}(1994)\citenamefont{Pocanic}
  \emph{et~al.}}]{Pocanic:1993mp}
\bibinfo{author}{\bibnamefont{Pocanic}, \bibfnamefont{D.}}, \emph{et~al.},
  \bibinfo{year}{1994}, \bibinfo{journal}{Phys. Rev. Lett.}
  \textbf{\bibinfo{volume}{72}}, \bibinfo{pages}{1156}.

\bibitem[{\citenamefont{Polchinski and Strassler}(2002)}]{Polchinski:2001tt}
\bibinfo{author}{\bibnamefont{Polchinski}, \bibfnamefont{J.}}, and
  \bibinfo{author}{\bibfnamefont{M.~J.} \bibnamefont{Strassler}},
  \bibinfo{year}{2002}, \bibinfo{journal}{Phys. Rev. Lett.}
  \textbf{\bibinfo{volume}{88}}, \bibinfo{pages}{031601}.

\bibitem[{Prakhov \emph{et~al.}(2004{\natexlab{a}})\citenamefont{Prakhov}
  \emph{et~al.}}]{Prakhov:2004an}
\bibinfo{author}{\bibnamefont{Prakhov}, \bibfnamefont{S.}}, \emph{et~al.},
  \bibinfo{year}{2004}{\natexlab{a}}, \bibinfo{journal}{Phys. Rev.}
  \textbf{\bibinfo{volume}{C70}}, \bibinfo{pages}{034605}.

\bibitem[{Prakhov \emph{et~al.}(2004{\natexlab{b}})\citenamefont{Prakhov}
  \emph{et~al.}}]{Prakhov:2004zv}
\bibinfo{author}{\bibnamefont{Prakhov}, \bibfnamefont{S.}}, \emph{et~al.},
  \bibinfo{year}{2004}{\natexlab{b}}, \bibinfo{journal}{Phys. Rev.}
  \textbf{\bibinfo{volume}{C69}}, \bibinfo{pages}{045202}.

\bibitem[{Prakhov \emph{et~al.}(2004{\natexlab{c}})\citenamefont{Prakhov}
  \emph{et~al.}}]{Prakhov:2004ri}
\bibinfo{author}{\bibnamefont{Prakhov}, \bibfnamefont{S.}}, \emph{et~al.},
  \bibinfo{year}{2004}{\natexlab{c}}, \bibinfo{journal}{Phys. Rev.}
  \textbf{\bibinfo{volume}{C69}}, \bibinfo{pages}{042202}.

\bibitem[{Prakhov \emph{et~al.}(2005)\citenamefont{Prakhov}
  \emph{et~al.}}]{Prakhov:2005qb}
\bibinfo{author}{\bibnamefont{Prakhov}, \bibfnamefont{S.}}, \emph{et~al.},
  \bibinfo{year}{2005}, \bibinfo{journal}{Phys. Rev.}
  \textbf{\bibinfo{volume}{C72}}, \bibinfo{pages}{015203}.

\bibitem[{\citenamefont{\protect{L\"oring}}
  \emph{et~al.}(2001{\natexlab{a}})\citenamefont{\protect{L\"oring},
  Kretzschmar, Metsch, and Petry}}]{Loring:2001kv}
\bibinfo{author}{\bibnamefont{\protect{L\"oring}}, \bibfnamefont{U.}},
  \bibinfo{author}{\bibfnamefont{K.}~\bibnamefont{Kretzschmar}},
  \bibinfo{author}{\bibfnamefont{B.~C.} \bibnamefont{Metsch}}, and
  \bibinfo{author}{\bibfnamefont{H.~R.} \bibnamefont{Petry}},
  \bibinfo{year}{2001}{\natexlab{a}}, \bibinfo{journal}{Eur. Phys. J.}
  \textbf{\bibinfo{volume}{A10}}, \bibinfo{pages}{309}.

\bibitem[{\citenamefont{\protect{L\"oring}}
  \emph{et~al.}(2001{\natexlab{b}})\citenamefont{\protect{L\"oring}, Metsch,
  and Petry}}]{Loring:2001kx}
\bibinfo{author}{\bibnamefont{\protect{L\"oring}}, \bibfnamefont{U.}},
  \bibinfo{author}{\bibfnamefont{B.~C.} \bibnamefont{Metsch}}, and
  \bibinfo{author}{\bibfnamefont{H.~R.} \bibnamefont{Petry}},
  \bibinfo{year}{2001}{\natexlab{b}}, \bibinfo{journal}{Eur. Phys. J.}
  \textbf{\bibinfo{volume}{A10}}, \bibinfo{pages}{395}.

\bibitem[{\citenamefont{\protect{L\"oring}}
  \emph{et~al.}(2001{\natexlab{c}})\citenamefont{\protect{L\"oring}, Metsch,
  and Petry}}]{Loring:2001ky}
\bibinfo{author}{\bibnamefont{\protect{L\"oring}}, \bibfnamefont{U.}},
  \bibinfo{author}{\bibfnamefont{B.~C.} \bibnamefont{Metsch}}, and
  \bibinfo{author}{\bibfnamefont{H.~R.} \bibnamefont{Petry}},
  \bibinfo{year}{2001}{\natexlab{c}}, \bibinfo{journal}{Eur. Phys. J.}
  \textbf{\bibinfo{volume}{A10}}, \bibinfo{pages}{447}.

\bibitem[{\citenamefont{Ratti}(2003)}]{Ratti:2003ez}
\bibinfo{author}{\bibnamefont{Ratti}, \bibfnamefont{S.~P.}},
  \bibinfo{year}{2003}, \bibinfo{journal}{Nucl. Phys. Proc. Suppl.}
  \textbf{\bibinfo{volume}{115}}, \bibinfo{pages}{33}.

\bibitem[{\citenamefont{Regge}(1959)}]{Regge:1959mz}
\bibinfo{author}{\bibnamefont{Regge}, \bibfnamefont{T.}}, \bibinfo{year}{1959},
  \bibinfo{journal}{Nuovo Cim.} \textbf{\bibinfo{volume}{14}},
  \bibinfo{pages}{951}.

\bibitem[{\citenamefont{Regge}(1960)}]{Regge:1960zc}
\bibinfo{author}{\bibnamefont{Regge}, \bibfnamefont{T.}}, \bibinfo{year}{1960},
  \bibinfo{journal}{Nuovo Cim.} \textbf{\bibinfo{volume}{18}},
  \bibinfo{pages}{947}.

\bibitem[{\citenamefont{Reinders} \emph{et~al.}(1985)\citenamefont{Reinders,
  Rubinstein, and Yazaki}}]{Reinders:1984sr}
\bibinfo{author}{\bibnamefont{Reinders}, \bibfnamefont{L.~J.}},
  \bibinfo{author}{\bibfnamefont{H.}~\bibnamefont{Rubinstein}}, and
  \bibinfo{author}{\bibfnamefont{S.}~\bibnamefont{Yazaki}},
  \bibinfo{year}{1985}, \bibinfo{journal}{Phys. Rept.}
  \textbf{\bibinfo{volume}{127}}, \bibinfo{pages}{1}.

\bibitem[{\citenamefont{Rho} \emph{et~al.}(1992)\citenamefont{Rho, Riska, and
  Scoccola}}]{Rho:1992yy}
\bibinfo{author}{\bibnamefont{Rho}, \bibfnamefont{M.}},
  \bibinfo{author}{\bibfnamefont{D.~O.} \bibnamefont{Riska}}, and
  \bibinfo{author}{\bibfnamefont{N.~N.} \bibnamefont{Scoccola}},
  \bibinfo{year}{1992}, \bibinfo{journal}{Z. Phys.}
  \textbf{\bibinfo{volume}{A341}}, \bibinfo{pages}{343}.

\bibitem[{\citenamefont{Richard}(1981)}]{Richard:1980tg}
\bibinfo{author}{\bibnamefont{Richard}, \bibfnamefont{J.~M.}},
  \bibinfo{year}{1981}, \bibinfo{journal}{Phys. Lett.}
  \textbf{\bibinfo{volume}{B100}}, \bibinfo{pages}{515}.

\bibitem[{\citenamefont{Richard}(1992)}]{Richard:1992uk}
\bibinfo{author}{\bibnamefont{Richard}, \bibfnamefont{J.~M.}},
  \bibinfo{year}{1992}, \bibinfo{journal}{Phys. Rept.}
  \textbf{\bibinfo{volume}{212}}, \bibinfo{pages}{1}.

\bibitem[{\citenamefont{Richard and Taxil}(1983)}]{Richard:1983mu}
\bibinfo{author}{\bibnamefont{Richard}, \bibfnamefont{J.~M.}}, and
  \bibinfo{author}{\bibfnamefont{P.}~\bibnamefont{Taxil}},
  \bibinfo{year}{1983}, \bibinfo{journal}{Ann. Phys.}
  \textbf{\bibinfo{volume}{150}}, \bibinfo{pages}{267}.

\bibitem[{Ripani \emph{et~al.}(2003)\citenamefont{Ripani}
  \emph{et~al.}}]{Ripani:2002ss}
\bibinfo{author}{\bibnamefont{Ripani}, \bibfnamefont{M.}}, \emph{et~al.},
  \bibinfo{year}{2003}, \bibinfo{journal}{Phys. Rev. Lett.}
  \textbf{\bibinfo{volume}{91}}, \bibinfo{pages}{022002}.

\bibitem[{\citenamefont{Rosner}(2007)}]{Rosner:2006jz}
\bibinfo{author}{\bibnamefont{Rosner}, \bibfnamefont{J.~L.}},
  \bibinfo{year}{2007}, \bibinfo{journal}{J. Phys.}
  \textbf{\bibinfo{volume}{G34}}, \bibinfo{pages}{S127}.

\bibitem[{Sadler \emph{et~al.}(2004)\citenamefont{Sadler}
  \emph{et~al.}}]{Sadler:2004yq}
\bibinfo{author}{\bibnamefont{Sadler}, \bibfnamefont{M.~E.}}, \emph{et~al.},
  \bibinfo{year}{2004}, \bibinfo{journal}{Phys. Rev.}
  \textbf{\bibinfo{volume}{C69}}, \bibinfo{pages}{055206}.

\bibitem[{Santoro \emph{et~al.}(2008)\citenamefont{Santoro}
  \emph{et~al.}}]{Santoro:2008ai}
\bibinfo{author}{\bibnamefont{Santoro}, \bibfnamefont{J.~P.}}, \emph{et~al.},
\bibinfo{year}{2008},
  \bibinfo{journal}{Phys. Rev.} \textbf{\bibinfo{volume}{C78}},
  \bibinfo{pages}{025210}.

\bibitem[{\citenamefont{Sarantsev} \emph{et~al.}(2005)\citenamefont{Sarantsev,
  Nikonov, Anisovich, Klempt, and Thoma}}]{Sarantsev:2005tg}
\bibinfo{author}{\bibnamefont{Sarantsev}, \bibfnamefont{A.~V.}},
  \bibinfo{author}{\bibfnamefont{V.~A.} \bibnamefont{Nikonov}},
  \bibinfo{author}{\bibfnamefont{A.~V.} \bibnamefont{Anisovich}},
  \bibinfo{author}{\bibfnamefont{E.}~\bibnamefont{Klempt}}, and
  \bibinfo{author}{\bibfnamefont{U.}~\bibnamefont{Thoma}},
  \bibinfo{year}{2005}, \bibinfo{journal}{Eur. Phys. J.}
  \textbf{\bibinfo{volume}{A25}}, \bibinfo{pages}{441}.

\bibitem[{Sarantsev \emph{et~al.}(2008)\citenamefont{Sarantsev}
  \emph{et~al.}}]{Sarantsev:2007bk}
\bibinfo{author}{\bibnamefont{Sarantsev}, \bibfnamefont{A.~V.}}, \emph{et~al.},
  \bibinfo{year}{2008}, \bibinfo{journal}{Phys. Lett.}
  \textbf{\bibinfo{volume}{B659}}, \bibinfo{pages}{94}.

\bibitem[{\citenamefont{Sasaki and Sasaki}(2005)}]{Sasaki:2005ug}
\bibinfo{author}{\bibnamefont{Sasaki}, \bibfnamefont{K.}}, and
  \bibinfo{author}{\bibfnamefont{S.}~\bibnamefont{Sasaki}},
  \bibinfo{year}{2005}, \bibinfo{journal}{Phys. Rev.}
  \textbf{\bibinfo{volume}{D72}}, \bibinfo{pages}{034502}.

\bibitem[{\citenamefont{Sasaki} \emph{et~al.}(2005)\citenamefont{Sasaki,
  Sasaki, and Hatsuda}}]{Sasaki:2005ap}
\bibinfo{author}{\bibnamefont{Sasaki}, \bibfnamefont{K.}},
  \bibinfo{author}{\bibfnamefont{S.}~\bibnamefont{Sasaki}}, and
  \bibinfo{author}{\bibfnamefont{T.}~\bibnamefont{Hatsuda}},
  \bibinfo{year}{2005}, \bibinfo{journal}{Phys. Lett.}
  \textbf{\bibinfo{volume}{B623}}, \bibinfo{pages}{208}.

\bibitem[{\citenamefont{Schneider} \emph{et~al.}(2006)\citenamefont{Schneider,
  Krewald, and Mei{\ss}ner}}]{Schneider:2006bd}
\bibinfo{author}{\bibnamefont{Schneider}, \bibfnamefont{S.}},
  \bibinfo{author}{\bibfnamefont{S.}~\bibnamefont{Krewald}}, and
  \bibinfo{author}{\bibfnamefont{U.-G.} \bibnamefont{Mei{\ss}ner}},
  \bibinfo{year}{2006}, \bibinfo{journal}{Eur. Phys. J.}
  \textbf{\bibinfo{volume}{A28}}, \bibinfo{pages}{107}.

\bibitem[{\citenamefont{Schumacher}(2006)}]{Schumacher:2006ii}
\bibinfo{author}{\bibnamefont{Schumacher}, \bibfnamefont{R.}},
  \bibinfo{year}{2006}, \eprint{$9^{\rm th}$ International Conference on
  Hypernuclear and Strange Particle Physics (HYP 2006), Mainz, Germany, 10-14
  Oct 2006. nucl-ex/0611035}.

\bibitem[{Schwille \emph{et~al.}(1994)\citenamefont{Schwille}
  \emph{et~al.}}]{Schwille:1994vg}
\bibinfo{author}{\bibnamefont{Schwille}, \bibfnamefont{W.~J.}}, \emph{et~al.},
  \bibinfo{year}{1994}, \bibinfo{journal}{Nucl. Instrum. Meth.}
  \textbf{\bibinfo{volume}{A344}}, \bibinfo{pages}{470}.

\bibitem[{Seftor \emph{et~al.}(1989)\citenamefont{Seftor}
  \emph{et~al.}}]{Seftor:1989er}
\bibinfo{author}{\bibnamefont{Seftor}, \bibfnamefont{C.~J.}}, \emph{et~al.},
  \bibinfo{year}{1989}, \bibinfo{journal}{Phys. Rev.}
  \textbf{\bibinfo{volume}{D39}}, \bibinfo{pages}{2457}.

\bibitem[{\citenamefont{Semay} \emph{et~al.}(2001)\citenamefont{Semay, Brau,
  and Silvestre-Brac}}]{Semay:2001th}
\bibinfo{author}{\bibnamefont{Semay}, \bibfnamefont{C.}},
  \bibinfo{author}{\bibfnamefont{F.}~\bibnamefont{Brau}}, and
  \bibinfo{author}{\bibfnamefont{B.}~\bibnamefont{Silvestre-Brac}},
  \bibinfo{year}{2001}, \bibinfo{journal}{Phys. Rev.}
  \textbf{\bibinfo{volume}{C64}}, \bibinfo{pages}{055202}.

\bibitem[{\citenamefont{Sengl} \emph{et~al.}(2007)\citenamefont{Sengl, Melde,
  and Plessas}}]{Sengl:2007yq}
\bibinfo{author}{\bibnamefont{Sengl}, \bibfnamefont{B.}},
  \bibinfo{author}{\bibfnamefont{T.}~\bibnamefont{Melde}}, and
  \bibinfo{author}{\bibfnamefont{W.}~\bibnamefont{Plessas}},
  \bibinfo{year}{2007}, \bibinfo{journal}{Phys. Rev.}
  \textbf{\bibinfo{volume}{D76}}, \bibinfo{pages}{054008}.

\bibitem[{Sevior \emph{et~al.}(1991)\citenamefont{Sevior}
  \emph{et~al.}}]{Sevior:1990yp}
\bibinfo{author}{\bibnamefont{Sevior}, \bibfnamefont{M.~E.}}, \emph{et~al.},
  \bibinfo{year}{1991}, \bibinfo{journal}{Phys. Rev. Lett.}
  \textbf{\bibinfo{volume}{66}}, \bibinfo{pages}{2569}.

\bibitem[{\citenamefont{Shifman and Vainshtein}(2008)}]{Shifman:2007xn}
\bibinfo{author}{\bibnamefont{Shifman}, \bibfnamefont{M.}}, and
  \bibinfo{author}{\bibfnamefont{A.}~\bibnamefont{Vainshtein}},
  \bibinfo{year}{2008}, \bibinfo{journal}{Phys. Rev.}
  \textbf{\bibinfo{volume}{D77}}, \bibinfo{pages}{034002}.

\bibitem[{\citenamefont{Shifman} \emph{et~al.}(1979)\citenamefont{Shifman,
  Vainshtein, and Zakharov}}]{Shifman:1978bx}
\bibinfo{author}{\bibnamefont{Shifman}, \bibfnamefont{M.~A.}},
  \bibinfo{author}{\bibfnamefont{A.~I.} \bibnamefont{Vainshtein}}, and
  \bibinfo{author}{\bibfnamefont{V.~I.} \bibnamefont{Zakharov}},
  \bibinfo{year}{1979}, \bibinfo{journal}{Nucl. Phys.}
  \textbf{\bibinfo{volume}{B147}}, \bibinfo{pages}{385}.

\bibitem[{\citenamefont{Shklyar}
  \emph{et~al.}(2005{\natexlab{a}})\citenamefont{Shklyar, Lenske, and
  Mosel}}]{Shklyar:2005xg}
\bibinfo{author}{\bibnamefont{Shklyar}, \bibfnamefont{V.}},
  \bibinfo{author}{\bibfnamefont{H.}~\bibnamefont{Lenske}}, and
  \bibinfo{author}{\bibfnamefont{U.}~\bibnamefont{Mosel}},
  \bibinfo{year}{2005}{\natexlab{a}}, \bibinfo{journal}{Phys. Rev.}
  \textbf{\bibinfo{volume}{C72}}, \bibinfo{pages}{015210}.

\bibitem[{\citenamefont{Shklyar}
  \emph{et~al.}(2005{\natexlab{b}})\citenamefont{Shklyar, Lenske, Mosel, and
  Penner}}]{Shklyar:2004ba}
\bibinfo{author}{\bibnamefont{Shklyar}, \bibfnamefont{V.}},
  \bibinfo{author}{\bibfnamefont{H.}~\bibnamefont{Lenske}},
  \bibinfo{author}{\bibfnamefont{U.}~\bibnamefont{Mosel}}, and
  \bibinfo{author}{\bibfnamefont{G.}~\bibnamefont{Penner}},
  \bibinfo{year}{2005}{\natexlab{b}}, \bibinfo{journal}{Phys. Rev.}
  \textbf{\bibinfo{volume}{C71}}, \bibinfo{pages}{055206}.

\bibitem[{\citenamefont{Shklyar} \emph{et~al.}(2007)\citenamefont{Shklyar,
  Lenske, and Mosel}}]{Shklyar:2006xw}
\bibinfo{author}{\bibnamefont{Shklyar}, \bibfnamefont{V.}},
  \bibinfo{author}{\bibfnamefont{H.}~\bibnamefont{Lenske}}, and
  \bibinfo{author}{\bibfnamefont{U.}~\bibnamefont{Mosel}},
  \bibinfo{year}{2007}, \bibinfo{journal}{Phys. Lett.}
  \textbf{\bibinfo{volume}{B650}}, \bibinfo{pages}{172}.

\bibitem[{\citenamefont{Shuryak and Rosner}(1989)}]{Shuryak:1988bf}
\bibinfo{author}{\bibnamefont{Shuryak}, \bibfnamefont{E.~V.}}, and
  \bibinfo{author}{\bibfnamefont{J.~L.} \bibnamefont{Rosner}},
  \bibinfo{year}{1989}, \bibinfo{journal}{Phys. Lett.}
  \textbf{\bibinfo{volume}{B218}}, \bibinfo{pages}{72}.

\bibitem[{\citenamefont{Shyam}
\emph{et~al.}(2009)\citenamefont{Shyam, Scholten and
  Lenske}}]{Shyam:2009za}
\bibinfo{author}{\bibnamefont{Shyam}, \bibfnamefont{R.}}, and
   \bibinfo{author}{\bibfnamefont{O.}~\bibnamefont{Scholten}},
 \bibinfo{author}{\bibfnamefont{H.}~\bibnamefont{Lenske}},
  \bibinfo{year}{2009}, \eprint{arXiv:0911.3351 [hep-ph]}.

\bibitem[\citenamefont{Sibirtsev} \emph{et~al.}(2004)]{Sibirtsev:2004bg}
\bibinfo{author}{\bibnamefont{Sibirtsev}, \bibfnamefont{A.}}, \emph{et~al.},
  \bibinfo{year}{2004}, \bibinfo{journal}{Phys. Lett.}
  \textbf{\bibinfo{volume}{B 599}}, \bibinfo{pages}{230}.

\bibitem[\citenamefont{Sibirtsev, Haidenbauer, and Mei{\ss}ner}(2007)]{Sibirtsev:2006ia}
\bibinfo{author}{\bibnamefont{Sibirtsev}, \bibfnamefont{A.}},
  \bibinfo{author}{\bibfnamefont{J.}~\bibnamefont{Haidenbauer}}, and
  \bibinfo{author}{\bibfnamefont{U.~G.} \bibnamefont{Mei{\ss}ner}},
  \bibinfo{year}{2007}, \bibinfo{journal}{Phys. Rev. Lett.}
  \textbf{\bibinfo{volume}{98}}, \bibinfo{pages}{039101}.

\bibitem[\citenamefont{Sibirtsev}\emph{et~al.}(2007)]{Sibirtsev:2007wk}
\bibinfo{author}{\bibnamefont{Sibirtsev}, \bibfnamefont{A.}}, \emph{et~al.},
  \bibinfo{year}{2007}, \bibinfo{journal}{Eur. Phys. J.}
  \textbf{\bibinfo{volume}{A34}}, \bibinfo{pages}{49}.

\bibitem[\citenamefont{Sibirtsev}\emph{et~al.}(2009{\natexlab{a}})]{Sibirtsev:2009bj}
\bibinfo{author}{\bibnamefont{Sibirtsev}, \bibfnamefont{A.}}, \emph{et~al.},
  \bibinfo{year}{2009{\natexlab{a}}}, \bibinfo{journal}{Eur. Phys. J.}
  \textbf{\bibinfo{volume}{A40}}, \bibinfo{pages}{65}.

\bibitem[\citenamefont{Sibirtsev}\emph{et~al.}(2009{\natexlab{b}})]{Sibirtsev:2009kw}
\bibinfo{author}{\bibnamefont{Sibirtsev}, \bibfnamefont{A.}}, \emph{et~al.},
  \bibinfo{year}{2009{\natexlab{b}}}, \bibinfo{journal}{Eur. Phys. J.}
  \textbf{\bibinfo{volume}{A41}}, \bibinfo{pages}{71}.

\bibitem[{Solovieva \emph{et~al.}(2005)\citenamefont{Solovieva}
  \emph{et~al.}}]{Solovieva:2008fw}
\bibinfo{author}{\bibnamefont{Solovieva}, \bibfnamefont{E.}}, \emph{et~al.},
  \bibinfo{year}{2005}, \bibinfo{journal}{Phys. Lett.}
  \textbf{\bibinfo{volume}{B672}}, \bibinfo{pages}{1}.

\bibitem[{Stanley and Robson}(1980)\citenamefont{Stanley}]{Stanley:1980fe}
\bibinfo{author}{\bibnamefont{Stanley}, \bibfnamefont{D.~P.}}, and
  \bibinfo{author}{\bibfnamefont{D.}~\bibnamefont{Robson}},
  \bibinfo{year}{1980}, \bibinfo{journal}{Phys. Rev. Lett.}
  \textbf{\bibinfo{volume}{45}}, \bibinfo{pages}{235}.

\bibitem[{Starostin \emph{et~al.}(2005)\citenamefont{Starostin}
  \emph{et~al.}}]{Starostin:2005pd}
\bibinfo{author}{\bibnamefont{Starostin}, \bibfnamefont{A.}}, \emph{et~al.},
  \bibinfo{year}{2005}, \bibinfo{journal}{Phys. Rev.}
  \textbf{\bibinfo{volume}{C72}}, \bibinfo{pages}{015205}.

\bibitem[{Stepanyan \emph{et~al.}(2003)\citenamefont{Stepanyan}
  \emph{et~al.}}]{Stepanyan:2003qr}
\bibinfo{author}{\bibnamefont{Stepanyan}, \bibfnamefont{S.}}, \emph{et~al.},
  \bibinfo{year}{2003}, \bibinfo{journal}{Phys. Rev. Lett.}
  \textbf{\bibinfo{volume}{91}}, \bibinfo{pages}{252001}.

\bibitem[{Strauch \emph{et~al.}(2005)\citenamefont{Strauch}
  \emph{et~al.}}]{Strauch:2005cs}
\bibinfo{author}{\bibnamefont{Strauch}, \bibfnamefont{S.}}, \emph{et~al.},
  \bibinfo{year}{2005}, \bibinfo{journal}{Phys. Rev. Lett.}
  \textbf{\bibinfo{volume}{95}}, \bibinfo{pages}{162003}.

\bibitem[{Sumihama \emph{et~al.}(2006)\citenamefont{Sumihama}
  \emph{et~al.}}]{Sumihama:2005er}
\bibinfo{author}{\bibnamefont{Sumihama}, \bibfnamefont{M.}}, \emph{et~al.},
  \bibinfo{year}{2006}, \bibinfo{journal}{Phys. Rev.}
  \textbf{\bibinfo{volume}{C73}}, \bibinfo{pages}{035214}.

\bibitem[{\citenamefont{Suzuki} \emph{et~al.}(2009)\citenamefont{Suzuki, Sato,
  and Lee}}]{Suzuki:2008rp}
\bibinfo{author}{\bibnamefont{Suzuki}, \bibfnamefont{N.}},
  \bibinfo{author}{\bibfnamefont{T.}~\bibnamefont{Sato}}, and
  \bibinfo{author}{\bibfnamefont{T.~S.~H.} \bibnamefont{Lee}},
  \bibinfo{year}{2009}, \bibinfo{journal}{Phys. Rev.}
  \textbf{\bibinfo{volume}{C79}}, \bibinfo{pages}{025205}.

\bibitem[{\citenamefont{Tang and Norbury}(2000)}]{Tang:2000tb}
\bibinfo{author}{\bibnamefont{Tang}, \bibfnamefont{A.}}, and
  \bibinfo{author}{\bibfnamefont{J.~W.} \bibnamefont{Norbury}},
  \bibinfo{year}{2000}, \bibinfo{journal}{Phys. Rev.}
  \textbf{\bibinfo{volume}{D62}}, \bibinfo{pages}{016006}.

\bibitem[{\citenamefont{Tariq}(2007)}]{Tariq:2007ck}
\bibinfo{author}{\bibnamefont{Tariq}, \bibfnamefont{A.~S.~B.}},
  \bibinfo{year}{2007}, \bibinfo{journal}{PoS}
  \textbf{\bibinfo{volume}{LAT2007}}, \bibinfo{pages}{136}.

\bibitem[{\citenamefont{de~Teramond and Brodsky}(2005)}]{deTeramond:2005su}
\bibinfo{author}{\bibnamefont{de~Teramond}, \bibfnamefont{G.~F.}}, and
  \bibinfo{author}{\bibfnamefont{S.~J.} \bibnamefont{Brodsky}},
  \bibinfo{year}{2005}, \bibinfo{journal}{Phys. Rev. Lett.}
  \textbf{\bibinfo{volume}{94}}, \bibinfo{pages}{201601}.

\bibitem[{\citenamefont{de~Teramond and Brodsky}(2009)}]{deTeramond:2009qx}
\bibinfo{author}{\bibnamefont{de~Teramond}, \bibfnamefont{G.~F.}}, and
  \bibinfo{author}{\bibfnamefont{S.~J.} \bibnamefont{Brodsky}},
  \bibinfo{year}{2009}, \eprint{0903.4922}.

\bibitem[{Thoma \emph{et~al.}(2008)\citenamefont{Thoma}
  \emph{et~al.}}]{Thoma:2007bm}
\bibinfo{author}{\bibnamefont{Thoma}, \bibfnamefont{U.}}, \emph{et~al.},
  \bibinfo{year}{2008}, \bibinfo{journal}{Phys. Lett.}
  \textbf{\bibinfo{volume}{B659}}, \bibinfo{pages}{87}.

\bibitem[{\citenamefont{Thomas} \emph{et~al.}(1981)\citenamefont{Thomas,
  Theberge, and Miller}}]{Thomas:1981vc}
\bibinfo{author}{\bibnamefont{Thomas}, \bibfnamefont{A.~W.}},
  \bibinfo{author}{\bibfnamefont{S.}~\bibnamefont{Theberge}}, and
  \bibinfo{author}{\bibfnamefont{G.~A.} \bibnamefont{Miller}},
  \bibinfo{year}{1981}, \bibinfo{journal}{Phys. Rev.}
  \textbf{\bibinfo{volume}{D24}}, \bibinfo{pages}{216}.

\bibitem[{\citenamefont{T{\"o}rnqvist and Zenczykowski}(1984)}]{Tornqvist:1984fy}
\bibinfo{author}{\bibnamefont{T{\"o}rnqvist}, \bibfnamefont{N.~A.}}, and
  \bibinfo{author}{\bibfnamefont{P.}~\bibnamefont{Zenczykowski}},
  \bibinfo{year}{1984}, \bibinfo{journal}{Phys. Rev.}
  \textbf{\bibinfo{volume}{D29}}, \bibinfo{pages}{2139}.

\bibitem[{\citenamefont{T{\"o}rnqvist and Zenczykowski}(1986)}]{Tornqvist:1985fi}
\bibinfo{author}{\bibnamefont{T{\"o}rnqvist}, \bibfnamefont{N.~A.}}, and
  \bibinfo{author}{\bibfnamefont{P.}~\bibnamefont{Zenczykowski}},
  \bibinfo{year}{1986}, \bibinfo{journal}{Z. Phys.}
  \textbf{\bibinfo{volume}{C30}}, \bibinfo{pages}{83}.

\bibitem[{\citenamefont{Trilling}(2008)}]{Trilling:2006wg}
\bibinfo{author}{\bibnamefont{Trilling}, \bibfnamefont{G.}},
  \bibinfo{year}{2008}, \bibinfo{note}{{Pentaquark update}, in
  \cite{Yao:2006px}}.

\bibitem[{Ungaro \emph{et~al.}(2006)\citenamefont{Ungaro}
  \emph{et~al.}}]{Ungaro:2006df}
\bibinfo{author}{\bibnamefont{Ungaro}, \bibfnamefont{M.}}, \emph{et~al.},
  \bibinfo{year}{2006}, \bibinfo{journal}{Phys. Rev. Lett.}
  \textbf{\bibinfo{volume}{97}}, \bibinfo{pages}{112003}.

\bibitem[{\citenamefont{Valcarce} \emph{et~al.}(1996)\citenamefont{Valcarce,
  Fernandez, Gonzalez, and Vento}}]{Valcarce:1995dm}
\bibinfo{author}{\bibnamefont{Valcarce}, \bibfnamefont{A.}},
  \bibinfo{author}{\bibfnamefont{F.}~\bibnamefont{Fernandez}},
  \bibinfo{author}{\bibfnamefont{P.}~\bibnamefont{Gonzalez}}, and
  \bibinfo{author}{\bibfnamefont{V.}~\bibnamefont{Vento}},
  \bibinfo{year}{1996}, \bibinfo{journal}{Phys. Lett.}
  \textbf{\bibinfo{volume}{B367}}, \bibinfo{pages}{35}.

\bibitem[{\citenamefont{Van~Dyck} \emph{et~al.}(2008)\citenamefont{Van~Dyck,
  Van~Cauteren, Ryckebusch, Metsch, and Petry}}]{VanDyck:2008zz}
\bibinfo{author}{\bibnamefont{Van~Dyck}, \bibfnamefont{A.}},
  \bibinfo{author}{\bibfnamefont{T.}~\bibnamefont{Van~Cauteren}},
  \bibinfo{author}{\bibfnamefont{J.}~\bibnamefont{Ryckebusch}},
  \bibinfo{author}{\bibfnamefont{B.~C.} \bibnamefont{Metsch}}, and
  \bibinfo{author}{\bibfnamefont{H.~R.} \bibnamefont{Petry}},
  \bibinfo{year}{2008}, \bibinfo{journal}{Prog. Part. Nucl. Phys.}
  \textbf{\bibinfo{volume}{61}}, \bibinfo{pages}{175}.

\bibitem[{\citenamefont{Varga} \emph{et~al.}(1999)\citenamefont{Varga,
  Genovese, Richard, and Silvestre-Brac}}]{Varga:1998wp}
\bibinfo{author}{\bibnamefont{Varga}, \bibfnamefont{K.}},
  \bibinfo{author}{\bibfnamefont{M.}~\bibnamefont{Genovese}},
  \bibinfo{author}{\bibfnamefont{J.-M.} \bibnamefont{Richard}}, and
  \bibinfo{author}{\bibfnamefont{B.}~\bibnamefont{Silvestre-Brac}},
  \bibinfo{year}{1999}, \bibinfo{journal}{Phys. Rev.}
  \textbf{\bibinfo{volume}{D59}}, \bibinfo{pages}{014012}.

\bibitem[{\citenamefont{Viehhauser}(2000)}]{Viehhauser:2000cu}
\bibinfo{author}{\bibnamefont{Viehhauser}, \bibfnamefont{G.}},
  \bibinfo{year}{2000}, \bibinfo{journal}{Nucl. Instrum. Meth.}
  \textbf{\bibinfo{volume}{A446}}, \bibinfo{pages}{97}.

\bibitem[{\citenamefont{Vijande} \emph{et~al.}(2008)\citenamefont{Vijande,
  Valcarce, and Fernandez}}]{Vijande:2008zn}
\bibinfo{author}{\bibnamefont{Vijande}, \bibfnamefont{J.}},
  \bibinfo{author}{\bibfnamefont{A.}~\bibnamefont{Valcarce}}, and
  \bibinfo{author}{\bibfnamefont{F.}~\bibnamefont{Fernandez}},
  \bibinfo{year}{2008}, \eprint{0810.4988}.

\bibitem[{\citenamefont{Vijande} \emph{et~al.}(2007)\citenamefont{Vijande,
  Valcarce, and Richard}}]{Vijande:2007ix}
\bibinfo{author}{\bibnamefont{Vijande}, \bibfnamefont{J.}},
  \bibinfo{author}{\bibfnamefont{A.}~\bibnamefont{Valcarce}}, and
  \bibinfo{author}{\bibfnamefont{J.~M.} \bibnamefont{Richard}},
  \bibinfo{year}{2007}, \bibinfo{journal}{Phys. Rev.}
  \textbf{\bibinfo{volume}{D76}}, \bibinfo{pages}{114013}.

\bibitem[{\citenamefont{Vrana} \emph{et~al.}(2000)\citenamefont{Vrana, Dytman,
  and Lee}}]{Vrana:1999nt}
\bibinfo{author}{\bibnamefont{Vrana}, \bibfnamefont{T.~P.}},
  \bibinfo{author}{\bibfnamefont{S.~A.} \bibnamefont{Dytman}}, and
  \bibinfo{author}{\bibfnamefont{T.~S.~H.} \bibnamefont{Lee}},
  \bibinfo{year}{2000}, \bibinfo{journal}{Phys. Rept.}
  \textbf{\bibinfo{volume}{328}}, \bibinfo{pages}{181}.

\bibitem[{\citenamefont{Walcher}(1988)}]{Walcher:1989qp}
\bibinfo{author}{\bibnamefont{Walcher}, \bibfnamefont{T.}},
  \bibinfo{year}{1988}, \bibinfo{journal}{Ann. Rev. Nucl. Part. Sci.}
  \textbf{\bibinfo{volume}{38}}, \bibinfo{pages}{67}.

\bibitem[{\citenamefont{Weber}(1990)}]{Weber:1989fv}
\bibinfo{author}{\bibnamefont{Weber}, \bibfnamefont{H.~J.}},
  \bibinfo{year}{1990}, \bibinfo{journal}{Phys. Rev.}
  \textbf{\bibinfo{volume}{C41}}, \bibinfo{pages}{2783}.

\bibitem[{\citenamefont{Weigel}(1986)}]{Weigel:2008zz}
\bibinfo{author}{\bibnamefont{Weigel}, \bibfnamefont{H.}},
  \bibinfo{year}{2008}, \bibinfo{journal}{Lect. Notes Phys.}
  \textbf{\bibinfo{volume}{743}}, \bibinfo{pages}{1-274}.

\bibitem[{Weis \emph{et~al.}(2008)\citenamefont{Weis}
  \emph{et~al.}}]{Weis:2007kf}
\bibinfo{author}{\bibnamefont{Weis}, \bibfnamefont{M.}},
\emph{et~al.}, \bibinfo{year}{2008}, \bibinfo{journal}{Eur.
  Phys. J.} \textbf{\bibinfo{volume}{A38}}, \bibinfo{pages}{27}.

\bibitem[{Wieland \emph{et~al.}(in preparation)\citenamefont{Wieland}
  \emph{et~al.}}]{Wieland:2010aaa}
\bibinfo{author}{\bibnamefont{Wieland}, \bibfnamefont{F. W.}}, \emph{et~al.},
  \bibinfo{year}{in preparation}, \bibinfo{journal}{Eur. Phys. J.}

\bibitem[{\citenamefont{Wilczek}(2004)}]{Wilczek:2004im}
\bibinfo{author}{\bibnamefont{Wilczek}, \bibfnamefont{F.}},
  \bibinfo{year}{2004}, \eprint{Solicited contribution to the Ian Kogan
  Memorial volume, ed. M. Shifman. Published in *Ann Arbor 2004, Deserfest*
  322-338, hep-ph/0409168}.

\bibitem[{Williams \emph{et~al.}(2009{\natexlab{a}})\citenamefont{Williams}
  \emph{et~al.}}]{Williams:2009yj}
\bibinfo{author}{\bibnamefont{Williams}, \bibfnamefont{M.}}, \emph{et~al.},
  \bibinfo{year}{2009}, \bibinfo{journal}{Phys. Rev.}
  \textbf{\bibinfo{volume}{C 80}}, \bibinfo{pages}{045213}.

\bibitem[{Williams \emph{et~al.}(2009{\natexlab{c}})\citenamefont{Williams}
  \emph{et~al.}}]{Williams:2009rb}
\bibinfo{author}{\bibnamefont{Williams}, \bibfnamefont{M.}}, \emph{et~al.},
  \bibinfo{year}{2009}, \eprint{arXiv:0908.2910 [nucl-ex]}.

\bibitem[{Williams \emph{et~al.}(2009{\natexlab{b}})\citenamefont{Williams}
  \emph{et~al.}}]{Williams:2009rc}
\bibinfo{author}{\bibnamefont{Williams}, \bibfnamefont{M.}}, \emph{et~al.},
  \bibinfo{year}{2009}, \eprint{arXiv:0908.2911 [nucl-ex]}.

\bibitem[{\citenamefont{Wohl}(2008{\natexlab{a}})}]{Wohl:2008gw}
\bibinfo{author}{\bibnamefont{Wohl}, \bibfnamefont{C.~G.}},
  \bibinfo{year}{2008}{\natexlab{a}}, \bibinfo{journal}{Charmed Baryons, in
  \protect\cite{Amsler:2008zz}} .

\bibitem[{\citenamefont{Wohl}(2008{\natexlab{b}})}]{Wohl:2008st}
\bibinfo{author}{\bibnamefont{Wohl}, \bibfnamefont{C.~G.}},
  \bibinfo{year}{2008}{\natexlab{b}}, \bibinfo{journal}{{Pentaquarks}, in
  \protect\cite{Amsler:2008zz}} .

\bibitem[{\citenamefont{Workman}(1999)}]{Workman:1998vh}
\bibinfo{author}{\bibnamefont{Workman}, \bibfnamefont{R.}},
  \bibinfo{year}{1999}, \bibinfo{journal}{Few Body Syst. Suppl.}
  \textbf{\bibinfo{volume}{11}}, \bibinfo{pages}{94}.

\bibitem[{\citenamefont{Workman \emph{et~al.}}(2004)\citenamefont{Workman \emph{et~al.}}}]{Workman:2004im}
\bibinfo{author}{\bibnamefont{Workman}, \bibfnamefont{R.~L.}}, \emph{et~al.},
  \bibinfo{year}{2006}, \bibinfo{journal}{Phys. Atom. Nucl.}
  \textbf{\bibinfo{volume}{69}}, \bibinfo{pages}{90}.

\bibitem[{\citenamefont{Workman and Arndt}(2008)}]{Workman:2008iv}
\bibinfo{author}{\bibnamefont{Workman}, \bibfnamefont{R.~L.}}, and
  \bibinfo{author}{\bibfnamefont{R.~A.} \bibnamefont{Arndt}},
  \bibinfo{year}{2008}, \eprint{0808.2176}.

\bibitem[{Wright \emph{et~al.}(1995)\citenamefont{Wright}
  \emph{et~al.}}]{Wright:1995em}
\bibinfo{author}{\bibnamefont{Wright}, \bibfnamefont{A.~M.}}, \emph{et~al.},
  \bibinfo{year}{1995}, \bibinfo{note}{7th International
  Conference on the Structure of Baryons, Santa Fe, New Mexico, 3-7 Oct 1995}.

\bibitem[{Wu \emph{et~al.}(2005)\citenamefont{Wu} \emph{et~al.}}]{Wu:2005wf}
\bibinfo{author}{\bibnamefont{Wu}, \bibfnamefont{C.}}, \emph{et~al.},
  \bibinfo{year}{2005}, \bibinfo{journal}{Eur. Phys. J.}
  \textbf{\bibinfo{volume}{A23}}, \bibinfo{pages}{317}.

\bibitem[{Yao \emph{et~al.}(2006)\citenamefont{Yao} \emph{et~al.}}]{Yao:2006px}
\bibinfo{author}{\bibnamefont{Yao}, \bibfnamefont{W.~M.}}, \emph{et~al.},
\bibinfo{year}{2006},
  \bibinfo{journal}{J. Phys.} \textbf{\bibinfo{volume}{G33}},
  \bibinfo{pages}{1}.

\bibitem[{\citenamefont{Yndur{\'a}in}(1999)}]{Yndurain:679369}
\bibinfo{author}{\bibnamefont{Yndur{\'a}in}, \bibfnamefont{F.~J.}},
  \bibinfo{year}{1999}, \emph{\bibinfo{title}{The theory of quark and gluon
  interactions; 3rd ed.}}, Texts and monographs in physics
  (\bibinfo{publisher}{Springer}, \bibinfo{address}{Berlin}).

\bibitem[{Zabrodin \emph{et~al.}(1999)\citenamefont{Zabrodin}
  \emph{et~al.}}]{Zabrodin:1999sq}
\bibinfo{author}{\bibnamefont{Zabrodin}, \bibfnamefont{A.}}, \emph{et~al.},
  \bibinfo{year}{1999}, \bibinfo{journal}{Phys. Rev.}
  \textbf{\bibinfo{volume}{C60}}, \bibinfo{pages}{055201}.

\bibitem[{Zegers \emph{et~al.}(2003)\citenamefont{Zegers}
  \emph{et~al.}}]{Zegers:2003ux}
\bibinfo{author}{\bibnamefont{Zegers}, \bibfnamefont{R.~G.~T.}}, \emph{et~al.},
  \bibinfo{year}{2003}, \bibinfo{journal}{Phys. Rev. Lett.}
  \textbf{\bibinfo{volume}{91}}, \bibinfo{pages}{092001}.

\bibitem[{Zhu \emph{et~al.}(2003)\citenamefont{Zhu \emph{et~al.}}}]{Zhu:2002su}
\bibinfo{author}{\bibnamefont{Zhu}, \bibfnamefont{L.~Y.}}, \emph{et~al.},
  \bibinfo{year}{2003}, \bibinfo{journal}{Phys. Rev. Lett.}
  \textbf{\bibinfo{volume}{91}}, \bibinfo{pages}{022003}.

\bibitem[{Zhu \emph{et~al.}(2005)\citenamefont{Zhu \emph{et~al.}}}]{Zhu:2004dy}
\bibinfo{author}{\bibnamefont{Zhu}, \bibfnamefont{L.~Y.}}, \emph{et~al.},
  \bibinfo{year}{2005}, \bibinfo{journal}{Phys. Rev.}
  \textbf{\bibinfo{volume}{C71}}, \bibinfo{pages}{044603}.

\bibitem[{\citenamefont{Zou}(2008)}]{Zou:2007mk}
\bibinfo{author}{\bibnamefont{Zou}, \bibfnamefont{B.~S.}},
  \bibinfo{year}{2008}, \bibinfo{journal}{Eur. Phys. J.}
  \textbf{\bibinfo{volume}{A35}}, \bibinfo{pages}{325}.



\end{thebibliography}

\end{document}